\documentclass[acmsmall,screen]{acmart}

\AtBeginDocument{%
  }

\setcopyright{acmlicensed}
\copyrightyear{2018}
\acmYear{2018}
\acmDOI{XXXXXXX.XXXXXXX}

\usepackage{balance}
\usepackage{soul}
\usepackage{url}
\usepackage[utf8]{inputenc}
\usepackage{graphicx}
\usepackage{amsmath}
\usepackage{mathtools}
\usepackage{amsthm}
\usepackage{xspace}
\usepackage{xcolor}
\usepackage{booktabs}
\usepackage{algorithm}
\usepackage{algorithmic}
\usepackage{thm-restate}
\usepackage{tikz}
\usepackage{enumerate}
\usetikzlibrary{positioning}
\usetikzlibrary {shapes.geometric} 
\newtheorem{claim}{Claim}
\newtheorem{remark}{Remark}[section]
\newcommand{\updownarrows}{\uparrow\downarrow}
\newcommand{\mn}[1]{\ensuremath{\mathsf{#1}}}

\newcommand{\Amf}{\mathfrak{A}}

\begin{document}

\title{Extremal Fitting Problems for Conjunctive Queries}

\author{Balder ten Cate}
\email{b.d.tencate@uva.nl}
\orcid{0000-0002-2538-5846}
\affiliation{%
  \institution{ILLC, University of Amsterdam}
  \country{The Netherlands}
}

\author{V{\'i}ctor Dalmau}
\email{victor.dalmau@upf.edu}
\orcid{0000-0002-9365-7372}
\affiliation{%
  \institution{Universitat Pompeu Fabra}
  \country{Spain}}

\author{Maurice Funk}
\email{mfunk@informatik.uni-leipzig.de}
\orcid{0000-0003-1823-9370}
\affiliation{%
  \institution{Universität Leipzig}
  \institution{ScaDS.AI Center Dresden/Leipzig}
  \country{Germany}
}

\author{Carsten Lutz}
\email{clu@informatik.uni-leipzig.de}
\orcid{0000-0002-8791-6702}
\affiliation{%
  \institution{Universität Leipzig}
  \institution{ScaDS.AI Center Dresden/Leipzig}
  \country{Germany}
}

\begin{abstract}
  The \emph{fitting problem} for conjunctive queries (CQs) is the
  problem to construct a CQ that fits a given set of labeled data
  examples. When a fitting CQ exists, it is in general not
  unique. This leads us to proposing natural refinements of the notion
  of a fitting CQ, such as \emph{most-general fitting CQ},
  \emph{most-specific fitting CQ}, and \emph{unique fitting CQ}. We
  give structural characterizations of these notions in terms of
  (suitable refinements of) homomorphism dualities, frontiers, and
  direct products, which enable the construction of the refined
  fitting CQs when they exist. We also pinpoint the complexity of the
  associated existence and verification problems, and determine the
  size of fitting CQs. We study the same problems for UCQs and for the
  more restricted class of tree %
  CQs.
\end{abstract}




\maketitle

\section{Introduction}

A fundamental challenge across many areas of computer science is to infer formal specifications, such as functions, programs, or queries, that are consistent with (or, ``fit'') a given set of examples. In machine learning tasks such as classification and regression, a model that achieves zero loss on the training data can be described as a perfect fit. 
In inductive program synthesis, the aim is to generate code that fits given examples
of input-output behavior. And in database management, finding queries
that fit a given set of data examples
lies at the heart of the classic \emph{Query-By-Example} paradigm
that aims to assist users in query formation and query refinement. We refer to such problems as \emph{fitting
problems}, highlighting the goal of constructing a formal object that exactly fits a set of relevant examples.

In this paper, we study the \emph{fitting problem} for database queries, which is the
problem to construct a query $q$ that fits a given set of labeled data
examples, meaning that $q$ returns all positive examples as an answer
while returning none of the negative examples.
This fundamental
problem has a long history in database research. It
has been intensively
studied for conjunctive queries (CQs)~\cite{Tran2014:reverse,Li2015:qfe,Barcelo017} and other types of database
queries, e.g.\ in~\cite{Bonifati2015:learning,Arenas2016:reverse,Cohen2016:complexity}. The fitting problem is also central to
\emph{Inductive Logic Programming}~\cite{Cropper2021:ilp,GottlobLS99}, and has close connections to fitting problems for
\emph{schema mappings}~\cite{Alexe2011:designing,CateD15}.  Apart from Query-By-Example, more recent motivation
comes from \emph{automatic feature generation in machine learning with
  relational
  data}~\cite{KimelfeldRe2018,Barcelo21:regularizing}. Here, the 
fitting problem arises because a database query that separates 
positive from negative examples in (a sufficiently large subset of) a labeled
dataset is a natural
contender for being added as an input feature to the
model~\cite{Barcelo21:regularizing}. 
In addition, there has been significant recent interest in fitting queries in knowledge representation, typically in the
presence of an ontology~\cite{DBLP:journals/ml/LehmannH10,DBLP:journals/ws/BuhmannLW16,Funk2019:when,DBLP:journals/fgcs/RizzoFd20,DBLP:conf/kr/JungLPW20,DBLP:conf/kr/JungLPW21}.

\begin{figure}
    \centering
    \begin{tabular}{|lll|}
    \hline
    \multicolumn{3}{|c|}{EmpInfo} \\
    \hline
EMP & DEPT & MGR \\
\hline
Hilbert & Math & Gauss \\
Turing & Computer Science & von Neumann \\
Einstein & Physics &  Gauss\\
    \hline
\end{tabular}
\caption{Example database instance}
\label{fig:intro-example}
\end{figure}

\begin{example}\label{ex:intro-example}
Consider the database instance $I$ depicted in Figure~\ref{fig:intro-example}, and suppose we are given the following positive and negative examples (in reference to this given database instance): (Hilbert,+), (Turing,-), (Einstein,+). For these labeled examples, a fitting database query is
\[ q_1(x) \colondash \texttt{EmpInfo}(x,y,\texttt{Gauss})\]
which returns all employees managed by Gauss. Other  fitting database queries are 
\[ q_2(x) \colondash \texttt{EmpInfo}(x,\texttt{Math},z) \lor \texttt{EmpInfo}(x,\texttt{Physics},z) \]
which returns all employees who work in the Math or Physics department, and
\[ q_3(x) \colondash \texttt{EmpInfo}(x,y,z), \texttt{not}(z= \texttt{von Neumann}) \]
which returns all employees managed by someone other than von Neumann.
\end{example}

In this paper, we study variants of the fitting problem for \emph{conjunctive queries} (CQs), a 
class of relational database queries that is of fundamental importance in data management. They correspond to 
the ``select-from-where'' fragment of SQL and to the ``select-project-join'' (SPJ) fragment of relational algebra. They are
also precisely the queries that can be expressed in first-order logic without using disjunction, negation or universal quantification.

When a fitting CQ exists, in general it need not be unique up to
equivalence. In fact, there may be infinitely many pairwise non-equivalent fitting CQs, even in very simple and practically relevant cases. However, the fitting CQs form a
\emph{convex set}: whenever two CQs $q_1, q_2$ fit a set of labeled
examples, then the same holds for every CQ $q$ with
$q_1\subseteq q \subseteq q_2$, where ``$\subseteq$'' denotes query
containment (i.e., 
$q\subseteq q'$ holds if $q(I)\subseteq q'(I)$ for all database instances $I$). Maximal elements of this convex set can be viewed as
``most-general'' fitting CQs, while minimal elements can be viewed as
``most-specific'' fitting CQs. The set of all most-general and all
most-specific fitting CQs (when they exist) can thus be viewed as natural
representatives of the entire set of fitting CQs, similarly to the
version-space representation theorem used in machine
learning~\cite[Chapter 2.5]{Mitchell97}. 
In the context of automatic feature generation
mentioned above, it would thus be natural to compute all extremal fitting CQs
and add them as features, especially when
infinitely many fitting CQs exist.
Likewise, in query refinement tasks
where the aim is to construct a modified query that excludes unwanted answers
or includes missing answers  (cf.~\cite{Tran2014:reverse}),
it is also natural to ask for a most-general, respectively, most-specific fitting query.

In this paper we embark on a systematic study of %
extremal fitting CQs. 
To the best of
our knowledge, we are the first to do so.  
Besides CQs, we perform the same study also for two other natural classes
of database queries, namely 
\emph{UCQs} (unions of conjunctive queries) and \emph{tree CQs} (unary CQs that are Berge-acyclic and connected, over a schema that consists of unary and binary relations). The latter class holds significance as it 
corresponds precisely to 
the concept language of $\mathcal{ELI}$, 
a description logic
that is prominent 
in knowledge representation. 

As our first contribution, we observe that the intuitive concept of most-general fitting queries can be formalized in multiple ways.
Let $E$ be a collection of labeled examples and $\mathcal{Q}$  a query language (e.g., CQs). Then
\begin{itemize}
    \item $q\in \mathcal{Q}$ is a
    \emph{weakly most-general fitting query} for $E$ with respect to $\mathcal{Q}$ if 
    $q$ fits $E$ and there does not exist a $q'\in \mathcal{Q}$ that fits $E$ such that $q\subsetneq q'$.
    \item $q\in \mathcal{Q}$ is a
    \emph{strongly most-general fitting query} for $E$ with respect to $\mathcal{Q}$ if 
    $q$ fits $E$ and for all $q'\in \mathcal{Q}$ that fit $E$, it holds that  $q'\subseteq q$.
    \item $q_1, \ldots, q_n\in \mathcal{Q}$ form a (finite)
    \emph{basis of most-general fitting queries} for $E$ with respect to $\mathcal{Q}$ if each
    $q_i$ fits $E$ and for all $q'\in \mathcal{Q}$ that fit $E$, it holds that  $q'\subseteq q_i$ for some $i\leq n$.
\end{itemize}
In the same way, the intuitive concept of
most-specific fitting queries can be formalized in multiple ways:
\begin{itemize}
    \item $q\in \mathcal{Q}$ is a
    \emph{weakly most-specific fitting query} for $E$ with respect to $\mathcal{Q}$ if 
    $q$ fits $E$ and there does not exist a $q'\in \mathcal{Q}$ that fits $E$ such that $q'\subsetneq q$.
    \item $q\in \mathcal{Q}$ is a
    \emph{strongly most-specific fitting query} for $E$ with respect to $\mathcal{Q}$ if 
    $q$ fits $E$ and for all $q'\in \mathcal{Q}$ that fit $E$, it holds that  $q\subseteq q'$.
    \item $q_1, \ldots, q_n\in \mathcal{Q}$ form a (finite)
    \emph{basis of most-specific fitting queries} for $E$ with respect to $\mathcal{Q}$ if each
    $q_i$ fits $E$ and for all $q'\in \mathcal{Q}$ that fit $E$, it holds that  $q_i\subseteq q'$ for some $i\leq n$.
\end{itemize}

\begin{table*}\scriptsize
\begin{tabular}{@{}l|lll@{}}
               & Verification  &  Existence  & Construction and size  \\ \hline 
Any Fitting    & DP-c (Thm~\ref{thm:any-verification})      
               & coNExpTime-c~\cite{Willard10,CateD15}                  
               & In ExpTime~\cite{Willard10}; \\&&&
                 Exp size lower bound (Thm~\ref{thm:unique-size-lowerbound})\\  
Most-Specific  & NExpTime-c (Thm~\ref{thm:most-specific-verification},~\ref{thm:nexptime-hard-many}) 
               & coNExpTime-c~\cite{Willard10,CateD15}                  
               & In ExpTime~\cite{Willard10}; \\&&& Exp size lower bound (Thm~\ref{thm:unique-size-lowerbound})\\  
Weakly Most-General   & %
                  NP-c (Thm~\ref{thm:weakly-most-general-verification})
               & ExpTime-c (Thm~\ref{thm:wmg-existence},~\ref{thm:wmg-hardness})                
               & In 2ExpTime (Thm~\ref{thm:wmg-existence}); \\&&& Exp size lower bound (Thm~\ref{thm:unique-size-lowerbound}) \\  
Basis of Most-General 
               & NExpTime-c (Thm~\ref{thm:verification-basis},~\ref{thm:nexptime-hard-many})
               & NExpTime-c (Thm~\ref{thm:existence-basis},~\ref{thm:nexptime-hard-many})
               & In 3ExpTime (Thm~\ref{thm:basis-construction}); \\&&& 2Exp size lower bound (Thm~\ref{thm:basis-size-lowerbound}) \\  
Unique         & NExpTime-c (Thm~\ref{thm:unique-verification},~\ref{thm:nexptime-hard-many}) 
               & NExpTime-c (Thm~\ref{thm:unique-existence},~\ref{thm:nexptime-hard-many})  
               & In ExpTime~\cite{Willard10}; \\&&& Exp size lower bound (Thm~\ref{thm:unique-size-lowerbound}) \\
\end{tabular}

\caption{Summary of results for CQs}
\label{tab:cq-results}

\end{table*}

\begin{table*}\scriptsize
\begin{tabular}{@{}l|lll@{}}
               & Verification  &  Existence  & Construction and size  \\ \hline 
Any Fitting    & DP-complete (Thm~\ref{thm:ucq-complexity-results-a})    
               & coNP-complete (Thm~\ref{thm:ucq-complexity-results-a})               
               & in PTime (Thm~\ref{thm:ucq-complexity-results-a}) \\  
Most-Specific  & DP-complete (Thm~\ref{thm:ucq-complexity-results-a}) 
               & coNP-complete (Thm~\ref{thm:ucq-complexity-results-a})                   
               & in PTime (Thm~\ref{thm:ucq-complexity-results-a}) \\  
Most-General   & \textsc{HomDual}-equivalent (Thm~\ref{thm:ucq-complexity-results-b})
               & NP-c (Thm~\ref{thm:ucq-complexity-results-a}) 
               & in 2ExpTime (Thm~\ref{thm:ucq-complexity-results-a}) \\  
Unique         & \textsc{HomDual}-equivalent (Thm~\ref{thm:ucq-complexity-results-b})
               & \textsc{HomDual}-equivalent (Thm~\ref{thm:ucq-complexity-results-b})
               & in PTime (Thm~\ref{thm:ucq-complexity-results-a}) \\
\end{tabular}

\caption{Summary of results for UCQs}
\label{tab:ucq-results}

\end{table*}

\begin{table*}\scriptsize
\begin{tabular}{@{}l|lll@{}}
              & Verification      &  Existence     & Construction and size  \\
\hline 
Any Fitting    & PTime  (Thm~\ref{thm:arbfittreeverptime})
               &  ExpTime-c \cite{Funk2019:when}      
                                                   & In 2ExpTime (Thm~\ref{thm:sizetreeCQarbupper}); \\&&& %
                                                     2Exp size lower bound (Thm~\ref{thm:sizetreelower})\\
Most-Specific  & ExpTime-c  (Thm~\ref{thm:mostspecificexphardtrees},~\ref{thm:mostspecific-exptime-hard-trees})
               & ExpTime-c  (Thm~\ref{thm:mostspecificexphardtrees},~\ref{thm:mostspecific-exptime-hard-trees})
               & In 2ExpTime (Thm~\ref{thm:sizetreeCQmsupper}) \\ %
Weakly Most-General   
               & PTime (Thm~\ref{thm:veri-weak-most-general-tree})      
               & ExpTime-c (Thm~\ref{thm:wmg-existence-tree},~\ref{thm:mauriceexpapp})
               &  In 2ExpTime (Thm~\ref{thm:wmg-existence-tree}) \\ 
Basis of Most-General  
               & ExpTime-c (Thm~\ref{thm:tree-cq-basis-verification},~\ref{thm:basesmostgeneralexphardtrees}) %
               & ExpTime-c
                 (Thm~\ref{thm:tree-cq-basis-existence},~\ref{thm:basesmostgeneralexphardtrees}) &
                                                                       In 3ExpTime \\&&&(2Exp upper bound on size of members) (Thm~\ref{thm:tree-cq-basis-existence}) \\
Unique         & ExpTime-c  (Thm~\ref{thm:uniqueexphardtrees},~\ref{thm:basesmostgeneralexphardtrees})
               & ExpTime-c  (Thm~\ref{thm:uniqueexphardtrees},~\ref{thm:basesmostgeneralexphardtrees})
               &  in 2ExpTime (Thm~\ref{thm:sizetreeCQarbupper})  
\end{tabular}

\caption{Summary of results for tree CQs}
\label{tab:tree-cq-results}

\end{table*}

Our second contribution lies in 
giving  structural characterizations of each of the above notions.
A well-known and fundamental fact in database theory~\cite{CM77} states that each (Boolean) CQ $q$ can be identified with a corresponding database instance, namely its \emph{canonical instance} (or, as we will call it in this paper, \emph{canonical example}) $I_q$, 
and that $q\subseteq q'$ holds
precisely if there is a \emph{homomorphism} from $I_{q'}$ to
$I_q$.  A natural approach for studying the above notions, 
is therefore to recast them in terms of 
properties of the \emph{homomorphism pre-order}, 
that is, the pre-order consisting of all database instances ordered by homomorphism.
This allows us to take advantage of the existing
literature on the order-theoretic and combinatorial properties of this pre-order. 
For instance, it is known that 
the \emph{direct product} $I\times J$ of two database instances $I,J$ is a greatest lower bound of $I$ and $J$
in the homomorphism pre-order. There is also extensive existing literature about \emph{finite dualities} in the homomorphism pre-order, where a finite duality
is a pair $(F,D)$ of finite sets of database instances
such that the upward closure of $F$ is equal to the complement of the downward closure of $D$ (with respect to the homomorphism pre-order) Such finite dualities can be \emph{relativized} in a natural way to a 
given database instance $J$, by restricting attention to only those database instances that are below $J$. Finally, a \emph{frontier} for a database instance $I$  is a finite set of instances $\{J_1, \ldots, J_n\}$
that are strictly below $I$ in the homomorphism pre-order and that
separate $I$ from all instances strictly below $I$. By the aforementioned correspondence between CQs and database instances, 
these same notions apply also to CQs.
Making use of these notions, we are able to structurally characterize each of the above six types of extremal fitting CQs. We illustrate this with a few example characterizations. For simplicity, 
we restrict attention to Boolean CQs here in the introduction.  
Let $E$ be a collection of labeled examples (i.e, database instances labeled as positive or negative),
and let $q$ be a Boolean CQ that fits $E$.
We show that:
\begin{itemize}
    \item $q$ is weakly most-general fitting for $E$ with respect to all CQs if and only if 
    $q$ has a frontier $\{q_1, \ldots, q_n\}$ consisting of CQs that  do not fit $E$. (Indeed, whenever $q$ is a weakly most-general fitting CQ  for $E$, the negative examples in $E$ induce such a frontier for $q$).
    \item $q$ is strongly most-general fitting for $E$ with respect to all CQs if and only
    $(\{I_q\},E^-)$ is a finite duality in the homomorphism pre-order,
    relative to $J$, where $J$ is the direct product of the positive examples in $E$ and where $E^-$ is the set of negative examples in $E$.
    \item $q$ is strongly (equivalently: weakly) most-specific fitting for $E$ with respect to all CQs if and only if 
    the $J$ maps homomorphically to $I_q$, where
    $J$ is the direct product of the positive examples in $E$.
\end{itemize}
Similar structural characterizations are obtained for
all flavors of extremal fitting CQs (of arbitrary arity),
as well as for the analogous notions in the case of
UCQs and tree CQs.
 
As our third contribution, we use the aforementioned structural
characterizations to 
obtain effective algorithms
for the associated verification, existence, and construction problems, and we
establish upper and lower bounds on the complexity of these problems as well as on the size of the extremal fitting CQs.
Our algorithms are based on a combination
of techniques from automata theory and from the area of constraint satisfaction problems.
The
main complexity results and size bounds for CQs, UCQs, and tree CQs are
summarized in Tables~\ref{tab:cq-results},
\ref{tab:ucq-results}, and~\ref{tab:tree-cq-results}. 
As shown in the table, we are able to pinpoint the exact complexity of all decision problems involved, except for three, which we prove to be ``\textsc{HomDual}-equivalent''. That is, 
these three problems are all equivalent, up to 
polynomial-time reductions, to the problem of 
testing whether a given pair of finite sets of database instances is a finite duality in the homomorphism pre-order --- a problem whose complexity lies between
NP-hard and ExpTime.

Note that, since the classical (non-extremal)
fitting problem for CQs is already coNExpTime-complete~\cite{Willard10,CateD15},
it is not surprising that many of the problems we consider here turn out to be of similarly high complexity. The complexities for UCQs 
tend to be significantly simpler, often by one exponential, than those for CQs. Transitioning from CQs to tree CQs also tends to reduce the complexity, although in a much milder way: often, a non-deterministic complexity class is replaced with its deterministic counterpart. For the case of arbitrary fittings, similar 
observations were already made in~\cite{Barcelo017}, we discuss
this in more depth in Section~\ref{sec:conclusion}. 
It is also interesting that verification and existence
of \emph{(weakly) most-general} fitting CQs is often computationally easier
than for arbitrary fittings or any other kind of fitting.





\subsection*{Related Work}

The (non-extremal) fitting problem for CQs and UCQs, as well as for bounded-treewidth CQs and UCQs, was studied in~\cite{Willard10,CateD15,Barcelo017,Alexe2011:designing}. Note that,
from a fitting point of view, the \emph{GAV schema mappings} and \emph{LAV schema mappings} studied in \cite{Alexe2011:designing} correspond in a precise way to UCQs and CQs, respectively, cf.~\cite{CateD15}.
The fitting problem for tree CQs 
(equivalently, \emph{$\mathcal{ELI}$-concept expressions}) was studied in 
\cite{Funk2019:when}, and with only positive examples in~\cite{DBLP:conf/aaai/JungLW20}. 
The fitting problem for CQs is also closely related to the consistency problem
for Horn clauses, studied in inductive logic programming (ILP)~\cite{GottlobLS99}, although the latter differs in assuming that the size of the fitting query is bounded by a number given a an input.

The notion of a \emph{most-specific fitting query} appears in several
places in this literature, largely because of the fact that some of the most natural fitting algorithms automatically produce such fittings. We are not aware of any prior work studying the 
verification or construction of \emph{most-general} fitting queries or \emph{unique} fitting queries, although~\cite{tCD2022:conjunctive} studies the inverse problem, namely
the existence and construction of uniquely characterizing examples for a query,
and we build on results from~\cite{tCD2022:conjunctive}.

The problem of deriving queries from data examples has also been studied from the perspective of computational learning theory, a central question being whether and when a derived query generalizes well to unseen examples. For more information, we refer the reader to the related work sections in \cite{cate2022nonefficient,tCD2022:conjunctive} and the 
column~\cite{DBLP:journals/sigmod/CateFJL23}.

Fitting algorithms for database queries can be viewed as a special case of the 
broader problem of \emph{program synthesis}. 
In program synthesis, one is interested in the
automatic construction of programs. This broad field can be further subdivided into \emph{deductive} and \emph{inductive} program synthesis. In deductive program synthesis, the input is a high-level description of the intended program behavior, while in inductive program synthesis (also known as \emph{programming by examples}), the input is a set of input-output examples~\cite{ProgramSynthesisReview}. 
A classic example of inductive program synthesis is Flashfill, which automatically synthesizes spreadsheet formulas from few input–output examples~\cite{Gulwani2012:spreadsheet,Gulwani2015:inductive}. We refer to the monograph \cite{Gulwani2017:program} for more information on program synthesis.
Our setting differs from the general setting in program synthesis in that there is a natural pre-order on the concept space. This is due to the fact that database queries, when evaluated on database instances, yield \emph{sets} of tuples, and hence they can be compared by query containment. 
Note that, for programs with an arbitrary output space, it is not clear how to define ``most general'' or ``most specific'' solutions. 

\emph{Inductive Logic Programming} (ILP) also operates on a hypothesis space  that can be ordered by generality (namely, logical clauses). The generality pre-order used there is typically based on $\theta$-subsumption, an approximation of logical entailment that also gives rise to a lattice, as  first described by Plotkin~\cite{Plotkin1971:Automatic}. To systematically explore the hypothesis space, ILP algorithms often employ \emph{refinement operators} that traverse the lattice in a controlled manner. By tuning the refinement strategy (upward vs.\ downward) and employing biases (like language constraints or search heuristics), ILP systems can target either most-specific hypotheses or most-general hypotheses (with respect to $\theta$-subsumption), as needed for the learning task. We refer to 
\cite{NienhuysCheng1997:Foundations} for more details, including a comparison between the $\theta$-subsumption and logical entailment hierarchies.

\medskip\par\noindent
A preliminary version of this paper, without detailed proofs, appeared in \cite{extremalFittingPODS23}. The present paper expands on~\cite{extremalFittingPODS23} by 
containing detailed proofs. In addition, the paper has been polished and expanded to make it  accessible for a broader audience. In particular, Section~\ref{sec:pre-order} was added, which gives an overview of relevant properties of the query containment pre-order and the closely related homomorphism pre-order. 










\section{Preliminaries}
\label{sec:prel}

\subsection{Basic Database Theory Concepts}
\label{sect:basicdatabasetheory}

\subsection*{Schema, Instance, Conjunctive Query} 
A \emph{schema} (or relational signature) is a finite set of relation symbols $\mathcal{S}=\{R_1, \ldots, R_n\}$, where each relation symbol $R_i$ has an associated arity $\text{arity}(R_i) \geq 1$. 
A \emph{fact} over $\mathcal{S}$ is an expression $R(a_1, \ldots, a_n)$, where
$a_1, \ldots, a_n$ are \emph{values}, $R\in\mathcal{S}$, and $\text{arity}(R)=n$.
An \emph{instance over $\mathcal{S}$} is a finite set $I$ of facts over~$\mathcal{S}$.
The \emph{active domain} of $I$ (denoted $\text{adom}(I)$) is the set of all values
occurring in facts of $I$. In Section~\ref{sec:tree-cq}, in some of our proofs, it will be convenient to consider infinite instances.
   We will always clearly mark when considering possibly infinite instances, and unless
   explicitly specified otherwise, instances are assumed to be finite.

Let $k\geq 0$.
A $k$-ary \emph{conjunctive query (CQ)} $q$ over a schema $\mathcal{S}$ is an expression of the form ~
$q(\textbf{x}) \colondash \alpha_1\land\cdots\land\alpha_n$ ~
where $\textbf{x}=x_1, \ldots, x_k$ is a sequence of variables, and each $\alpha_i$ is an atomic formula using a relation from $\mathcal{S}$. Note that $\alpha_i$ may use variables from $\textbf{x}$ as well as other variables.
The variables in $\textbf{x}$ are called \emph{answer variables}, and
the other variables \emph{existential variables}.
Each answer variable is required to occur in at least one
conjunct $\alpha_i$. This requirement is known as the \emph{safety} condition.
A CQ of arity 0 is called a \emph{Boolean CQ}.
\looseness=-1

If $q$ is a $k$-ary CQ and $I$ is an instance over the same schema as $q$, we denote by $q(I)$ the
set of all $k$-tuples of values from the active domain of $I$ that satisfy the query $q$ in $I$.
We write $q\subseteq q'$ if $q$ and $q'$ are queries over the same schema and of the 
same arity such that $q(I)\subseteq q'(I)$ holds for all instances~$I$. We say that
$q$ and $q'$ are \emph{equivalent} (denoted $q\equiv q'$) if $q\subseteq q'$ and $q'\subseteq q$ both hold. %

By the \emph{degree} of a CQ $q$ we mean the largest number $n$ for which there is a variable occuring $n$ times in (the body of) $q$. Equivalently, it is the maximum degree of  variables in the incidence graph of $q$, as defined below, cf.~Definition~\ref{def:cacyclicity}. 

\subsection*{Pointed Instance, Data Example, Fitting Problem} %
A \emph{pointed instance} for schema $\mathcal{S}$ is a 
pair $(I,\textbf{a})$ where $I$ is an instance over $\mathcal{S}$, and
$\textbf{a}$ is a tuple of values. The values in $\textbf{a}$ are 
typically elements of $\text{adom}(I)$, but we also admit here values 
from outside of $\text{adom}(I)$ as this allows us to simplify
some definitions and proofs. If the tuple $\textbf{a}$ consists of 
$k$ values, then we  call $(I,\textbf{a})$ a \emph{$k$-ary}
pointed instance. We refer to $\textbf{a}$ as the
\emph{distinguished elements} of the pointed instance.
A pointed instance $(I,\textbf{a})$, with $\textbf{a}=a_1, \ldots, a_k$,
has the \emph{Unique Names Property (UNP)} if $a_i\neq a_j$ for all $i\neq j$.

A $k$-ary \emph{data example} for schema $\mathcal{S}$ (for $k\geq 0$) is 
 a pointed instance $e=(I,\textbf{a})$ where
$\textbf{a}$ is a $k$-tuple of values from $\text{adom}(I)$.
A data example $(I,\textbf{a})$ is said to be a \emph{positive example} for a 
query $q$ (over the same schema and of the same arity) if 
$\textbf{a}\in q(I)$, and a \emph{negative example} otherwise.
By a \emph{collection of labeled examples} we  mean a pair 
$E=(E^+, E^-)$ of finite sets of data examples. The size of a data example $e$ (as measured by the number of facts) is denoted
by $|e|$, and the combined size of a set of data examples $E$ by
$||E||=\Sigma_{e\in E}|e|$.

We say that 
$q$ \emph{fits} $E$ if each data example in $E^+$ is a positive example
for $q$ and each data example in $E^-$ is a negative example
for $q$.
The \emph{fitting problem} (for CQs) is the problem, given as 
input a collection of 
labeled examples, to decide if a fitting CQ exists.

A special case is where the input
examples involve a single database instance $I$, 
and hence can be given jointly as $(I,S^+,S^-)$, where $S^+, S^-$ are sets of tuples. We focus on the general version of the fitting problem here, but note that the aforementioned special case typically carries the same complexity (cf.~\cite[Theorem~2]{CateD15}).

\subsection*{Homomorphism, Core, Canonical Example, Canonical CQ}

Given two instances $I,J$ over the same schema, a 
\emph{homomorphism} $h\colon I\to J$ is a map from $\text{adom}(I)$ to $\text{adom}(J)$ that  preserves all facts. When such a homomorphism exists, we   say that $I$ ``homomorphically maps to'' $J$ and write
$I\to J$. 
We say that $I$ and $J$ are \emph{homomorphically equivalent} if
$I\to J$ and $J\to I$.
It is well known that every instance $I$ has a unique (up to isomorphism)
minimal subinstance to which it is homomorphically equivalent, known as the \emph{core} of $I$. Furthermore, two instances are 
homomorphically equivalent if and only if their cores are isomorphic.

The definitions of a homomorphism and of cores naturally extend to pointed instances.
More precisely a homomorphism $h\colon (I,\textbf{a})\to(J,\textbf{b})$ is a 
map from $\text{adom}(I)\cup\{\textbf{a}\}$ to $\text{adom}(J)\cup\{\textbf{b}\}$
that maps every fact of $I$ to a fact of $J$, and that maps every 
distinguished element $a_i$ to the corresponding distinguished element $b_i$ (where $\textbf{a}=a_1,\ldots, a_k$
and $\textbf{b}=b_1,\ldots, b_k$).

There is a natural correspondence between $k$-ary CQs over a schema 
$\mathcal{S}$ and $k$-ary data examples over $\mathcal{S}$.
In one direction, the \emph{canonical example} of a CQ $q(\textbf{x})$ is the pointed instance $e_q = (I_q,\textbf{x})$, where the active domain of $I_q$ is the set of variables
occurring in $q$ and the facts of $I_q$ are the conjuncts of $q$. Note that
every distinguished element of $e_q$ does indeed belong to the active domain (i.e.\ occurs in a fact), due to the safety condition of CQs, and therefore, $e_q$ is a well-defined data example.
Conversely, the \emph{canonical CQ} of a data example $e=(I,\textbf{a})$ with 
$\textbf{a}=a_1, \ldots, a_k$ is the CQ $q(x_{a_1}, \ldots, x_{a_k})$ 
 that has a variable $x_a$ for every value $a\in\text{adom}(I)$, 
 and a conjunct for every fact of $I$. Here, it is important that all distinguished 
 elements belong to the active domain (cf.~the definition of data examples), to ensure that $q$ satisfies the safety condition. 

 We  write $q\to q'$ when there is a homomorphism $h\colon e_q\to e_{q'}$ and 
  $q\to e$ when $e_q\to e$.
 By the classic Chandra-Merlin Theorem~\cite{CM77},  then,
 a tuple $\textbf{a}$ belongs to $q(I)$ if and only if $q\to e$ holds, where $e=(I,\textbf{a})$; and
 $q\subseteq q'$ holds if and only if $e_{q'}\to e_q$. 

We say that a CQ has the \emph{Unique Names Property (UNP)} its canonical example does. In other words, a CQ has the UNP there are no repeated variables in its list of answer variables. 

\subsection{Structure of the query containment pre-order and the homomorphism pre-order}
\label{sec:pre-order}

Let us denote by 
$CQ[\mathcal{S},k]$ the set of all 
$k$-ary CQs over schema $\mathcal{S}$. 
The query containment relation $\subseteq$ forms a pre-order 
on the set $CQ[\mathcal{S},k]$ (i.e., a reflexive and transitive relation). We will refer to this pre-order as the 
\emph{query containment pre-order}.
Closely related to it is the \emph{homomorphism pre-order}
on data examples: 
let $Ex[\mathcal{S},k]$ be the set of all
$k$-ary data examples over $\mathcal{S}$.
Then, by the \emph{homomorphism pre-order} 
we mean the relation $\to$ where 
$e_1\to e_2$ denotes the
existence of a homomorphism from $e_1$ to $e_2$.
The function that maps every CQ $q$ to its canonical example $e_q$ can be seen as a (surjective)
isomorphic embedding
of $(CQ[\mathcal{S},k],\subseteq)$ into
$(Ex[\mathcal{S},k],\to)$, except that
the direction of the order gets reversed:
$q\subseteq q'$ if and only if $e_{q'}\to e_q$. Thus, both pre-orders are isomorphic, except for a reversal of the direction of the pre-order.
In this section, we will review the relevant order-theoretic and combinatorial properties of these pre-orders, which will play a central role in this paper.

In the special case where $k=0$, the
pre-order $(Ex[\mathcal{S},k],\to)$ is simply the set of all finite relational structures ordered by homomorphism, which has been 
studied extensively and is the topic of
several textbooks such as
\cite{HellNesetril2004} and~\cite{nesetril2012sparsity}. Results for $k=0$ can typically be adapted to 
$k>0$ by suitably refining the constructions. A somewhat simpler in-between case consists of $k$-ary data examples with $k>0$ that have the UNP. Note that every data example with $k=0$ trivially has the UNP.

For the remainder, let us fix a schema $\mathcal{S}$ and an arity $k\geq 0$. 

\subsection*{Least upper bounds}
Let $(I,\textbf{a})$ and $(J,\textbf{a})$
be data examples with the UNP, where
both data examples have the \emph{same} tuple of distinguished elements. 
Furthermore, assume that $\text{adom}(I)\cap \text{adom}(J)\subseteq \{\textbf{a}\}$. Then the \emph{disjoint union}  $(I,\textbf{a})\uplus (J,\textbf{a})$ is the data example $(I\cup J, \textbf{a})$, where the facts of $I\cup J$ are the union of the facts of $I$ and $J$. 

This construction generalizes to arbitrary pairs of data examples with the UNP, by taking suitable isomorphic copies of the
input instances (to ensure that they have the same tuple of distinguished elements,
and are disjoint otherwise).
This operation also naturally generalizes to finite
sets of data examples with the UNP
(note that the operation is associative, in the sense that $(e_1\uplus e_2) \uplus e_3$
is isomorphic to $e_1\uplus(e_2\uplus e_3)$).

\begin{figure}
    \begin{center}
        \begin{tabular}{ccccc}
\begin{tikzpicture}[->, shorten >=1pt, auto, node distance=2cm]
      \draw[blue!20!black,fill=blue!20,rounded corners=5,thick]
     (1.5,-2.5) rectangle (2.5,0.5);
    \node (A3) {$a_3$};
    \node (A1) [right of=A3] {$a_1$};
    \node (A2) [below of=A1] {$a_2$};

    \path[->] (A3) edge (A1)
              (A1) edge (A2)
              (A2) edge (A3);
\end{tikzpicture}
&&
\begin{tikzpicture}[->, shorten >=1pt, auto, node distance=2cm]
      \draw[blue!20!black,fill=blue!20,rounded corners=5,thick]
     (-0.5,-2.5) rectangle (0.5,0.5);
    \node (B1) {$b_1$};
    \node (B2) [below of=B1] {$b_2$};
    \node (B3) [right of=B2] {$b_3$};
    \node (B4) [right of=B1] {$b_4$};

    \path[->] (B2) edge (B3)
              (B3) edge (B4)
              (B4) edge (B1);
\end{tikzpicture}
&&\begin{tikzpicture}[->, shorten >=1pt, auto, node distance=2cm]
      \draw[blue!20!black,fill=blue!20,rounded corners=5,thick]
     (1.5,-2.5) rectangle (2.5,0.5);
    \node (A3) {$a_3$};
    \node (A1) [right of=A3] {$c_1$};
    \node (A2) [below of=A1] {$c_2$};
    \node (B3) [right of=A2] {$b_3$};
    \node (B4) [right of=A1] {$b_4$};

    \path[->] (A3) edge (A1)
              (A1) edge (A2)
              (A2) edge (A3);
    \path[->] (A2) edge (B3)
              (B3) edge (B4)
              (B4) edge (A1);
\end{tikzpicture}
\\ \\
        $e_1$ && $e_2$ && $e_1\uplus e_2$
        \end{tabular}
    \end{center}
    \caption{Example of disjoint union}
    \label{fig:ex-du}
\end{figure}
\begin{example}
    Consider the following two data examples with a binary relation $R$ and with $k=2$:
    \begin{itemize}
        \item $e_1=(I,\langle a_1,a_2\rangle)$ where $I$ consists of the facts $R(a_1,a_2), R(a_2,a_3), R(a_3,a_1)$
        \item $e_2=(J,\langle b_1,b_2\rangle)$ where $J$ consists of the facts $R(b_2,b_3), R(b_3,b_4), R(b_4,b_1)$.
    \end{itemize}
    The disjoint union $e_1\uplus e_2$ is depicted in Figure~\ref{fig:ex-du}. It is the union of $e_1$ and $e_2$ in which the corresponding distinguished elements have been identified.
\end{example}

\begin{proposition} For all data examples $e_1$ and $e_2$ with the UNP,  
$e_1\uplus e_2$ is a least upper bound for 
$e_1$ and $e_2$ in the homomorphism pre-order, in the following sense:

    \begin{enumerate}
        \item $e_1\to e_1\uplus e_2$,
        \item $e_2\to e_1\uplus e_2$, and
        \item For all data examples $e'$, if 
        $e_1\to e'$ and $e_2\to e'$, then $e_1\uplus e_2\to e'$
    \end{enumerate}
    \end{proposition}

For CQs $q_1, q_2$ with the UNP, we define $q_1\uplus q_2$ 
to be the canonical query of $e_{q_1}\uplus e_{q_2}$. It then follows that $q_1\uplus q_2$ acts as 
a \emph{greatest lower bound} for $q_1$ and $q_2$ in the query containment pre-order (note the order reversal, as discussed in the beginning of this section).

The restriction to data examples and CQs with \emph{UNP} in the above propositions is not essential: similar results hold for the more general case without the UNP, except that the
 disjoint union construction needs to be further refined (it then becomes necessary to quotient by an equivalence relation induced by the equalities among the distinguished variables in the two data examples). Since it will not be relevant for us we will not go into the details here.

In fact, the above disjoint union operation can be applied to arbitrary pointed instances, not only data examples 
(recall that a data example is a pointed instance with the additional requirement that the distinguished elements belong to the active domain).
It just  happens that if the input pointed instances are data examples, then so is the output of the disjoint union operation. 

We
say that a pointed instance is \emph{connected} if it cannot be represented as the disjoint union of two or more non-empty pointed instances. Every pointed instance, and hence, also, every data example $e$, is 
a disjoint union of connected pointed instances. The latter will be referred to as the \emph{connected components} of $e$. Note that, by this definition, the connected components of a data example are not necessarily data examples.

\begin{example}
    Consider the data example $e=(I,\langle a,b\rangle)$, where
    $I$ consists of the facts $R(a,b)$, $S(a,c)$, $S(c,b)$, and $P(b)$. 
    Then $e$ has three connected components:
    \begin{itemize}
        \item $e_1=(I_1,\langle a,b\rangle)$ with $I_1$ consisting of $R(a,b)$,
        \item $e_2=(I_2,\langle a,b\rangle)$ with $I_2$ consisting of $S(a,c), S(c,b)$, and
        \item $e_3=(I_3,\langle a,b\rangle)$ with $I_3$ consisting of $P(a)$,
    \end{itemize}
    Indeed, it can be confirmed that $e=e_1\uplus e_2\uplus e_3$. Note however, that $e_3$ is not a data example: it is a pointed instance where one of the distinguished elements does not belong to the active domain.
\end{example}

\begin{proposition}\label{prop:du}
    Let $e, e'$ be data examples. Then $e\to e'$ holds if and only if every connected component of $e$ homomorphically maps to $e'$.
\end{proposition}

Indeed, in one direction, every homomorphism $h\colon e\to e'$ maps every connected component of $e$ to $e'$, while, in the other direction, the connected components have disjoint domains except for distinguished elements, and the homomorphisms from the connected components of $e$ to $e'$ must, by definition, all agree on the distinguished elements. 

\subsection*{Greatest lower bounds}
In the same way that least upper bounds for data examples can be constructed using disjoint union, 
\emph{greatest lower bounds} for data examples can be constructed using \emph{direct products}.
The \emph{direct product} of two pointed instances  $(I,\textbf{a})$ and $(J,\textbf{b})$, where $\textbf{a}=\langle a_1, \ldots a_k\rangle$ and $\textbf{b}=\langle b_1, \ldots, b_k\rangle$ is the $k$-ary pointed instance
$(I\times J, \langle (a_1,b_1), \ldots, (a_k,b_k)\rangle)$, where 
$I\times J$ consists of all facts $R((c_1,d_1), \ldots, (c_n,d_n))$ such that $R(c_1, \ldots, c_n)$ is a fact of $I$ and 
$R(d_1, \ldots, d_n)$ is a fact of $J$. 
This construction extends naturally to finite sets of
pointed instances (where the direct product of an empty set of pointed instances is, by convention, the 
pointed instance $(I,\langle a, \ldots, a\rangle)$ where
$I$ consists of all possible facts over the singleton domain $\{a\}$).

    \begin{figure}
    \begin{center}
    \begin{tabular}{ccc}
    \begin{tikzpicture}[thick, >=stealth]
    \node(a) at (0,0)   {$a$};	
    \node(b) at (2,0)   {$b$};	
    \draw[->] (a) -- (b) node[midway,above] {$R$};
    \draw[->] (-0.2,0.2) arc (45:330:0.2) node[midway, left] {$S$};
    \draw[->] (2.2,0.2) arc (135:-130:0.2) node[midway, right] {$S$};
    \end{tikzpicture}  
    &
    \begin{tikzpicture}[thick, >=stealth]
    \node(c) at (0,0)   {$c$};	
    \node(d) at (0,2)   {$d$};	
    \draw[->] (c) -- (d) node[midway,right] {$S$};
    \draw[->] (-0.2,0.2) arc (45:330:0.2) node[midway, left] {$R$};
    \draw[->] (-0.2,2.2) arc (45:330:0.2) node[midway, left] {$R$};
    \end{tikzpicture}  
    &
    \begin{tikzpicture}[thick, >=stealth]
    \node(ac) at (0,0)   {$\langle a,c\rangle$};	
    \node(ad) at (0,2)   {$\langle a,d\rangle$};	
    \node(bc) at (2,0)   {$\langle b,c\rangle$};	
    \node(bd) at (2,2)   {$\langle b,d\rangle$};	
    \draw[->] (ac) -- (ad) node[midway,left] {$S$};
    \draw[->] (ac) -- (bc) node[midway,above] {$R$};
    \draw[->] (bc) -- (bd) node[midway,right] {$S$};
    \draw[->] (ad) -- (bd) node[midway,above] {$R$};
\end{tikzpicture}  
\\ \\
$e_1$ & $e_2$ & $e_1\times e_2$
\end{tabular}
\end{center}
\caption{Example of direct product}
\label{fig:ex-product}
\end{figure}

\begin{example}
    Consider the following data examples with binary relations $R, S$ and $k=0$:
    \begin{itemize}
        \item $e_1$ consists of the facts $R(a,b), S(a,a), S(b,b)$,
        \item $e_2$ consists of the facts $S(c,d), R(c,c), R(d,d)$.
    \end{itemize}
    The direct product $e_1\times e_2$ has active domain $\{\langle a,c\rangle, \langle a,d\rangle, \langle b,c\rangle, \langle b,d\rangle\}$ and is depicted in Figure~\ref{fig:ex-product}.
\end{example}

One problem that arises is that the direct product of two data examples may not be a data example: although the direct product is always a pointed instance, it is not guaranteed that the designated elements occur in the active domain.

\begin{example}
    Consider the following data examples, with unary relations $P,Q$ and binary relation $R$, and where $k=1$:
    \begin{itemize}
        \item $e_1=(I,a)$, where $I$ consists of the facts $P(a), R(c,d)$, and
        \item $e_2=(J,b)$, where $J$ consists of the facts $Q(b), R(c,d)$.
    \end{itemize}
    In this case, the direct product $e_1\times e_2$ consists only of the fact
    $R(\langle c,c\rangle, \langle d,d\rangle)$, and has distinguished element $\langle a,b\rangle$. Since the distinguished element does not belong to the active domain, $e_1\times e_2$ is not a data example.
\end{example}

The following fundamental fact about direct products (cf.~\cite{HellNesetril2004}) will be used in many proofs.

\begin{proposition}\label{prop:instance-products}
For all data examples $e_1$ and $e_2$, if
$e_1\times e_2$ is a  data example, then it is a greatest lower bound for $e_1$ and $e_2$ in the homomorphism pre-order, in the following sense:
\begin{enumerate}
    \item $e_1\times e_2 \to e_1$,
    \item $e_1\times e_2\to e_2$, and
    \item  For all data examples $e'$, if 
   $e'\to e_1$ and $e'\to e_2$, then $e'\to e_1\times e_2$.
\end{enumerate}
Furthermore, whenever there exists a data example $e'$ such that $e'\to e_1$ and $e'\to e_2$, then $e_1\times e_2$ is a  data example.
\end{proposition}

\begin{proof} 
    Item (1), (2) and (3) are well known. Indeed,
    the homomorphism $h\colon e_1\times e_2\to e_i$ (for $i=1,2$) 
    is simply the natural projection that maps each pair 
    $(a_1, a_2)$ to $a_i$, while, from homomorphisms
    $h_1 \colon e'\to e_1$ and $h_2 \colon e'\to e_2$ one can construct
    a homomorphism $h \colon e'\to e_1\times e_2$ by setting
    $h(x)=(h_1(x),h_2(x))$. It therefore remains only to show that,
    if $e'\to e_1$ and $e'\to e_2$, then $e_1\times e_2$ is a data example.
    Indeed, by (3), 
    in this case, 
    there is a homomorphism $h$
    from $e'$ to the 
    pointed instance $e_1\times e_2$. 
    By definition, each distinguished element $y$ of $e_1\times e_2$,
    is the $h$-image
    of a distinguished 
    element $x$ of $e'$. Since $e'$ is a data example, $e'$ occurs in a fact. It follows that $y$ occurs in the $h$-image of this fact.
\end{proof}

The same operation of direct product can also be applied to CQs: for CQs
$q_1, q_2$, we define $q_1\times q_2$ 
to be the canonical CQ of $e_{q_1}\times e_{q_2}$ (note that this is well-defined only if the pointed instance $e_{q_1}\times e_{q_2}$ is indeed 
a data example).
In this way, the following proposition, which we will use in several places, follows immediately from Proposition~\ref{prop:instance-products}:

\begin{proposition}\label{prop:cq-products}
For all CQs $q, q_1, q_2$, the following are equivalent:
\begin{enumerate}
    \item $q\to q_1$ and $q\to q_2$,
    \item $q_1\times q_2$ is a well-defined CQ and $q\to q_1\times q_2$.
\end{enumerate}
\end{proposition}
%

\subsection*{Frontiers}
A frontier for an element $x$ in a pre-ordered set $(X,\leq)$ is a finite set $\{y_1, \ldots, y_n\}$ of elements that are strictly below $x$ and that separate $x$ from all other elements strictly below $x$. When such a frontier exists, 
this can be viewed as a local failure of density of the pre-order.
\footnote{To make this precise, recall that, an element
$x$ \emph{covers} an element $y$ in a pre-order $\leq$ if
$y\lneq x$ and there does not exist an element $z$
with $y\lneq z\lneq x$. Such pairs $(x,y)$ are also known as 
\emph{gap pairs}. It is not difficult to see that if $F=\{y_1, \ldots, y_n\}$ is a minimal frontier for $x$ (i.e., 
$F$ is a frontier for $x$ and no strict subset of $F$ is a frontier for $x$), then $x$ covers each $y_i$.}
In the homomorphism pre-order, it turns out, some data example have a frontier while others do not. Furthermore, whether a data example has a frontier can be characterized in terms of a suitable notion of \emph{acyclicity}. 

Formally, a \emph{frontier} for a
data example $e$ in the homomorphism pre-order is a finite
set of data examples $\{e_1, \ldots, e_n\}$ such that 
\begin{enumerate}
    \item for all $i\leq n$, 
    $e_i\to e$ and $e\nrightarrow e_i$.
    \item for all data examples $e'$, if $e'\to e$ and $e\nrightarrow e'$, then $e'\to e_i$ for some $i\leq n$.
\end{enumerate}
This definition can be relativized to a class $\mathcal{C}$ of data examples: 
 a finite
set of data examples $\{e_1, \ldots, e_n\}$  is a \emph{frontier for $e$ with respect to $\mathcal{C}$} if
the above two conditions hold, where, in the second condition, $e'$ ranges over members of $\mathcal{C}$ only (note that $e_1, \ldots, e_n$ are not required to belong to $\mathcal{C}$).

The same notion can, of course,  also be phrased in terms of CQs instead of data examples. Thus, we can speak about a frontier for a CQ, which is a finite set of CQs, and we can speak about a frontier for a CQ with respect to a restricted class $\mathcal{C}$ of CQs.

\begin{example}\label{ex:frontier}
    Fix a schema consisting of a single binary relation $R$, and let $k=0$. Consider the 
    following data examples:
    \begin{itemize}
        \item $e_1$ consists of the facts $R(a,b), R(b,c), R(c,d)$.  That is, $e_1$ is a directed path of length 3. 
        \item $e_2$ consists of the fact $R(a,a)$.
    \end{itemize}
    It can be shown that $e_1$ has a frontier, namely the singleton set containing the data example consisting of the facts $R(a,b), R(b,c), R(b',c), R(b',c'), R(c',d')$.
    The data example $e_2$, on the other hand, does not have a frontier. Indeed, a frontier for $e_2$ would have to be a finite set of data examples into which arbitrarily large cliques can be homomorphically mapped. No such finite set exists.
\end{example}
The existence of a frontier for a data example without distinguished elements (i.e., for $k=0$) 
was characterized in \cite{NesetrilTardif2000} in terms of a suitable notion of acyclicity. As observed in~\cite{AlexeCKT2011,tCD2022:conjunctive}, the same holds for arbitrary data examples (i.e., $k\geq 0$) as long as we use the right notion of acyclicity.

\begin{definition}[c-acyclicity]
\label{def:cacyclicity}
The \emph{incidence graph} of a data example $e$ is the bipartite multi-graph consisting of the 
  active domain elements and the facts of $e$, and such that there is a distinct edge between an element and a fact for each occurrence of the element in the fact.  
A data example $e$ is \emph{c-acyclic} if every cycle in the incidence graph (including every cycle of length 2 consisting of different edges that connect the same pair of nodes)  passes through a distinguished element.
\end{definition}

\begin{example}
Of the two data examples given in Example~\ref{ex:frontier}, $e_1$ is c-acyclic while $e_2$ is not.
\end{example}

\begin{theorem}[\cite{NesetrilOssona2008,NesetrilTardif2005,tCD2022:conjunctive}]\label{thm:cacyclic-frontier}
~\begin{enumerate}
    \item A data example has a frontier if and only if its core is c-acyclic.
    \item For fixed $k \geq 0$, a frontier for a c-acyclic $k$-ary data example can be constructed in polynomial time.
\end{enumerate}
\end{theorem}

The same notion applies to CQs: a CQ is c-acyclic if its canonical  example is. The above theorem can therefore again be equivalently phrased in terms of CQs.

\begin{example}
Consider the CQs
\[\begin{array}{lll}
q_1(x) &\colondash& R(x,y), R(y,z)\\
q_2(x) &\colondash& R(x,x), S(u,v), S(v,w)\\
q_3(x) &\colondash& R(x,y), R(y,y)
\end{array}\]
Of these CQs, the first two are c-acyclic while the third is not. 
We will not prove it here, 
but a frontier for $q_1$ is
$\{q'_1\}$ where $q'_1(x)\coloneq R(x,y), R(u,y), R(u,v), R(v,w)$, while 
a frontier for $q_2$ is
$\{q'_2, q''_2\}$ where
$q'_2(x) \colondash R(x,x), S(u,v)$
and
$q''_2(x) \colondash R(x,y), R(y,x), R(y,y), S(u,v), S(v,w)$.
The CQ $q_3$ does not have a frontier.
\end{example}

The polynomial-time frontier construction from~\cite{tCD2022:conjunctive} referred to above
will be described in Section~\ref{sect:mostgenfitCQs}, where we  also observe that it can, in some sense, be implemented in the form of a finite tree automaton. 

\subsection*{Homomorphism Dualities}

Another important concept, closely related to frontiers, is that of
\emph{homomorphism dualities}. By a homomorphism duality, we  mean, here, a pair of finite sets of data examples $(F,D)$ such that
$$
\begin{array}{l}
\{e\mid e \text{ is a data example  and } e\to e' \text{ for some } e'\in D\} = \\[1mm]
\{e\mid e \text{ is a data example  and } e'\nrightarrow e \text{ for all } e'\in F\}.
\end{array}
$$

In other words, if $(F,D)$ is a homomorphism duality, then every data example is either above an element of $F$ or below an element of $D$ (and not both) in the homomorphism pre-order.

We illustrate the notion of homomorphism dualities with two examples.

\begin{example} \label{ex:gallai}
This example concerns directed graphs, viewed as a data example over a schema with one binary relation, where $k=0$. For $n\geq 1$,
let $e_n$ be a directed path of length $n$,
and let $e'_n$ be the linear order of length $n$ (i.e., the directed graph with nodes $\{1,\dots n\}$ and edges $\{(i,j) \mid i<j\}$). Then, for $n>1$, 
$(\{e_n\},\{e'_{n-1}\})$ is a homomorphism duality. This fact is known as the Gallai-Hasse-Roy-Vitaver Theorem, as it was proved and published independently by each of these four researchers (each in a different language) in the 1960s.
\end{example}

\begin{example} 
Consider the following data examples
with unary relations $P, Q, R$, and with $k=0$: 
\begin{itemize}
    \item $e_1$ consists of the facts $P(a), Q(b)$,
    \item $e_2$ consists of the facts  $P(a), R(a)$,
    \item $e_3$ consists of the facts  $Q(a), R(a)$.
\end{itemize}
It is not hard to see that $(\{e_1\}, \{e_2, e_3\})$ is a homomorphism duality. Indeed, 
a data example admits a homomorphism from $e_1$ precisely if it contains both a $P$-fact and a $Q$-fact, which holds precisely if the data example in question fails to map homomorphically to $e_2$ and $e_3$.
\end{example}

Note that the above definition of homomorphism dualities requires that $F$ and $D$ are \emph{finite} sets. These are not the only interesting type of homomorphism dualities, but they will be the relevant ones for us in this paper.

Homomorphism dualities have been 
studied extensively in the literature on constraint satisfaction problems and graph combinatorics (e.g.,~\cite{NesetrilTardif2000,FoniokNT08, Larose2007:characterization}),
and have also found several applications in database theory (e.g., for schema mapping design~\cite{AlexeCKT2011}, ontology-mediated data access~\cite{Bienvenu2015:obda}, and query learning~\cite{tCD2022:conjunctive}).
In particular, in~\cite{AlexeCKT2011,tCD2022:conjunctive}, 
homomorphism dualities were used as a tool
for studying the unique characterizability,
and exact learnability, of schema mappings and of conjunctive queries. 

It turns out that there is an intimate relationship between homomorphism dualities and frontiers. Indeed, a frontier for a data example $e$ can be constructed out of a 
homomorphism duality of the form $(\{e\},D)$,
while a homomorphism duality $(F,D)$ can be 
constructed from frontiers for the data examples in $F$ \cite{NesetrilTardif2000,FoniokNT08} (see also \cite[Lemma 3.4]{tCD2022:conjunctive}). As a consequence of this,
the data examples that are the left-hand side of a homomorphism duality are precisely the
data examples that have a frontier. As we have seen already, the latter are precisely
the data examples whose core is c-acyclic.
This leads to the following characterization:

\begin{theorem}[\cite{NesetrilTardif2000,NesetrilTardif2005,Alexe2011:designing,tCD2022:conjunctive}]\label{thm:cacyclic-duality}
For all data examples $e$,
\begin{enumerate}
    \item There exists a homomorphism duality $(\{e\},D)$ if and only if the core of $e$ is c-acyclic.
\item A set $D$ as in (1) can be computed in exponential time for a c-acyclic data example.
\end{enumerate}
\end{theorem}

It may not yet be clear to the reader why homomorphism dualities are relevant for the study of  extremal fitting problems.
As we will see, in extremal fitting problems for CQs (Section~\ref{sec:cq})  and tree-CQs (Section~\ref{sec:tree-cq}), suitable refinements of homomorphism dualities 
provide a key to understanding when, for a 
given collection of labeled examples, there is a finite basis of most-general fitting queries. In our study of extremal fitting UCQs (Section~\ref{sec:ucq}), 
we will see that there is an even more 
transparent and easy-to-grasp connection between homomorphism dualities and 
most-general fitting queries.

\subsection{Tree Automata}

Tree automata will play an important role in some parts of this paper. Specifically, in Section~\ref{sect:mostgenfitCQs} and
in Section~\ref{sec:tree-cq}, where we will be concerned with tree-shaped CQs. In this section, we recall some basic definitions and facts about non-deterministic tree automata.

\newcommand*{\transition}[1]{\xRightarrow{~#1~}}

\begin{definition}[$\Sigma$-labeled $d$-ary trees]
  Fix a finite alphabet $\Sigma$ and a $d>0$. We denote by $\{1, \ldots, d\}^*$ the set of all finite sequences of elements of $\{1, \ldots, d\}$.
  A \emph{$\Sigma$-labeled $d$-ary tree} is a pair $(T, Lab)$ where $T$ is a non-empty
  prefix-closed finite subset of $\{1, \ldots, d\}^*$, and $Lab\colon T\to\Sigma$.
  By abuse of notation, we will sometimes use the symbol $T$ to refer to the pair 
  $(T,Lab)$ and we will write $Lab^T$ for $Lab$. 
\end{definition}

Note that the empty sequence $\varepsilon$
  belongs to every tree representing its root. Each sequence $(i_1, \ldots, i_n)\in T$ represents a node of the tree $T$, and, more 
  specifically, it is the $i_n$-th successor of the node represented by $(i_1, \ldots, i_{n-1})$. We will denote $(i_1, \ldots, i_n)$ also as 
  $Suc_{i_n}((i_1, \ldots, i_{n-1}))$. 
The above definition permits a node to have an ``$i$-th successor'' without 
having a ``$j$-th successor'' for~$j<i$. 

\begin{definition}[Non-Deterministic Tree Automaton; Accepting Run; Acceptance]
  A (bottom-up) $d$-ary non-deterministic tree automaton (NTA) is a tuple $\mathfrak{A}=(Q,\Sigma,\Delta,F)$ where
  \begin{itemize}
      \item $Q$ is a finite set of \emph{states}
      \item $\Sigma$ is a finite alphabet
      \item $\Delta\subseteq (Q\cup\{\bot\})^{\{1, \ldots, d\}}\times \Sigma\times Q$ is the \emph{transition relation}
      \item $F\subseteq Q$ is the set of \emph{accepting states}.
  \end{itemize}

  An \emph{accepting run} of $\mathcal{A}$ on a $\Sigma$-labeled $d$-ary tree $T$
  is a mapping $\rho:T\to Q$ such that:
  \begin{itemize}
      \item $\rho(\varepsilon) \in F$, and
      \item for each $t \in T$,  the tuple $\langle \rho_1(t), \ldots, \rho_d(t),Lab^T(t), \rho(t)\rangle$ belongs to $\Delta$, where $\rho_i(t)=\rho(t\cdot i))$  if $(t\cdot i) \in T$, and $\rho_i(t)=\bot$ otherwise. 
  \end{itemize}
  When such an accepting run exist, we say that $\mathfrak{A}$ \emph{accepts} $T$. The \emph{tree language recognized by $\mathfrak{A}$}, denoted by $L(\mathcal{A})$, is the set of all $\Sigma$-labeled $d$-ary trees accepted by $\mathfrak{A}$.
\end{definition}

When specifying an NTA, 
for the sake of readability we will use the notation
$\langle q_1, \ldots, q_d\rangle\transition{\sigma} q$ for transitions,
instead of writing $\langle q_1, \ldots, q_d, \sigma, q\rangle\in\Delta$.

The following theorem lists a number of well-known facts about NTAs.%

\begin{theorem}\label{thm:nta-basic-facts}\ 
\begin{enumerate}
\item Given an NTA $\mathfrak{A}$, it can be decided in PTime whether $L(\mathfrak{A})$ is non-empty. 
\item Given an NTA $\mathfrak{A}$ for which $L(\mathfrak{A})$ is non-empty, we can 
compute in polynomial time a succinct representation, in the form of a directed acyclic graph (DAG), of a tree $T$ of minimal size accepted by $L(\mathfrak{A})$.
\item Given NTAs $\mathfrak{A}_1,\dots,\mathfrak{A}_n$ and an expression $e$ that combines the symbols $L_1,\dots,L_n$ in terms of the operators
$\setminus, \cap,\cup$, we can construct in single exponential time
an NTA $\mathfrak{A}$ with $L(\mathfrak{A})$  the language obtained
by replacing each $L_i$ in $e$ with $L(\mathfrak{A}_i)$ and applying the operators.
\item Given a constant number of NTAs $\mathfrak{A}_1,\dots,\mathfrak{A}_n$, we can construct in polynomial
  time an NTA $\mathfrak{A}$ with $L(\mathfrak{A})=L(\mathfrak{A}_1) \cap \cdots \cap L(\mathfrak{A}_n)$.
\end{enumerate}
\end{theorem}

Here, the alphabet $\Sigma$ and arity $d$ of the NTAs is not treated as a fixed in the complexity analyses, but as part of the input. In Point (2), with the succinct representation of a tree $T$ as a DAG, we mean the DAG that is obtained from $T$ by identifying isomorphic subtrees. 
Such a  representation can be of size logarithmic in the
size of the original tree .
The facts listed in 
Theorem \ref{thm:nta-basic-facts} can be found in any standard textbook on tree automata, possibly with the exception of Point (2). Point (2), however, can be shown by a straightforward dynamic programming algorithm that computes, starting with the final states,
for each state $s$ of the automaton a pair $(T,n)$ where 
$T$ is a succinct representation of a tree accepted by the automaton from starting state $s$, and $n$ is the size of the corresponding non-succinct representation. 

\section{The Case of Conjunctive Queries}
\label{sec:cq}

In this section, we study the fitting problem for CQs. We first review
 results for the case where the fitting CQ needs not satisfy any further properties. After that, we introduce and study extremal fitting CQs, including most-general, most-specific,
 and unique fittings. For these, we first concentrate on characterizations and upper bounds, deferring lower bounds to Section~\ref{sec:cq-lowerbounds}.\looseness=-1

To simplify presentation, when we speak of a CQ $q$ in the context of
a collection of labeled examples $E$, we mean that $q$ 
ranges over CQs that have the same schema and arity as the data examples in $E$.
\looseness=-1

\subsection{Arbitrary Fitting CQs}
\label{sect:arbfittingcqs}
We first consider the verification problem for fitting CQs:
\emph{given a collection of labeled examples $E$ and a CQ $q$, does $q$ fit $E$?}
This problem naturally
falls in the complexity class DP (i.e., it can be expressed as the
intersection of a problem in NP and a problem in coNP). Indeed: 
\looseness=-1

\begin{restatable}{theorem}{thmanyverification}
\label{thm:any-verification}
The verification problem for fitting CQs  is DP-complete.
The lower bound holds for a schema consisting of a single binary relation, a fixed collection of labeled examples, and Boolean CQs. \looseness=-1
\end{restatable}

\begin{proof}
Clearly it is equivalent to a conjunction of problems that are in NP or coNP.
For the lower bound we can reduce from exact-4-colorability, i.e., testing 
that a graph is 4-colorable and not 3-colorable \cite{Rothe2001:exact}. Fix a schema consisting of a single 
binary relation $R$. Let $K_3$ be the 3-clique (viewed as an instance with a symmetric, irreflexive relation), and let $K_4$ be the 4-clique. 
Let $E^- = \{K_3\}$ and $E^+=\{K_4\}$. 
Then a graph $G$ is exact-4-colorable if and only if the canonical CQ of $G$ fits $(E^+,E^-)$. 
\end{proof}

The \emph{existence} problem for fitting CQs is: \emph{given a collection of labeled examples $E$, is there a CQ that fits $E$?} It
was studied in~\cite{Willard10,CateD15}.

\begin{theorem}[\cite{Willard10,CateD15}] \label{thm:icdt2015}
The existence problem for fitting CQs is coNExpTime-complete. 
The lower bound holds already for Boolean CQs over a fixed schema consisting of a single binary relation.
\end{theorem}

\begin{theorem}[\cite{Willard10}] \label{thm:icdt2015-b}
If any CQ fits a collection of labeled examples $E=(E^+,E^-)$, then the canonical CQ of the direct product $\Pi_{e\in E^+}(e)$ is well-defined and fits $E$.
\end{theorem}

\begin{proof}
We include a proof as it provides intuition for the role
of direct products in the
construction of fitting CQs.
Let $q$ be a fitting CQ for
$(E^+,E^-)$.
By definition, this means that $e_q$
maps homomorphically into each data example in $E^+$ and not into any data example in $E^-$. 
It follows by Proposition~\ref{prop:instance-products} that
$\Pi_{e\in E^+}$ is a
well-defined data example and that $e_q\to \Pi_{e\in E^+}(e)$. Furthermore, it follows that
$\Pi_{e\in E^+}(e)$ does not
map homomorphically to any
data example in $E^-$, for otherwise, by transitivity, 
$e_q$ would map homomorphically to the same
data example, contradicting the fact that $q$ fits $(E^+,E^-)$. We conclude
that the canonical CQ of 
$\Pi_{e\in E^+}(e)$ is a 
fitting CQ for $(E^+,E^-)$.
\end{proof}

When we are promised that a fitting CQ exists, we can therefore construct
one in (deterministic) single exponential time. We will see in Section~\ref{sec:size-lowerbounds} that
this is optimal, as there is a matching lower bound.

\begin{remark}\label{rem:sicos} \rm
A special case of the existence problem for  fitting CQs is \emph{CQ definability}, where
the input is a pair $(I,S)$ with $I$ an instance and $S\subseteq \text{adom}(I)^k$
a $k$-ary relation, and the task is to decide whether there exists a CQ $q$
such that $q(I)=S$. Note that this problem is meaningful only for $k\geq 1$.
For fixed $k\geq 1$, this polynomially reduces to fitting existence, namely
by choosing $E=(E^+,E^-)$ with $E^+=\{(I,\textbf{a})\mid \textbf{a}\in S\}$, and
$E^-=\{(I,\textbf{a})\mid \textbf{a}\in \text{adom}(I)^k\setminus S\}$. 
In other words, CQ definability can be viewed as the special case of  CQ fitting existence
where all input examples share the same instance $I$ and where the $k$-tuples appearing
in the positive and negative examples cover the complete set of all $k$-tuples over
$\text{adom}(I)$.
The lower bound in Theorem~\ref{thm:icdt2015}
was shown to hold already for the CQ definability problem \cite{Willard10,CateD15}.

The lower bound for the verification problem in Theorem~\ref{thm:any-verification}, also already holds for
the restricted case where the labeled examples cover the entire tuple space of
a single instance. For the second variant of the lower bound in Theorem~\ref{thm:any-verification}  this can be seen as follows: we let $I$
be the disjoint union $K_3\uplus K_4$,
let $E^+$ be the set of all triples $(I,a)$ where  $a$ lies on the 4-clique, 
and let $E^-$ be the set of all data examples $(I,a)$ where $a$ lies
 on the 3-clique. Then, for any connected graph $G$, if $a$ 
 is an arbitrarily chosen vertex of $G$,  then we have that  $G$ is exact-4-colorable if and only if the canonical unary CQ of $(G,a)$ fits $E$.
Disconnected graphs $G$  can be handled similarly: in this case, we can linearly order its connected components and add a connecting edge from each component to the next one, without affecting the 3-colorability or 4-colorability of the graph. 
\end{remark}

\subsection{Most-Specific Fitting CQs (Characterizations and Upper Bounds)}

There are two natural ways to define \emph{most-specific} fitting CQs. This is analogous 
to the distinction between ``minimum'' and 
``minimal'' elements of a partial order: we may be
interested in a fitting CQ that is at least as specific
as \emph{any} fitting CQ or, weaker than that, in 
a fitting CQ $q$ such that there is no fitting CQ that
is strictly more specific than $q$.

\begin{definition}\label{def:mostspec} \ 
\begin{itemize}
    \item A CQ $q$ is a \textbf{strongly most-specific fitting} CQ for a collection of labeled examples $E$ if $q$ fits $E$ and
    for every CQ  $q'$ that fits $E$, we have $q\subseteq q'$.
    \item A CQ $q$ is a \textbf{weakly most-specific fitting} CQ for a collection of labeled examples $E$ if $q$ fits $E$ and
    for every CQ $q'$ that fits $E$, $q'\subseteq q$ implies $q \equiv q'$.
\end{itemize}
\end{definition}

It follows from Theorem~\ref{thm:icdt2015-b} that the above two notions  coincide:

\begin{restatable}{proposition}{propmostspecific}
\label{prop:most-specific}
For all CQs $q$ and collections of labeled examples $E=(E^+,E^-)$, the following are equivalent:
\begin{enumerate}
    \item  $q$ is strongly most-specific fitting for $E$,
    \item  $q$ is weakly most-specific fitting for $E$,
    \item  $q$ fits $E$ and $q$ is  equivalent to the 
    canonical CQ of $\Pi_{e\in E^+}(e)$.
    \footnote{In particular, in this case, the 
    canonical CQ of $\Pi_{e\in E^+}(e)$ is well-defined.}
\end{enumerate}
\end{restatable}

\begin{proof}
(1 $\Rightarrow$ 2) follows directly from Definition~\ref{def:mostspec}.

(2 $\Rightarrow$ 3):
Since $q$ fits $E$, by Theorem~\ref{thm:icdt2015-b}, the canonical CQ $q^*$ of $\Pi_{(I,\textbf{a})\in E^+}(I,\textbf{a})$ is well defined and fits $E$. It then follows by Proposition~\ref{prop:instance-products} that $q^*$ is a strongly most-specific fitting CQ for $E$:
if $q'$ is any fitting CQ, then $q'$ has a homomorphism to each positive example, and hence,
$q'$ has a homomorphism to their product. It follows that $q\to q^*$, and, hence, $q^*\subseteq q$. Since $q$ is 
weakly most-specific fitting, it cannot be the case that $q^*\subsetneq q$. We can conclude that $q$ and $q^*$ are equivalent.

(3 $\Rightarrow$ 1): 
Let $q^*$ be the canonical CQ of $\Pi_{e\in E^+}(e)$.
As we already pointed out, if $q^*$ fits $E$ then it is a strongly most-specific fitting CQ for $E$.
\end{proof}

In light of Proposition~\ref{prop:most-specific}, we simply speak of \emph{most-specific fitting CQs}, dropping ``weak'' and ``strong''. 

\begin{example}
  Let $\mathcal{S}=\{R,P\}$, where $R$ is a ternary relation and $P$ is a unary relation. Consider the collection of 
  labeled examples $E=(E^+=\{I_1,I_2\},E^-=\{I_3\})$,
  where $I_1=\{R(a,a,b),P(a)\}$, $I_2=\{R(c,d,d),P(c)\}$, 
  and $I_3=\emptyset$. The Boolean CQs 
  $q_1() \colondash  R(x,y,z)$ and 
  $q_2() \colondash R(x,y,z)\land P(x)$
  both fit $E$, but $q_2$ is more specific
  than $q_1$. Indeed, $q_2$ is the most-specific fitting CQ for $E$, as it is equivalent to the canonical query of $I_1\times I_2$. 
  Note that $I_1\times I_2$ here consists of the facts
  $R(\langle a,c\rangle, \langle a,d\rangle,\langle b,d\rangle)$ and $P(\langle a,b\rangle)$.
\end{example}

It follows from Proposition~\ref{prop:most-specific} and Theorem~\ref{thm:icdt2015} that the existence problem for 
most-specific fitting CQs coincides with that for arbitrary fitting CQs,
and hence, is coNExpTime-complete; and that we can construct in exponential time a CQ $q$ (namely, the canonical CQ of $\Pi_{e\in E^+}(e)$), with the property that, if there is a most-specific fitting CQ, then $q$ is one.
For the \emph{verification} problem, finally,  
Theorem~\ref{thm:icdt2015-b}, with Theorem~\ref{thm:any-verification}, implies:

\begin{restatable}{theorem}{thmmostspecificverification}
\label{thm:most-specific-verification}
The verification problem for most-specific fitting CQs 
is in NExpTime.
\end{restatable}

\begin{proof} 
We first verify that $q$ fits $E$ (in DP by Theorem~\ref{thm:any-verification}).  If this test succeeds, we apply
by Theorem~\ref{thm:icdt2015-b} and Proposition~\ref{prop:most-specific}, 
we test that $q$ is equivalent to the canonical CQ of
$\Pi_{e\in E^+}(e)$. This puts us in NExpTime because $\Pi_{e\in E^+}(e)$
can be computed in exponential time,
and equivalence can be tested by guessing homomorphisms in both directions.
\end{proof}

\subsection{Most-General Fitting CQs (Characterizations and Upper Bounds)}
\label{sect:mostgenfitCQs}

For \emph{most-general fitting} CQs, there are again two natural definitions.

\begin{definition}\label{def:mostgenfitCQs}\ 
\begin{itemize}
\item
A CQ $q$ is a \textbf{strongly most-general fitting} CQ for a collection of labeled examples $E$ if $q$ fits $E$ and 
for all CQs $q'$ that fit $E$, we have $q'\subseteq q$.
    \item 
A CQ $q$ is a \textbf{weakly most-general fitting} CQ for a collection of labeled examples $E$ if $q$ fits $E$ and for every CQ $q'$ that fits $E$,
$q\subseteq q'$ implies $q\equiv q'$
\end{itemize}
\end{definition}

Unlike in the case of most-specific fitting CQs, as we will see, these two notions do \emph{not} coincide. In fact, there is a third:

\begin{definition}\label{def:basismostgenfitCQs}
A finite set of CQs $\{q_1, \ldots, q_n\}$ is a \textbf{basis of most-general fitting CQs}
for  $E$ if each $q_i$ fits $E$ and 
for all CQs $q'$ that fit $E$, we have $q'\subseteq q_i$ for some $i\leq n$.
If, in addition, no strict subset of $\{q_1, \ldots, q_n\}$ is a 
basis of most-general fitting CQs for $E$, we say that
$\{q_1, \ldots, q_n\}$ is a \emph{minimal} basis. 
\end{definition}

Each member of a minimal basis is indeed guaranteed to be 
weakly most-general fitting. The same does not necessarily
hold for non-minimal bases. We could have 
included this as an explicit requirement in the definition,
but we decided not to, in order to simplify the statement 
of the characterizations below.

It is easy to see that minimal bases are unique up to homomorphic equivalence.
Also, a strongly most-general fitting CQ is simply a basis of size 1. We will therefore consider the notions of \emph{weakly most-general fitting CQs} 
and \emph{basis of most-general fitting CQs}, only.

\begin{example}
Let $\mathcal{S} = \{R,P,Q\}$, where $R$ is a binary relation and $P, Q$ are unary
relations. The following examples pertain to Boolean CQs. 
Let $K_2$ be the 2-element clique, i.e., $K_2=\{R(a,b), R(b,a)\}$. 
Furthermore, let  $I_{P}$, $I_{Q}$, and $I_{PQ}$
be the instances consisting of the set of facts $\{P(a)\}$, $\{Q(a)\}$, and 
$\{P(a), Q(a)\}$, respectively. 
\begin{enumerate}
    \item The collection of labeled examples $(E^+=\emptyset, E^-=\{I_{PQ}\})$ has a strongly
    most-general fitting CQ, namely $q() \colondash R(x,y)$.
    \item The collection of labeled examples $E=(E^+=\emptyset, E^-=\{I_{P},I_{Q}\})$ has a basis
    of most-general fitting CQs of size two, consisting of 
    $q_1() \colondash  R(x,y)$ and $q_2() \colondash P(x)\land Q(y)$.
    In particular, each of these two CQs is weakly most-general fitting for $E$.
\item 
If we restrict the schema to consist only of the binary relation $R$, then
the collection of labeled examples $E=(E^+=\emptyset, E^-=\{K_2\})$ does not have
a weakly most-general fitting CQ. 
Note that a  CQ~$q$  fits $E$
if and only if $q$ is not two-colorable. The latter, in turn, holds if and only
if $q$ contains a cycle of odd length (where, by a \emph{cycle of length $n$}, we mean a sequence of variables
$(v_i)_{i=1\ldots n+1}$ where
$a_{n+1}=a_1$ and where, for each $1\leq i\leq n$, either $R(v_i,v_{i+1})$
or $R(v_{i+1},v_i)$ is an atom of $q$). Take a fitting CQ $q$ and let $k$ be the size of the smallest cycle of odd length in $q$, and let
$(v_i)_{i=1\ldots n+1}$ be the cycle in question.
We can construct a CQ $q'$ consisting of a cycle of length $3k$ obtained by ``blowing up'' the cycle
$(v_i)_{i=1\ldots n+1}$ so that it becomes 3 times as long. In this
way, we have that $q'$ fits $E$, 
$q'\to q$ and $q\nrightarrow q'$. It follows that $q\subsetneq q'$, 
and therefore $q$ is not 
a weakly most-general fitting CQ for $E$.
\item 
The collection of labeled examples $(E^+=\emptyset, E^-=\{ K_2, I_P, I_Q\})$
has a weakly most-general fitting CQ, namely $q() \colondash P(x)\land Q(y)$.
By the same reasoning as in (3), there is no  basis of most-general fitting CQs.
\end{enumerate}
\end{example}

\subsection*{Weakly most-general fitting CQs}
As it turns out, weakly most-general fitting CQs can be 
 characterized in terms of frontiers.

\begin{restatable}{proposition}{propweaklymostgeneral} 
\label{prop:weakly-most-general}
The following are equivalent for all collections of labeled examples $E=(E^+,E^-)$ and all CQs $q$:
\begin{enumerate}
    \item $q$ is weakly most-general fitting for $E$,
    \item $q$ fits $E$, $q$ has a  frontier and every element 
    of the frontier has a homomorphism to an example in $E^-$,
    \item $q$ fits  $E$ and $\{q \times q_e\mid e\in E^- \text{ and } q\times q_e \text{ is a well-defined CQ}\}$ is a frontier for $q$,
\end{enumerate}
where $q_e$ is the canonical CQ of $e$. 
\end{restatable}

\begin{proof}
(1 $\Rightarrow$ 3): If $\{q\times q_e\mid e\in E^- \text{ and $q\times q_e$ is a well-defined CQ}\}$ is not a frontier for $q$, then there exists a query $q'$ that is homomorphically strictly weaker than $q$ but that does not map to $\{q\times q_e\mid e\in E^-  \text{ and $q\times q_e$ is a well-defined CQ}\}$.
It follows that $q'$ has no homomorphism to any example in $e\in E^-$
(for, if it did, then, by Proposition~\ref{prop:cq-products}, we would have that $q'\to q\times q_e$ and $q\times q_e$ would be well-defined). Hence, $q'$ is 
a fitting CQ that is strictly weaker than $q$, showing that $q$ is not a weakly most-general fitting CQ.

(3 $\Rightarrow$ 2) is trivial.

(2 $\Rightarrow$ 1): let $q'$ be homomorphically strictly weaker 
than $q$. Then $q'$ maps to the frontier of $q$ and hence to a negative example.
Therefore $q'$ does not fit $E$.
\end{proof}

Note that if a CQ $q$ has frontiers
$F$ and $F'$, then 
these are equivalent 
up to homomorphism, in the following sense: for each
$q_i\in F$ there is a $q_j\in F$ 
such that $q_i\to q_j$, and,
conversely, for every $q_i\in F'$ there is a $q_j\in F$ such that
$q_i\to q_j$. This justifies the fact that item (2) of Proposition~\ref{prop:weakly-most-general} refers to ``the frontier'': the condition in question does not depend on the choice of frontier.

Using Theorem~\ref{thm:cacyclic-frontier}, we can now show:

\begin{restatable}{theorem}{thmweaklymostgeneralverification}
\label{thm:weakly-most-general-verification}
Fix $k\geq 0$.
The verification problem for weakly most-general fitting $k$-ary CQs is %
NP-complete.
In fact, it remains NP-complete even if the examples are fixed suitably and, in addition,
the input query is assumed to fit the examples.
\end{restatable}

\begin{proof}

The algorithm showing NP-membership is as follows. We check by means of a non-deterministic guess that $q$ is  equivalent to a c-acyclic CQ $q'$
(which can be assumed to be of polynomial size by~Theorem~\ref{thm:cacyclic-frontier}).
Then we check whether $q'$ fits. This can be done in polynomial time since $q'$ is c-acyclic by means of an easy dynamic programming argument. Finally, we compute the frontier of $q'$
(which we can do in polynomial time by Theorem~\ref{thm:cacyclic-frontier} since $k$ is fixed), and we 
check each member of the frontier has a 
homomorphism to a negative example. 

Let us now turn our attention to the NP-hardness.  For every graph $T$, let $CSP(T)$ be the problem consisting to determine whether an input graph $G$ is homomorphic to $T$. It is known (\cite{GutjahrWW92}) that there exists some directed trees $T$ such that $CSP(T)$ is NP-complete.
Fix any such directed tree $T$, and let us turn it into a pointed instance $(T,\textbf{a})$ by selecting $\textbf{a}$ to be any tuple of $k$ (non necessarily different) values from $T$. 

Since $(T,\textbf{a})$ is  c-acyclic, 
it has a frontier $F$. Now, given any graph $G$, we have that $G$ is homomorphic to $T$ if and only if 
$F$ is a frontier for $(T,\textbf{a})\uplus G$. This,
in turn, holds if and only if
the canonical CQ of $(T,\textbf{a})\uplus G$ is a weakly most-general fitting CQ for $(E^+=\emptyset, E^-=F)$.
To see that this is the case, note
that if $G$ is homomorphic to $T$, then $(T,\textbf{a})\uplus G$ is homomorphically equivalent
to $(T,\textbf{a})$ itself, whereas if $G$ is not homomorphic to $T$, then $(T,\textbf{a})\uplus G$ is strictly
greater than $(T,\textbf{a})$ in the homomorphism order.
\end{proof}

The next theorem addresses the \emph{existence} and \emph{construction} problems for weakly most-general fitting CQs.

\begin{restatable}{theorem}{thmwmgexistence}
\label{thm:wmg-existence}
The existence problem for weakly most-general fitting CQs
    is in ExpTime. Moreover, if such a CQ exists, then
\begin{enumerate}
    \item 
there is one of doubly exponential size and 
    \item  we can produce one in time $2^{poly(n)} + poly(m)$ where $n=||E||$ and $m$ is the size of the smallest weakly most-general fitting CQ.
    \end{enumerate}
\end{restatable}

The proof of Theorem~\ref{thm:wmg-existence} uses 
tree automata. More precisely, we  will show that,
given a collection of labeled examples $E=(E^+,E^-)$, 
    (i)~if there is a weakly most-general fitting CQ for $E$, then there is one 
      that is c-acyclic and has a degree at most $||E^-||$; and
    (ii)~we can construct in ExpTime a non-deterministic 
tree automaton $\mathfrak{A}_E$ that accepts precisely the (suitably encoded)
      c-acyclic weakly most-general fitting CQs for $E$ of degree at most 
      $||E^-||$.
  The existence problem for weakly most-general fitting CQs then reduces to the emptiness problem for the corresponding automaton, and hence can be solved in ExpTime. Similarly, the other 
  claims in Theorem~\ref{thm:wmg-existence} follow by Theorem~\ref{thm:nta-basic-facts}.
  Note that, by Proposition~\ref{prop:weakly-most-general} and Theorem~\ref{thm:cacyclic-frontier}, the core of a weakly most-general fitting CQ is always c-acyclic, and hence we can restrict attention 
  to c-acyclic CQs here. In fact,
  we will restrict attention to c-acyclic CQs with the Unique Names Property (UNP).
  We will explain afterwards how to lift the UNP restriction.

The following proposition provides the starting point for our encoding. It tells us that we can restrict attention to c-acyclic CQs and it provides an upper bound on the required degree of the tree encoding.

\newcommand{\up}{\textsf{up}}
\newcommand{\down}{\textsf{down}}
\newcommand{\ans}{\textsf{ans}}
\newcommand{\arity}{\textup{arity}}
\newcommand{\maxarity}{\textup{max-arity}}

 \begin{restatable}{proposition}{propdegree}
\label{prop:degree}
Let $E=(E^+,E^-)$ be a collection of $k$-ary labeled examples over a schema $\mathcal{S}$, and let $d= \max\{\maxarity(\mathcal{S}), ||E^-||\}$
If there exists a weakly most-general fitting CQ with UNP for $E$, then there is one satisfying the following
additional properties:
\begin{enumerate}
    \item $q$ is c-acyclic
    \item $q$ has at most $|E^-|+k$ many connected components, and 
    \item every existential variable of $q$ occurs in at most $||E^-||+k$ facts. 
\end{enumerate}
In particular, the incidence graph of $q$ has degree at most $d=\max\{||E^-||+k,\maxarity(\mathcal{S})\}$. 
\end{restatable}

\begin{proof} 
Let $q(\textbf{x})$ be a weakly most-general fitting CQ with UNP for $E$. 
Without loss of generality, we may assume that $q$ is minimal, in the sense
that no strict sub-query of $q$ fits $E$. In particular, then, $q$ is a core, and,
it follows from Proposition~\ref{prop:weakly-most-general} and Theorem~\ref{thm:cacyclic-frontier}
that $q$ is c-acyclic.

We know that $q$ fits the negative examples, which means
that for each negative example, there is a connected component of $q$ that does not homomorphically map to that negative example (cf.~Proposition~\ref{prop:du}). Therefore, there
is a subquery of $q$ 
with at most $d+k$ connected components that already fits the negative examples (note the $k$ here to ensure the safety condition for the subquery).
It follows from the minimality assumption that the number of connected components of $q$ is at most $|E^-|\leq d+k$. 

Next, we tackle the degree. Let $y$ be any existential variable in $q$,
and let $f_1, \ldots, f_n$ be the facts in which $y$ occurs.
We have already argued that 
$q$ is c-acyclic. Therefore, the neighborhood of $y$ in the incidence graph of $q$ can be depicted as follows (omitting edges between facts and answer variables):

\begin{center}
	\begin{tikzpicture}[thick, >=stealth]
		\node(y) at (2,2)   {$y$};	
		\node(fone) at (2,4)   {$f_1$};	
		\node(ftwo)    at (0.5,3)   {$f_2$};	
		\node(fthree)   at (0.33,1.33)   {$f_3$};	
		\node(rest) at (2,1)   {$\ldots$};	
		\node(fn)    at (3.5,1)   {$f_n$};	
		\draw[-] (y) -- (fone) ;
		\draw[-] (y) -- (ftwo) ;
		\draw[-] (y) -- (fthree) ;
		\draw[-] (y) -- (fn) ;
            \draw[dashed] (1,3.5) to[out=-20,in=-70] (0,2.5);
            \draw[dashed] (1.5,4) to[out=-80,in=-70] (2.5,4);
            \draw[dashed] (0,2) to[out=0,in=30] (1,1);
            \draw[dashed] (4,1.5) to[out=130,in=130] (2.7,1);
	\end{tikzpicture} 
\end{center}

We can correspondingly partition the connected component of $q$ that contains $y$, into $n$ ``subcomponents''  overlapping only in the variable $y$. 

To simplify the following 
argument, let us assume that
none of the subcomponents 
includes a fact containing a distinguished element. 
We will afterwards explain how to remove this assumption.
We will show that, then, $n\leq ||E^-||$.

Let  us denote by $q_i$ the subquery of $q$ obtained by removing all $n$ subcomponents except the one containing $f_i$. 
Note that, by our assumption, this each $q_i$ still satisfies the safety condition and hence is a well-defined CQ.
Let
\[S_i = \{(e,d)\mid e=(I,\textbf{c})\in E^-, d\in \text{adom}(I), (q_i,\textbf{x},y)\to (I,\textbf{c},d)\}~.\]
In other words, $S_i$ is set of all values from the negative examples, to which $y$ can be
mapped by a homomorphism from $q_i$. Let 
$T_i = \bigcap_{j=1\ldots i} S_i$. Clearly, the sequence $T_1, T_2, \ldots$ is decreasing
in the sense that $T_{i+1}\subseteq T_i$. Furthermore, we know that $T_n=\emptyset$, 
because $q$ fits $E$ and hence $q$ does not have a homomorphism to any of the negative
examples. We claim that the sequence $T_1, T_2, \ldots$ must be \emph{strictly} 
decreasing, and hence, its length is bounded by $||E^-||$. Suppose, for the sake of a
contradiction, that $S_i=S_{i+1}$. In particular, then
$\bigcap_{j=1\ldots,i,i+2\ldots n}S_j=\emptyset$. It follows that the
subquery $q'$ be obtained from $q$ by removing the subcomponent containing $f_{i+1}$
fits $E$. 
This contradicts our initial minimality assumption on $q$. 

In the general case, 
where some of the subcomponents may contain 
distinguished variables, 
the same argument applies,
except that we must take
care to keep, for each 
distinguished element, at least one subcomponent 
containing this distinguished element,
so that the subquery satisfies
the safety condition. 
It follows by the same reasoning that, in this
case, $n\leq ||E^-||+k$.
\end{proof}

To simplify the presentation in the remainder of this subsection,
let us fix a schema $\mathbf{S}$ and arity $k\geq 0$. Furthermore
choose some $d>\maxarity(\mathbf{S})$ (Proposition~\ref{prop:degree} tells us more precisely how to choose $d$). We will encode $k$-ary c-acyclic CQs over $\mathbf{S}$ by trees over the
alphabet $$\Sigma = \{\langle R,\pi\rangle\mid R\in \mathbf{S} \text{ and } \pi\in(\{\up, \down, \ans_1,\ldots, \ans_k\})^{\arity(R)})\}\cup \{\nu\}$$ where $\nu$ is a new symbol.
The intuition behind this choice of alphabet is as follows: each node of the tree,
other than the root node,
represents an existentially quantified variable or a fact (i.e., an atomic 
conjunct) of the query. The nodes at even distance from the root represent
existentially quantified variables while the nodes at odd distance from the root
represent facts. Each node of the tree that represents a fact has a label of the form $\langle R,\pi\rangle$,  where $R$ indicates the relation of the fact, and $\pi$ describes the arguments of the fact (cf.~Figure~\ref{fig:example-tree-encoding}).

\begin{definition}[Proper $\Sigma$-labeled $d$-ary tree] \label{def:proper-trees}
We say that a $\Sigma$-labeled $d$-ary tree $(T,Lab)$ is \emph{proper} if the following hold
for every $t\in T$ (where $|t|$ denotes the length of the sequence $t$):
\begin{enumerate}
    \item If $|t|$ is even, then $Lab(t)=\nu$. In particular, $Lab(\varepsilon)=\nu$
    \item If $|t|=1$, then $Lab(t)$ is of the form $\langle R,\pi\rangle$ for some $\pi$  not containing $\up$.
    \item If $|t|$ is odd and $|t|>1$, then $Lab(t)$ is of the form $\langle R,\pi\rangle$ for some $\pi$ that includes exactly one occurrence of $\up$.
    \item If 
    $Lab(t)=\nu$ and $(t\cdot i)\in T$, then
    $(t\cdot j)\in T$ for all $j<i$.
    \item If $Lab(t)=\langle R,\pi\rangle$ with $\pi=dir_1\ldots dir_n$, then
    we have that
    \begin{enumerate}
        \item for all $i\leq n$, $dir_i=\down$ if and only if $(t\cdot i)\in T$, and 
        \item for all $i>n$, $(t\cdot i)\not\in T$.
    \end{enumerate} 
    \item For each $i\leq k$, there exists at least one node whose label
    is of the form $\langle R,\pi\rangle$ where $\pi$ contains $\ans_i$.
    \end{enumerate}
\end{definition}

\begin{definition}[The CQ encoded by a proper tree]
Let $T=(T,Lab)$ be a proper $\Sigma$-labeled $d$-ary tree. We construct
a corresponding CQ $q_T(x_1, \ldots, x_k)$ as follows:
\begin{itemize}
    \item For each non-root $t\in T$ labeled $\nu$, $q_T$ contains an
       existentially quantified variable $y_t$. 
    \item For each $t\in T$ labeled $\langle R,\pi\rangle$ with $\pi=dir_1, \ldots, dir_n$, $q_T$ contains a conjunct of the form 
    $R(u_1, \ldots, u_n)$ where $u_i=x_j$ if $dir_i=\ans_j$;
    $u_i=y_{(t\cdot i)}$ if $dir_i=\down$; and $u_i=y_{t'}$ if $dir_i=\up$ and $t'$ is the parent of $t$ in $T$.
\end{itemize}
\end{definition}

We leave it to the reader to verify that $q_T$ is indeed a well-defined CQ. 
In particular the 5th item in the above definition of proper trees guarantees
that $q_T$ satisfies the safety condition.
Note that the size of $q_T$, as counted by the number of existential quantifiers plus the number of conjuncts, is at most the number of nodes of $T$ minus 1.

\begin{figure}
{\small
\begin{tabular}{ccc}
\multicolumn{3}{c}{C-acyclic query:  $q(x_1, x_2) \colondash R(x_1,z)\land R(z,z')\land R(x_1,z')\land P(x_2)$} \\ \\ \\
	\begin{tikzpicture}[thick, >=stealth, baseline={(xone.base)}]
		\node(xone) at (0,4)   {$x_1$};	
		\node(z)    at (0,2)   {$z$};	
		\node(zz)   at (0,0)   {$z'$};	
		\node(xtwo) at (2,4)   {$x_2$};	
		\node(fa) at (0,3)  {$R(x_1,z)$};
		\node(fb) at (0,1)  {$R(z,z')$};
		\node(fc) at (0,-1)  {$R(x_1,z')$};
		\node(fd) at (2,3)  {$P(x_2)$};
		\draw[-,dashed] (xone) -- (fa) ;
		\draw[-] (z) -- (fa) ;
		\draw[-] (z) -- (fb) ;
		\draw[-] (zz) -- (fb) ;
		\draw[-] (zz) -- (fc) ;
		\draw[-,dashed] (xone) to[out=190,in=170] (fc) ;
		\draw[-,dashed] (xtwo) -- (fd) ;
	\end{tikzpicture}  &\mbox{\hspace{15mm}}&
	\begin{tikzpicture}[thick, >=stealth, text width=.5cm, baseline={(root.base)}]
		\node(root)   at (0,3)   {$~~~ \textcolor{gray}{\varepsilon} {~~~~~} \nu$};	
		\node(i)      at (-2,2)   {$\textcolor{gray}{\langle 1\rangle} ~~~ \langle R,\ans_1,\down\rangle$};	
		\node(ii)     at (-2,1)   {$\textcolor{gray}{\langle 1,1\rangle} ~~~ \nu$};	
		\node(iii)    at (-2,0)   {$\textcolor{gray}{\langle 1,1,1\rangle} ~~~ \langle R,\up,\down\rangle$};	
		\node(iiii)   at (-2,-1)   {$\textcolor{gray}{\langle 1,1,1,1\rangle} ~~~ \nu$};	
		\node(iiiii)  at (-2,-2)   {$\textcolor{gray}{\langle 1,1,1,1,1\rangle} ~~~ \langle R,\ans_1,\up\rangle$};	
		\node(t)    at (2,2)    {$\textcolor{gray}{\langle 2\rangle} ~~~ \langle P, \ans_2\rangle$};	
		\draw[-] (root) -- (i) ;
		\draw[-] (i) -- (ii) ;
		\draw[-] (ii) -- (iii) ;
		\draw[-] (iii) -- (iiii) ;
		\draw[-] (iiii) -- (iiiii) ;
		\draw[-] (root) -- (t) ;
	\end{tikzpicture} \\ \\
Incidence graph of $q$ && Tree encoding
of $q$ \end{tabular}
}
\caption{Example of a c-acyclic CQ and its encoding as a $\Sigma$-labeled $d$-ary tree}
\label{fig:example-tree-encoding}
\end{figure}

An example of a tree encoding of a c-acyclic CQ is given in Figure~\ref{fig:example-tree-encoding}.

Let us say that a CQ is \emph{encodable by a $\Sigma$-labeled $d$-ary tree} if there is a 
  proper $\Sigma$-labeled $d$-ary tree $T$ such that $q_T$ is equal to $q$ (up to a one-to-one renaming of variables).

\begin{restatable}{proposition}{propencodable}
  A $k$-ary CQ $q$ over $\mathbf{S}$ is encodable by a $\Sigma$-labeled $d$-ary tree 
  (for $d\geq \maxarity(\mathbf{S})$)
  if and only if the following hold:
  \begin{enumerate}
      \item $q$ is c-acyclic,
      \item $q$ has the UNP,
      \item $q$ has at most $d$ connected components, and
      \item every existential variable of $q$ occurs in at most $d+1$ facts.
  \end{enumerate}
  Indeed, if a c-acyclic CQ $q$ meets these conditions, then $q$ is encodable by a 
  $\Sigma$-labeled $d$-ary tree $T$ where the number of nodes of $T$ is equal to the 
  number of existential variables of $q$ plus the number of conjuncts of $q$ plus 1.
\end{restatable}

The proof is given in 
Appendix~\ref{app:wmg-existence}.

Next, we show how to construct an NTA that
accept proper trees encoding a fitting CQ 
for a given collection of labeled examples.

\begin{lemma}
Given a schema, $k\geq 0$, and $d>0$, 
we can construct in polynomial time a
$d$-ary NTA $\mathfrak{A}_{proper}$ that accepts precisely the proper $\Sigma$-labeled $d$-ary trees.
\end{lemma}

\begin{proof} 
To simplify the presentation, we describe the automaton that accepts a 
$\Sigma$-labeled $d$-ary tree $T$ if and only if $T$ satisfies the first five of the five conditions
of Definition~\ref{def:proper-trees}. Our automaton has  
four states: $Q=\{ q_{\text{root}}, q_{\text{root-fact}}, q_{\text{fact}}, q_{\text{exvar}}\}$, where $F=\{ q_{\text{root}}\}$.
The transition relation
$\Delta$ consists of all transitions of the form
\begin{itemize}
\item $\langle q_1, \ldots, q_d\rangle \transition{\nu} q_{\text{root}}$ where $\langle q_1, \ldots, q_d\rangle \in \{q_{\text{root-fact}}\}^+\{\bot\}^*$
\item $\langle q_1, \ldots, q_d\rangle\transition{\langle R,dir_1, \ldots, dir_\ell\rangle} q_{\text{root-fact}}$, where, for each $i\leq \ell$, either $dir_i=\down$ and $q_i=q_{\text{exvar}}$, or  $dir_i\in \{\ans_1, \ldots, \ans_k\}$ and $q_i=\bot$; furthermore $q_i=\bot$ for $i>\ell$
\item $\langle q_1, \ldots, q_d\rangle\transition{\nu} q_{\text{exvar}}$ where  $\langle q_1, \ldots, q_d\rangle\in \{q_{fact}\}^* \{\bot\}^*$
\item $\langle q_1, \ldots, q_d\rangle\transition{\langle R,dir_1, \ldots, dir_\ell\rangle} q_{\text{fact}}$, where, for each $i\leq \ell$, either $dir_i=\down$ and $q_i=q_{\text{exvar}}$, or $dir_i\in \{\up,\ans_1, \ldots, \ans_k\}$ and $q_i=\bot$; furthermore, $q_i=\bot$ for $i>\ell$, and there is exactly one $i\leq \ell$ for which $dir_i=up$.
\end{itemize}
It is easily verified that this automaton accepts precisely the $d$-trees that satisfy
the first five conditions of Definition~\ref{def:proper-trees}.
The automaton can be extended in a straightforward way to 
test the sixth condition as well. 
To do this, we use state set
$Q' = Q\times \wp(\{1, \ldots, k\}$. In other words, states are now pairs $(q,X)$ where $q$ is as before and
$X\subseteq\{1, \ldots, k\}$ is used to represent which $\ans_i$ occur within the node labels of nodes in the subtree at hand.
\end{proof}

\begin{restatable}{lemma}{lemautomatonpositive}
\label{lem:automaton-positive}
Given a schema, $k\geq 0$, $d>0$ 
and a $k$-ary data example $e=(I,\textbf{a})$,
we can construct in polynomial time, an NTA
$\mathfrak{A}_e$, such that, for all proper $\Sigma$-labeled $d$-ary trees $T$,
we have $T\in L(\mathfrak{A}_e)$ if and only if $q_T$ fits $e$ as a positive example (i.e.,  $\textbf{a}\in q_T(I)$).
\end{restatable}

\begin{proof}
The set of states $Q$ of the automaton consists of:
\begin{itemize}
    \item an accepting state $q^{\text{root}}$,
    \item a state $q^{\text{root-fact}}_{R(b_1, \ldots, b_n)}$ for each fact $R(b_1, \ldots, b_n)$ of $I$, 
    \item a state $q^{\text{fact}}_{R(b_1, \ldots, b_n),j}$ for each fact $R(b_1, \ldots, b_n)$ of $I$, and $j\leq n$,
    \item a state $q^{\text{exvar}}_{b}$ for each $b\in \text{adom}(I)$.
\end{itemize} 

The transition relation $\Delta$
contains all transitions of the form:
\begin{itemize}
    \item $\langle q_1, \ldots, q_d\rangle\transition{\nu} q^{\text{root}}$ where each $q_i\in\{\bot, ~ q^{\text{root-fact}}_{R(b_1, \ldots, b_n)}\mid\text{$R(b_1, \ldots, b_n)$ is a fact of $I$}\}$,
    \item $\langle q_1, \ldots, q_d\rangle\transition{\nu} q_a$ where each 
    $q_i\in\{\bot, ~ q^{\text{fact}}_{R(b_1, \ldots, b_n),j}\mid\text{$R(b_1, \ldots, b_n)$ is a fact of $I$ and $b_j = a$}\}$ 
    \item $\langle q_1, \ldots, q_n, \bot, \ldots, \bot\rangle \transition{\langle S, dir_1, \ldots, dir_n\rangle} q^{\text{root-fact}}_{S(b_1, \ldots, b_n)}$ where for each $i\leq n$, either $dir_i=\ans_\ell$ for some $\ell\leq k$ and $b_i=a_\ell$,
    or else $dir_i=\down$ and $q_i=q^{\text{exvar}}_{b_i}$
    \item $\langle q_1, \ldots, q_n, \bot, \ldots, \bot\rangle \transition{\langle S, dir_1, \ldots, dir_n\rangle} q^{\text{fact}}_{(S(b_1, \ldots, b_n),i)}$ where
    $dir_i=\up$, and, for each $j\leq n$ with $j\neq i$, either $dir_j=\ans_\ell$ for some $\ell\leq k$ and $b_j=a_\ell$, or else $dir_j=\down$ and $q_j=q_{b_j}$
\end{itemize}

By construction, every accepting run of $\mathfrak{A}_e$ on a tree encoding $T$ corresponds to a homomorphism from $q_T(\textbf{a})$ to $e$, and vice versa. In 
particular, $\mathfrak{A}_e$ accepts $T$ if and only if $q_T$ fits $e$ as a positive example.
\end{proof}

From the above two lemmas, together with Theorem~\ref{thm:nta-basic-facts},
we immediately get:

\begin{theorem}\label{thm:automaton-fitting}
Given a schema, $k\geq 0$, $d>0$, and
a collection of labeled examples $E=(E^+,E^-)$, we can construct in exponential time an NTA $\mathfrak{A}_E$ that defines the tree language 
consisting of all proper $\Sigma$-labeled $d$-ary trees $T$ for which it holds that  $q_T$ 
fits $E$.
\end{theorem}

Recall that every c-acyclic CQ has a frontier. 
Our next aim is to show that we can create, for a given collection of labeled examples $E$, an automaton that accepts (tree encodings of) those c-acyclic CQs with the UNP whose frontier
consists of queries that \emph{do not fit} $E$. 
Combining this with Theorem~\ref{thm:automaton-fitting}, by 
Propositions~\ref{prop:weakly-most-general}, we then obtain an automaton
that accepts precisely weakly most-general fitting c-acyclic CQs with the UNP.

We make use of a frontier construction from~\cite{tCD2022:conjunctive}.
The presentation given here is slightly different  to the one in~\cite{tCD2022:conjunctive} (because it is phrased in terms of conjunctive
queries instead of finite structures), but it is equivalent.
We will denote the set of
answer variables of a CQ $q$ by $\text{ANSVAR}_q$ and we will denote the set of 
existential 
variables by $\text{EXVAR}_q$. We denote the set of facts of $q$ by $\text{FACTS}_q$.

\begin{definition}[$F(q)$]\label{def:F}
  Let $q$ be any connected c-acyclic CQ with the UNP. Then $F(q)$
  is the possible-unsafe CQ defined as follows: 
\begin{itemize}
\item $\text{ANSVAR}_{F(q)} = \text{ANSVAR}_q$
\item $\text{EXVAR}_{F(q)} = \{u_{(y,f)}\mid \text{$y\in\text{EXVAR}_q$ occurs in $f\in \text{FACTS}_q$}\}\cup\{u_{x}\mid x\in \text{ANSVAR}_q\}$. We will call each variable of the form $u_{y,f}$ a \emph{replica} of the existential variable $y$. 
By the replicas of an answer variable $x\in\text{ANSVAR}_k$ we will mean $u_x$ and
$x$ itself.
\item $\text{FACTS}_{F(q)}$ consists of all 
  acceptable instances of facts in $\text{FACTS}_q$, where 
  an \emph{acceptable instance} of 
 a fact $f = R(z_1, \ldots, z_n)\in \text{FACTS}_q$ is a fact
of the form $R(z'_1, \ldots, z'_n)$ where each $z'_i$ is a replica of $z_i$, and for some $i\leq n$, either $z'_i$ is of the form $u_{(z_i,f')}$ with $f'\neq f$, 
  or $z'_i$ is of the form $u_{z_i}$. 
\end{itemize}
\end{definition}

\begin{definition}[Frontier construction for c-acyclic CQs with the UNP~\cite{tCD2022:conjunctive}]
  Let $q(\textbf{x})$ be a c-acyclic CQ with the UNP with $m$ connected components.
  Then $\mathcal{F}_q = \{ F^i(q)\mid i\leq m\}$, where $F^i(q)$ denotes the possibly-unsafe CQ obtained from $q$
  by performing the $F(\cdot)$ operation on the $i$-th connected component 
  (and leaving all other connected components unchanged).
\end{definition}

\begin{proposition}[\cite{tCD2022:conjunctive}]\label{prop:frontier}
Let $q(\textbf{x})$ be a c-acyclic CQ with the UNP.
\begin{enumerate}
\item
Each query in $\mathcal{F}_q$ maps homomorphically to $q$, and
\item
If $q$ is a core, then $\mathcal{F}_q$ is a frontier for $q$.
\end{enumerate}
\end{proposition}

It is worth pointing out that $F(q)$ as constructed above is not necessarily c-acyclic. Furthermore, it may in fact not satisfy the safety condition that is part of the definition of
CQs. Indeed, consider the (connected) c-acyclic CQ $q(x) \colondash P(x)$. Then, $F(q)$ is the query  $q'(x) \colondash P(y)$, which is an unsafe CQ. 
Consequently, $\mathcal{F}_q$ in general includes unsafe CQs. This will however not be a problem for what follows, because the characterization of weakly most-general fitting CQs in terms of frontiers (Proposition~\ref{prop:weakly-most-general}) applies also if the frontier is taken to include unsafe CQs.
\footnote{More precisely, 
let $q$ be any CQ and $F$ a
set of possibly-unsafe CQs $q'$, and
let $\mathcal{F}'$ be the set of safe CQs in $\mathcal{F}$. It is easy to see that if $\mathcal{F}$ is a frontier for $q$ (in the homomorphism pre-order of possibly-unsafe CQs)  then so is $\mathcal{F}'$ (in the ordinary sense).
}

\begin{restatable}{lemma}{lemautomatonfrontier}
\label{lem:automaton-frontier}
Given a schema, $k\geq 0$, $d>0$, and a set $E$ of
$k$-ary data examples, 
we can construct, in exponential time, 
for every $i\leq d$, 
an NTA
$\mathfrak{A}^{\text{frontier}}_{E,i}$, such that, for all  proper $d$-ary 
$\Sigma$ trees $T$, we have that 
$T\in L(\mathfrak{A}^{\text{frontier}}_{E,i})$ if and only if either $q_T$ has less than $i$ connected components or else $F^i(q_T)$ admits a homomorphism to an example in $E$.
\end{restatable}

The high-level
intuition behind the construction is as follows:
recall that, in the proof
of Lemma~\ref{lem:automaton-positive}, we constructed an automaton, based on a data example $e$, that takes an input tree $T$ and tests
if $q_T$ has a homomorphism to $e$. The automaton had 
a state corresponding to every domain element of $e$ and a state corresponding to every fact of $e$, and an accepting run of the automaton (being a function from the nodes of $T$ to the states of the automaton) corresponded to a homomorphism from $q_T$ to $e$. The proof of Lemma~\ref{lem:automaton-frontier} is similar except that the queries $q'$ in the frontier of $q$ can be substantially larger.
Indeed, every existential variable of $q$ may 
have up to $d+1$ many replicas in $q'$, where $d$ is the degree of $q$. To specify a homomorphism from $q'$ to $e$ we must send each such replica to an element of $e$. To make this work, 
we construct an automaton that has a different state for every 
 $d+1$-length vector of values $e$. The details are in 
Appendix~\ref{app:wmg-existence}. 

By combining Theorem~\ref{thm:automaton-fitting} with Lemma~\ref{lem:automaton-frontier}, we get:

\begin{theorem}
Let a schema $\mathcal{S}$ and a collection of labeled examples $E=(E^+,E^-)$ be given. We can construct in exponential time
an NTA $\mathfrak{A}$ such that 
\begin{enumerate}
    \item Each $T\in L(\mathfrak{A})$ is a proper $\Sigma$-labeled $d$-ary tree for which $q_T$ is a weakly most-general fitting CQ (with UNP) for $E$.
    \item      $L(\mathfrak{A})$ is non-empty if there is a CQs with UNP  that is weakly most-general fitting for $E$.
\end{enumerate}
\end{theorem}

\begin{proof}
    Let $d=\max\{||E^-||+k,\maxarity(\mathcal{S})\}$.
By taking the intersection of the NTA $\mathfrak{A}_E$ from Theorem~\ref{thm:automaton-fitting} with the NTAs $\mathfrak{A}^{frontier}_{E^-,i}$ (for $i=1, \ldots, d$) from Lemma~\ref{lem:automaton-frontier}, we  can construct in exponential time
an NTA $\mathfrak{A}$ such that
$L(\mathfrak{A})$ consists of proper $\Sigma$-labeled $d$-ary trees $T$ 
for which 
    it holds that $q_T$ is weakly most-general fitting for $E$.
    This establishes the first item, and the second item follows from Proposition~\ref{prop:degree}.
\end{proof}

By Theorem~\ref{thm:nta-basic-facts}, 
this yields:

\begin{corollary}\label{cor:wmg-with-unp}
Given a collection of labeled examples $E$, we can decide in ExpTime the existence
of a CQ with UNP that is weakly most-general fitting for $E$, and we can produce in ExpTime a succinct DAG-representation of a  minimal-size such CQ if it exists.
\end{corollary}

In the above, for simplicity we restricted attention to CQs with the UNP.
However, the same  techniques apply in the general case, allowing us to
remove the UNP requirement from 
the statement of Corollary~\ref{cor:wmg-with-unp}:

\begin{restatable}{corollary}{corwmgwithoutunp}
\label{cor:wmg-without-unp}
Given a collection of labeled examples $E$, we can decide in ExpTime the existence
of a CQ that is weakly most-general fitting for $E$, and we can produce in ExpTime a succinct DAG-representation of a  minimal-size such CQ if it exists.
\end{restatable}

The proof of Corollary~\ref{cor:wmg-without-unp} requires
us to revisit our previous 
automata constructions and refine them
to take into account
the different possible equality types of  data examples and CQs, where by 
the equality type of a data
example $(I,\textbf{a})$ with $\textbf{a}=\langle a_1, \ldots, a_k\rangle$ we mean the equivalence relation on the set $\{1,\ldots, k\}$ induced by the repeat-occurrences in the tuple $\textbf{a}$ (and 
likewise for the equality-type of a CQs).
The details are in 
Appendix~\ref{app:wmg-existence}.

Corollary~\ref{cor:wmg-without-unp} 
immediately implies 
Theorem~\ref{thm:wmg-existence},
due to the following general facts 
about DAG-representations of trees:
if $d$ is a DAG-representation 
of a tree $t$, then the size of $t$ is 
bounded exponentially in the size of $d$,  and we can compute
an explicit representation of $t$ from $d$ in time $O(m)$ where $m$ is the size of $t$.

\subsection*{Bases of most-general fitting CQs}
In the same way that the weakly most-general fitting CQs are
characterized in terms of frontiers, \emph{bases of most-general
  fitting CQs} admit a characterization in terms of \emph{homomorphism
  dualities}.  To spell this out, we need a refinement of this
concept,  \emph{relativized} homomorphism dualities. 

\begin{definition}[Relativized homomorphism dualities] 
A pair of finite sets of data examples $(F,D)$ forms a homomorphism duality
relative to a data example $p$,  
if for all data examples $e$ with $e\to p$, the following are equivalent:
\begin{enumerate}
    \item $e$ homomorphically maps to a data example in $D$,
    \item No data example in $F$ homomorphically maps to $e$.
\end{enumerate}
\end{definition}

\begin{restatable}{proposition}{propbasisvsduality}
\label{prop:basis-vs-duality}
For all collections of labeled examples $E=(E^+,E^-)$, the following are equivalent for all CQs $q_1, \ldots, q_n$:
\begin{enumerate}
    \item $\{q_1, \ldots, q_n\}$ is a basis of most-general fitting CQs for $E$,
    \item each $q_i$ fits $E$ and $(\{e_{q_1}, \ldots, e_{q_n}\}, E^-)$ is 
a homomorphism duality relative to $p$,
\end{enumerate}
where $p=\Pi_{e\in E^+}(e)$. \end{restatable}

Before we give the proof, we remark that, intuitively, the condition that $(\{e_{q_1}, \ldots, e_{q_n}\}, E^-)$ is 
a homomorphism duality relative to $p$ can be understood as saying: every data
example that does
not already satisfy 
one of the queries $q_1, \ldots, q_n$ is 
below a data example in $E^-$ in the homomorphism pre-order and hence must necessarily be a negative example for any fitting CQ for $(E^+,E^-)$.

\begin{proof} 
(1 $\Rightarrow$ 2): By assumption, each $q_i$  fits. Let $e$ be any data example such that
$e\to p$. We need to show that $e_{q_i}\to e$ for some $i\leq n$ if and only if $e$ does not map to any data example in $E^-$. First, assume $e_{q_i}\to e$, and assume for the sake of a
contradiction that $e$ has a homomorphism to a data example in $E^-$. Then, by transitivity, $e_{q_i}$ also has a homomorphism to the
same negative example, contradicting the fact that $q_i$ fits $E$. For the converse
direction, assume
$e$ does not have a homomorphism to a data example in $E^-$. Since $e\to p$, 
by Proposition~\ref{prop:instance-products}, $e$ has a homomorphism to every 
data example in $E^+$. Therefore, the canonical CQ $q_e$ of $e$ fits $E$. Hence, we have $q_e\subseteq q_i$ for some $q_i$, and therefore, $e_{q_i}\to e$.

(2 $\Rightarrow$ 1): Let $q'$ be any CQ that fits $(E^+, E^-)$. 
Then, by Proposition~\ref{prop:instance-products}, $e_{q'}\to p$, and $e_{q'}$ does not map to any negative example in $E^-$. It follows
that some $e_{q_i}$ maps to $e_{q'}$, and hence, $q_i\to q'$,
which means that $q'\subseteq q_i$.
\end{proof}

While homomorphism dualities have been studied extensively,
we are not aware of the relativized variant having been
considered.

\begin{restatable}{theorem}{thmrelativedualities}
\label{thm:relative-dualities}\ 
\begin{enumerate}
    \item 
The following is NP-complete:
given a finite set of data examples~$D$ and a data example $p$,
is there a finite set of data examples $F$ such that
$(F, D)$ is a homomorphism duality relative to $p$? 
     \item 
Given a finite set of data examples~$D$ and a data example $p$,
if there is a finite set of data examples $F$ such that
$(F, D)$ is a homomorphism duality relative to $p$, then we
can compute in 2ExpTime 
such a set $F$, where each $e\in F$ is of
size $2^{O(||D||^2\cdot \log||D||\cdot |p|)}$.
\end{enumerate}
\end{restatable}

We provide a brief description of the key ingredients of the proof. We
first observe that the
general case reduces to the case where $D$ consists of a single instance $e$, and then address this case by adapting known techniques.
Larose et al. \cite{Larose2007:characterization} showed that, under the assumption that $e$ has no distinguished elements and is a core, there is a combinatorial characterization of the existence of a finite set of data examples $F$ such that $(F,\{e\})$ is a homomorphism duality. Notably, this  characterization can be tested by a simple algorithm: start with $e^2$ and iteratively remove non-diagonal elements (i.e., of the form $(a,b)$ with $a\neq b$) that are `dominated' in a certain sense from its domain until no further removal is possible. If only diagonal elements remain then accept and, otherwise, reject.  This can be carried out in polynomial time implying a NP algorithm for an arbitrary (non necessarily core) $e$ since, additionally, one needs to non-deterministically guess the core.  Further, as a byproduct of the above-mentioned combinatorial characterization it follows that such $F$, if exists, consists of trees with a diameter bounded by $adom(e)^2$. Since there are exponentially many minimal non-isomorphic such trees it follows that $F$ can be computed in 2ExpTime by brute force.

The main ingredient of our proof is a generalization of the previous combinatorial characterization that includes the case when $e$ might have distinguished elements and where the duality is relativized to some instance $p$. Our point of departure is an alternative proof of the above-mentioned combinatorial characterization given in~\cite{Briceno2021:dismantlability}. However, although it would have been possible to directly generalize the  proof from~\cite{Briceno2021:dismantlability} to our more general setting this would have resulted in a very lengthy and technical development.
The reason is that the proof  in~\cite{Briceno2021:dismantlability} is obtained via a long chain of equivalences, because the main motivation there was to connect dualities with certain  topological properties of the solution set of constraint satisfaction problems that are not relevant to us. Hence, reusing ideas
from~\cite{Briceno2021:dismantlability} we 
devise a new, simpler and more streamlined proof that links directly the combinatorial and duality conditions, and  adapt it to our more
general setting. Details are given in Appendix~\ref{app:relative-dualities}. 


As a consequence, we get: 

\begin{restatable}{theorem}{thCBmostgeneral}
\label{thm:existence-basis}
\label{thm:verification-basis}
The existence and verification problems for bases of most-general fitting CQs
is in NExpTime. 
\end{restatable}

\begin{proof}
For the existence problem,
let a collection of labeled examples $E=(E^+,E^-)$ be given.
Let $p=\Pi_{e\in E^+}(e)$.
We claim that the following are equivalent:
\begin{enumerate}
    \item a basis of most-general fitting CQs exists for $E$,
    \item there exists a finite set of data examples $F$ such that
$(F,E^-)$ is a homomorphism duality relative to $p$.
\end{enumerate}
The direction from (1) to (2) is immediate from
Proposition~\ref{prop:basis-vs-duality}. For the 
converse direction, let $Q$ be the set of all 
canonical CQs of data examples in $F$ that fit $E$. Then
$Q$ is a basis of most-general fitting CQs: let $q'$
be any query that fits $E$. Then, by Proposition~\ref{prop:instance-products}, $e_{q'}\to p$. Furthermore $e_{q'}$ does not have a homomorphism to any
data example in $E^-$. Therefore, we have $e\to e_{q'}$ for some $e\in F$.
Since $e\to e_{q'}$ and $e_{q'}\to p$, $e$ has a homomorphism to every 
data example in $E^+$. Therefore, $q_{e}$ fits $E$, and hence, belongs to $Q$.

This puts the problem in NExpTime, 
since $p$ can be computed in exponential time, and, by
Theorem~\ref{thm:relative-dualities}, (2) can be tested in NP given $p$.
\looseness=-1

For the verification problem, 
let $E=(E^+,E^-)$ and $\{q_1, \ldots, q_n\}$ be given. 
We may assume that each $q_i$ fits $E$ (as this can be checked in DP by Theorem~\ref{thm:any-verification}).
We may also assume that $q_1, \ldots, q_n$ are pairwise homomorphically
incomparable (if not, we can select a minimal subset, with the property
that the queries in this subset homomorphically map into all others),
and core.
It is then straightforward to see that, in order for $\{q_1, \ldots, q_n\}$ to be 
a basis of most-general fitting CQs, each $q_i$ must be a weakly 
most-general fitting CQ. By Proposition~\ref{prop:weakly-most-general}
and Theorem~\ref{thm:cacyclic-frontier}, for this to be the case, 
each $q_i$ must be c-acyclic.

Since, at this point, we have that $q_1, \ldots, q_n$ are c-acyclic, 
by Theorem~\ref{thm:cacyclic-duality}, we can compute, in single exponential
time, for each $q_i$, a set of data examples $D_{q_{i}}$, such that
$(\{e_{q_i}\},D_{q_i})$ is a homomorphism
duality. Let
$D = \{ e_1\times \cdots\times e_n\mid e_i\in D_{q_i}\}$. It is 
easy to see (using Proposition~\ref{prop:instance-products})
that $(\{e_{q_1}, \ldots, e_{q_n}\},D)$ is a homomorphism duality. 

Finally, we claim that the following are equivalent:
\begin{enumerate}
    \item $\{q_1, \ldots, q_n\}$ is a basis of most-general fitting CQs for $E$,
    \item $(\{e_{q_1}, \ldots, e_{q_n}\}, E^-)$ is a homomorphism duality relative to $p$,
    where $p=\Pi_{e\in E^+}(e)$,
    \item For each $e\in D$, there is $e'\in E^-$ such that $e\times p \to e'$.
\end{enumerate}
The equivalence of 1 and 2 is given by Proposition~\ref{prop:basis-vs-duality}.

(2 $\Rightarrow$ 3): let $e\in D$. Since  $(\{e_{q_1}, \ldots, e_{q_n}\},D)$ is a homomorphism duality and $e\in D$, we have $e_{q_i}\nrightarrow e$ for all $i\leq n$. Hence, by Proposition~\ref{prop:instance-products}, also $e_{q_i}\nrightarrow e\times p$. Therefore, since $e\times p\to p$, we have that $e\times p\to e'$ for some $e'\in E^-$.

(3 $\Rightarrow$ 2):
Let $e$ be any data example such that $e\to p$. 
If some $e_{q_i}\to e$, then, since $q_i$ fits $E$, we know that $e\nrightarrow e'$ for all $e'\in E^-$. If, on the other hand, no $e_{q_i}$ has a homomorphism to $e$, then $e\to e'$ for some $e'\in D$. Hence, since $e\to p$, by Proposition~\ref{prop:instance-products}, we have that 
$e\to e'\times p$, and therefore $e\to e''$ for some $e''\in E^-$.

This concludes the proof since (3) can be tested in NExpTime.
\end{proof}

\begin{restatable}{theorem}{thmcriticalobstructiondiameter} \label{thm:basis-construction}
Let $E=(E^+,E^-)$ be a collection of labeled examples, for which a 
basis of most-general fitting CQs exists. Then we can compute
 a minimal such basis in 3ExpTime, consisting of CQs of size $2^{poly(||E^-||)\cdot 2^{O(||E^+||)}}$. 
\end{restatable}

\begin{proof}
It follows from Proposition~\ref{prop:basis-vs-duality} together with Theorem~\ref{thm:relative-dualities}(2) that
there exists a basis consisting of CQs of size $2^{poly(||E^-||)\cdot 2^{O(||E^+||)}}$. 
Trivially, this means that the set of all fitting such CQs is a basis.
Since the fitting problem is in DP by Theorem~\ref{thm:any-verification}, 
this basis can be enumerated, and, subsequently, minimized, in 3ExpTime.
\end{proof}

\subsection{Unique Fitting CQs (Characterizations and Upper Bounds)}

By a \emph{unique fitting CQ} for a collection of labeled examples $E$, we mean a 
fitting CQ $q$ with the property that \emph{every  CQ that fits $E$ is  equivalent to $q$}.

\begin{example} \label{ex:sicos-with-unique-fitting-cq}
Let $\mathcal{S}$ consist of a single binary relation $R$, and 
let $I$ be the instance consisting of the facts $R(a,b)$, $R(b,a)$, and $R(b,b)$. 
Let $E=(E^+,E^-)$, where $E^+=\{(I,b)\}$ and $E^-=\{(I,a)\}$.

The query $q(x) \colondash R(x,x)$ is a unique fitting CQ for $E$. 
Indeed, $q$ fits $E$, and it is 
easy to see that if $q'(x)$ is any CQ that fits $E$, then $q'$ must contain the conjunct $R(x,x)$ (in order to fit $E^-$). From this, it is 
easy to see that $q$ and $q'$ admit homomorphisms to each other.
\end{example}

\begin{proposition}\label{prop:unique-fitting}
For every CQ $q$ and collection of labeled examples
$E=(E^+, E^-)$ the following are equivalent:
\begin{enumerate}
    \item $q$ is a unique fitting CQ for $E$,
    
    \item $q$ is a most-specific and 
    weakly most-general fitting CQ for $E$,
    \item $q$ is equivalent to $\Pi_{e\in E^+}(q_e)$ and 
    $\{q\times e \mid 
     e\in E^- \text{ and $q\times e$ is a well-defined CQ}\}$ is a  frontier for $q$.
\end{enumerate}
\end{proposition}

Our previous results on most-specific fitting CQs and weakly most-general fitting CQs now  imply:

\begin{theorem}\label{thm:unique-verification}\label{thm:unique-existence}
The verification and existence problems for unique fitting CQs are in NExpTime. 
When a unique fitting CQ exists, it can be computed in exponential time.
\end{theorem}

Indeed, we can verify that a CQ is unique fitting by verifying that it is both most-specific fitting (using Theorem~\ref{thm:most-specific-verification}) and weakly most-general fitting (using Theorem~\ref{thm:weakly-most-general-verification}), while 
the existence of a unique fitting CQ for $E=(E^+,E^-)$ can be tested simply by
checking that $\Pi_{e\in E^+}(e)$ is weakly most-general fitting for $E$ (using Theorem~\ref{thm:weakly-most-general-verification}).

\subsection{Lower Bounds}
\label{sec:cq-lowerbounds}
\label{sec:size-lowerbounds}

The lower bound proofs below involve reductions from 
the \emph{Product Homomorphism Problem} (PHP)~\cite{CateD15}.
The PHP takes as input a set of instances $I_1, \ldots, I_n$ and an instance $J$, and asks
whether the direct product $I_1\times\cdots\times I_n$ admits a homomorphism to $J$.
This problem is NExpTime-complete~\cite{Willard10,CateD15}.
We need a refinement of this:

\begin{restatable}{theorem}{thmphpcacyclic}
\label{thm:php-cacyclic}
Let $\mathcal{S}$ consist of a single binary relation. There is a fixed $k$
for which the following problem is NExpTime-complete: given $k$-ary data examples
$e_1, \ldots, e_n$ and $e'$ with the UNP, where $e'$ is c-acyclic, is it the case that $\Pi_i (e_i)\to e'$? 
\end{restatable}

\begin{proof}
It was shown in \cite[Theorem 1(2)]{CateD15} that,
for some fixed constant $N>0$, the following
problem is NExpTime-complete:
\begin{itemize}
    \item[(*)] Given instances $I_1, \ldots, I_n$ and $J$ over a schema consisting
    of a single relation, each of domain size at most $N$, decide if 
    $\Pi_i (I_i)\to J$.
\end{itemize}
Note that the arity of the relation is not bounded.
Furthermore, it was shown in \cite[Proof of Theorem 1(3)]{CateD15} that
(*) reduces to the 
product homomorphism problem for digraphs. We describe this reduction 
here, as we will build on it.
Let $I_1, \ldots, I_n$ and $J$ be given,
each of size at most $N$, 
and let $r$ be the arity of the relevant relation $R$. We may assume without loss of generality that for each instance $C$ among
$I_1, \ldots, I_n, J$, the projection of $R^C$ to the first coordinate is the entire active domain of domain $C$. This is
because we can always replace the $r$-ary relation $R$ by the $(r + 1)$-ary relation $adom(C) \times R$. This
transformation can be carried out in polynomial time and it does not affect the existence or
non-existence of a homomorphism from $\Pi_i(I_i)\to J$.

Next, for every $i\in \{1, \ldots, n\}$, we denote by $G(I_i)$ the digraph defined as follows. The nodes
of $G(I_i)$ include all elements of $adom(I_i)$. Furthermore, for every tuple $t\in R^{I_i}$, $G(I_i)$ contains $r$
additional nodes, which we denote $t_1. \ldots, t_r$. Furthermore, we include the following edges:
\begin{itemize}
    \item $(t_j, t_{j+1})$ for every $1\leq j < r$,
    \item
$(t[j], t_j)$ for every $1\leq j\leq r$
(where $t[j]$ refers to the $j$-th element of the tuple $t$).
\end{itemize}

We define $G(J)$ as the digraph obtained from $J$ in the same way, except that we further
add $r-1$ additional elements $s_1, \ldots, s_{r-1}$
called \emph{sink nodes}. We also add edge $(s_j, s_{j+1})$ for
every $1 \leq j < r-1$. Furthermore we add an edge from every element in $J$ to every sink node.
It was proved in \cite[Proof of Theorem 1(3)]{CateD15} that, then, $\Pi_i(I_i)\to I$ 
if and only if $\Pi_i (G(I_i))\to G(J)$.
This completes our review of the proof in \cite{CateD15}.
We now require a further modification to the reduction, in order to 
turn $G(J)$ into a c-acyclic data example. 
Observe that the incidence graph of $G(J)$ contain cycles, but that each
such cycle passes through one of the (at most $N$) elements of $J$. We can therefore
turn $G(J)$ into a c-acyclic data example, by selecting the elements of $J$ as designated elements. More precisely, let $e'$ be the data example $(G(J),\langle b_1, \ldots, b_k\rangle)$, where $k\leq N$.
Since we have now augmented $G(J)$ with $k$ designated elements, we must now do the same
with each $G(I_i)$.
For $i\leq n$, let $e_i$ be
the data example obtained 
from $G(I_i)$ by adding $k$ fresh
isolated edges (using fresh vertices) $(a_1, b_1), \ldots, (a_k,b_k)$ and selecting 
as designated elements $\langle a_1, \ldots, a_k\rangle$. 
By construction, we then have that
$\Pi_i (e_i)]\to e'$ if and only if $\Pi_i G(I_i)\to G(J)$.
\end{proof}

In addition, we will make use of the following lemma:

\begin{lemma}\label{lem:product-hom}
Let $e_1, \ldots, e_n$ and $e'$ be data examples
of the same arity and with the UNP. Then $\Pi_i (e_i) \to e'$ if and only if $\Pi_i(e_i\uplus e') \to e'$.
\end{lemma}

\begin{proof} 
A homomorphism from $\Pi_i (e_i)$ to $e'$, 
can be extended to a homomorphism from $\Pi_i (e_i\uplus  e')$ to $e'$ by sending every $k$-tuple that contains at least one value from $e'$, to the first element of the $k$-tuple in question that is a value from $e'$.
The converse direction is trivial.
\end{proof}

Using the above, we obtain the following results:

\begin{restatable}{theorem}{thmnexptimehardmany}
\label{thm:nexptime-hard-many}
The following problems are NExpTime-hard:
\begin{enumerate}
    \item The verification problem for most-specific fitting CQs.
    \item The verification problem for unique fitting CQs.
    \item The existence problem for unique fitting CQs.
    \item The verification problem for bases of most-general fitting CQs.
    \item The existence problem for bases of most-general fitting CQs.
\end{enumerate}
Each problem is NExpTime-hard already for a fixed schema and arity, and, in the case of 
the verification problems, when restricted to inputs where the input CQ fits the examples, or, in the case of the existence problems, 
when restricted to inputs where a fitting CQ exists.
\end{restatable}

\begin{proof}
(1) By a reduction from the PHP, which is NExpTime-hard already for
$k=0$ and over a schema consisting of a single binary relation~\cite{CateD15}. Given $I_1, \ldots, I_n$ and $J$,
let $E$ be the set of labeled examples that contains $(I_1\uplus J), \ldots, (I_n \uplus J)$ as positive examples,
and that does not contain any negative examples. Let $q$ be the canonical query of $J$.
It is clear from the construction that $q$ fits $E$. Furthermore, 
$q$ is a most-specific fitting CQ for $E$ if and only if 
(by Proposition~\ref{prop:most-specific}) $\Pi_i(I_i\uplus J)\to J$ if and only if 
(by Lemma~\ref{lem:product-hom}) $\Pi_i I_i \to J$.

(2--3)
By reduction from Theorem~\ref{thm:php-cacyclic}.
Let $e_1, \ldots, e_n$ and $e'$ be  data examples of the same arity, where $e'$ is c-acyclic.
Let $q$ be the canonical CQ of  $e'$, and let 
$F$ be the frontier of $e'$ (which can be computed in polynomial time, since $e'$ is c-acyclic).
It follows from Proposition~\ref{prop:unique-fitting} and Lemma~\ref{lem:product-hom} that then the following are equivalent:
\begin{enumerate}[(i)]
    \item $e_1\times \cdots\times e_n\to e'$,
    \item $q$ is a unique fitting CQ for $(E^+=\{e_i\uplus e'\mid 1\leq i\leq n\}, E^-=F)$,
    \item There is a unique fitting CQ for $(E^+=\{e_i\uplus e'\mid 1\leq i\leq n\}, E^-=F)$.
    \end{enumerate}
    Therefore, the verification and the existence problem for unique fitting CQs are both NExpTime-hard.

(4--5)
Next, we show how to modify the above reduction to also show hardness for the verification
and existence problems for \emph{bases} of most-general CQs. 
For any $k$-ary pointed instance $(C,\textbf{a})$ over schema $\mathcal{S}$,
we will denote by $(C,\textbf{a})^*$ the 
$k+1$-ary pointed instance over schema $\mathcal{S}^*=\mathcal{S}\cup \{R,P\}$,
that extends $(C,\textbf{a})$ with a fresh designated 
element $a_{k+1}$ and a non-designated element $d$, and with facts
$R(a_{k+1},d)$ and $P(d)$. In addition, we denote by $(C_{sink},\textbf{c})$
the pointed instance with distinguished elements $\textbf{c}=c_1, \ldots, c_{k+1}$ and
non-distinguished element $d$, consisting of all possible $\mathcal{S}$-facts
over $\{c_1,\ldots, c_{k+1}\}$ and all possible $\mathcal{S}^*$ facts over $\{d\}$.
We claim that the following are equivalent
(where $F$ is the frontier of $(J,\textbf{b})^*$, which can be computed in polynomial time, since $(J,\textbf{b})^*$ is c-acyclic, and where $q$ is the canonical CQ of $(J,\textbf{b})^*$): 
\begin{enumerate}[(i)]
    \item $(I_1,\textbf{a}_1)\times\cdots\times (I_n,\textbf{a}_n)\to (J,\textbf{b})$
    \item $(I_1,\textbf{a}_1)^* \times\cdots\times (I_n,\textbf{a}_n)^* \to (J,\textbf{b})^*$
    \item $q$ is a unique fitting CQ for $(E^+=\{(I_i,\textbf{a}_i)^*\uplus (J,\textbf{b})^* \mid 1\leq i\leq n\}, E^-=F\cup\{(C_{sink},\textbf{c})\})$
    \item There is a basis of most-general fitting CQ for $(E^+=\{(I_i,\textbf{a}_i)^*\uplus (J,\textbf{b})^* \mid 1\leq i\leq n\}, E^-=F\cup\{(C_{sink},\textbf{c})\})$
    \end{enumerate}
    Therefore, the verification and the existence problem for unique fitting CQs are both NExpTime-hard.

The equivalence between (1) and (2) is easy to see. Note that 
$\Pi_i ((I_i,\textbf{a}_i)^*)$ is homomorphically equivalent to $(\Pi_i (I_i,\textbf{a}_i))^*$.
The argument for the implication from (2) to (3) is similar to the
one we gave above with the simpler reduction (note that
$q$ clearly does not map homomorphically to $(C_{sink},\textbf{c})$).
The implication from (3) to (4) is trivial, because a unique fitting CQ,
by definition, constitutes a singleton basis of most-general fitting CQs.
It therefore remains only to show that (4) implies (2).

Let $\{q_1, \ldots, q_m\}$ be a basis of most-general 
fitting CQs for $(E^+,E^-)$ and assume towards a contradiction that (2) fails. Let $p=\Pi_{e\in E^+}(e)$. 
It is not hard to see that $p$ is homomorphically
equivalent to $\Pi_i((I_i,\textbf{a}_i)\uplus (J,\textbf{b}))^*$. With a slight abuse 
of notation, in what follows we will identify $p$ with 
$\Pi_i((I_i,\textbf{a})\uplus (J,\textbf{b}))^*$, so that we can speak about the 
unique $R$-edge in $p$.
For $i\geq 1$, let $p'_i$ be obtained from $p$ by replacing this
$R$-edge by a zig-zag path of length $i$, i.e., an oriented path
of the form $\rightarrow (\leftarrow ~ \rightarrow)^i$.
Clearly, $p'_i\to p$. Therefore, in particular $p'_i$ fits the 
positive examples $E^+$. Also, clearly, $p'_i$ fits the 
negative example $(C_{sink},\textbf{c})$. It also fits the other negative
examples: Suppose $p'_i$ had a homomorphism $h$ to $e\in F$. 
Since $F$ is a frontier for $(J,\textbf{b})^*$, this implies that 
 $p'_i\to (J,\textbf{b})^*$. Since $(J,\textbf{b})^*$ has a unique $R$-edge it
 follows that every $R$-edge in the zigzag of $p'_i$ must necessarily be mapped to it by $h$. In turn,
 this implies that $p$ is homomorphic to $(J,\textbf{b})^*$, contradiction the fact that (2) fails.
Thus, each $p'_i$ fits $(E^+,E^-)$. Hence, some member
$q_j$ of the basis must homomorphically map to infinitely many $p'_i$. 
It follows that $q_j$ does not contain any finite undirected $R$-path
from a designated element to a value satisfying the unary $P$
(for if such a path existed, of length $\ell$, then $q_j$ would not 
map to $p'_i$ for any $i>\ell$). It follows that $q_j$ maps to 
$(C_{sink},\textbf{c})$, a contradiction.
\end{proof}

For the existence of weakly most-general fitting CQs, we prove
an ExpTime lower bound by adapting a reduction from the word problem for certain
alternating Turing machines used in
\cite{DBLP:journals/iandc/HarelKV02} to prove hardness of a 
product simulation problem for transition systems. 
Unlike the previous reductions, it does not apply to the 
restricted case where a fitting CQ is promised to exist.

\begin{restatable}{theorem}{thmwmghardness}
\label{thm:wmg-hardness}
    The existence problem for weakly most-general fitting CQs is
    ExpTime-hard. %
\end{restatable}

The proof of Theorem~\ref{thm:wmg-hardness} is given in 
Appendix~\ref{app:lowerboundstreecqs}, where the result is 
established simultaneously for CQs and for \emph{tree CQs}
(cf.~Section~\ref{sec:tree-cq}).  

The following results provide lower bounds on the \emph{size} of fitting CQs:

\begin{restatable}{theorem}{thmcqsizelowerbound}
\label{thm:cq-size-lowerbound}
Fix a schema consisting of a single binary relation.
For $n>0$, we can construct a collection of Boolean data examples
 of combined size polynomial in $n$ such that a fitting CQ exists, but not one of size less than $2^n$.
\end{restatable}

\begin{proof}
For $i\geq 1$, let $C_{p_i}$ denote the directed cycle of length $p_i$, with $p_i$ the $i$-th prime number (where $p_1 = 2$). Note that, by the prime number theorem, $C_{p_i}$ is of size
$O(i\log i)$.
Let $E_n^+ = \{ C_{p_i}\mid i=2, \ldots, n\}$
and let $E_n^- = \{ C_{p_1} \}$.
Then it is easy to see that a fitting CQ for $(E^+_n, E^-_n)$ exists
(namely any cycle whose length is a common multiple of the lengths of the cycles in $E_n^+$). Furthermore, every fitting CQ
must necessarily contain a cycle of odd length (in order not to fit the negative example), and the length of this cycle must be a common multiple of the prime numbers $p_2, \ldots, p_n$ (in order to fit
the positive examples). This shows that the query must have size
at least $2^n$.
\end{proof}

\begin{remark}\rm
Continuing from Remark~\ref{rem:sicos},
the above proof can be adapted to also apply
 to the \emph{CQ definability} problem:
let $I_n=\biguplus_{i=1\ldots n} C_{p_i}$,
let $E_n^+ = \{(I_n,a)\mid \text{$a$ lies on the cycle of length 2}\}$ 
and 
$E_n^- = \{(I_n,a)\mid \text{$a$ lies on a cycle of length greater than $2$}\}$.
By the same reasoning as before, there is a unary CQ that fits 
these examples, but every unary CQ that fits must have size at least
$2^n$.
\end{remark}

Theorem~\ref{thm:cq-size-lowerbound} provides a size lower bound for arbitrary fitting CQs, and, consequently, also for most-specific fitting CQs. However, it does not provide
a lower bound on the size of
most-general fitting CQs or unique CQs when they exist, since there may not exist such fitting CQs for the labeled examples constructed in the proof. The next theorem provides a size lower bound that pertains to unique fitting CQs and therefore also to most-general fitting CQs (but not with a fixed schema).

\begin{restatable}{theorem}{thmuniquesizelowerbound}
\label{thm:unique-size-lowerbound}
For $n\geq 0$, we can construct a schema with $O(n)$ unary and binary relations and a collection of labeled 
examples without designated elements, of combined size polynomial in $n$  such that
\begin{enumerate}
    \item There is a unique fitting (Boolean) CQ. 
    \item Every fitting CQ contains at least $2^n$ variables.
\end{enumerate}
\end{restatable}

\begin{proof}
The follow construction is inspired by the lower bound arguments in~\cite{CateD15}.
The schema
contains unary relations $T_1, \ldots, T_n$, $F_1, \ldots, F_n$ (used to describe bit-strings of length $n$, where $T_i$ stands for ``the $i$-th bit is set to 1'', and $F_i$ stands for ``the $i$-th bit is set to 0'') and binary relations
$R_1, \ldots, R_n$ where the intended interpretation of $R_i$ is ``the successor relation on bit-strings of length $n$, restricted to pairs of bit-strings where the 
$i$-th bit is the one that flips to from 0 to 1''. Note that the 
union of these $R_i$'s is precisely the ordinary successor relation on bit-strings.

Next, we describe the positive examples.
There will be $n$ positive examples, 
$P_1, \ldots, P_n$, each of which has domain
$\{0,1\}$. The direct product $P_1\times \cdots\times P_n$, will be the intended
instance as described above. Specifically,
for each $i\leq n$, let $P_i$ be the two-element instance that has domain $\{0,1\}$ and contains the following facts:
\begin{itemize}
    \item $F_i(0)$ and $T_i(1)$
    \item all facts involving unary relations $T_j$ and $F_j$ for $j\neq i$
    \item all facts $R_j(0,0)$ and $R_j(1,1)$ for $j<i$
    \item all facts $R_i(0,1)$ 
    \item all facts $R_j(1,0)$ for $j>i$
\end{itemize}

Let $P$ be the direct product $P_1\times \cdots\times P_n$.
By construction, the domain of $P$ consists
of bit-strings of length $n$. Indeed,
it can easily be verified that $P$ is a directed path of length $2^n$, starting with $\langle 0,\ldots,0\rangle$ and ending with $\langle 1,\ldots,1\rangle$, where the unary and binary relations  have the intended interpretation as described above. For example, if $n=2$, then the instance $P$ can
be depicted as follows:

\[ \langle 0,0\rangle \xrightarrow{R_2} 
   \langle 0,1\rangle \xrightarrow{R_1} 
   \langle 1,0\rangle \xrightarrow{R_2} 
   \langle 1,1\rangle \]

Our negative example is the instance $N$, 
whose domain consists of $3n$ values, $a_1, \ldots, a_n$ and $b_1, \ldots, b_n$,
and $c_1, \ldots, c_n$, and such that $N$ contains the following facts:
\begin{itemize} 
\item All facts over domain $A=\{a_1, \ldots, a_n\}$ except $T_i(a_i)$ for $i\leq n$;
\item All facts over domain $B=\{b_1, \ldots, b_n\}$ except $F_i(b_i)$ for $i\leq n$;
\item All facts over domain $C=\{c_1, \ldots, c_n\}$ except $T_i(c_i)$ and $F_i(c_i)$ for $i\leq n$;
\item All binary facts $R_j(x,y)$ where $x\in B$ and $y\in A$
\item All binary facts $R_j(x,y)$ where $x\in C$ or $y\in C$
\end{itemize}
In particular, note that there are no directed edges going from the $A$ cluster to the $B$ cluster.

We claim that $P$ does not map to $N$. 
Indeed, the only values in $N$ that satisfy $F_1(x)\land\cdots\land F_n(x)$ 
(and hence, to which the value $\langle 0,\ldots, 0\rangle$ of $P$ could be mapped)
are $a_1, \ldots, a_n$, while the only values in $N$ that satisfy 
$T_1(x)\land\cdots\land T_n(x)$ (and hence,
to which the value $\langle 1,\ldots, 1\rangle$ of $P$ could be mapped), are 
$b_1, \ldots, b_n$. By construction, the latter cannot be reached from  the former
by a path consisting of forward edges only, except if the path goes through the $C$ cluster.
Note that the values in the $C$ cluster are not viable candidates for the homomorphism
because each fails to satisfy $T_i\lor F_i$ for some $i$.
It follows that the canonical CQ of $P$ is a 
fitting CQ, and, indeed, a most-specific fitting CQ. In the remainder of the proof, we 
show that it is  in fact a unique fitting CQ.

Let $q'$ be any fitting CQ and let $I$ be its canonical example. 
Then $I$ maps to $P$ and not to $N$.
Consider any connected component
of $I$ that does not map to $N$. 

The component in question must contain a node $a$ satisfying all 
$F_1, \ldots, F_n$ (otherwise the entire component could be mapped
to the $\{b_1, \ldots, b_n\}$-subinstance of $N$). Similarly,
it must contain a node $b$ satisfying all 
$T_1, \ldots, T_n$ (otherwise the entire component could be mapped
to the $\{a_1, \ldots, a_n\}$-subinstance of $N$.

We can now distinguish three cases:
\begin{itemize}
    \item There is no directed path in $I$ from a node $a$ satisfying $F_1,\ldots,F_n$
    to a node $b$ satisfying $T_1, \ldots, T_n$. In this case, let $X$ be the set of all
    values of $I$ that are reachable by a directed path from an value satisfying
    $F_1,\ldots,F_n$. Take any map $h$ that sends every $x\in X$ to some $a_j$ (where $j$ is chosen so that $x$ omits $T_j$)
    and that sends every value $y$ outside $X$ to a $b_j$ of $N$ (where $j$ is chosen so that $y$ omits $F_j$).
    Then $h$ is a homomorphism from $I$ to $N$, contradicting the fact that $q'$ is a fitting CQ.
    Therefore, this cannot happen.
    
    \item Every directed path in $I$ from a node $a$ satisfying $F_1,\ldots,F_n$ to a
         node $b$ satisfying $T_1, \ldots, T_n$, contains a ``bad'' node, by which we mean
         a value that, for some $j\leq n$, fails
         to satisfy either $T_j$ or $F_j$.
         In this case, let $X$ be the set of all values of $I$ that can be reached
         from a node satisfying $F_1,\ldots, F_n$ by a path that does not contain bad nodes.
         We construct a homomorphism from $I$ to $N$ by sending all the values in $X$
         to a suitable $a_j$; all bad nodes to $c_j$ (where $j$ is such that the bad node in question fails to satisfy $T_j$ or $F_j$); all other nodes to suitable $b_j$.
         Therefore, again, we have a contradiction, showing that this cannot happen.
         
    \item There are nodes $a$ and $b$ satisfying $F_1,\ldots,F_n$ and $T_1, \ldots, T_n$, respectively, such that there is a 
     directed path from $a$ to $b$ that does not contain any bad node.
    In this case, we can easily see that the homomorphism from $I$ to $P$ maps this path 
    bijectively to $P$ and hence its inverse contains a homomorphism from $P$ to $I$.
    It follows that $I$ and $P$ are homomorphically equivalent, and hence, $q'$ is 
    equivalent to $q$.
\end{itemize}
\end{proof}

\begin{restatable}{theorem}{thmbasissizelowerbound}
\label{thm:basis-size-lowerbound}
For $n\geq 0$, we can construct a schema with $O(n)$ unary and binary relations and a collection of labeled
examples without distinguished elements, of combined size polynomial in $n$ such that
\begin{enumerate}
    \item There is a basis of most-general fitting (Boolean) CQs.
    \item Every such basis contains at least $2^{2^n}$ CQs.
\end{enumerate}
\end{restatable}

\begin{proof}
It suffices to make minor changes to the construction used in the proof of Theorem~\ref{thm:unique-size-lowerbound}. Specifically,
(i) we expand the schema with unary relation symbols $Z_0$ and $Z_1$,
(ii) we extend the positive examples and the negative example with all possible $Z_0$- and $Z_1$-facts over their domain, and (iii) we extend the negative example $N$
with one further value $z$ where $z$ satisfies all possible unary facts except $Z_0(z)$ and $Z_1(z)$, as well as all binary facts 
$R_j(x,y)$ for which it holds that $z\in \{x,y\}$.

Let $P$ be the direct product $P_1\times \cdots \times P_n$.
It can again be verified that $P$ is a directed path of length $2^n$,
starting with $\langle 0,\ldots,0\rangle$ and ending with $\langle
1,\ldots,1\rangle$, where the unary relations $T_1, \ldots, T_n,
F_1,\ldots, F_n$  and the binary relations $R_1, \ldots R_n$ have the
intended interpretation, and  such that the unary relation symbols $Z_0$ and $Z_1$ are true everywhere.

Let $X$ be the set containing all subinstances of $P$ obtained by removing, for each node $x$ in its domain, exactly one of the facts $Z_0(x)$ or $Z_1(x)$. We shall show that $(X,\{N\})$ is a homomorphism duality relative to $P$. 

First, note that no instance in $X$ is  homomorphic to $N$. Now, let $Q$ be any instance satisfying $Q\rightarrow P$ and $Q\nrightarrow N$.  We need to show that $Q$ admits an homomorphism from some instance in $X$. To do so, let $Q'$ be any connected component of $Q$ such that $Q'\nrightarrow N$.

By the same arguments as in the proof of Theorem~\ref{thm:unique-size-lowerbound}, $Q'$ contains nodes $a$ and $b$
satisfying $F_1,\dots,F_n$ and $T_1,\dots,T_n$ respectively and there a directed path from $a$ to $b$ containing no bad nodes (that is a value that for some $j\leq n$, fails to satisfy either $T_j$ or $F_j$). Mimicking the same arguments it is immediate to show that, additionally,
every node in this directed path satisfies $Z_0$ or $Z_1$. Since
$Q\rightarrow P$ there is an homomorphism $h$ from this directed path to $P$. It is easy to see that $h$ must be bijective.
Since every node $x$ in the path satisfies $Z_0$ or $Z_1$ it follows
that the inverse of $h$ defines an homomorphism from some instance in $X$ to $Q$.

Finally, note that $X$ has $2^{2^n}$ values and that every pair of instances in $X$ is not homomorphically equivalent. By Proposition~\ref{prop:basis-vs-duality}, the set containing the canonical queries of instances in $X$, which clearly fits, is a minimal basis of most general fitting CQs.
\end{proof}

\section{The Case of UCQs}
\label{sec:ucq}

A $k$-ary \emph{union of conjunctive
queries} (UCQ) over a schema $\mathcal{S}$ 
is an expression of the form
$q_1\cup\cdots\cup q_n$, where $q_1, \ldots, q_n$ are $k$-ary CQs over $\mathcal{S}$.
Each $q_i$ is called a \emph{disjunct} of $q$. Logically, $q$ is interpreted
as the disjunction of $q_1, \ldots, q_n$. That is, 
$q(I)=\bigcup_i q_i(I)$.
All the notions and 
problems
considered in Section~\ref{sec:cq} now naturally generalize to UCQs.
The correspondence between query containment and homomorphisms that holds for
CQs extends to UCQs as well:
for two UCQs $q, q'$, we say that $q$ \emph{maps homomorphically to} $q'$
(written: $q\to q'$) if, for every disjunct $q'_i$ of $q'$, there
is a disjunct $q_j$ of $q$ such that $q_j\to q'_i$. %
Under this definition, as for CQs we have $q\to q'$  precisely if
$q' \subseteq q$.

The following example shows that sometimes a fitting UCQ exists when a fitting CQ does not exist.

\begin{example} \label{ex:ucq}
Consider a schema consisting of the unary relations $P,Q,R$, and let $k=0$.
Let $E$ consist of positive examples $\{P(a),Q(a)\}$ and $\{P(a),R(a)\}$, and negative example $\{P(a), Q(b), R(b)\}$. Let $q$ be the union of the following two CQs:
\[\begin{array}{lll}
q_1 &\colondash& P(x)\land Q(x) \\
q_2 &\colondash& P(x)\land R(x) 
\end{array}\]

Clearly, $q$ 
fits. Indeed, it can be shown that 
$q$ is a unique fitting UCQ for $E$.
However, there is no fitting CQ
for $E$, as the direct product of the positive examples maps
to the first negative example. \end{example}

The following proposition tells us when a fitting UCQ exists for a given collection of labeled examples.

\begin{proposition}
For all collections of labeled examples $E=(E^+,E^-)$ the following
are equivalent:
\begin{enumerate}
    \item There is a fitting UCQ for $E$,
    \item There is no $e\in E^+$ and $e'\in E^-$ with $e\to e'$.
    \item The UCQ $\bigcup_{e\in E^+}q_e$ fits $E$,
\end{enumerate}
\end{proposition}

\begin{proof}
    The implication from (2) to (3) and
    from (3) to (1) are immediate.
    The implication from (1) to (2), 
    can be seen as follows: let $q=\bigcup_i q_i$ be a UCQ that fits $E$, and let $e\in E^+$ and 
    $e'\in E^-$. By definition, 
    there must be a homomorphism from some $q_i$ to $e$. If there were a homomorphism from $e$ to $e'$, then,
    by transitivity, $q_i$ would have a homomorphism to $e'$, contradicting the fact that $q$ fits $E$.
\end{proof}

This shows that the fitting existence problem for UCQs is in coNP (cf.~also Theorem~\ref{thm:ucq-complexity-results-a} below). Moreover, it shows that there is a PTime-computable 
``canonical candidate'' fitting UCQ, namely $\bigcup_{e\in E^+}(q_e)$. 
That is, computing a fitting UCQ under the
promise that one exists, is in PTime.
While this can be viewed as a positive result,
it is somewhat disappointing as the UCQ in question does
nothing more than enumerate the positive examples. It
does not ``compress'' or ``generalize from'' 
the input examples in a meaningful way. 
This, it turns out, is unavoidable:
it was shown in \cite{CateDK13:learning} that 
there does not exist an ``efficient Occam algorithm'' for UCQs, i.e., 
a PTime algorithm taking as input a collection of labeled examples $E$
for which a fitting UCQ is promised to exists and producing a fitting UCQ
of size $O(m)^\alpha \cdot poly(n)$ where $m$ is the size of the input,
$n$ is the size of the smallest fitting UCQ, and $\alpha<1$.

Next, we next give characterizations for most-specific fitting UCQs,
most-general fitting UCQs, and unique fitting UCQs, in the style of
the characterization for CQs provided in Section~\ref{sec:cq}. For
most-general fitting UCQs, the weak and the strong version turn
out to coincide, unlike for CQs. 

\begin{restatable}{proposition}{propucqmostspecific}
(Implicit in \cite{Alexe2011:designing}.)
\label{prop:ucq-most-specific}
For all collections of labeled examples $E=(E^+,E^-)$ and UCQs $q$, the following are 
equivalent:
\begin{enumerate}
    \item $q$ is a strongly most-specific fitting UCQ for $E$,
    \item $q$ is a weakly most-specific fitting UCQ for $E$,
    \item $q$ fits $E$ and is  equivalent to 
         $\bigcup_{e\in E^+} q_e$.
 \end{enumerate}
\end{restatable}

\begin{proof}
The implication from 1 to 2 is trivial. 

For the implication from 2 to 3, suppose that $q$ is a weakly most-specific fitting  UCQ for $E$. Let $q'=\bigcup_{e\in E^+} q_e$. Since $q$ fits $E$, we have 
$q\to q'$. Furthermore, $q'$ fits $E$. Indeed, $q'$ fits $E^+$ by construction, and
if there was a homomorphism from a disjunct of $q'$ to a negative example, then,
since $q\to q'$, also $q$ would fail to fit the same negative example. Thus, 
$q'\subseteq q$ and $q$ fits $E$. Therefore, by the definition of ``weakly most-specific'',
we have that $q\equiv q'$.

For the implication from 3 to 1, let $q'=\bigcup_i q'_i$ be any UCQ that fits $E$.
We must show that $q\subseteq q'$. Consider any disjunct $q_i$ of $q$. Since $q$ is
homomorphically equivalent to $\bigcup_{e\in E^+}q_e$, we know that, for some
$e\in E^+$, $q_e\to q_i$. Furthermore, since $q'$ fits $E^+$, for some disjunct
$q'_j$ of $q'$ we have $q'_j\to q_e$. Therefore, by composition, $q'_j\to q_i$, which means that $q_i\subseteq q'_j$.
\end{proof}

This also shows that the notion of a
\emph{most-specific fitting query} 
is perhaps not very interesting in the case of UCQs, as it
trivializes to something that is essentially a disjunction of complete descriptions of positive examples.

\begin{restatable}{proposition}{propucqmostgeneral}
\label{prop:ucq-most-general}
For all collections of labeled examples $E=(E^+,E^-)$ and UCQs $q=q_1\cup \cdots\cup q_n$,
the following are equivalent:
\begin{enumerate}
    \item $q$ is a strongly most-general fitting UCQ for $E$,
    \item $q$ is a weakly most-general fitting UCQ for $E$,
    \item $q$ fits $E$ and 
          $(\{e_{q_1}, \ldots, e_{q_n}\}, E^-)$ is a homomorphism duality.
\end{enumerate}
\end{restatable}

\begin{proof} 
The implication from 1 to 2 is trivial. 

For the implication from 2 to 3, assume
that $q=\bigcup_i q_i$ is weakly most-general fitting for $E$,
and let $e$ be any data example. If $e_{q_i}\to e$, then
$e\nrightarrow E^-$, as otherwise, by transitivity, we would have
that $e_{q_i}\to E^-$, which we know is not the case because $q$ fits $E$.
Conversely, if $e\nrightarrow E^-$, then the UCQ $q' = q\cup q_{e}$ 
fits $E$. Since $q$ is contained in $q'$,
it follows by the definition of ``weakly most-general''
that $q\equiv q'$, which means that $q$ and $q'$ are homomorphically equivalent. In particular,
some $q_i$ maps to $q_e$, and hence, $e_{q_i}\to e$.

Finally, for the implication from 3 to 1,
suppose $q'$ is a fitting UCQ for $E$.
Consider any disjunct $q'_i$ of $q'$. Then $q'_i$ does not map to $E^-$.
Hence, since $(\{e_{q_1}, \ldots, e_{q_n}\}, E^-)$ is a homomorphism duality,
we have that some $e_{q_j}$ maps to $e_{q'_i}$, and hence,
$q_j\to q'_i$. This shows that $q$ homomorphically maps to $q'$.
\end{proof}

We can now revisit Example~\ref{ex:ucq}, and see can that 
(i) $q$ is equal to the union of the canonical CQs of the positive examples, therefore $q$ is a most-specific fitting UCQ; and (ii) 
$(\{e_{q_1},e_{q_2}\},E^-)$ is a 
homomorphism duality (we leave it to the reader to verify this) and therefore
$q$ is also a most-general fitting UCQ.

Combining the above two propositions, we obtain:

\begin{restatable}{proposition}{propucqunique}
\label{prop:ucq-unique}
For all collections of labeled examples $E=(E^+,E^-)$ and UCQs $q$,
the following are equivalent:
\begin{enumerate}
    \item $q$ is a unique fitting UCQ for $E$,
    \item $q$ fits $E$ and 
         the pair $(E^+, E^-)$ is a homomorphism duality,
    \item $q$ is equivalent to $\bigcup_{e\in E^+}q_e$ and 
         $(E^+, E^-)$ is a homomorphism duality.
\end{enumerate}
\end{restatable}

\begin{proof}
From 1 to 2, suppose $q$ is a unique fitting UCQ for $E$. 
We must show that 
$(E^+,E^-)$ is a homomorphism duality. Let $e$ be any data example.
If a positive example maps to $e$, then $q$ maps to $e$. Since
$q$ does not map to any negative example, it follows that $e$
does not map to any negative example either. If, on the other hand,
no positive example maps to $e$, then $e$ must map to 
a negative example. For, otherwise, $q'=q\cup q_{e}$ would be
a fitting UCQ that is not homomorphically equivalent (and hence not equivalent) to $q$.

For the implication from 2 to 3, it suffices to show that $q$ is homomorphically equivalent 
to $q'=\bigcup_{e\in E^+} q_e$. The direction $q\to q'$ is immediate from the 
fact that $q$ fits $E^+$. For the other direction, let $q_i$ be any disjunct of $q$.
Since $q_i$ fits $E^-$ and $(E^+,E^-)$ is a homomorphism duality, we know that,
for some positive example $e\in E^+$, $q_e\to q_i$. This shows that $q'\to q$.

For the implication from 3 to 1, it follows from Proposition~\ref{prop:ucq-most-specific} 
that $q$ is a most-specific fitting UCQ for $E$, and from Proposition~\ref{prop:ucq-most-general} that $q$ is a most-general fitting UCQ for $E$.
Hence, $q$ is a unique fitting UCQ for $E$.
\end{proof}

Based on these characterizations we obtain:

\begin{restatable}{theorem}{thmucqcomplexityresultsa} \ 
\label{thm:ucq-complexity-results-a}
\begin{enumerate}
\item 
The existence problem for fitting UCQs (equivalently, for most-specific fitting UCQs) is coNP-complete; if a fitting UCQ exists, a most-specific fitting UCQ can be computed in PTime.
\item
The existence problem for most-general fitting UCQs is NP-complete; 
if a most-general fitting UCQ exists, one can be computed in 2ExpTime.
\item 
The verification problem for fitting UCQs is DP-complete.
\item
The verification problem for most-specific fitting UCQs is DP-complete.
\end{enumerate}
\end{restatable}

\begin{proof}
1. By Proposition~\ref{prop:ucq-most-specific}, it suffices to test that 
the UCQ $\bigcup_{e\in E^+} q_e$ fits. It fits the positive examples by definition.
Therefore, it is enough to test the non-existence of a homomorphism to $E^-$, 
which can be done in coNP. The lower bound is by reduction from graph homomorphism:
$G\to H$ holds if and only if there is no fitting UCQ for $(E^+=\{G\}, E^-=\{H\})$.

2. The upper bound follows immediately from
Proposition~\ref{prop:ucq-most-general} together with Theorem~\ref{thm:relative-dualities} (one can choose $p$ to be the single-element instance containing all possible facts). 
The NP-hardness follows directly from~\cite{Larose2007:characterization}
(we can choose $E^+=\emptyset$). Finally, it follows from Proposition~\ref{prop:ucq-most-general} and Theorem~\ref{thm:relative-dualities}(2) (again choosing $p$ to be the single-element instance containing all possible facts) that the union of the canonical queries of all instances of size at most $2^{O(poly(||E^{-}||))}$ 
not homomorphic to any instance in $E^-$ is a most-general fitting UCQ, provided it exists. Hence, such $q$ can be constructed 
in 2ExpTime. 

3. To test if $q$ fits $E=(E^+,E^-)$, we test that (i) for each $e\in E^+$, some disjunct of $q$ maps to it, and (ii) no disjunct of $q$ maps to any $e\in E^-$. This clearly shows that the problem is in DP. For the lower bound, we reduce from exact 4-colorability~\cite{Rothe2001:exact}: a graph $G$ is exact 4-colorable if and only if 
the canonical CQ of
$G$ fits $(E^+=\{K_4\}, E^-=\{K_3\})$, where $K_n$ is the $n$-element clique.

4. To verify that $q$ is a most-specific fitting UCQ for $E=(E^+,E^-)$, by Proposition~\ref{prop:ucq-most-specific}, it suffices to test that (i) $q$ is 
homomorphically equivalent to $\bigcup_{e\in E^+}q_e$, and (ii) fits $E^-$. 
This clearly places the problem in DP. For the lowerbound, we reduce again from exact 4-colorability~\cite{Rothe2001:exact}: a graph $G$ is exact 4-colorable if and only if 
the canonical CQ of
$G$ is a most-specific fitting UCQ for $(E^+=\{K_4\times G\}, E^-=\{K_3\})$.
\end{proof}

In order to state the remaining complexity results, let \textsc{HomDual} be the problem of testing
if a given pair $(F,D)$ of finite sets of data examples is a homomorphism duality. The precise complexity of this
problem is not known.

\begin{restatable}{proposition}{prophomdual}
\label{prop:homdual}
\textsc{HomDual} is in %
ExpTime
and NP-hard. 
\end{restatable}

\begin{proof}
We first prove the upper bound.
For this, we make use of the following algorithm:

\begin{quote}
  There is polynomial time algorithm (arc consistency) that given instances $e',e$ with possible distinguished elements as input determines whether it is true that for each $c$-acyclic $t$, $t\rightarrow e'$ implies $t\rightarrow e$.
\end{quote}
  The constraint literature includes several sightly different algorithms under the name `arc consistency' so we give a reference for the sake of concreteness \cite{chen2011arc}. It is well known that the arc-consistency algorithm, which has been defined only for instances without designated elements, verifies precisely the condition stated.
  To extend it to instances with distinguished elements, the only modification that is needed is to initialize the algorithm such that each designated element in $e'$ is mapped to the corresponding designated element in $e$.

Let us now proceed with the ExpTime upperbound. Let $(F,D)$ be given.
We may assume that $F$ consists of pairwise homomorphically incomparable instances: if not, then we can take a minimal subset $F'\subseteq F$ with the property that for every $e\in F$, there is $e'\in F'$
such that $e'\to e$. Similarly, we can assume that $D$ consists
of pairwise homomorphic incomparable instances: again, we can take a
minimal subset $D'\subseteq D$, with the property that for every $e\in D$, there is $e'\in D'$
such that $e\to e'$.

It is easy to see that $(F,D)$ is a 
homomorphism duality if and only if $(F',D')$ is. We may also assume 
that each $e\in F\cup D$ is a core.

We claim that, in order for $(F,D)$ to be a homomorphism duality, each
$e\in F$ must be c-acyclic. By Theorem~\ref{thm:cacyclic-frontier},
it suffices to show that, if $(F,D)$ is a homomorphism duality, then
each $e\in F$ has a frontier (since, core instances that have a frontier
are c-acyclic). Indeed, if $(F,D)$ is a homomorphism duality, then
 $Fr_e = \{e'\times e\mid e'\in D\}$ is a frontier for $e$:
 since $e\nrightarrow e'$,  we have that 
 $e'\times e\to e$ and $e\nrightarrow e\times e'$ (by Proposition~\ref{prop:instance-products}).
 Furthermore, suppose $e''\to e$ and $e\nrightarrow e''$. 
 Since $F$ consists of pairwise homomorphically incomparable data examples,
 it follows that there is no data example in $F$ has a homomorphism to $e''$.
 Hence, $e''\to e'$ for some $e'\in D$. Therefore, 
 $e''\to e'\times e$. The latter belongs to $Fr_e$ by construction.

 Next, we therefore test that $F$ consists of c-acyclic instances.
 If this test succeeds, then, by Theorem~\ref{thm:cacyclic-duality}, we can 
 compute, for each $e\in F$, a finite set $D_e$ such that $(\{e\},D_e)$
 is a homomorphism duality. Let $D'= \{e_1\times \cdots\times e_n\mid
 (e_1, \ldots, e_n)\in \Pi_{e\in F}D_e\}$. It is straightforward to show
 that $(F,D')$ is a homomorphism duality. 
 
 It follows that $(F,D)$ is a homomorphism duality if and only $D$ and $D'$ are homomorphically equivalent, in the sense that (i) for each $e\in D$, there is $e'\in D'$ such that $e\rightarrow e'$, and (ii) vice versa:
 for each $e'\in D$, there is a $e\in D$, such that $e'\to e$. 
 
Condition (i) can be tested in polynomial time since it is equivalent to the fact that $e\nrightarrow e'$ for every $e\in F$ and $e'\in D$ 
(and $e$ is guaranteed to be $c$-acyclic).
For condition (ii) we first check whether there is some set $F'$ of instances such that $(F',D)$ is a homomorphism duality. It follows from Lemmas \ref{le:reductiontosingleexample} and \ref{le:characterization} that this can be done by verifying 
that every $e\in D$ satisfies condition (1) in Lemma \ref{le:characterization} (choosing $P$ to be the instance with only one element and having all possible facts). This check can be done clearly in NP (and, indeed, in polynomial time if $e$ is a core although this is not needed here). 

Next, let $e'\in D'$ and $e\in D$. We shall show that $e'\rightarrow e$ is equivalent to the condition checked by the arc-consistency algorithm we mentioned at the beginning of this proof. 
We note that since $D'$ has at most exponentially many instances and all of them have size bounded above exponentially then this implies that (ii) can be verified in ExpTime by an iterative application of
 arc-consistency.
 
Let us proof our claim. Let $e'\in D'$ and $e\in D$.
If $e'\to e$, then, clearly (by composition of homomorphisms),
$t\to e'$ implies $t\to e$, for all c-acyclic instances $t$.
For
the converse, assume that $t\rightarrow e'$ implies $t\rightarrow e$ for every c-acyclic instance $t$. Since $e$ satisfies 
Lemma~\ref{le:characterization}(1) it follows that there exists some set $T$ of instances such that $(T,\{e\})$ is a homomorphism duality. 
Moreover, by~\cite{AlexeCKT2011}, there is such a set $T$ consisting of c-acyclic
instances. Then, for
every $t\in T$, we have that $t\nrightarrow e'$ since otherwise
$t\rightarrow e$, which is impossible as $(T,\{e\})$ is a homomorphism duality. Again, using the fact that $(T,\{e\})$ is a homomorphism duality it follows that $e'\rightarrow e$.

For the lower bound, we use an argument that was also used in \cite{Larose2007:characterization} to show that FO definability of a CSP is NP-hard:
we reduce from 3-SAT. 
Fix a schema consisting of a single binary relation $R$.
Let $F=\{P_{n+1}\}$ where $P_{n+1}$ is the path of length $n+1$,
and let $D=\{T_n\}$ where $T_{n}$ is the transitive tournament (i.e.,
total linear order) of length $n$.
It is well known that $(F,D)$ is a homomorphism duality (cf.~Example~\ref{ex:gallai}).
Now, consider any 3-SAT input
\[ \phi = \bigwedge_{i=1\ldots n} \bigvee_{j=1,2,3} L_{ij} \]
Let $H$ be the instance with domain $\{1, \ldots, n\}\times \{1,2,3\}$,
with an atom $R(\langle i,j\rangle, \langle i',j'\rangle)$ whenever
$i<j$ and $L_{i'j'}$ is not the negation of $L_{ij}$. 
We claim that the following are equivalent:
\begin{enumerate}
    \item $\phi$ is satisfiable
    \item $H$ is homomorphically equivalent to $T_n$
    \item $(F,\{H\})$ is a homomorphism duality
\end{enumerate}
The equivalence of 2 and 3 is obvious. Therefore, it suffices
only to show that 1 and 2 are equivalent.
By construction, $H\to T_n$.
From 1 to 2,  a 
homomorphism from $T_n$ to $H$ may be constructed
out of a satisfying assignment by mapping the $i$-th element of $T_n$
to any true literal from the $i$-th clause of $\phi$. Conversely,
any homomorphism from $T_n$ to $H$ clearly induces a satisfying
truth assignment for $\phi$.
\end{proof}


\begin{restatable}{theorem}{thmucqcomplexityresultsb}
\label{thm:ucq-complexity-results-b}
The following problems are 
computationally equivalent  to \textsc{HomDual} (via polynomial conjunctive reductions):
\begin{enumerate}
    \item The existence  problem for unique fitting UCQs,
    \item The verification  problem for unique fitting UCQs, 
    \item The verification problem for most-general fitting UCQs.
\end{enumerate}
\end{restatable}

\begin{proof}
Recall that a conjunctive reduction takes an instance of the first
problem and produces one or more instances of the second problem,
such that the input instance is a \emph{Yes} instance for the first problem if and only if
each output instance is a \emph{Yes} instance of the second problem.
It follows immediately from Proposition~\ref{prop:ucq-unique} and  Proposition~\ref{prop:ucq-most-general}, together with the NP-hardness
of \textsc{HomDual} that these problems polynomially conjunctively reduce to \textsc{HomDual}. The converse
direction is immediate as well (where, for the verification problems, it suffices to choose $q=\bigcup_{e\in E^+}q_{e}$).
\end{proof}

In summary, we see that 
the complexity of the various 
extremal fitting tasks
tends to be lower for UCQs than for CQs, and that \emph{most-general fitting UCQs} admit a simple characterization in terms of homomorphisms dualities. At the same time, Proposition~\ref{prop:ucq-most-specific} shows that 
\emph{most-specific fitting UCQs} are not very interesting as they essentially boil down to
disjunctions of complete descriptions of the given positive examples.

\section{The Case of Tree CQs}
\label{sec:tree-cq}

We now consider the fitting problem for CQs that take the form of a tree. Formally, a
unary CQ $q(x)$ is a
\emph{tree CQ} if its incidence graph is acyclic and connected. 
This form of acyclicity is also known as Berg\'e-acyclicity. 
Without further notice, we also restrict our attention to schemas that consist only of
unary and binary relations, which we refer to as \emph{binary}
schemas. Note that binary schemas are at the core of prominent web data
formalisms such as RDF and OWL. 
Apart from being 
natural per se, tree CQs hold significance as they correspond to
concept expressions in the description logic $\mathcal{ELI}$, see e.g.~\cite{DBLP:conf/lpar/KrisnadhiL07}. An example for a tree
CQ is $q(x)  \colondash R(x,y) \wedge S(x,z) \wedge A(z)$
and a non-example is  $q(x)  \colondash R(x,y) \wedge S(x,y)$.

Tree CQs behave differently from
unrestricted CQs in several ways which enables different techniques than those
used in Section~\ref{sec:cq}, often leading to lower computational complexity.
For instance, it is well-known that given an instance
$I$, an $a \in \text{adom}(I)$ and a tree CQ $q$, it
can be decided in PTime  whether $a \in q(I)$, details are given below. This results in lower complexity of fitting verification.
An important observation is that tree CQs are by definition c-acyclic, which means that they always have a frontier (c.f.\ Theorem~\ref{thm:cacyclic-frontier}). This affects the
characterization of weakly most-general fitting tree CQs and  enables algorithms based on tree automata.
Another difference that is worth pointing out is
that  greatest lower bounds in the homomorphism order of tree CQs do
not always exist. Note that existence does no longer follow from  Proposition~\ref{prop:cq-products} because the direct product $q_1 \times q_2$ is not guaranteed to be
equivalent to a tree CQ. In fact, as discussed below we will
be interested in the simulation lattice rather than
in the homomorphism lattice, but there greatest lower bounds are not guaranteed to exist either. 
%
This also
means that, in contrast to the unrestricted case, most-specific fitting tree CQs are not guaranteed to exist.

We next introduce the notions of
simulation and unraveling, which are closely linked to tree CQs. We also introduce two-way alternating parity automata which will be an important tool throughout this section.

\subsection*{Simulations}

While homomorphisms are closely linked to CQs, they are too strong a 
notion for tree CQs. To see this recall that, by Theorem~\ref{thm:icdt2015-b}, a collection of labeled examples $E=(E^+,E^-)$  has a fitting CQ if and only if the canonical CQ of the direct product $\Pi_{e\in E^+}(e)$ is well-defined and does not admit a homomorphism to any of the negative examples. This is no longer true for tree CQs.
\begin{example}
\label{ex:treeCQfirst}
Consider a schema that consists of a single binary relation $R$.
  Let $E^+$ consist of the example $(I,a)$ with $I=\{ R(a,a) \}$ and let $E^-$ consist of the example $(J,a)$ with $J=\{R(a,b),R(b,a)\}$. Then the canonical CQ of $(I,a)$ does not admit a homomorphism to $(J,a)$, yet (it follows from the characterizations below that) there is no tree CQ that fits $(E^+,E^-)$.
\end{example}

%
%


Given two instances $I,J$ over the same binary schema, a \emph{simulation of $I$
in} $J$ is a relation $S \subseteq \text{adom}(I) \times \text{adom}(J)$ that
satisfies the following properties:
\begin{enumerate}

    \item if $A(a) \in I$ and $(a,a') \in S$, then $A(a') \in J$;
    
    \item if $R(a,b) \in I$ and $(a,a') \in S$, then there is an
      $R(a',b') \in J$ with $(b,b') \in S$;
    
    \item if $R(a,b) \in I$ and $(b,b') \in S$, then there is an
      $R(a',b') \in J$ with $(a,a') \in S$.
    
\end{enumerate}
We write $(I,a) \preceq (J,b)$ if there exists a simulation $S$ of $I$ in $J$ with $(a,b) \in S$. 
Note that $(I, a) \to (J, b)$ implies $(I, a) \preceq (J, b)$, as every
homomorphism can be viewed as a simulation, but the reverse implication does not
hold in general.

Informally, simulations can be thought of as a relaxed version of homomorphisms that can map a single value
of the source instance to multiple values of the target instance instead of only a single one.
\begin{example}
  Consider again the pointed instances $(I,a)$ and $(J,b)$ from Example~\ref{ex:treeCQfirst}. While clearly $(I,a) \not\rightarrow (J,b)$, the simulation
  $\{(a,a),(a,b)\}$ witnesses $(I,a) \preceq (J,b)$.
\end{example}
A simulation
of a tree CQ $q(x)$ in an instance $I$ is a simulation of $I_q$ in
$I$, and we write $q \preceq (I,a)$ as shorthand for
$(I_q,x) \preceq (I,a)$, and likewise for $(I,a) \preceq
q$.
It is
well-known that if $I$ is a tree, then $(I,a) \preceq (J,b)$ if and only if 
$(I, a) \to (J, b)$. We thus have the following.
\begin{lemma}
\label{lem:simbasic}
  For all instances $I$, $a \in \text{adom}(I)$, and tree CQs $q(x)$, 
  $I \models q(a)$ if and only if $q \preceq (I,a)$. Moreover,
  for all tree CQs $q_1(x_1)$ and $q_2(x_2)$, $q_1 \subseteq q_2$
  if and only if $(I_{q_2},x_2) \preceq (I_{q_1},x_1)$.
\end{lemma}

Like homomorphisms, simulations induce a pre-order on data examples which
shares many of the properties of the homomorphism pre-order. Greatest lower bounds and least upper bounds are
obtained exactly as for the homomorphism pre-order by product and disjoint union, respectively.
Since we will only make use of the former, we shall not formally state the latter.
%
%
\begin{proposition}\label{prop:instance-products-sim}
For all data examples $e_1$ and $e_2$, if
$e_1\times e_2$ is a  data example, then it is a greatest lower bound for $e_1$ and $e_2$ in the simulation pre-order, in the following sense:
\begin{enumerate}
    \item $e_1\times e_2 \preceq e_1$,
    \item $e_1\times e_2\preceq e_2$, and
    \item  for all data examples $e'$, if $e'\preceq e_1$ and $e'\preceq e_2$, then $e'\preceq e_1\times e_2$.
\end{enumerate}
\end{proposition}
An important difference to the case of unrestricted CQs is that, there, CQs and examples are in a sense interchangeable since every CQ gives rise to a corresponding canonical example and every example gives rise to a corresponding canonical CQ. For tree CQs, the latter is no longer the case because we do not require examples to be trees.
One consequence is that the simulation pre-order on 
data examples is no longer (up to reversal) isomorphic to the simulation pre-order on tree CQs.
Since CQs are trees while the
instances in data examples are not, however, we are in a sense not interested in either of these pre-orders, but rather in a careful mix of them.
In particular, we shall use carefully
adapted
definitions of frontiers and dualities. Regarding frontiers,
we will consider exclusively tree CQs. As a consequence, a frontier of a tree CQ as defined in Section~\ref{sect:basicdatabasetheory} w.r.t.\ the set of all CQs is also a frontier in the more carefully defined sense, but not necessarily vice versa. Regarding dualities, we shall 
consider \emph{simulation} dualities in place of homomorphism dualities. These are
defined on data examples, thus not on trees. We defer the precise definitions to the subsequent sections where these notions are actually applied.

%

\subsection*{Unraveling}

Unraveling is an operation that is closely linked to 
simulations and transforms an instance into a tree instance. The resulting instance may be infinite and in fact
we shall frequently consider infinite instances in the
context of unraveling. All relevant notions from the
preliminaries such as homomorphisms and disjoint
unions naturally apply to infinite instances as well.

Let $I$ be an instance over a binary schema. A \emph{role}
is a binary relation symbol $R$ or its \emph{converse}~$R^-$.
For an instance~$I$, we may write $R^-(a,b) \in I$ to mean $R(b,a) \in I$.
A \emph{path} in $I$ is a sequence
$p=a_1R_1a_2R_2\cdots R_{k-1}a_k$, $k \geq 1$, where $a_1,\dots,a_k \in \text{adom}(I)$
and $R_1,\dots,R_{k-1}$ are roles
such that $R_i(a_i,a_{i+1}) \in I$ for $1 \leq i< k$.
We say that $p$ is of \emph{length} $k$, \emph{starts} at $a_1$ and \emph{ends}
at $a_k$.
The \emph{unraveling of $I$ at} $a \in \text{adom}(I)$
is the instance $U$ with active domain $\text{adom}(U)$ that
consists of all paths 
starting at $a$. It contains the fact
\begin{itemize}

\item $A(p)$ for every path $p \in \text{adom}(U)$ that ends with some
  $b \in \text{adom}(I)$ such that $A(b) \in I$, and

\item $R(p,pRb)$ for every path $pRb \in \text{adom}(U)$.

\end{itemize}
For all $m \geq 1$, the \emph{$m$-unraveling of $I$ at} $a$
is the (finite) restriction of the unraveling of $I$ to all paths of length at most $m$.

For all instances $I$ and $a \in \text{adom}(I)$, the unraveling of $I$ at $a$
is a (potentially infinite) tree, and for every $m$, the $m$-unraveling of $I$
at $a$ is a finite tree. This means that for every $m$, the canonical CQ of the
$m$-unraveling of $I$ at $a$ is a tree CQ, which will be important later.
Unraveling turns out to be a central operation in characterizing fitting tree
CQs. One of its main properties is the following.

\begin{restatable}{lemma}{lemunravbasics}
\label{lem:unravbasics}
  Let $(I,a)$ and $(J,b)$ be pointed instances and $U$ the unraveling of $I$ at $a$. Then 
  \begin{enumerate}
   \item $(I,a) \preceq (J,b)$ if and only if $(U,a) \preceq (J,b)$. Moreover, if $I$ and $J$ are finite, then this is the case if and only if $(U_m,a)
   \preceq (J,b)$ for all $m$-unravelings $U_m$ of $I$ at $a$.
    \item $(J,b) \preceq (I,b)$ if and only if $(J,b) \preceq (U,a)$, provided that $J$ is a tree.
  \end{enumerate}
\end{restatable}
\begin{proof}
We start with Item~(1).
For the `if' direction, assume that $(U,a) \preceq (J,b)$ is witnessed
by simulation $S$. Then $S'=\{(a',b') \mid (p,b') \in S \text{ and $p$ ends in } a' \}$ is a simulation that witnesses $(I,a) \preceq (J,b)$.
For the `only if' direction, assume that $(I,a) \preceq (J,b)$ is witnessed
by simulation $S$. Then $S'=\{(p,b') \mid (a',b') \in S \text{ and $p$ ends in } a' \}$ is a simulation that witnesses $(U,a) \preceq (J,b)$.

For the `moreover' part, the `only if' direction follows directly from the fact that each $U_m$ is a restriction of $U$.
For the `if' direction,
assume that $(U_m,a) \preceq (J,b)$ for all $m$-unravelings $U_m$ of $I$ at $a$.
Then also $(U_{m}, a) \to (J, b)$ for all $m \geq 1$ via homomorphisms $h_{1},h_{2},\dots$ with $h_m(a)=b$. 
To show that there is a simulation of the (infinite) unraveling $U$ in $J$, it
suffices to manipulate this sequence so that for all $i, j \geq 1$ and $\widehat a \in \text{adom}(U)$
\begin{equation}\label{eqn:hiahja}
    h_i(\widehat a)=h_j(\widehat a) \text{ whenever $h_i(\widehat a),h_j(\widehat a)$ are both defined} \tag{$*$}
\end{equation}
 as then $\bigcup_{i \geq 1} h_i$ is a homomorphism from $U$ to $J$ that maps
 $a$ to $b$. Then the first part of the lemma yields $(I,a) \preceq (J,b)$, as
 required.

To achieve (\ref{eqn:hiahja}), we start with
$h_{1}$ and observe that since $U_1$ and $J$ are finite, there are
only finitely many homomorphisms $h$ from $U_1$ to $J$. Some such
homomorphism must occur infinitely often in the restrictions of
$h_1,h_{2},\dots$ to $\text{adom}(U_1)$ and thus we find a subsequence
$h'_1,h'_{2},\dots$ of $h_1,h_{2},\dots$ such that the restriction of each $h'_i$ to $\text{adom}(U_1)$ is identical. We may assume w.l.o.g.\ that each $h'_i$ is a homomorphism from $U_i$ to $J$
and can thus replace $h_1,h_2,\dots$ with $h'_1,h'_2,\dots$.
We proceed in the same way for the restrictions of the sequence $h_{2},h_{3},\dots$ to $\text{adom}(U_{2})$, 
then for the restrictions of the sequence $h_{3},h_{4},\dots$ to $\text{adom}(U_{3})$, and so on. In the limit, we obtain a sequence
that satisfies~(\ref{eqn:hiahja}).

\smallskip
For Item~(2), it suffices to observe that any simulation $S$ which witnesses $(J,b) \preceq (I,b)$ gives rise to a simulation $S'=\{ (c,p) \mid (c,d) \in S \text{ and $p$ is a path in $I$ that starts at $a$ and ends at $d$} \}$
that witnesses
 $(J,b) \preceq (U,a)$. Conversely,  any simulation $S$ which witnesses $(J,b) \preceq (U,a)$ gives rise to a simulation $S'=\{ (c,d) \mid (c,p) \in S \text{ and $p$ is a path in $I$ that starts at $a$ and ends at $d$} \}$
that witnesses  $(J,b) \preceq (I,b)$.  
\end{proof}

\subsection*{Two-Way Alternating Tree Automata}

In the proof of Theorem~\ref{thm:wmg-existence}, we used NTAs to show an upper
bound for the existence and construction problems for weakly most-general
fitting CQs. For tree CQs, we will also use automata based techniques, but instead of NTAs whenever possible we prefer to 
use two-way alternating tree automata (TWAPAs)
because they often result in simpler constructions.
We remark that it does not seem possible to use
TWAPAs in the proof of Theorem~\ref{thm:wmg-existence} because we would have to construct TWAPAs with only polynomially many states and it is not clear how to achieve that.

For any set $X$, let $\mathcal{B}^+(X)$ denote the set of all positive Boolean
formulas over $X$, i.e., formulas built using conjunction and disjunction over
the elements of $X$ used as propositional variables, and where the special
formulas $\mn{true}$ and $\mn{false}$ are admitted as well. 

An \emph{infinite path} $P$ in a $\Sigma$-labeled $d$-ary tree $T$ is a
prefix-closed set $P \subseteq T$ such that for every $i \geq 0$, there is a
unique $x \in P$ with $|x|=i$, where $|x|$ denotes the length of word $x$.

\begin{definition}[Two-way alternating parity Automaton]
  A $d$-ary \emph{two-way alternating parity automaton 
    (TWAPA)} is a tuple
  $\mathfrak{A}=(S,\Sigma,\delta,s_0,c)$ where 
  \begin{itemize}
    \item $S$ is a finite set of \emph{states},
    \item $\Sigma$ is a finite alphabet,
    \item $\delta\colon S \times
  \Sigma \rightarrow \mathcal{B}^+(\mn{tran}(\mathfrak{A}))$ is the \emph{transition
    function} with $\mn{tran}(\mathfrak{A}) = \{ \langle i\rangle s, \ [i] s
  \mid -1 \leq
  i \leq d \text{ and } s \in S \}$ the set of
  \emph{transitions} of $\mathfrak{A}$,
  \item $s_0 \in S$ is the \emph{initial state},
  \item and $c \colon S \rightarrow \mathbb{N}$ is the \emph{parity condition} that 
  assigns to each state a \emph{priority}.
  \end{itemize}  
\end{definition}
Intuitively, a transition $\langle i \rangle s$ with $i>0$ means that
a copy of the automaton in state $s$ is sent to the $i$-th successor
of the current node, which is then required to exist. Similarly,
$\langle 0 \rangle s$ means that the automaton stays at the current
node and switches to state $s$, and $\langle -1 \rangle s$ indicates
moving to the predecessor of the current node, which is then required
to exist. Transitions $[i] s$ mean that a copy of the automaton in
state $s$ is sent to the relevant successor/predecessor if it exists,
which is then not required. %
\begin{definition}[Run, Acceptance]
  Let $\mathfrak{A} = (S,\Sigma,\delta,s_0,c)$ be a $d$-ary TWAPA.
  A \emph{run} of $\mathfrak{A}$ on a finite $\Sigma$-labeled $d$-ary tree
  $(T,Lab)$ is a $T \times S$-labeled $d$-ary tree $(T_r,r)$
  such that the following conditions are satisfied:
  \begin{enumerate}

  \item $r(\varepsilon) = ( \varepsilon, s_0)$;

  \item if $y \in T_r$, $r(y)=(x,s)$, and $\delta(s,Lab(x))=\varphi$, then
    there is a set $S \subseteq \mn{tran}(\mathfrak{A})$ 
    that (viewed as a propositional valuation over
    the set of symbols $\mn{tran}(\mathfrak{A})$) satisfies $\varphi$ as
    well as the following conditions:
    \begin{enumerate}

    \item if $\langle i \rangle s' \in S$, then $x \cdot i$ is defined and 
      there is a node $y \cdot j \in T_r$ such that $r(y \cdot j)=(x 
      \cdot i,s')$;

    \item if $[i]s' \in S$ and $x \cdot i$ is defined and a node in $T$, then
      there is a
      node $y \cdot j \in T_r$ such that $r(y \cdot j)=(x \cdot
      i,s')$.

    \end{enumerate}

  \end{enumerate}
  We say that $(T_r,r)$ is \emph{accepting} if on all infinite paths
  $\varepsilon = y_1 y_2 \cdots$ in $T_r$, the maximum priority that
  appears infinitely often is even.  A finite $\Sigma$-labeled $d$-ary
  tree
  $(T,Lab)$ is \emph{accepted} by $\mathfrak{A}$ if there is an accepting run of
  $\mathfrak{A}$ on $(T,Lab)$. We use $L(\mathfrak{A})$ to denote the set of all finite
  $\Sigma$-labeled $d$-ary trees accepted by $\mathfrak{A}$.
\end{definition}
We remark that most of the automata we are going to construct use the acceptance
condition in a trivial way, that is, every state has priority 0 and thus every
run is accepting. Note that this does \emph{not} imply that every input tree is
accepted because the TWAPA encounter transitions to \mn{false}.
Also, when we apply complementation to a TWAPA with a trivial acceptance condition, the acceptance
condition of the resulting TWAPA is no longer trivial.
%
%
The following properties of TWAPAs are
well-known, see for instance~\cite{DBLP:journals/tcs/MullerS87,DBLP:conf/icalp/Vardi98}.
\begin{theorem}[\cite{DBLP:journals/tcs/MullerS87,DBLP:conf/icalp/Vardi98}]\label{thm:TWAPAstuff}
  \mbox{}
  \begin{enumerate}

  \item Given a TWAPA $\mathfrak{A} = (S,\Sigma,\delta,s_0,c)$,
    it can be decided in time single exponential in $|S|$ and
    the maximum priority used by $c$, and 
    polynomial in $|\mathfrak{A}|$, whether $L(\mathfrak{A})$ is
    empty.

  \item Given a TWAPA $\mathfrak{A} = (S,\Sigma,\delta,s_0,c)$, one
    can compute in polynomial time a TWAPA
    $\mathfrak{A} = (S,\Sigma,\delta',s_0,c')$ such that
    $L(\mathfrak{A}')=\overline{L(\mathfrak{A})}$.

  \item Given TWAPAs
    $\mathfrak{A_i} = (S_i,\Sigma,\delta_i,s_{0,i},c_i)$,
    $i \in \{1,2\}$, one can compute in polynomial time a TWAPA
    $\mathfrak{A}=(S_1 \uplus S_2 \uplus \{ s_0 \},\Sigma,
    \Delta,s_0,c)$ with $L(\mathfrak{A})=L(\mathfrak{A}_1) \cap
    L(\mathfrak{A}_2)$.

  \item Given a TWAPA $\mathfrak{A} = (S,\Sigma,\delta,s_0,c)$, one
    can compute in single exponential time an NTA
    $\mathfrak{A'}=(S',\Sigma,\Delta,F)$ with
    $L(\mathfrak{A})=L(\mathfrak{A}')$.
    
  \item Given a $d$-ary TWAPA $\mathfrak{A} = (S,\Sigma,\delta,s_0,c)$ with \mbox{$L(\mathfrak{A}) \neq \emptyset$}, one can
    compute in single exponential time a succinct representation (in
    the form of a directed acyclic graph) of a tree with a minimal
    number of nodes accepted by $\mathfrak{A}$. The number of nodes in
    that tree is at most $d^{2^{p(|S|)}}$, $p$ a polynomial.

  \item Given a $d$-ary TWAPA $\mathfrak{A} = (S,\Sigma,\delta,s_0,c)$, it can be decided in time single exponential in
    $|S|$ and the maximum priority used by $c$, and polynomial in
    $|\mathfrak{A}|$, whether $L(\mathfrak{A})$ is infinite.
    Moreover, if $L(\mathfrak{A})$ is finite, then the size of every
    tree in it is at most $d^{2^{p(|S|)}}$, $p$ a polynomial.

  \end{enumerate}
\end{theorem}
Note that Point~(5) follows from Point~(4) together with Point~2 of
Theorem~\ref{thm:nta-basic-facts}. To obtain the decision procedure
promised by Point~(6), one would first convert the TWAPA
$\mathfrak{A}$ to an equivalent NTA $\mathfrak{A}'$ via Point~(4). Let
the state set of $\mathfrak{A}'$ be $Q$ and let $m$ be the cardinality
of $Q$. We then use the fact that $L(\mathfrak{A}')$ is infinite if and only if
$L(\mathfrak{A}')$ contains a tree whose depth exceeds the number $m$
of states of~$\mathfrak{A}'$. More concretely, we convert
$\mathfrak{A}'$ into an NTA $\mathfrak{A}''$ that accepts exactly the
trees in $L(\mathfrak{A}')$ that are of depth exceeding $m$ by using
as the new states $\{0,\dots,m\} \times Q$ and implementing a counter
on the first component of the states and simulation $\mathfrak{A}'$ on
the second component. It remains to decide the non-emptiness of
$L(\mathfrak{A}'')$ via Point~(1) of
Theorem~\ref{thm:nta-basic-facts}.

\subsection{Arbitrary Fitting Tree CQs}
\label{sect:arbMSfittreeCQ}

As in the non-tree case, we begin our investigation by considering the
verification and existence problem for arbitrary fitting tree CQs.
By the characterization of query answers of tree CQs in
Lemma~\ref{lem:simbasic}, we can decide the verification problem for arbitrary
fitting tree CQs by checking the existence of simulations to positive and
negative examples.
Since the existence of simulations can be decided in
PTime~\cite{DBLP:conf/focs/HenzingerHK95}, we obtain the following.
\begin{theorem}
  \label{thm:arbfittreeverptime}
  The verification problem for fitting tree CQs is in PTime. 
\end{theorem}
The following was proved in~\cite{Funk2019:when} in the setting
of the description logic $\mathcal{ELI}$, see Theorem~12 of that paper.
\begin{theorem}[\cite{Funk2019:when}]
\label{thm:arbfit} 
The existence problem for fitting tree CQs is ExpTime-complete. The
lower bound already holds for a fixed schema.
\end{theorem}
For our results, it is convenient reprove the upper bound in Theorem~\ref{thm:arbfit} using an
approach based on TWAPAs.  We gain from this an upper bound on 
the size of fitting tree CQs and the size of their succinct
representations as a DAG.

\begin{restatable}{theorem}{thmsizetreeCQarbupper}
  \label{thm:sizetreeCQarbupper}
If any tree CQ fits a collection of labeled examples  
$E=(E^+,E^-)$, then we can produce a succinct representation as a DAG of a fitting  
tree CQ with a minimal number of variables in single exponential  
time and the size of such a tree CQ is at most double exponential.  
\end{restatable}
To prove Theorems~\ref{thm:arbfit} and~\ref{thm:sizetreeCQarbupper} we
show how to construct, given a collection $(E^+,E^-)$ of labeled
examples, a TWAPA $\mathfrak{A}$ that accepts exactly the tree CQs
that fit $(E^+,E^-)$.
The automaton $\mathfrak{A}$ has polynomially many
states and the upper bound in Theorem~\ref{thm:arbfit} can be obtained
by constructing $\mathfrak{A}$ and checking its non-emptiness.
Theorem~\ref{thm:sizetreeCQarbupper} then follows from
Theorem~\ref{thm:TWAPAstuff}. 
To achieve this, we need to represent tree CQs as
$\Sigma$-labeled $d$-ary trees for some suitable alphabet $\Sigma$ and $d > 0$.
The encoding is simpler than the one used in Section~\ref{sect:mostgenfitCQs},
as we only consider tree CQs rather than c-acyclic CQs. 

Assume that we are concerned with a collection of labeled examples over
some binary schema~$\mathcal{S}$. Then the symbols in $\Sigma$ are the
sets $\sigma$ that contain at most one $\mathcal{S}$-role (binary
relation symbol from $\mathcal{S}$ or converse thereof) and any number
of unary relation symbols from $\mathcal{S}$. A $\Sigma$-labeled $d$-ary tree
is \emph{tree-proper} if the symbol $\sigma$ that labels the root contains
no $\mathcal{S}$-role while any other label used contains exactly one
$\mathcal{S}$-role. Nodes in the tree correspond to variables in the
tree CQ and the $\mathcal{S}$-role in a node/variable label determines
how the predecessor links to the variable. If, for example, $(T,L)$ is
a $\Sigma$-labeled $d$-ary tree, $122 \in T$ with $R^- \in L(122)$, $x$ is
the variable that corresponds to node 122 in the tree CQ and $y$
its predecessor (corresponding to node 12), then the tree CQ contains
the atom $R(x,y)$. In this way, tree CQs of  degree at most $d$
correspond to tree-proper $\Sigma$-labeled $d$-ary trees in an obvious way, and vice versa. In the
following, we do often not explicitly distinguish between tree CQs
and their encoding as a $\Sigma$-labeled $d$-ary tree and say, for example,
that an automaton \emph{accepts} a tree CQ. Note that tree-properness
can always be ensured by intersecting with a straightforward two-state TWAPA
which ensures that the input tree is tree-proper.

Since TWAPAs run on trees of bounded degree, we also need to bound the degree of fitting
tree CQs. The following lemma shows that this can be done in many relevant
cases. We define extremal
fitting tree CQs in exact analogy with extremal fitting unrestricted CQs.
For instance, a tree CQ $q$ is a 
\emph{strongly most-general fitting} CQ for a collection of labeled examples $E$ if $q$ fits $E$ and 
for all tree CQs $q'$ that fit $E$, we have $q'\subseteq q$.

\begin{restatable}{lemma}{lemdegreebnd}
  \label{lem:degreebnd}
  Let $(E^+,E^-)$ be a collection of labeled examples.  If there is a
  tree CQ $q$ that is a fitting of $(E^+,E^-)$, then there is such a $q$
  of degree at most $||E^-||$. The same is true for
  weakly and strongly most-general fittings, %
  for unique fittings,
  and for all these kinds of
  fittings with a minimal number of variables. 
\end{restatable}

We show Lemma~\ref{lem:degreebnd} by observing that every fitting tree CQ can be turned into a fitting
tree CQ of degree at most $||E^-||$ by dropping atoms. Dropping atoms does not
affect fitting with regard to the examples in $E^+$, and it can be shown that
degree $||E^-||$ always suffices to fit the negative examples in $E^-$. A full
proof of this lemma is available in Appendix~\ref{app:tree-degree-bound}.
Now we are ready to show Theorem~\ref{thm:sizetreeCQarbupper}.

\begin{proof}[Proof of Theorem~\ref{thm:sizetreeCQarbupper}]
Let $(E^+,E^-)$ be a collection of labeled examples over
schema~$\mathcal{S}$. We construct a TWAPA $\mathfrak{A}$ that accepts
exactly the fittings for $(E^+,E^-)$ of degree at most $||E^-||$ and
whose number of states is only polynomial in $||E^+ \cup E^-||$. We
may then use an emptiness check on $\mathfrak{A}$ to reprove the
ExpTime upper bound from Theorem~\ref{thm:arbfit}. What is more
important, we might use Point~5 of Theorem~\ref{thm:TWAPAstuff} to
extract from $\mathfrak{A}$ in single exponential time the DAG
representation of a fitting for $(E^+,E^-)$ with a minimal number of
variables, and to show that the number of variables in that fitting is
at most $2^{2^{p(||E^+ \cup E^-||)}}$, $p$ a polynomial. By
Lemma~\ref{lem:degreebnd}, the number of variables in the resulting
fitting is minimal not only among the fittings of degree at most $||E^-||$, but
also among all fittings.

To obtain the desired TWAPA $\mathfrak{A}$, we construct a TWAPA
$\mathfrak{A}_e$ for each example $e \in E^+ \cup E^-$, ensuring that
it accepts exactly the tree CQs of degree at most $m$ that admit a
 homomorphism to $e$,\footnote{Since the canonical example $(I_q,x)$ of a tree CQ $q(x)$ is a tree, there is a homomorphism from $(I_q,x)$ to $e$ if and only if there is a simulation of $(I_q,x)$ in $e$; we find it more natural to work with homomorphisms here.} then complement the
TWAPA $\mathfrak{A}_e$ if $e \in E^-$, and finally take the
intersection of all obtained TWAPAs. Building the TWAPA
$\mathfrak{A}_e$ is very simple, we only give a sketch.  Let
$e=(I,c_0)$. The set of states $S$ of $\mathfrak{A}_e$ contains the
pairs $(c,R)$ such that $I$ contains a fact of the form $R(c',c)$, with $R$ a potentially inverse role, or
$c \in \text{adom}(I)$ and $R$ is the special symbol~`$-$'. The
initial state is $(c_0,{-})$. All that $\mathfrak{A}_e$ does is
repeatedly sending a copy of itself to every successor node in the
input tree, guessing a homomorphism target in $I$ for that node.
Since the connecting role is only made explicit at that successor, we
also guess that role and then verify that the guess was correct once
that we are at the successor. More precisely, for $(c,R) \in S$ and
$\sigma \in \Sigma$, we set
$$
\delta((c,R),\sigma) = \displaystyle\bigwedge_{1 \leq i \leq m} \ \bigvee_{R' \ \mathcal{S}\text{-role and }
  R'(c,d) \in I} [i] (d,R')
$$
if $R \in \sigma \cup \{ {-}
  \}$ and $A \in \sigma$ implies $A(c) \in I$,
  and otherwise put $\delta((c,R),\sigma) = \text{false}$. Every state
  gets
  priority~0.
\end{proof}

A lower bound that matches the upper bounds on query size given in
Theorem~\ref{thm:sizetreeCQarbupper}, both with succinct representation and without, is established in
Section~\ref{sect:treeCQlower}. Note that fitting tree CQs of minimal size can
thus be double exponential in size, which is exponentially larger than fitting unrestricted CQs (c.f.\ Section~\ref{sect:arbfittingcqs}). However, this has little
impact on the complexity of deciding fitting existence as one may resort to succinct
representation as a DAG, bringing down the size by one exponential.

\subsection{Most-Specific Fitting Tree CQs}
\label{sect:mostspectreecqs}

We study a strong and a weak version of most-specific fitting
tree CQs, defined in exact analogy with Definition~\ref{def:mostspec}.
For instance, a tree CQ $q$ is a \emph{strongly most-specific fitting} tree CQ for a collection of labeled examples $E$ if $q$ fits $E$ and
    for every tree CQ  $q'$ that fits $E$, we have $q\subseteq q'$.
Unlike in the non-tree case, the existence of (strongly or weakly) most-specific
fitting tree CQs does not coincide with the existence of arbitrary fitting tree
CQs.
\begin{example}
  Consider a schema that contains the single binary relation $R$, the instance $I = \{ R(a, a) \}$, the single positive example $(I, a)$, 
  and the empty set of negative examples. Then $q(x) \colondash R(x,y)$ is a fitting tree
  CQ and $p(x) \colondash R(x,x)$ is a strongly most-specific fitting CQ, but
  there is no strongly most-specific fitting tree CQ. In fact, any $m$-unraveling
  of $p(x)$ at $x$ is a fitting tree CQ and there is no (finite)
  fitting tree CQ that is more specific than all these.
\end{example}

Indeed, we observe the following counterpart of Proposition~\ref{prop:most-specific}.
\begin{restatable}{proposition}{propmostspecifictree}
  \label{prop:most-specific-tree}
For all tree CQs $q$ and collections of labeled examples $E=(E^+,E^-)$, the following are equivalent:
\begin{enumerate}
    \item $q$ is a strongly most-specific fitting for $E$,
    \item $q$ is a weakly most-specific fitting for $E$,
    \item $q$ fits $E$ and $\Pi_{e\in E^+}(e) \preceq q$.
\end{enumerate}
\end{restatable}

Note that in contrast to the characterization for CQs in Proposition~\ref{prop:most-specific}, Point~3 does not
require $q$ to be equivalent to $\Pi_{e \in E^+}(e)$.
The equivalence between Points~1 and~3  has already been observed in \cite{Jung2020:least} in the absence of negative examples.

\begin{proof}
 $(1 \Rightarrow 2)$ is immediate from the definition.
For $(3 \Rightarrow 1)$, assume that $q$ fits~$E$ and $\Pi_{e\in E^+}(e) \preceq q$. Further, let $p$ be a tree CQ that fits~$E$.
We have to show that $q\subseteq p$. Since $p$ fits $E$, we
have $p \preceq (I,a)$ for all
$(I,a) \in E^+$ and thus also $p \preceq \Pi_{e\in E^+}(e)$.
From $\Pi_{e\in E^+}(e) \preceq q$, we obtain $p \preceq q$, thus
$q\subseteq p$ as required.

For $(2 \Rightarrow 3)$, assume that $q$ is a weakly most-specific
fitting for~$E$. Then $q \preceq e$ for all $e \in E^+$ and
thus also $q \preceq \Pi_{e\in E^+}(e)$. Let $m$ be the depth of
$q$. Then also $q \preceq (U_m, \bar e)$ where $U_m$ is the $m$-unraveling
of $\Pi_{e\in E^+}(e)$ at the tuple $\bar e$ that consists
of the distinguished elements of the examples in $E^+$.  Since $q$ is
a fitting for $E$, $q \not\preceq e$ for all $e \in E^-$ and
thus $q \preceq (U_m, \bar e)$ implies that $(U_m, \bar e) \not\preceq e$ for all
$e \in E^-$. Consequently, the canonical CQ of $(U_m, \bar e)$ is a fitting for $E$.  It easily
follows that, for all $m' \geq m$, the canonical CQ of the $m'$-unraveling $U_{m'}$ of
$\Pi_{e\in E^+}(e)$ at $\bar e$ is also a fitting for $E$.
Since $q$ is weakly most-specific and $q \preceq (U_{m'}, \bar e)$,
we must have $(U_{m'}, \bar a) \preceq q$ for all $m' \geq m$
(for, otherwise, the canonical CQ of $(U_{m'}, \bar e)$ would
a fitting tree CQ that is strictly more specific than $q$).
Then 
also $(U_{i}, \bar e) \preceq q$ for all $i \geq 1$. Thus
Lemma~\ref{lem:unravbasics} yields $\Pi_{e\in E^+}(e) \preceq q$, as
required.
\end{proof}
As in the non-tree case, we thus simply speak of \emph{most-specific} fitting
tree CQs, dropping the \emph{weak} and \emph{strong} qualifications. 

\medskip

In~\cite{Jung2020:least} it is shown that verification and existence
of a most-specific fitting tree CQ are in ExpTime and
PSpace-hard when there are only positive examples, but no negative
examples. Here, we consider the more general case with negative examples and
show that both problems are in ExpTime using the characterization in Proposition~\ref{prop:most-specific-tree}.
In fact, as we will see later in Section~\ref{sect:treeCQlower}, both are
ExpTime-complete. 
%
\begin{restatable}{theorem}{thmmostspecificexphardtrees}
\label{thm:mostspecificexphardtrees}
  Verification and existence of most-specific fitting tree CQs is in
  ExpTime.
\end{restatable}
\begin{proof}
  By Proposition~\ref{prop:most-specific-tree}, we may verify whether a tree CQ $q$
  is a most-specific fitting for some $E=(E^+,E^-)$ by checking whether $q$
  fits $E$ based on Theorem~\ref{thm:arbfittreeverptime}, then constructing
  $\Pi_{e\in E^+}(e)$ and deciding in PTime whether $\Pi_{e\in E^+}(e) \preceq
  q$. This gives the desired ExpTime upper bound.
  
  Now for existence. An ExpTime upper bound is proved in
  \cite{Jung2020:least} for the case where there are only positive
  examples, but no negative examples. We extend the same argument to negative
  examples. Given a collection of labeled examples $E=(E^+,E^-)$, we may first
  decide whether $E$ has a fitting tree CQ based on Theorem~\ref{thm:arbfit},
  answer `no' if this is not the case, and then use the algorithm from~\cite{Jung2020:least}
  to decide whether $(E^+,\emptyset)$ has a strongly
  most-specific fitting tree CQ and return the result. 
  
  We have to argue that this is correct. This is clearly the case if $E$ has no
  fitting tree CQ. Thus assume that there is such a CQ $q_0$. 
  First assume that the answer returned by the second check is `no'. Assume to
  the contrary of what we have to prove that $E$ has a  most-specific fitting
  tree CQ $q$. Then Point~3 of Proposition~\ref{prop:most-specific-tree} yields
  $\Pi_{e\in E^+}(e) \preceq q$. Since $q$ fits also $(E^+,\emptyset)$, again
  from Point~3 we obtain that $q$ is also a most-specific fitting for
  $(E^+,\emptyset)$, a contradiction.
  Now assume that the answer returned by the second check is `yes'. It suffices
  to show that any (strongly) most-specific fitting tree CQ $q^+$ for
  $(E^+,\emptyset)$ satisfies $q^+ \not\preceq e$ for all $e \in E^-$. But this
  follows from the existence of $q_0$  since we
  know that $q_0 \preceq q^+$ (since $q_0$ also fits $(E^+,\emptyset)$ and
  $q^+$ is strongly most-specific for $(E^+,\emptyset)$) and $q_0 \not\prec e$
  for all $e \in E^-$.
\end{proof}

Next, we consider the problem of constructing most-specific fitting tree CQs and establish upper bounds on their size.
We build on the ideas used in 
the decision procedure for the
existence of weakly most-specific fitting tree CQs with only positive
examples given in \cite{Jung2020:least}. The general idea is that we
characterize most-specific fitting tree CQs in terms of certain initial pieces of the unraveling of 
 $\prod_{e \in E^+}(e)$.  We next make this precise. 

Let $E=(E^+,E^-)$ be a collection of labeled examples, and let $U$ be the
unraveling of~$\Pi_{e \in E^+}(e)$ at the tuple $\bar e$ that consists of the
distinguished elements of the examples in $E^+$.
An \emph{initial piece} $U'$
of $U$ is a connected instance that is obtained as the restriction of
$U$ to some finite non-empty subset of $\text{adom}(U)$ that contains $\bar e$.
We say that $U'$ is \emph{complete} if for all paths
$pRa \in \text{adom}(U)$ with
$p \in \text{adom}(U')$ and $pRa \notin \text{adom}(U')$,
there is an $R(p,c) \in U'$ with $(U,pRa) \preceq (U,c)$. 
\begin{lemma}
  \label{lem:inipiececomplete}
  If $U'$ is a complete initial piece of $U$, then $(U, \bar e) \preceq (U', \bar e)$.
\end{lemma}
\begin{proof}
  Let $U'$ be a complete initial piece of $U$. It can be verified
  that $
  S= \{ (p,p') \mid p \in \text{adom}(U), p' \in \text{adom}(U') \text{ and }
  (U,p) \preceq (U,p') \}
  $
  is a simulation of $U$ in $U'$ with $(\bar e, \bar e) \in S$.
\end{proof}

In the following, slightly abusing notation we do not
distinguish between $U$ and its canonical (tree)
CQ with answer variable $\bar e$,
and likewise for initial pieces $U'$ of $U$. 
The following proposition links complete initial
pieces tightly to most-specific fittings. In particular, it implies
that if there is a most-specific fitting, then there is a complete
initial piece that is a most-specific fitting. 
\begin{proposition}
    \label{prop:most-specific-tree-initial-piece}
  Let $E=(E^+,E^-)$ be a collection of labeled examples. Then
  \begin{enumerate}
  \item any complete initial piece of the unraveling $U$ of
    $\Pi_{e \in E^+}(e)$ that fits $E$ is a most-specific fitting tree CQ for $E$
    and, conversely,

  \item any  most-specific fitting tree CQ for $E$ is  equivalent to some
  and every complete initial piece of~$U$.
    
  \end{enumerate}
\end{proposition}
\begin{proof}
  For Point~1, let $U'$ be a complete initial piece of $U$ that fits
  $E$. By Lemma~\ref{lem:inipiececomplete}, $(U, \bar e) \preceq (U', \bar e)$, and thus
  also $\Pi_{e \in E^+}(e) \preceq (U', \bar e)$. By
  Proposition~\ref{prop:most-specific-tree}, $U'$ is thus a most-specific
  fitting for~$E$.

  For Point~2, let $q$ be a most specific fitting tree CQ for $E$. Since $q$ fits the positive examples, we must have $q \preceq \Pi_{e \in E^+}(e)$. From Proposition~\ref{prop:most-specific-tree}, we additionally get $\Pi_{e \in E^+}(e) \preceq q$ and thus it follows
  from Lemma~\ref{lem:simbasic}
  that $q$ is
   equivalent to the canonical CQ of $\Pi_{e \in E^+}(e)$. Since $q$ is a tree CQ and by Lemma~\ref{lem:unravbasics}, this
  implies that $q$ is also equivalent to $U$. By
Lemma~\ref{lem:inipiececomplete}, $U$ is  equivalent to
  every complete initial piece of $U$.  It therefore remains to show
  that a complete initial piece of $U$ exists.

  We first observe that there is an $m \geq 1$ such that
  $(U, \bar e) \preceq (U_m, \bar e)$ with $U_m$ the $m$-unraveling of~$U$.
  In fact, equivalence of $U$ and $q$ implies that 
  there
  is a homomorphism $h_1$ from  $(U,\bar a)$ to $(I_q,x)$ and $h_2$ from $(I_q,x)$ to
  $(U,\bar a)$.  Since $q$ is finite, the composition $h=h_2 \circ h_1$ is
  a homomorphism from $(U,\bar e)$ into some~$(U_m,\bar e)$.
   %
   Let $U'$ be the initial piece of $U$ such that
  $\text{adom}(U')$ is the range of $h$. We argue that $U'$ is
  complete.  Take any path $pRa \in \text{adom}(U)$ with
  $p \in \text{adom}(U')$ and $pRa \notin \text{adom}(U')$.
  Then $h(pRa) \in \text{adom}(U')$ and $R(p,h(pRa)) \in U'$.
  Moreover, $h$ witnesses that $(U,pRa) \preceq (U',h(pRa))$, thus
  $(U,pRa) \preceq (U,h(pRa))$ as desired.
\end{proof}

Using the characterization in
Proposition~\ref{prop:most-specific-tree-initial-piece}, we can now show that
most-specific fitting CQs can be computed in exponential time, if they exist.

\begin{restatable}{theorem}{thmsizetreeCQmsupper}
  \label{thm:sizetreeCQmsupper}
  If a collection of labeled examples $E=(E^+,E^-)$ admits a most-specific
  fitting tree CQ, then one can construct a succinct representation as a DAG of such
  a tree CQ with a minimal number of variables in single exponential
  time and the size of the tree CQ in question is at most double exponential.
\end{restatable}
To prove Theorem~\ref{thm:sizetreeCQmsupper},
we start with establishing an upper
bound on the degree of most-specific fitting tree CQs. 
It is exponential rather than polynomial because
we derive it from (the unraveling of) $\Pi_{e \in E^+}(e)$. A proof is in Appendix~\ref{app:ms-fit-tree-cq}.
%
\begin{restatable}{lemma}{lemdegreebndtwo}
  \label{lem:degreebndtwo}
  Let $(E^+,E^-)$ be a collection of labeled examples.  If there is a
  tree CQ $q$ that is a most-specific fitting of $(E^+,E^-)$, then there
  is such a $q$ of degree at most $2^{||E^+||}$. The same is
  true with a minimal number of variables. Moreover, any most-specific
  fitting with a minimal number of variables must be isomorphic to a
  complete initial piece of the unraveling of $\Pi_{e \in E^+}(e)$.
\end{restatable}

We next show how to construct, given a collection
$(E^+,E^-)$ of labeled examples over schema $\mathcal{S}$, an NTA $\mathfrak{A}$ that
exactly the most-specific tree CQ fittings for $(E^+,E^-)$ which have degree at most
$m:=2^{||E^+||}$ and are isomorphic to a complete initial piece of the unraveling
$U$ of $\Pi_{e \in E^+}(e)$.   Note that the latter conditions are
without loss of generality due to
Proposition~\ref{prop:most-specific-tree-initial-piece} and
Lemma~\ref{lem:degreebndtwo}, no matter whether we want to decide the
existence of most-specific fittings or construct a most-specific
fitting with a minimal number of variables.

In contrast to the
proofs of Theorems~\ref{thm:arbfit} and~\ref{thm:sizetreeCQarbupper},
we use an NTA rather than a
TWAPA because for checking the input to be isomorphic to a
complete initial piece of the unraveling of $\Pi_{e \in E^+}(e)$,
there seems no way around having at least one state for each element
of $\Pi_{e \in E^+}(e)$. To obtain the desired results via TWAPAs,
however, we would need to have only polynomially many
states. 

%

We may use an emptiness check on $\mathfrak{A}$ to reprove the
ExpTime upper bound from Theorem~\ref{thm:mostspecificexphardtrees}.
Moreover, we might use Point~2 of Theorem~\ref{thm:nta-basic-facts} to
extract from $\mathfrak{A}$ in single exponential time the DAG
representation of a most-specific fitting for $(E^+,E^-)$ with a
minimal number of variables, and to show that the number of variables
in that fitting is at most $2^{2^{p(||E^+ \cup E^-||)}}$, $p$ a polynomial. 

The NTA $\mathfrak{A}$ is constructed as the intersection of two NTAs
$\mathfrak{A}_1$ and $\mathfrak{A}_2$, c.f.\ Point~4 of
Theorem~\ref{thm:nta-basic-facts}. The first NTA checks that the input
tree CQ is a fitting for $(E^+,E^-)$. It is obtained by converting the
TWAPA used in the proof of Theorem~\ref{thm:arbfit} into an NTA as per
Point~4 of Theorem~\ref{thm:TWAPAstuff}. The second NTA verifies that
the input tree CQ is a complete initial piece of the unraveling $U$ of
$\Pi_{e \in E^+}(e)$ and thus, by
Proposition~\ref{prop:most-specific-tree-initial-piece}, a most-specific
fitting.

We define $\mathfrak{A}_2=(Q,\Gamma,\Delta,F)$ where $\Gamma$ is the
alphabet for tree CQs over schema $\mathcal{S}$. The states in $Q$
take the form $(aR,b)$ with $a,b \in \text{adom}(\Pi_{e \in E^+}(e))$
and $R$ an $\mathcal{S}$-role.  As a special case, $aR$ can be `$-$'.
Informally, state $(aR,b)$ means that we are currently visiting a path
in $U$ that ends with $aRb$. In the transition relation, we verify
that all successors required for the initial piece to be complete are
present. It is convenient to view $\mathfrak{A}_2$ as a top-down
automaton and $F$ as a set of initial states. We set $F=\{(-,e_0)\}$
where $e_0 \in \text{adom}(\Pi_{e \in E^+}(e))$ is the root of the
unraveling $U$ of $\Pi_{e \in E^+}(e)$, that is, the tuple that
consists of all the selected points in the data examples in $E^+$.  We
then include in $\Delta$ all transitions
$$\langle q_1,\dots,q_m\rangle\transition{\sigma} (aR,b),$$
where each $q_i$ can also be $\bot$,
such that the following
conditions are satisfied:
\begin{enumerate}

\item $R \in \sigma$ (unless $aR={-}$);

\item $A \in \sigma$ if and only if $A(b) \in \Pi_{e \in E^+}(e)$, for all
  unary $A \in \mathcal{S}$;

\item if $q_i=(bS_i,c_i)$, then $S_i(b,c_i) \in \Pi_{e \in E^+}(e)$
  for $1 \leq i \leq m$;
  
\item if $b$ has an $S$-successor $c$ in $U$, then one of
  the following holds:
  \begin{itemize}

  \item there is an $i$ such that $q_i=(bS,c_i)$ and
        $(\Pi_{e \in E^+}(e),c) \preceq (\Pi_{e \in E^+}(e),c_i)$;

  \item $R=S^-$ and
    $(\Pi_{e \in E^+}(e),c) \preceq (\Pi_{e \in E^+}(e),a)$;

  \end{itemize}

\item all of $q_1,\dots,q_m$ that are not $\bot$ are pairwise
  distinct.
  
\end{enumerate}
It can be verified that the automaton recognizes precisely the
intended language. Note that to construct the NTA, we need to know
about simulations between values in $\Pi_{e \in E^+}(e)$.  We
determine these by first constructing $\Pi_{e \in E^+}(e)$ in single
exponential time and then computing in PTime the maximal simulation on
it.


\subsection{Weakly Most-General and Unique Fitting Tree CQs}

We define weakly and strongly most-general tree CQs exactly as in the case of unrestricted CQs, 
and likewise for bases of most-general fitting tree CQs and unique
fitting tree CQs, see Definitions~\ref{def:mostgenfitCQs}
and~\ref{def:basismostgenfitCQs}. As in the non-tree case, the notions of weakly
most-general and strongly most-general fitting tree CQ do not coincide.
\begin{example}
Consider the schema that consists of a binary relation $R$ and two unary relations $P$ and $Q$. 
  Let $E^+ = \{ (I,a) \}$ and $E^-=\{ (J_1,a),(J_2,a) \}$ with  
  $I=\{ P(a), R(a,b), Q(b) \}$,
  $J_1=\{ P(a), R(a,b) \}$,
  and $J_2=\{ R(a,b), R(c,b), R(c,d), Q(d) \}$. Then
  $q(x) \colondash R(x,y) \wedge Q(y)$
  is a weakly most-general fitting tree CQ, but there is no strongly most-general fitting tree CQ, and  not even a basis of most-general fitting tree CQs. 
  This is due to the fact
  that the following tree CQs fit,
  for all $i \geq 0$:
  $$
    q_i(x) \colondash P(x) \wedge R(x,y_0) \wedge Q(y_i) \wedge
    \bigwedge_{1 \leq j \leq i}
    R(z_i,y_{i-1}) \wedge R(z_i,y_i).
  $$
\end{example}
The following example illustrates
that the existence of weakly most-general tree CQs does not coincide
with the existence of weakly most-general CQs.
\begin{example} \label{ex:most-general-tree}
Consider the schema that consists of a binary relation $R$ and a unary relation $P$.
  Let $E^+=\emptyset, E^-=\{ (\{P(a)\}, a), (\{R(a,a)\}, a)\}$. Then there
  are no weakly most-general fitting tree CQs.  To see this, let
  $q(x)$ be a tree CQ that fits the examples.  Clearly, $q(x)$ must
  contain both an $R$-atom and a $P$-atom.  Take a  shortest path in the graph of $q$, from $x$ to some $y$ that satisfies
  $P$, and let $\pi$ be its label, that is,  a
  sequence of roles $R$ and~$R^-$.  If $\pi$ is empty, then $q$ must contain some atom
  $R(x,y)$ or $R(y,x)$. In the first case,
    the tree CQ
  $q'(x)\colondash R(x,y)\land R(y,z)\land P(z)$ is homomorphically strictly weaker
  than~$q$, but still fits $E$. In the
  second case, we may use 
  $q'(x) \colondash R(y,x)\land R(z,y)\land P(z)$
  instead.
  Finally if $\pi$ is empty, then we also find a tree CQ $q'$ that is homomorphically
  weaker than $q$, but fits $E$: the graph of
  $q'$ is a path starting at $x$ and ending
  at a variable $y$ with $P(y)$ an atom in $q'$, and the label of the path is $\pi$,
  followed by its reversal $\pi^-$, followed
  again by $\pi$. Thus $q$ is not a weakly
  most-general fitting tree CQ.

  However, weakly most-general fitting CQs exist that are not
  tree CQs. In fact, we obtain a complete basis of most-general
  fitting CQs by taking the CQs
  $q(x)\colondash \alpha(x) \land R(y,z) \land P(u)$ where
  $\alpha(x)$ is $P(x)$ or $R(x,v)$ or $R(v,x)$,
  serving purely to make the CQ safe.
\end{example}
Similarly to the non-tree case, we may characterize weakly most-general
fitting tree CQs using frontiers. As a weakly most-general fitting tree CQ only
needs to be most-general among all fitting tree CQs, it suffices to consider
frontiers w.r.t.\ the class of all tree CQs. 
For the reader's convenience, we spell
out the definition  in detail: a \emph{frontier} for a
tree CQ $q$ \emph{w.r.t.\ all tree CQs}  is a finite
set of tree CQs $\{q_1, \ldots, q_n\}$ such that 

\begin{enumerate}
    \item for all $i\leq n$, 
    $q \subseteq q_i$ and $q_i \not\subseteq q$.
    \item for all tree CQs $q'$, if $q \subseteq q'$ and $q' \not\subseteq q$, then $q_i\subseteq q'$ for some $i\leq n$.
\end{enumerate}
Based on this, we obtain the
following characterization. The
proof is straightforward and 
uses the same arguments as the proof of Proposition~\ref{prop:weakly-most-general}, details are omitted.
\begin{proposition} \label{prop:weakly-most-general-tree-cq}
The following are equivalent for all collections of labeled examples
$E=(E^+,E^-)$ and tree CQs $q$:
\begin{enumerate}
    \item $q$ is a weakly most-general fitting for $E$,
    \item $q$ fits $E$ and every element of the frontier for $q$
      w.r.t.\ all tree CQs is simulated in an example in~$E^-$.
\end{enumerate}
\end{proposition}

Proposition~\ref{prop:weakly-most-general-tree-cq} allows us to directly
determine the complexity of the verification problem for weakly most-general
fitting tree CQs.

\begin{theorem}
  \label{thm:veri-weak-most-general-tree}
Verification of weakly most-general fitting tree CQs is in PTime.
\end{theorem}

\begin{proof}
As every tree CQ is c-acyclic, by Theorem~\ref{thm:cacyclic-frontier} every tree
CQ has a frontier w.r.t.\ the class of all CQs (and thus also w.r.t.\ all tree
CQs) that can be computed in polynomial time. 
Based on Proposition~\ref{prop:weakly-most-general-tree-cq},
Theorem~\ref{thm:veri-weak-most-general-tree} is then easy to prove. Given a tree CQ
$q$ and a
collection of labeled examples $(E^+,E^-)$, we may first verify in
polynomial time that $q$ fits $E$. We then compute in polynomial time
a frontier $\mathcal{F}$ of $q$ w.r.t.\ tree CQs and then check
that none of the CQs in $\mathcal{F}$ simulates into a negative
example.  
\end{proof}




Proposition~\ref{prop:weakly-most-general-tree-cq} also serves as 
the basis for a decision procedure for the existence of weakly 
most-general fitting tree CQs. In principle, we could use the 
same approach as in the proof of Theorem~\ref{thm:wmg-existence}
and use NTAs that accept exactly the weakly most-general fitting tree CQs.
However, the NTA construction used in the proof of Theorem~\ref{thm:wmg-existence} is 
already complex and we would have to replace homomorphisms with
simulations, which results in additional technicalities.
This led us to working with a simpler frontier construction tailored towards
tree CQs, presented in \cite{tCD2022:conjunctive}, that `only' yields a frontier
w.r.t.\ tree CQs, and with TWAPAs. It runs in polynomial time and yields a frontier with polynomially many elements.
Using this approach, we obtain the same results regarding the size and
computation of weakly most-general fitting tree CQs as for arbitrary fitting tree CQs and most-specific fitting tree CQs in
Sections~\ref{sect:arbMSfittreeCQ} and~\ref{sect:mostspectreecqs}.
\begin{restatable}{theorem}{thmwmgexistencetree}
  \label{thm:wmg-existence-tree}
  Existence of weakly most-general fitting tree CQs is in ExpTime.
  Moreover, if a collection of labeled examples $E=(E^+,E^-)$ admits a weakly
  most-general tree CQ fitting, then we can construct a succinct
  representation as a DAG of such a fitting with a minimal number of variables
  in single exponential time and the size of such a tree CQ is at most
  double exponential.
\end{restatable}
We only give an overview of the proof,
details are in Appendix~\ref{app:wmg-tree-cq}.
For a collection of labeled examples $(E^+,E^-)$ over
binary schema~$\mathcal{S}$ and $m := ||E^-||$, we  construct a
TWAPA $\mathfrak{A}$ with polynomially many states that accepts
exactly the weakly most-general tree CQ fittings for $(E^+,E^-)$ which
have degree at most $m$.  The latter restriction is justified by
Lemma~\ref{lem:degreebnd}. To prove Theorem~\ref{thm:wmg-existence-tree}, it then
remains to invoke the results about TWAPA in Point~(1) and~(5) of
Theorem~\ref{thm:TWAPAstuff}.
The details of the construction of $\mathfrak{A}$ rely on 
 Proposition~\ref{prop:weakly-most-general-tree-cq}. Based on that proposition, we may
construct $\mathfrak{A}$ as the intersection of two TWAPAs
$\mathfrak{A}_1$ and $\mathfrak{A}_2$ where $\mathfrak{A}_1$ verifies
that the $q$ fits $(E^+,E^-)$ and $\mathfrak{A}_2$ verifies that every element
of the frontier $\mathcal{F}$ for $q$ w.r.t.\ tree CQs is simulated in an
example in $E^-$. For $\mathfrak{A}_1$, we can use the TWAPA from
the proof of Theorem~\ref{thm:sizetreeCQarbupper}.
The TWAPA $\Amf_2$ implements, in a certain sense, the mentioned frontier construction w.r.t.\ tree CQs from 
 \cite{tCD2022:conjunctive}. It is constructed so that, for every tree CQ $p$ in the frontier of the tree CQ $q$ represented by the input tree, an accepting run must contain a subtree that is isomorphic to $p$ and encodes, via the states, a simulation into a negative example.
 
\medskip
We now turn to
uniquely fitting tree CQs, first observing that, as in the non-tree case, a fitting
tree CQ is a unique fitting if and only if it is both a most-specific
and a weakly most-general fitting. This immediately gives an ExpTime
upper bound for verification, from the ExpTime upper bounds for
verifying most-specific and weakly most-general tree CQs in Theorem~\ref{thm:mostspecificexphardtrees}.
We obtain an 
ExpTime upper bound for the existence of uniquely fitting tree CQs by 
combining the automata constructions for those two cases.
\begin{restatable}{theorem}{thmuniqueexphardtrees}
\label{thm:uniqueexphardtrees}
  Verification and existence of unique fitting tree CQs is in ExpTime.
\end{restatable}

The upper bound for existence can be established as follows. In the
proof of Theorem~\ref{thm:sizetreeCQmsupper}, we have constructed,
given a collection of labeled examples $(E^+,E^-)$, an NTA
$\mathfrak{A}_1$ with single exponentially many states that accepts
exactly the fitting tree CQs for $(E^+,E^-)$ that are most-specific,
have degree at most $2^{||E^+||}$, and are isomorphic to a complete
initial piece of the unraveling of $\Pi_{e \in E^+}(e)$.  Recall that
the latter two conditions are w.l.o.g.\ in the sense that if a
most-specific fitting exists, then (i)~there exists one that satisfies
the conditions and even~(ii) the most-specific fitting with the
minimal number of variables satisfies the conditions. The
same is true for unique fittings because, if a unique fitting exists,
then the unique fittings and the most-specific fittings are
identical.  We can easily modify
$\mathfrak{A}_1$ so that it runs on trees of degree
$m := \max\{ 2^{||E^+||}, ||E^-||\}$.

Furthermore, in the proof of Theorem~\ref{thm:wmg-existence-tree}, we have constructed a TWAPA
$\mathfrak{A}_2$ with polynomially many states that accepts exactly
the fitting tree CQs for $(E^+,E^-)$ that are weakly most-general and
have degree at most $||E^-||$.   We can
easily modify $\Amf_2$ so that it runs on trees of degree $m$.
Moreover, we can convert it into an equivalent NTA with single
exponentially many states
and then intersect with $\Amf_1$ to obtain an NTA $\mathfrak{A}$ with
still single exponentially many states that accepts exactly the unique
fitting tree CQs for $(E^+,E^-)$ of degree at most $m$. It remains to
check emptiness of $\mathfrak{A}$ in polynomial time, which can be done by Theorem~\ref{thm:TWAPAstuff}.

\medskip

Regarding the computation and size of unique fitting tree CQs, we remark that Theorem~\ref{thm:sizetreeCQarbupper}, which provides an algorithm for computing arbitrary fitting tree CQs and establishes an upper bound on their size, trivially also applies
to unique fitting tree CQs. In particular, if a unique fitting tree CQ exists, then the
algorithm from the proof of Theorem~\ref{thm:sizetreeCQarbupper} clearly must 
compute it.

\subsection{Bases of Most-General Fitting Tree CQs}

We study bases of most-general fitting tree CQs.
As in the case of unrestricted CQs, this also (implicitly) yields results
for strongly most-general fitting tree CQs, which correspond to bases of most-general fitting tree CQs of size~1.
In  Proposition~\ref{prop:basis-vs-duality}, we have characterized bases of most-general
fitting CQs in terms of relativized homomorphism dualities. However,
Example~\ref{ex:most-general-tree} demonstrates that there are data examples for
which bases of most-general fitting CQs exist, but no weakly most-general
fitting tree CQs exist, and therefore also no bases of most-general fitting tree CQs.
Thus, the mentioned characterization does not apply to tree CQs.
We  instead give a characterization that is based on relativized \emph{simulation} dualities.

\begin{definition}[Relativized simulation dualities] 
A pair of finite sets of data examples $(F,D)$ forms a \emph{simulation duality}
if, for all data examples $e$, the following are equivalent:
\begin{enumerate}
    \item $e \preceq e'$ for some $e' \in D$,
    \item $e' \not\preceq e$ for all $e' \in F$.
\end{enumerate}
We say that $(F,D)$ forms a simulation duality
\emph{relative to a data example $p$} if 
the above conditions hold for all $e$ with $e\preceq p$.
\end{definition}

Using the notion of relativized simulation dualities, we obtain a characterization very close to Proposition~\ref{prop:basis-vs-duality}.

\begin{restatable}{proposition}{propbasisvsdualitytreecqs}
\label{prop:basis-vs-duality-tree-cqs}
For all collections of labeled examples $E=(E^+,E^-)$, the following are equivalent,
for $p=\Pi_{e\in E^+}(e)$:
\begin{enumerate}
    \item $\{q_1, \ldots, q_n\}$ is a basis of most-general fitting tree CQs for $E$,
    \item each $q_i$ fits $E$ and $(\{q_1, \ldots, q_n\}, E^-)$ is 
a simulation duality relative to $p$.
\end{enumerate}
\end{restatable}

\begin{proof}
(1 $\Rightarrow$ 2). Assume that
 $\{q_1, \ldots, q_n\}$ is a basis of most-general fitting tree CQs for $E$. Then 
 each $q_i$ fits $E$. Let $e$ be any data example such that
$e\preceq p$. We need to show that 
$e\preceq e'$ for some $e'\in E^-$
if and only if $q_i\not\preceq e$ for all $i\leq n$. 

``$\Rightarrow$''. Assume that $e\preceq e'$ for some $e'\in E^-$ and, to the contrary of what we have to show, $q_i\preceq e$ for some $i \leq n$. Then, by transitivity, $q_i\preceq e'$, 
and hence (since $q_i$ is a tree CQ), $q_i\to e'$, contradicting the fact that $q_i$ fits $E$.

``$\Leftarrow$''.
We prove the contraposition.
Assume that
$e\not\preceq e'$ for all $e'\in E^-$. 
Then by Lemma~\ref{lem:unravbasics},
there is some $m \geq 1$ such that $e^* \not\preceq e'$ for all $e'\in E^-$,
where $e^*$ is the $m$-unraveling $e^*$ of $e$ at the distinguished element of $e$. 
Thus the canonical CQ of $e^*$ fits $E^-$.
Recall that $e\preceq p$. This clearly
implies that the canonical CQ of   $e$ fits $E^+$ and by Lemma~\ref{lem:unravbasics} so
does the canonical CQ of $e^*$. So in 
summary the  canonical CQ of $e^*$ fits $E$.
Since $\{q_1, \ldots, q_n\}$ is a basis of most-general fitting tree CQs for $E$,
we obtain 
$q_i\preceq e^*$ for some $i\leq n$.
This clearly implies $q_i\preceq e$, as required. 

\smallskip

(2 $\Rightarrow$ 1). Assume that
each $q_i$ fits $E$ and $(\{q_1, \ldots, q_n\}, E^-)$ is 
a simulation duality relative to $p$.
Take any tree CQ
 $q'$  that fits $E$. We have to show
 that $q_i \subseteq q'$ for some $i\leq n$.
Since  $q'$   fits $E$, we have $q'\to p$ and hence $q' \preceq p$. Furthermore, 
for all $e\in E^-$, $q'\nrightarrow e$ and hence $q'\not\preceq e$ since $q'$ is a tree CQ. Since $(\{q_1, \ldots, q_n\}, E^-)$ is 
a simulation duality relative to $p$,
this implies $q' \preceq q_i$ for some $i\leq n$ which yields $q_i \subseteq q'$, as required.
\end{proof}
Based on Proposition~\ref{prop:basis-vs-duality-tree-cqs}, we establish an ExpTime upper bound for the verification problem 
for bases of most-general fitting tree CQs.

\begin{restatable}{theorem}{thmtreecqbasisverification}
  \label{thm:tree-cq-basis-verification}
The verification problem for bases of most-general fitting tree CQs is
in ExpTime.
\end{restatable}
 
\begin{proof}
  Let $E=(E^+,E^-)$ and $\{q_1, \ldots, q_n\}$ be
  given.  We first verify in PTime that each $q_i$ fits $E$.

By Theorem~\ref{thm:cacyclic-duality}, for each $q_i$ we can compute in single exponential
time a set of data examples~$D_{q_{i}}$ such that
$(\{e_{q_i}\},D_{q_i})$ is a homomorphism
duality.  Let $D = \{ e_1\times \cdots\times e_n\mid e_i\in
D_{q_i} \text{ for } 1 \leq i \leq n\}$. From the properties of direct products in
Proposition~\ref{prop:instance-products}, it follows that
$(\{e_{q_1}, \ldots, e_{q_n}\},D)$ is also a homomorphism duality. It is not difficult to
verify that $(\{e_{q_1}, \ldots, e_{q_n}\},D)$
is then also a simulation duality. 
In fact, this follows from the observation 
that the instances 
in $e_{q_1}, \ldots, e_{q_n}$ are trees
and the facts that (i)~every homomorphism is a simulation and (ii)~$e \preceq e'$ implies $e \rightarrow e'$ if the instance in $e$ is a tree.


We claim that the following are equivalent:
\begin{enumerate}
    \item $\{q_1, \ldots, q_n\}$ is a basis of most-general fitting
      tree CQs for $E$,
    \item $(\{e_{q_1}, \ldots, e_{q_n}\}, E^-)$ is a simulation duality relative to $p$,
    where $p=\Pi_{e\in E^+}(e)$,
    \item For each $e\in D$, there is $e'\in E^-$ such that $e\times p \preceq e'$.
\end{enumerate}
The equivalence of 1 and 2 is given by Proposition~\ref{prop:basis-vs-duality-tree-cqs}.

(2 $\Rightarrow$ 3). Let $e\in D$. Since  $(\{e_{q_1}, \ldots, e_{q_n}\},D)$ is a simulation duality and $e\in D$, we have $e_{q_i}\not\preceq e$ for all $i\leq n$. Hence, by Proposition~\ref{prop:instance-products-sim}, also $e_{q_i}\not\preceq e\times p$. Therefore, since $e\times p\preceq p$, we have that $e\times p\preceq e'$ for some $e'\in E^-$.

(3 $\Rightarrow$ 2). Let $e$ be any data example such that
$e\preceq p$.  If some $e_{q_i}\preceq e$, then, since $q_i$ fits $E$,
we know that $e\not\preceq e'$ for all $e'\in E^-$. If, on the other
hand, $e_{q_i} \not\prec e$ for all $e_{q_i}$, then $e\preceq e'$ for
some $e'\in D$. Hence, since $e\preceq p$, by
Proposition~\ref{prop:instance-products-sim},
we have that
$e\preceq e'\times p$, and therefore $e\preceq e''$ for some
$e''\in E^-$.

\smallskip
This concludes the proof since (3) can straightforwardly be tested in ExpTime.
\end{proof}

To approach the existence problem of bases of most-general fitting tree CQs, we characterize
the existence of relativized simulation dualities as follows.
Let $D$ be a finite collection of data examples. A tree CQ $q$ is a
\emph{critical tree obstruction} for $D$ if $q \not\preceq e$ for
all $e \in D$ and every tree CQ $q'$ that can be obtained from $q$ by
removing subtrees satisfies $q' \preceq e$ for some $e \in D$.
\begin{restatable}{proposition}{propcharsimdual}\label{prop:charsimdual}
  Let $D$ be a finite set of data examples and $\widehat e$ a data
  example. Then the following are equivalent:
  \begin{enumerate}

  \item there is a finite set of tree data examples $F$ such
    that $(F,D)$ is a simulation duality relative to $\widehat e$,

  \item the number of critical tree obstructions $q$ for $D$
    that satisfy $q \rightarrow \widehat e$ is finite (up to isomorphism). 
    
  \end{enumerate}
\end{restatable}
    
\begin{proof}
  $(1 \Rightarrow 2)$: Assume that $(F,D)$ is a
  simulation duality relative to $\widehat e$ with $F$ a set of tree
  examples. Let $n$ be the maximum number of variables of any example
  in $F$. To show that there are only finitely many critical tree
  obstructions $q$ for $D$ that satisfy $q \preceq \widehat e$, it
  suffices to show that each such $q$ has at most $n$ variables. So
  take a critical tree obstruction $q$ for $D$ with
  $q \preceq \widehat e$.  Then $q \not\preceq e$ for all $e \in D$
  and thus there is an $e' \in F$ with $e' \preceq q$.  Since $e'$ is
  a tree, this implies $e' \rightarrow q$. Now, the homomorphism
  witnessing the latter must be surjective as otherwise it gives rise
  to a tree CQ $q'$ that can be obtained from $q$ by dropping subtrees
  and that still satisfies $q' \not\preceq e$ for all $e \in D$
  (because $q' \preceq e$ would yield $e' \preceq e$ by composition of
  simulations) which
  contradicts the fact that $q$ is a critical tree obstruction for
  $D$. Consequently, the number of variables in $q$ is bounded by the
  number of variables in $e'$, thus by $n$.
  
  \smallskip $(2 \Rightarrow 1)$: Assume that there is a finite
  number of critical tree obstructions $q$ for $D$ that satisfy
  $q \rightarrow \widehat e$, up to isomorphism. Let $F$ be a set of
  tree CQs that contains one representative for every isomorphism
  class. Then $(F,D)$ is a simulation duality relative
  to~$\widehat e$. 

  To see this, first take a data example $e$ such that $e \preceq e'$
  for some $e' \in D$ and $e \preceq \widehat e$. Then
  $e'' \not \preceq e$ for all $e'' \in F$ because otherwise 
  we obtain $e'' \preceq e'$ be composing simulations, which
  contradicts the fact that $e''$ satisfies the first condition of
  critical tree obstructions.

  Now take a data example $e$ with $e' \not \preceq e$ for all
  $e' \in F$ and $e \preceq \widehat e$. Assume to the contrary of
  what we have to show that $e \not\preceq e''$ for all $e'' \in D$.
  By Lemma~\ref{lem:unravbasics}, there is then some $m$-unraveling
  $u$ of $e$ such that $u \not\preceq e''$ for all
  $e'' \in D$.  Moreover, $u \preceq \widehat e$. Let $u'$ be obtained
  from $u$ by dropping subtrees as long as $u' \not\preceq e''$ for
  all $e'' \in D$. Clearly, $u'$ is a critical tree obstruction. But
  then $F$ contains a CQ that is isomorphic to $u'$, contradicting our
  assumption that $e' \not \preceq e$ for all $e' \in F$.
\end{proof}
Let $(E^+,E^-)$ be a collection of labeled
examples. A fitting tree CQ $q$ for $(E^+,E^-)$ is \emph{critical} if no tree CQ
that can be obtained from $q$ by removing subtrees is a fitting. It is clear from the definitions that critical fitting tree CQs
for $(E^+, E^-)$ must be critical tree obstructions for $E^-$.
As a direct consequence of Propositions~\ref{prop:basis-vs-duality-tree-cqs}
and~\ref{prop:charsimdual},
we thus obtain the following.
\begin{lemma}
  \label{lem:basiscritfit}
  A collection $E=(E^+,E^-)$ of labeled examples has 
  a basis of most-general fitting tree CQs if and only if the number of critical fitting tree CQs for $E$ is finite (up to isomorphism). 
\end{lemma}

We also notice the following, using a proof similar to the one of Lemma~\ref{lem:degreebnd}. Full details are available in Appendix~\ref{app:wmg-tree-cq}. 

\begin{restatable}{lemma}{lemcritdeg}\label{lem:critdeg}
  If $q$ is a critical fitting tree CQ for a collection of labeled
  examples $(E^+,E^-)$, then the degree of $q$ is bounded
  by $||E^-||$.
\end{restatable}

We now use Lemmas~\ref{lem:basiscritfit} and \ref{lem:critdeg} to provide a reduction from the
existence
problem for  bases of most-general fitting tree CQs  to the
infinity problem for TWAPAs. This also yields bounds for the
size of such bases.
\begin{restatable}{theorem}{thmtreecqbasisexistence}
  \label{thm:tree-cq-basis-existence}
The existence problem for bases of most-general fitting tree CQs is
in ExpTime. Moreover, if a collection of labeled examples $E$ has a basis of
    most-general fitting tree CQs, then it has such a basis
    in which every tree CQ has size at most double exponential
    in $||E||$.
\end{restatable}

\begin{proof}
We construct, given a collection of labeled examples $E = (E^+,
E^-)$ a TWAPA $\Amf$ with polynomially many states that accepts exactly the
critical fitting tree CQs for $E$. By Lemma~\ref{lem:critdeg} it suffices to 
consider trees of degree at most $||E^-||$. The TWAPA $\Amf$ is the
intersection of three TWAPAs $\Amf_1$, $\Amf_2$, $\Amf_3$, where
\begin{itemize}
  \item $\Amf_1$ accepts those tree CQs that have no simulation in any
    negative example;
  \item $\Amf_2$ accepts those tree CQs that have a simulation in some negative
    example once any subtree is dropped;
  \item $\Amf_3$ accepts those tree CQs that have a simulation in all positive examples.
\end{itemize}

The TWAPA $\Amf_1$ and $\Amf_3$ are constructed in exactly the same way as the ones in the
proof of Theorem~\ref{thm:sizetreeCQarbupper}.
The TWAPA
$\Amf_2$ is a straightforward variation, we only sketch the
idea. The automaton first sends a copy of itself to every node in
the input tree except the root. It then verifies that, when the
subtree rooted at the node that it currently visits is dropped, then
the remaining input tree maps to some negative example. It does this
by traveling upwards one step to the predecessor and memorizing the
successor that it came from.  It also uses disjunction to guess an
$(I,c) \in E^-$ and an $a \in \mn{adom}(I)$ that the predecessor
maps to. It then behave essentially like $\Amf_3$, verifying the
existence of a simulation in~$I$, but avoiding the subtree at the
memorized successor. Once the root of the input tree is reached, the
automaton verifies that the constructed simulation uses $c$ as the
target.

  Since $\mathfrak{A}$ accepts exactly the critical fitting
  tree CQs for~$E$, by Lemma~\ref{lem:basiscritfit} it remains to
  complement $\mathfrak{A}$ and solve the infinity problem for the remaining TWAPA. By Points~(2) and~(6) of
  Theorem~\ref{thm:TWAPAstuff}, we obtain an ExpTime upper bound.

  We next observe that  if a collection of labeled examples $E$ has a basis of
    most-general fitting tree CQs, then it has such a basis
    in which every tree CQ has size at most double exponential
    in $||E||$.  If, in fact, $E$ has a basis of most-general fitting tree CQs, then
  we might assume w.l.o.g.\ that the basis contains only critical
  fitting tree CQs. By Lemma~\ref{lem:basiscritfit}, $E$ has only
  finitely many critical fitting tree CQs, and by the construction of
  $\Amf$ above and Point~(6) of
  Theorem~\ref{thm:TWAPAstuff}, every critical fitting tree CQ has
  size at most double exponential in $||E||$.
\end{proof}

  \subsection{Lower Bounds}
  \label{sect:treeCQlower}

  All the complexity upper bounds stated for tree CQs above are
  tight. 
  We establish matching ExpTime lower bounds by a polynomial time
  reduction from the product simulation problem into trees, as defined below, with one
  exception where we reduce from a related but different problem. Some of the reductions are
  rather technical and lengthy; the details of
  those are given in the appendix in order to not disrupt
  the flow of the paper.

\paragraph*{Product Simulation Problem Into Trees}

The \emph{product simulation problem} asks, for finite
pointed instances $(I_1,a_1),\ldots,(I_n,a_n)$ and $(J,b)$, whether
$\Pi_{1 \leq i \leq n} (I_i,a_i) \preceq (J,b)$. A variant of this
problem was shown to be ExpTime-hard in
\cite{DBLP:journals/iandc/HarelKV02} where simulations are replaced
with $\downarrow$-simulations, meaning that the third condition of
simulations is dropped, and certain transition systems are used in
place of instances. This result was transferred from transition systems to database instances in
\cite{Funk2019:when}, still using $\downarrow$-simulations. Here, we consider instead the \emph{product
  simulation problem into trees} where the target instance $J$ is
required to be a tree and full simulations  as defined at the beginning of Section~\ref{sec:tree-cq} are used in place of
$\downarrow$-simulations. For clarity, we may refer
to these as \emph{$\updownarrows$-simulations}.
We prove ExpTime-hardness of the product ($\updownarrows$-)simulation problem into trees by a
 reduction from the product $\downarrow$-simulation problem studied
in \cite{Funk2019:when}.

\begin{restatable}{theorem}{thmsimtreeexp}
\label{thm:simtreeexp}
  The product simulation problem
  into trees is ExpTime-hard, even for a fixed schema.
\end{restatable}
The rather technical proof is given in Appendix~\ref{app:product-sim-problem}, we only give a high-level overview.
Given an input $(I_1,a_1),\ldots,(I_n,a_n),(J,b)$
for the product $\downarrow$-simulation problem,
we construct an input $(I'_1,a'_1),\ldots,(I'_n,a'_n),(J',b')$ for the product $\updownarrows$-simulation problem into trees. The instances $(I'_i,a'_i)$ are obtained from the instances
$(I_i,a_i)$ by replacing every binary atom with a certain gadget. The intance $(J',b')$ takes the form of a tree of depth three that branches only at the root. There is one leaf for every pair of values in $J$. Intuitively, these are  potential targets that the values in a fact  of $\prod_{1 \leq i \leq n} I'$ are mapped to by a simulation. 
%
Note that the structure of $J'$ does not reflect the structure of $J$ (it cannot, because we want $J'$ to be a tree), but rather serves as a navigation gadget that must be traversed by any $\updownarrows$-simulation of $\prod_{1 \leq i \leq n} I'$ in $J'$
in a certain systematic way, due to the use of the
gadgets in the instances $I'_i$. Various issues are
induced by the facts that the gadgets in the instances $I'_i$ give rise to some subtructures in  $\prod_{1 \leq i \leq n} I'$ that we would rather not like to be there, and that we use $\updownarrows$-simulations
rather than $\downarrow$-simulations. We solve these issues by attaching several additional (tree-shaped) gadgets
to values in $J'$.


We remark that Theorem~\ref{thm:simtreeexp} improves a PSpace lower bound from \cite{Jung2020:least}
where, however, \emph{all} involved instances are required to be
trees. It is easy to prove an ExpTime upper bound for the product simulation problem into trees by computing
the product and then deciding the existence of a simulation in
polynomial time~\cite{DBLP:conf/focs/HenzingerHK95}.

We use Theorem~\ref{thm:simtreeexp} to prove lower bounds for the verification
and existence problems of unique fitting tree CQs, bases of most-general fitting
tree CQs, and most-specific fitting tree CQs.

\begin{restatable}{theorem}{thmstronglymostgeneralexphardtrees}
\label{thm:basesmostgeneralexphardtrees}
The verification and existence problems of unique fitting tree CQs and bases of
 most-general fitting tree CQs are ExpTime-hard.
This holds already when the tree CQ / all tree CQs in the
basis are
promised to fit resp.\ when a fitting tree CQ is promised to exist,
and with a fixed schema.
\end{restatable}
Before giving a formal proof of Theorem~\ref{thm:basesmostgeneralexphardtrees},
we describe the main ideas on an intuitive level.
Given an input $(I_1,a_1),\dots,(I_n,a_n),(J,b)$
for the product homomorphism problem into trees,
we construct a collection of labeled examples $E=(E^+, E^-)$ and a tree CQ $q'$ such that  (a)~if $\Pi_{1 \leq i \leq n} (I_i,a_i) \preceq
           (J,b)$, then $q'$ is a unique fitting tree CQ for $E$; and (b)~if $\Pi_{1 \leq i \leq n} (I_i,a_i) \not\preceq
           (J,b)$, then $E$ has no %
            basis of strongly most-general fitting tree CQs. The reader might want to verify that this proves ExpTime-hardness of all problems mentioned in Theorem~\ref{thm:basesmostgeneralexphardtrees}. We then use $J$ as~$q'$,  build the negative examples from the queries in the frontier of $q'$, and  have one positive example for each instance $(I_i,a_i)$ which 
            contains this instance and a copy of $J$.
            It can be seen that this achieves~(a). To also achieve~(b), we add to $q'$ and to each example a fresh root connected via a fresh binary relation $R$. This enables ever more general fittings if $\Pi_{1 \leq i \leq n} (I_i,a_i) \not\preceq
           (J,b)$ by replacing $R$ with a `zig-zag structure'.

\begin{proof}
  We prove the hardness results simultaneously, by
  reduction from the product simulation problem into trees.
  Assume that we are given
  finite pointed instances $(I_1,a_1),\dots,(I_n,a_n)$ and $(J,b)$
  with~$J$ a tree, and that we are to decide whether
  $\Pi_{1 \leq i \leq n} (I_i,a_i) \preceq (J,b)$. Let
  $I_1,\dots,I_n,J$ be formulated in binary schema $\mathcal{S}$.
  
  We construct a collection of labeled examples $E=(E^+, E^-)$, as
  follows. Assume w.l.o.g.\ that
  $\text{adom}(I_i) \cap \text{adom}(J) = \emptyset$ for
  $1 \leq i \leq n$. Let $R$ be a binary relation symbol  %
  that does
    not occur in $\mathcal{S}$. $E^+$
  contains the instances
  $(I'_1,a'_1),\dots,(I'_n,a'_n)$ where
  $(I'_i,a'_i)$ is obtained by starting with $(I_i,a_i)$ and adding
  the following facts:
  \begin{enumerate}
      
      \item $R(a'_i,a_i)$;
      
      \item  $R(a'_i,b)$ %
        and all facts from $J$ (we refer to this as the copy of $J$ in $I'_i$).
      
   \end{enumerate}
   Since $(J,b)$ is a tree, we may view it as a tree CQ $q(x)$.  Let
   $q'(x')$ be the tree CQ obtained from $q$ by making $x'$ the answer
   variable and adding the atom $R(x',x)$.
   It was shown in \cite{tCD2022:conjunctive} that every tree CQ has a
   frontier w.r.t.\ tree CQs which can be computed in polynomial
   time. We may thus compute such a frontier
   $\mathcal{F} = \{ p_1(x),\dots,p_k(x)\}$ for $q'$. $E^-$ contains
   the instances $(p_1,x),\dots,(p_k,x)$.
   It is easy to verify that $q'$ is a fitting for $E$.

    To establish all of the results in the theorem, it remains to show the following:
    \begin{enumerate}[(a)]

         \item If $\Pi_{1 \leq i \leq n} (I_i,a_i) \preceq
           (J,b)$, then $q'$ is a unique fitting tree CQ for $E$;
         
         \item If $\Pi_{1 \leq i \leq n} (I_i,a_i) \not\preceq
           (J,b)$, then $E$ has no %
            basis of strongly most-general fitting tree CQs.
        
    \end{enumerate}
    For Point~(a), assume that
    $\Pi_{1 \leq i \leq n} (I_i,a_i) \preceq (J,b)$ and let
    $\widehat q(x)$ be a fitting tree CQ for $E$.  We have to show
    that $\widehat q \preceq q'$ and $q' \preceq \widehat q$. The
    latter actually follows from the former since $\widehat q \preceq q'$
    and $q' \not\preceq \widehat q$ would imply that $\widehat q$
    simulates into some element of the frontier of $q'$, thus into a
    negative example, in contradiction to $\widehat q(x)$ being a
    fitting for $E$.
    
    To obtain $\widehat q \preceq q'$, in turn, it suffices to show
    $\Pi_{1 \leq i \leq n} (I'_i,a'_i) \preceq q'$ since $\widehat q$
    is a fitting and thus
    $\widehat q \preceq \Pi_{1 \leq i \leq n} (I'_i,a'_i)$.  Let $S$
    be a simulation witnessing
    $\Pi_{1 \leq i \leq n} (I_i,a_i) \preceq (J,b)$. We obtain $S'$
    from $S$ as follows:
    \begin{enumerate}[(i)]

    \item add $(\bar a',x')$ for $\bar a' =
          (a'_1,\dots,a'_n)$;

        \item for all
          $\bar a \in \text{adom}(\Pi_{1 \leq i \leq n} (I'_i,a'_i))$
          that contain an element $c$ of $\text{adom}(J)$, add
          $(\bar a,c)$.
        
    \end{enumerate}
    It can be verified that $S'$ is a simulation of
    $\Pi_{1 \leq i \leq n} (I'_i,a'_i)$ in $q'$. We have thus shown that
    $\Pi_{1 \leq i \leq n} (I'_i,a'_i) \preceq q'$, as desired.
    
    \smallskip

    For Point~(b), let
    $\Pi_{1 \leq i \leq n} (I_i,a_i) \not\preceq (J,b)$ and assume
    to the contrary of what we have to show that $E$ has a complete
    basis
    of strongly most-general fitting tree CQs. By
    Lemma~\ref{lem:unravbasics}, there is then also an $m \geq 1$
    such that $(U_m,\bar a) \not\preceq (J,b)$ where $U_m$ is 
    the $m$-unraveling of $\Pi_{1 \leq i \leq n} (I_i,a_i)$ 
    and $\bar a = a_1 \cdots a_n$.  
    Since $(U_m, \bar a)$ is a tree, we may view it as a tree CQ $p'(z)$.
    For all $i \geq 0$, let $p'_i(y_0)$ be
    obtained from $p'$ by making $y_0$ the answer variable and adding an initial $R$-zig-zag path, that is:
    $$
    R(y_0,z_0), R(y_1,z_0),R(y_1,z_1), R(y_2,z_1),R(y_2,z_2),\dots,
    R(y_{i},z_{i - 1}),R(y_{i},z)
    $$
    where $y_0,\dots,y_i$ and $z_0,\dots,z_{i - 1}$ are fresh variables.  We
    argue that each $p'_i$ is a fitting for $E$. By construction,
    $p'_i$ fits the positive examples. It does not fit any of the
    negative examples because any such example simulates into $q'$ and
    thus we would obtain $p'_i \preceq q'$ and any simulation witnessing
    this would also show
    $U_m \preceq (J,b)$. 

    Since $E$ has a basis of strongly most-general fitting
    tree CQs, there is some CQ $\widehat q$ that maps into
    infinitely many of the tree CQs $p'_i$. Since $\widehat q$ is
    connected and the length of the initial $R$-zig-zag path gets
    longer with increasing $i$, it follows that the query $\widehat q$
    simulates into an $R$-zig-zag path, and thus into the simple
    CQ $q_0(x) \colondash R(x,y)$. But then $\widehat q$ clearly
    simulates into a negative example, which is a contradiction.
\end{proof}
The proof of the subsequent lower bound is similar to that of
Theorem~\ref{thm:basesmostgeneralexphardtrees}, and in particular
it also uses a reduction from the product simulation problem into trees. This reduction is slightly more
intricate than the one used in the proof of Theorem~\ref{thm:basesmostgeneralexphardtrees}
because in the case that  $\Pi_{1 \leq i \leq n} (I_i,a_i) \not\preceq (J,b)$, it is more difficult to achieve that there is no most-specific fitting than it is to achieve that there is no stronlgy most-general fitting. In particular, adding a fresh root connected by a fresh binary relation no longer suffices. Details can be found in
Appendix~\ref{app:lowerboundstreecqs}.
\begin{restatable}{theorem}{thmmostspecificexptimehardtrees}
  \label{thm:mostspecific-exptime-hard-trees}
  Verification and existence of most-specific fitting tree CQs is
  ExpTime-hard. This holds already when the tree CQ / all tree CQs in
  the basis are promised to fit resp.\ when a fitting tree CQ is
  promised to exist, and with a fixed schema.
\end{restatable}
Next, we prove a matching lower bound for the existence of most-general fitting queries using a different technique. The proof applies simultaneously
to tree CQs and to arbitrary CQs. Here, we follow a different 
strategy than in the previous two lower bounds.  We adapt a proof 
of ExpTime-hardness of the simulation
 problem for concurrent transition systems
 from~\cite{DBLP:journals/iandc/HarelKV02} that proceeds by reducing
  the word problem for alternating, linear space bounded Turing
  machines.  In \cite{Funk2019:when}, a similar adaptation is used to show that deciding
  the existence of an arbitrary fitting tree CQ is ExpTime-hard, for a different notion of `tree CQ' that requires all edges in the tree to be directed away from the answer variable. In our adaptation, we
  need to be more careful since we are interested in the existence of weakly
  most-general fittings and work with a more liberal notion of tree CQ. Our proof does \emph{not} apply in the case that a weakly most-general fitting tree CQ is promised to exist. In fact, we leave the complexity of that case as an open problem. Our proof
  also does not use a fixed schema, although we conjecture that it can be modified to do so (at the cost of making it  more cumbersome than it already is).
\begin{restatable}{theorem}{thmmauriceexpapp}
  \label{thm:mauriceexpapp}
The existence problem is ExpTime-hard for weakly most-general fitting CQs and for weakly most-general fitting tree CQs.
\end{restatable}

The rather technical  proof it available in Appendix~\ref{app:lowerboundstreecqs}.
To close this section, we establish a double exponential lower bound on the size of
(arbitrary) fitting tree CQs.
\begin{restatable}{theorem}{thmsizetreelower}
  \label{thm:sizetreelower}
  For all $n \geq 0$, there is a collection of labeled examples of 
  combined size polynomial in $n$ such that a fitting tree CQ exists 
  and the size of every fitting tree CQ is at least $2^{2^n}$. This 
  even holds for a fixed schema. 
\end{restatable}

\begin{proof}
The following construction extends the one used in the proof of Theorem~\ref{thm:cq-size-lowerbound} with \emph{branching} to force every fitting tree CQ to have double exponential size.
Let $A$ be the unary relation of the schema and $R, L$ the binary relations.

First, we describe the positive examples which each will consist of a cycle of prime length where two facts, an $R$ fact and an $L$ fact, connect an element of the cycle to the next one.
Formally, for $j \geq 1$, let $D_j$ denote the instance with domain $\{0, \dots, j - 1\}$ and the following facts:
\begin{itemize}
    \item $R(k, k + 1), L(k, k + 1)$ for all $k < j - 1$,
    \item and $R(j - 1, 0), L(j - 1, 0), A(j - 1)$.
\end{itemize}
For $i \geq 1$, let $p_i$ denote the $i$-th prime number (where $p_1 = 2$).
Note that by the prime number theorem, $D_{p_i}$ is of size $O(i \log i)$.

    \begin{figure}
    \begin{center}
    \begin{tikzpicture}[thick, >=stealth]
    \node(11) at (2,2)  [circle, draw=black, fill=lightgray] {$11$};	
    \node(01) at (0,0)  [circle, draw=black, fill=lightgray] {$01$};	
    \node(10) at (4,0)  [circle, draw=black, fill=lightgray] {$10$};	
    \node(b) at (2,-2)  [circle, draw=black, fill=lightgray] {$b$};	

    \node(A1) at (2.5,2.3) {$A$};
    \node(A1) at (2.5,-2.3) {$A$};	

    \path [->] (b) edge [bend left = 10] node [midway,left] {$L,R$} (01);
    \path [->] (b) edge [bend right = 10] node [midway,right] {$L,R$} (10);
    \path [->] (01) edge [bend left] node [midway,above] {$L$} (10);

    \draw[->] (01) -- (11) node[midway,left] {$R$};
    \draw[->] (10) -- (11) node[midway,right] {$L$};
    \draw[->] (10) -- (01) node[midway,above] {$R$};   

    \draw[->] (2.1,2.4) arc (-45:240:0.2) node[midway, above] {$L,R$};
    \draw[->] (1.6,-1.9) arc (45:330:0.2) node[midway, left] {$L,R$};
    \draw[->] (-0.4,0.1) arc (45:330:0.2) node[midway, left] {$L$};
    \draw[->] (4.4,0.1) arc (135:-130:0.2) node[midway, right] {$R$};
    \end{tikzpicture}  
    \end{center}
    \caption{The instance $I$}
    \label{fig:instance-I}
    \end{figure}
    
For the negative examples, construct the instance $I$ with domain
$\{01, 10, 11, b\}$ and the following facts:
\begin{itemize}
    \item $L(10, 11)$, and $R(10, a)$ for all $a \in \{01, 10\}$,
    \item $R(01, 11)$, and $L(01, a)$ for all $a \in \{01, 10\}$,
    \item $R(b, b), L(b, b)$, $A(b)$, and $R(b, a), L(b, a)$ for all $a \in \{01, 10\}$,
    \item $L(11, 11)$, $R(11, 11)$, $A(11)$.
\end{itemize}
The instance $I$ is displayed in Figure~\ref{fig:instance-I}.

To establish the result of the theorem, we will show that there is a tree CQ that fits the examples
$E^+_n = \{ D_{p_i} \mid i = 1, \dots, n\}$ and
$E^-_n = \{ (I, a) \mid a \in \{01, 10\} \}$,
and that every fitting CQ has size at least $2^{2^n}$.
For this we will talk about a tree CQ $q$
as if it were a tree, i.e.\ use the notions of successors and predecessors of a variable as well as subtree below a variable.
Additionally, we will refer to a binary tree where $A$ is holds at every leaf and every non-leaf has exactly one direct $L$ successor and one direct $R$ successor as an $L, R, A$-tree.
We say that a tree CQ $q(x)$ contains an $L, R, A$-tree if there is a subset of atoms of $q$ that is an $L, R, A$-tree rooted at $x$.
The following claim holds for the negative examples:

\smallskip
\textit{Claim.} Let $q$ be a tree CQ over the schema $\{L, R, A\}$.
If $q$ does not contain an $L, R, A$-tree, then $q \preceq (I, a)$ for some $a \in \{10, 01\}$.

\smallskip
\textit{Proof of the claim.} We show this claim by induction on the height of $q(x)$. In the induction start $q$ has height $0$.
If $q$ contains no $L, R, A$ tree, then $q$ does not contain $A(x)$
and therefore $q \preceq (I, a)$ for any $a \in \{10, 01\}$.
Now let the claim hold for all tree CQs of height at most $i$ and let $q$ be a tree CQ of height $i + 1$. If $q$ contains no $L, R, A$-tree, then there is a $P \in \{L, R\}$ such that there is no direct $P$ successor $x'$ of $x$ that contains an $L, R, A$-tree.
Consider the case $P = L$, the case $P = R$ is analogous using the element $10$ instead of $01$. Then there is a simulation $S$ that witnesses $q \preceq (I, 01)$, that can be constructed as follows:
Start with $(x, 01) \in S$. If $q$ contains the atom $R(x, x')$ for some $x'$, map $x'$ and the entire subtree below $x'$ to $11$.
If $q$ contains the atom $L(x, x')$, then, by the induction hypothesis, there is an $a \in \{01, 10\}$ and a simulation from the subtree below $x'$ to $(I, a)$. Extend $S$ to the subtree below $x'$ according to this simulation.
If $q$ contains $R(x', x)$ or $L(x', x)$ map $x'$ and the entire subtree below $x'$ to $b$.
This completes the construction of $S$ and the proof of the claim.

\smallskip
Now, let $q$ be the full binary $L, R, A$-tree of depth $(\prod^n_{i = 1} p_i) - 1$.
Observe that $\prod^n_{i = 1} D_{p_i}$ is a double-linked cycle of size $\prod^n_{i = 1} p_i > 2^n$
and $A$ is only true at the \emph{last} element.
Therefore, $q \preceq D_{p_i}$ for all $i = 1, \dots, n$ and it can be shown that $q \not\preceq (I, a)$ for all $a \in \{01, 10\}$..
Thus, $q$ is a fitting tree CQ. The query $q$ is even a weakly most-general fitting of $E^+$ and $E^-$, as every element of its frontier no longer contains an $L, R, A$-tree.

Let $p$ be any fitting tree CQ over the schema. By the property of $I$ shown in the claim, 
$p$ must contain a $L, R, A$-tree, but since $p \preceq D_{p_i}$ for all $i = 1, \dots, n$ every $A$ in this $L, R, A$-tree must have distance $(\prod^n_{i = 1} p_i) - 1 > 2^n$ from the root.
Hence, $p$ must at least have size $2^{2^{n}}$.
\end{proof}

We do not
currently have a similar lower bound for any of the other types of
fitting tree CQs listed in Table~\ref{tab:tree-cq-results}.


\section{Conclusion}\label{sec:conclusion}

The characterizations and complexity results we presented, we believe,
give a fairly complete picture of extremal fitting
problems for CQs, UCQs, and tree CQs. Similar studies could be
performed, of course, for other query  and specification
languages (e.g., graph database queries, schema mappings). In
particular, the problem of computing fitting queries has received
considerable interest in knowledge representation, where, additionally,
background knowledge in the form of an ontology  is considered. The existence of a fitting $\mathcal{ELI}$ concept
(corresponding to a tree CQ) is undecidable in the presence of an
$\mathcal{ELI}$ ontology \cite{Funk2019:when}, but there are more restricted
settings, involving e.g.\ $\mathcal{EL}$ concept queries, that are
decidable and have received considerable interest
\cite{DBLP:conf/ilp/LehmannH09,DBLP:journals/ws/BuhmannLW16,Funk2019:when}.
\looseness=-1

Since the non-extremal fitting problem for CQs is
already coNExpTime-complete~\cite{Willard10,CateD15}, it is 
not surprising that many of our complexity bounds  are similarly
high. In~\cite{Barcelo017}, it was shown that the (non-extremal) fitting problem for
CQs can be made tractable by a combination
of two modifications to the problem: (i) ``desynchronization'', which effectively means to consider
UCQs instead of CQs, and (ii) replacing homomorphism tests by $k$-consistency tests, 
which effectively means to restrict attention to queries of bounded treewidth.
Similarly, in our results we also see improved complexity bounds when considering UCQs and tree CQs. While we have not studied
unions of tree CQs in this paper, based on results in~\cite{Barcelo017} one may expect that they
will exhibit a further reduction in the complexity of  fitting. We leave this as future work.
Another way to reduce the complexity is to consider
size-bounded versions of the  fitting problem,
an approach that also has  learning-related benefits~\cite{OurNewPaper}.
\looseness=-1

A question that we have not addressed so far is what to do if an extremal fitting query of interest does not exist. 
For practical purposes, in such cases (and possibly in general) it may be natural to consider relaxations where the fitting query is required to be, for instance, most-general, only as compared to other queries \emph{on some given (unlabeled) dataset}. It is easy to see that, under this relaxation,
a basis of most-general fitting queries always exists.

It would also be interesting to extend our extremal fitting analysis to allow for
approximate fitting, for instance using a threshold based approach  as in~\cite{Barcelo21:regularizing}
or an optimization-based approach as in~\cite{GottlobS10,tC2017:approximation}.


\begin{acks}
Balder ten Cate was supported by the \grantsponsor{EUH}{European Union's Horizon
2020 research and innovation programme}{} (\grantnum{EUH}{MSCA-101031081}),  Victor Dalmau was supported by the MICIN under
grants PID2019-109137GB-C22 and PID2022-138506NB-C22, and the Maria de Maeztu program (CEX2021-
001195-M), and 
Carsten Lutz by the DFG Collaborative Research Center 1320 EASE and by BMBF in DAAD
project 57616814 (SECAI).
\end{acks}

\bibliographystyle{ACM-Reference-Format}
\bibliography{bib}


\begin{thebibliography}{59}


\ifx \showCODEN    \undefined \def \showCODEN     #1{\unskip}     \fi
\ifx \showDOI      \undefined \def \showDOI       #1{#1}\fi
\ifx \showISBNx    \undefined \def \showISBNx     #1{\unskip}     \fi
\ifx \showISBNxiii \undefined \def \showISBNxiii  #1{\unskip}     \fi
\ifx \showISSN     \undefined \def \showISSN      #1{\unskip}     \fi
\ifx \showLCCN     \undefined \def \showLCCN      #1{\unskip}     \fi
\ifx \shownote     \undefined \def \shownote      #1{#1}          \fi
\ifx \showarticletitle \undefined \def \showarticletitle #1{#1}   \fi
\ifx \showURL      \undefined \def \showURL       {\relax}        \fi
\providecommand\bibfield[2]{#2}
\providecommand\bibinfo[2]{#2}
\providecommand\natexlab[1]{#1}
\providecommand\showeprint[2][]{arXiv:#2}

\bibitem[Alexe et~al\mbox{.}(2011a)]%
        {AlexeCKT2011}
\bibfield{author}{\bibinfo{person}{Bogdan Alexe}, \bibinfo{person}{Balder~ten
  Cate}, \bibinfo{person}{Phokion~G. Kolaitis}, {and}
  \bibinfo{person}{Wang-Chiew Tan}.} \bibinfo{year}{2011}\natexlab{a}.
\newblock \showarticletitle{Characterizing Schema Mappings via Data Examples}.
\newblock \bibinfo{journal}{\emph{ACM Trans. Database Syst.}}
  \bibinfo{volume}{36}, \bibinfo{number}{4} (\bibinfo{year}{2011}),
  \bibinfo{pages}{23:1--23:48}.
\newblock
\showISSN{0362-5915}
\urldef\tempurl%
\url{https://doi.org/10.1145/2043652.2043656}
\showDOI{\tempurl}


\bibitem[Alexe et~al\mbox{.}(2011b)]%
        {Alexe2011:designing}
\bibfield{author}{\bibinfo{person}{Bogdan Alexe}, \bibinfo{person}{Balder ten
  Cate}, \bibinfo{person}{Phokion~G. Kolaitis}, {and}
  \bibinfo{person}{Wang-Chiew Tan}.} \bibinfo{year}{2011}\natexlab{b}.
\newblock \showarticletitle{Designing and Refining Schema Mappings via Data
  Examples}. In \bibinfo{booktitle}{\emph{Proceedings of the 2011 ACM SIGMOD
  International Conference on Management of Data}} (Athens, Greece)
  \emph{(\bibinfo{series}{SIGMOD '11})}. \bibinfo{publisher}{Association for
  Computing Machinery}, \bibinfo{address}{New York, NY, USA},
  \bibinfo{pages}{133--144}.
\newblock
\showISBNx{9781450306614}
\urldef\tempurl%
\url{https://doi.org/10.1145/1989323.1989338}
\showDOI{\tempurl}


\bibitem[Arenas et~al\mbox{.}(2016)]%
        {Arenas2016:reverse}
\bibfield{author}{\bibinfo{person}{Marcelo Arenas}, \bibinfo{person}{Gonzalo~I.
  Diaz}, {and} \bibinfo{person}{Egor~V. Kostylev}.}
  \bibinfo{year}{2016}\natexlab{}.
\newblock \showarticletitle{Reverse Engineering SPARQL Queries}. In
  \bibinfo{booktitle}{\emph{Proceedings of the 25th International Conference on
  World Wide Web}} (Montr\'{e}al, Qu\'{e}bec, Canada)
  \emph{(\bibinfo{series}{WWW '16})}. \bibinfo{publisher}{International World
  Wide Web Conferences Steering Committee}, \bibinfo{address}{Republic and
  Canton of Geneva, CHE}, \bibinfo{pages}{239–249}.
\newblock
\showISBNx{9781450341431}
\urldef\tempurl%
\url{https://doi.org/10.1145/2872427.2882989}
\showDOI{\tempurl}


\bibitem[Barcel{\'{o}} and Romero(2017)]%
        {Barcelo017}
\bibfield{author}{\bibinfo{person}{Pablo Barcel{\'{o}}} {and}
  \bibinfo{person}{Miguel Romero}.} \bibinfo{year}{2017}\natexlab{}.
\newblock \showarticletitle{The Complexity of Reverse Engineering Problems for
  Conjunctive Queries}. In \bibinfo{booktitle}{\emph{20th International
  Conference on Database Theory, {ICDT} 2017, March 21-24, 2017, Venice,
  Italy}} \emph{(\bibinfo{series}{LIPIcs}, Vol.~\bibinfo{volume}{68})},
  \bibfield{editor}{\bibinfo{person}{Michael Benedikt} {and}
  \bibinfo{person}{Giorgio Orsi}} (Eds.). \bibinfo{publisher}{Schloss Dagstuhl
  - Leibniz-Zentrum f{\"{u}}r Informatik}, \bibinfo{pages}{7:1--7:17}.
\newblock
\urldef\tempurl%
\url{https://doi.org/10.4230/LIPIcs.ICDT.2017.7}
\showDOI{\tempurl}


\bibitem[Barceló et~al\mbox{.}(2021)]%
        {Barcelo21:regularizing}
\bibfield{author}{\bibinfo{person}{Pablo Barceló}, \bibinfo{person}{Alexander
  Baumgartner}, \bibinfo{person}{Victor Dalmau}, {and} \bibinfo{person}{Benny
  Kimelfeld}.} \bibinfo{year}{2021}\natexlab{}.
\newblock \showarticletitle{Regularizing conjunctive features for
  classification}.
\newblock \bibinfo{journal}{\emph{J. Comput. System Sci.}}
  \bibinfo{volume}{119} (\bibinfo{year}{2021}), \bibinfo{pages}{97--124}.
\newblock
\showISSN{0022-0000}
\urldef\tempurl%
\url{https://doi.org/10.1016/j.jcss.2021.01.003}
\showDOI{\tempurl}


\bibitem[Bienvenu et~al\mbox{.}(2014)]%
        {Bienvenu2015:obda}
\bibfield{author}{\bibinfo{person}{Meghyn Bienvenu}, \bibinfo{person}{Balder
  ten Cate}, \bibinfo{person}{Carsten Lutz}, {and} \bibinfo{person}{Frank
  Wolter}.} \bibinfo{year}{2014}\natexlab{}.
\newblock \showarticletitle{Ontology-Based Data Access: A Study through
  Disjunctive Datalog, CSP, and MMSNP}.
\newblock \bibinfo{journal}{\emph{ACM Trans. Database Syst.}}
  \bibinfo{volume}{39}, \bibinfo{number}{4} (\bibinfo{year}{2014}),
  \bibinfo{pages}{33:1--33:44}.
\newblock
\showISSN{0362-5915}
\urldef\tempurl%
\url{https://doi.org/10.1145/2661643}
\showDOI{\tempurl}


\bibitem[Bonifati et~al\mbox{.}(2015)]%
        {Bonifati2015:learning}
\bibfield{author}{\bibinfo{person}{Angela Bonifati}, \bibinfo{person}{Radu
  Ciucanu}, {and} \bibinfo{person}{Aur{\'e}lien Lemay}.}
  \bibinfo{year}{2015}\natexlab{}.
\newblock \showarticletitle{Learning Path Queries on Graph Databases}. In
  \bibinfo{booktitle}{\emph{EDBT}}. \bibinfo{pages}{109--120}.
\newblock
\urldef\tempurl%
\url{https://doi.org/10.5441/002/edbt.2015.11}
\showDOI{\tempurl}


\bibitem[Briceño et~al\mbox{.}(2021)]%
        {Briceno2021:dismantlability}
\bibfield{author}{\bibinfo{person}{Raimundo Briceño}, \bibinfo{person}{Andrei
  Bulatov}, \bibinfo{person}{Víctor Dalmau}, {and} \bibinfo{person}{Benoît
  Larose}.} \bibinfo{year}{2021}\natexlab{}.
\newblock \showarticletitle{Dismantlability, connectedness, and mixing in
  relational structures}.
\newblock \bibinfo{journal}{\emph{J. of Comb. Theory, Ser. {B}}}
  \bibinfo{volume}{147} (\bibinfo{year}{2021}), \bibinfo{pages}{37--70}.
\newblock
\showISSN{0095-8956}
\urldef\tempurl%
\url{https://doi.org/10.1016/j.jctb.2020.10.001}
\showDOI{\tempurl}


\bibitem[B{\"{u}}hmann et~al\mbox{.}(2016)]%
        {DBLP:journals/ws/BuhmannLW16}
\bibfield{author}{\bibinfo{person}{Lorenz B{\"{u}}hmann}, \bibinfo{person}{Jens
  Lehmann}, {and} \bibinfo{person}{Patrick Westphal}.}
  \bibinfo{year}{2016}\natexlab{}.
\newblock \showarticletitle{DL-Learner - {A} framework for inductive learning
  on the Semantic Web}.
\newblock \bibinfo{journal}{\emph{J. Web Semant.}}  \bibinfo{volume}{39}
  (\bibinfo{year}{2016}), \bibinfo{pages}{15--24}.
\newblock
\urldef\tempurl%
\url{https://doi.org/10.1016/j.websem.2016.06.001}
\showDOI{\tempurl}


\bibitem[Cate and Dalmau(2015)]%
        {CateD15}
\bibfield{author}{\bibinfo{person}{{Balder ten} Cate} {and}
  \bibinfo{person}{V{\'{\i}}ctor Dalmau}.} \bibinfo{year}{2015}\natexlab{}.
\newblock \showarticletitle{The Product Homomorphism Problem and Applications}.
  In \bibinfo{booktitle}{\emph{18th International Conference on Database
  Theory, {ICDT} 2015, March 23-27, 2015, Brussels, Belgium}}
  \emph{(\bibinfo{series}{LIPIcs}, Vol.~\bibinfo{volume}{31})},
  \bibfield{editor}{\bibinfo{person}{Marcelo Arenas} {and}
  \bibinfo{person}{Mart{\'{\i}}n Ugarte}} (Eds.). \bibinfo{publisher}{Schloss
  Dagstuhl - Leibniz-Zentrum f{\"{u}}r Informatik}, \bibinfo{pages}{161--176}.
\newblock
\urldef\tempurl%
\url{https://doi.org/10.4230/LIPIcs.ICDT.2015.161}
\showDOI{\tempurl}


\bibitem[Cate and Dalmau(2022)]%
        {tCD2022:conjunctive}
\bibfield{author}{\bibinfo{person}{{Balder ten} Cate} {and}
  \bibinfo{person}{Victor Dalmau}.} \bibinfo{year}{2022}\natexlab{}.
\newblock \showarticletitle{Conjunctive Queries: Unique Characterizations and
  Exact Learnability}.
\newblock \bibinfo{journal}{\emph{{ACM} Trans. Database Syst.}}
  \bibinfo{volume}{47}, \bibinfo{number}{4} (\bibinfo{year}{2022}),
  \bibinfo{pages}{14:1--14:41}.
\newblock
\urldef\tempurl%
\url{https://doi.org/10.1145/3559756}
\showDOI{\tempurl}


\bibitem[Cate et~al\mbox{.}(2023a)]%
        {extremalFittingPODS23}
\bibfield{author}{\bibinfo{person}{{Balder ten} Cate}, \bibinfo{person}{Victor
  Dalmau}, \bibinfo{person}{Maurice Funk}, {and} \bibinfo{person}{Carsten
  Lutz}.} \bibinfo{year}{2023}\natexlab{a}.
\newblock \showarticletitle{Extremal Fitting Problems for Conjunctive Queries}.
  In \bibinfo{booktitle}{\emph{Proceedings of the 42nd {ACM}
  {SIGMOD-SIGACT-SIGAI} Symposium on Principles of Database Systems, {PODS}
  2023, Seattle, WA, USA, June 18-23, 2023}},
  \bibfield{editor}{\bibinfo{person}{Floris Geerts}, \bibinfo{person}{Hung~Q.
  Ngo}, {and} \bibinfo{person}{Stavros Sintos}} (Eds.).
  \bibinfo{publisher}{{ACM}}, \bibinfo{pages}{89--98}.
\newblock
\urldef\tempurl%
\url{https://doi.org/10.1145/3584372.3588655}
\showDOI{\tempurl}


\bibitem[Cate et~al\mbox{.}(2013)]%
        {CateDK13:learning}
\bibfield{author}{\bibinfo{person}{{Balder ten} Cate},
  \bibinfo{person}{V{\'{\i}}ctor Dalmau}, {and} \bibinfo{person}{Phokion~G.
  Kolaitis}.} \bibinfo{year}{2013}\natexlab{}.
\newblock \showarticletitle{Learning schema mappings}.
\newblock \bibinfo{journal}{\emph{{ACM} Trans. Database Syst.}}
  \bibinfo{volume}{38}, \bibinfo{number}{4} (\bibinfo{year}{2013}),
  \bibinfo{pages}{28:1--28:31}.
\newblock
\urldef\tempurl%
\url{https://doi.org/10.1145/2539032.2539035}
\showDOI{\tempurl}


\bibitem[Cate et~al\mbox{.}(2022)]%
        {cate2022nonefficient}
\bibfield{author}{\bibinfo{person}{{Balder ten} Cate}, \bibinfo{person}{Maurice
  Funk}, \bibinfo{person}{Jean~Christoph Jung}, {and} \bibinfo{person}{Carsten
  Lutz}.} \bibinfo{year}{2022}\natexlab{}.
\newblock \bibinfo{title}{On the non-efficient PAC learnability of acyclic
  conjunctive queries}.
\newblock
\newblock
\urldef\tempurl%
\url{https://doi.org/10.48550/arXiv.2208.10255}
\showDOI{\tempurl}
\showeprint[arxiv]{2208.10255}~[cs.DB]


\bibitem[Cate et~al\mbox{.}(2023b)]%
        {DBLP:journals/sigmod/CateFJL23}
\bibfield{author}{\bibinfo{person}{{Balder ten} Cate}, \bibinfo{person}{Maurice
  Funk}, \bibinfo{person}{Jean~Christoph Jung}, {and} \bibinfo{person}{Carsten
  Lutz}.} \bibinfo{year}{2023}\natexlab{b}.
\newblock \showarticletitle{Fitting Algorithms for Conjunctive Queries}.
\newblock \bibinfo{journal}{\emph{{SIGMOD} Rec.}} \bibinfo{volume}{52},
  \bibinfo{number}{4} (\bibinfo{year}{2023}), \bibinfo{pages}{6--18}.
\newblock


\bibitem[Cate et~al\mbox{.}(2023c)]%
        {OurNewPaper}
\bibfield{author}{\bibinfo{person}{{Balder ten} Cate}, \bibinfo{person}{Maurice
  Funk}, \bibinfo{person}{Jean~Christoph Jung}, {and} \bibinfo{person}{Carsten
  Lutz}.} \bibinfo{year}{2023}\natexlab{c}.
\newblock \bibinfo{title}{{SAT}-Based {PAC} Learning of Description Logic
  Concepts}.
\newblock \bibinfo{howpublished}{Forthcoming}.
\newblock


\bibitem[Cate et~al\mbox{.}(2017)]%
        {tC2017:approximation}
\bibfield{author}{\bibinfo{person}{{Balder ten} Cate},
  \bibinfo{person}{Phokion~G. Kolaitis}, \bibinfo{person}{Kun Qian}, {and}
  \bibinfo{person}{Wang-Chiew Tan}.} \bibinfo{year}{2017}\natexlab{}.
\newblock \showarticletitle{Approximation Algorithms for Schema-Mapping
  Discovery from Data Examples}.
\newblock \bibinfo{journal}{\emph{ACM Trans. Database Syst.}}
  \bibinfo{volume}{42}, \bibinfo{number}{2} (\bibinfo{year}{2017}),
  \bibinfo{pages}{12:1--12:41}.
\newblock
\showISSN{0362-5915}
\urldef\tempurl%
\url{https://doi.org/10.1145/3044712}
\showDOI{\tempurl}


\bibitem[Chandra et~al\mbox{.}(1981)]%
        {chandra81}
\bibfield{author}{\bibinfo{person}{Ashok~K. Chandra}, \bibinfo{person}{Dexter
  Kozen}, {and} \bibinfo{person}{Larry~J. Stockmeyer}.}
  \bibinfo{year}{1981}\natexlab{}.
\newblock \showarticletitle{Alternation}.
\newblock \bibinfo{journal}{\emph{J. {ACM}}} \bibinfo{volume}{28},
  \bibinfo{number}{1} (\bibinfo{year}{1981}), \bibinfo{pages}{114--133}.
\newblock
\urldef\tempurl%
\url{https://doi.org/10.1145/322234.322243}
\showDOI{\tempurl}


\bibitem[Chandra and Merlin(1977)]%
        {CM77}
\bibfield{author}{\bibinfo{person}{Ashok~K. Chandra} {and}
  \bibinfo{person}{Philip~M. Merlin}.} \bibinfo{year}{1977}\natexlab{}.
\newblock \showarticletitle{{Optimal Implementation of Conjunctive Queries in
  Relational Data Bases}}. In \bibinfo{booktitle}{\emph{ACM Symposium on Theory
  of Computing (STOC)}}. \bibinfo{pages}{77--90}.
\newblock


\bibitem[Chen et~al\mbox{.}(2011)]%
        {chen2011arc}
\bibfield{author}{\bibinfo{person}{Hubie Chen}, \bibinfo{person}{Victor
  Dalmau}, {and} \bibinfo{person}{Berit Grußien}.}
  \bibinfo{year}{2011}\natexlab{}.
\newblock \bibinfo{title}{Arc Consistency and Friends}.
\newblock
\newblock
\urldef\tempurl%
\url{https://doi.org/10.48550/arXiv.1104.4993}
\showDOI{\tempurl}
\showeprint[arxiv]{1104.4993}~[cs.AI]


\bibitem[Cohen and Weiss(2016)]%
        {Cohen2016:complexity}
\bibfield{author}{\bibinfo{person}{Sara Cohen} {and} \bibinfo{person}{Yaacov~Y.
  Weiss}.} \bibinfo{year}{2016}\natexlab{}.
\newblock \showarticletitle{The Complexity of Learning Tree Patterns from
  Example Graphs}.
\newblock \bibinfo{journal}{\emph{ACM Trans. Database Syst.}}
  \bibinfo{volume}{41}, \bibinfo{number}{2} (\bibinfo{year}{2016}),
  \bibinfo{pages}{14:1--14:44}.
\newblock
\showISSN{0362-5915}
\urldef\tempurl%
\url{https://doi.org/10.1145/2890492}
\showDOI{\tempurl}


\bibitem[Cropper et~al\mbox{.}(2022)]%
        {Cropper2021:ilp}
\bibfield{author}{\bibinfo{person}{Andrew Cropper}, \bibinfo{person}{Sebastijan
  Duman\v{c}i\v{c}}, \bibinfo{person}{Richard Evans}, {and}
  \bibinfo{person}{Stephen~H. Muggleton}.} \bibinfo{year}{2022}\natexlab{}.
\newblock \showarticletitle{Inductive logic programming at 30}.
\newblock \bibinfo{journal}{\emph{Mach. Learn.}} \bibinfo{volume}{111},
  \bibinfo{number}{1} (\bibinfo{year}{2022}), \bibinfo{pages}{147--172}.
\newblock
\urldef\tempurl%
\url{https://doi.org/10.1007/s10994-021-06089-1}
\showDOI{\tempurl}


\bibitem[Feder and Vardi(1998)]%
        {FV98}
\bibfield{author}{\bibinfo{person}{Tom{\'a}s Feder} {and}
  \bibinfo{person}{Moshe~Y. Vardi}.} \bibinfo{year}{1998}\natexlab{}.
\newblock \showarticletitle{The Computational Structure of Monotone Monadic
  {SNP} and Constraint Satisfaction: A Study through {D}atalog and Group
  Theory}.
\newblock \bibinfo{journal}{\emph{{SIAM} J.~on Computing}}
  \bibinfo{volume}{28}, \bibinfo{number}{1} (\bibinfo{year}{1998}),
  \bibinfo{pages}{57--104}.
\newblock


\bibitem[Foniok et~al\mbox{.}(2008)]%
        {FoniokNT08}
\bibfield{author}{\bibinfo{person}{Jan Foniok}, \bibinfo{person}{Jaroslav
  Nesetril}, {and} \bibinfo{person}{Claude Tardif}.}
  \bibinfo{year}{2008}\natexlab{}.
\newblock \showarticletitle{Generalised dualities and maximal finite antichains
  in the homomorphism order of relational structures}.
\newblock \bibinfo{journal}{\emph{Eur. J. Comb.}} \bibinfo{volume}{29},
  \bibinfo{number}{4} (\bibinfo{year}{2008}), \bibinfo{pages}{881--899}.
\newblock
\urldef\tempurl%
\url{https://doi.org/10.1016/j.ejc.2007.11.017}
\showDOI{\tempurl}


\bibitem[Funk et~al\mbox{.}(2019)]%
        {Funk2019:when}
\bibfield{author}{\bibinfo{person}{Maurice Funk}, \bibinfo{person}{Jean Jung},
  \bibinfo{person}{Carsten Lutz}, \bibinfo{person}{Hadrien Pulcini}, {and}
  \bibinfo{person}{Frank Wolter}.} \bibinfo{year}{2019}\natexlab{}.
\newblock \showarticletitle{Learning Description Logic Concepts: When can
  Positive and Negative Examples be Separated?}. In
  \bibinfo{booktitle}{\emph{Proceedings of IJCAI 2019}}.
  \bibinfo{pages}{1682--1688}.
\newblock
\urldef\tempurl%
\url{https://doi.org/10.24963/ijcai.2019/233}
\showDOI{\tempurl}


\bibitem[Gottlob et~al\mbox{.}(1999)]%
        {GottlobLS99}
\bibfield{author}{\bibinfo{person}{Georg Gottlob}, \bibinfo{person}{Nicola
  Leone}, {and} \bibinfo{person}{Francesco Scarcello}.}
  \bibinfo{year}{1999}\natexlab{}.
\newblock \showarticletitle{On the Complexity of Some Inductive Logic
  Programming Problems}.
\newblock \bibinfo{journal}{\emph{New Generation Comput.}}
  \bibinfo{volume}{17}, \bibinfo{number}{1} (\bibinfo{year}{1999}),
  \bibinfo{pages}{53--75}.
\newblock
\urldef\tempurl%
\url{https://doi.org/10.1007/BF03037582}
\showDOI{\tempurl}


\bibitem[Gottlob and Senellart(2010)]%
        {GottlobS10}
\bibfield{author}{\bibinfo{person}{Georg Gottlob} {and} \bibinfo{person}{Pierre
  Senellart}.} \bibinfo{year}{2010}\natexlab{}.
\newblock \showarticletitle{Schema mapping discovery from data instances}.
\newblock \bibinfo{journal}{\emph{J. {ACM}}} \bibinfo{volume}{57},
  \bibinfo{number}{2} (\bibinfo{year}{2010}), \bibinfo{pages}{6:1--6:37}.
\newblock
\urldef\tempurl%
\url{https://doi.org/10.1145/1667053.1667055}
\showURL{%
\tempurl}


\bibitem[Gulwani et~al\mbox{.}(2012)]%
        {Gulwani2012:spreadsheet}
\bibfield{author}{\bibinfo{person}{Sumit Gulwani}, \bibinfo{person}{William~R.
  Harris}, {and} \bibinfo{person}{Rishabh Singh}.}
  \bibinfo{year}{2012}\natexlab{}.
\newblock \showarticletitle{Spreadsheet data manipulation using examples}.
\newblock \bibinfo{journal}{\emph{Commun. ACM}} \bibinfo{volume}{55},
  \bibinfo{number}{8} (\bibinfo{date}{Aug.} \bibinfo{year}{2012}),
  \bibinfo{pages}{97–105}.
\newblock
\showISSN{0001-0782}
\urldef\tempurl%
\url{https://doi.org/10.1145/2240236.2240260}
\showDOI{\tempurl}


\bibitem[Gulwani et~al\mbox{.}(2015)]%
        {Gulwani2015:inductive}
\bibfield{author}{\bibinfo{person}{Sumit Gulwani}, \bibinfo{person}{Jos{\'e}
  Hern{\'a}ndez-Orallo}, \bibinfo{person}{Emanuel Kitzelmann},
  \bibinfo{person}{Stephen~H. Muggleton}, \bibinfo{person}{Ute Schmid}, {and}
  \bibinfo{person}{Benjamin~G. Zorn}.} \bibinfo{year}{2015}\natexlab{}.
\newblock \showarticletitle{Inductive programming meets the real world}.
\newblock \bibinfo{journal}{\emph{Commun. ACM}} \bibinfo{volume}{58},
  \bibinfo{number}{11} (\bibinfo{year}{2015}), \bibinfo{pages}{90--99}.
\newblock


\bibitem[Gulwani et~al\mbox{.}(2017)]%
        {Gulwani2017:program}
\bibfield{author}{\bibinfo{person}{Sumit Gulwani}, \bibinfo{person}{Oleksandr
  Polozov}, {and} \bibinfo{person}{Rishabh Singh}.}
  \bibinfo{year}{2017}\natexlab{}.
\newblock \showarticletitle{Program Synthesis}.
\newblock \bibinfo{journal}{\emph{Foundations and Trends{\textregistered} in
  Programming Languages}} \bibinfo{volume}{4}, \bibinfo{number}{1-2}
  (\bibinfo{year}{2017}), \bibinfo{pages}{1--119}.
\newblock
\showISSN{2325-1107}
\urldef\tempurl%
\url{https://doi.org/10.1561/2500000010}
\showDOI{\tempurl}


\bibitem[Gutjahr et~al\mbox{.}(1992)]%
        {GutjahrWW92}
\bibfield{author}{\bibinfo{person}{Wolfgang Gutjahr}, \bibinfo{person}{Emo
  Welzl}, {and} \bibinfo{person}{Gerhard~J. Woeginger}.}
  \bibinfo{year}{1992}\natexlab{}.
\newblock \showarticletitle{Polynomial graph-colorings}.
\newblock \bibinfo{journal}{\emph{Discret. Appl. Math.}} \bibinfo{volume}{35},
  \bibinfo{number}{1} (\bibinfo{year}{1992}), \bibinfo{pages}{29--45}.
\newblock
\urldef\tempurl%
\url{https://doi.org/10.1016/0166-218X(92)90294-K}
\showDOI{\tempurl}


\bibitem[Harel et~al\mbox{.}(2002)]%
        {DBLP:journals/iandc/HarelKV02}
\bibfield{author}{\bibinfo{person}{David Harel}, \bibinfo{person}{Orna
  Kupferman}, {and} \bibinfo{person}{Moshe~Y. Vardi}.}
  \bibinfo{year}{2002}\natexlab{}.
\newblock \showarticletitle{On the Complexity of Verifying Concurrent
  Transition Systems}.
\newblock \bibinfo{journal}{\emph{Inf. Comput.}} \bibinfo{volume}{173},
  \bibinfo{number}{2} (\bibinfo{year}{2002}), \bibinfo{pages}{143--161}.
\newblock
\urldef\tempurl%
\url{https://doi.org/10.1006/inco.2001.2920}
\showDOI{\tempurl}


\bibitem[Hell and Ne\v{s}et\v{r}il(2004)]%
        {HellNesetril2004}
\bibfield{author}{\bibinfo{person}{Pavol Hell} {and} \bibinfo{person}{Jaroslav
  Ne\v{s}et\v{r}il}.} \bibinfo{year}{2004}\natexlab{}.
\newblock \bibinfo{booktitle}{\emph{Graphs and homomorphisms}}.
  \bibinfo{series}{Oxford lecture series in mathematics and its applications},
  Vol.~\bibinfo{volume}{28}.
\newblock \bibinfo{publisher}{Oxford University Press}.
\newblock
\showISBNx{978-0-19-852817-3}


\bibitem[Henzinger et~al\mbox{.}(1995)]%
        {DBLP:conf/focs/HenzingerHK95}
\bibfield{author}{\bibinfo{person}{Monika~Rauch Henzinger},
  \bibinfo{person}{Thomas~A. Henzinger}, {and} \bibinfo{person}{Peter~W.
  Kopke}.} \bibinfo{year}{1995}\natexlab{}.
\newblock \showarticletitle{Computing Simulations on Finite and Infinite
  Graphs}. In \bibinfo{booktitle}{\emph{36th Annual Symposium on Foundations of
  Computer Science, Milwaukee, Wisconsin, USA, 23-25 October 1995}}.
  \bibinfo{publisher}{{IEEE} Computer Society}, \bibinfo{pages}{453--462}.
\newblock
\urldef\tempurl%
\url{https://doi.org/10.1109/SFCS.1995.492576}
\showDOI{\tempurl}


\bibitem[Jacindha et~al\mbox{.}(2022)]%
        {ProgramSynthesisReview}
\bibfield{author}{\bibinfo{person}{S. Jacindha}, \bibinfo{person}{G. Abishek},
  {and} \bibinfo{person}{P. Vasuki}.} \bibinfo{year}{2022}\natexlab{}.
\newblock \showarticletitle{Program Synthesis---A Survey}. In
  \bibinfo{booktitle}{\emph{Computational Intelligence in Machine Learning}},
  \bibfield{editor}{\bibinfo{person}{Amit Kumar}, \bibinfo{person}{Jacek~M.
  Zurada}, \bibinfo{person}{Vinit~Kumar Gunjan}, {and} \bibinfo{person}{Raman
  Balasubramanian}} (Eds.). \bibinfo{publisher}{Springer Nature Singapore},
  \bibinfo{address}{Singapore}, \bibinfo{pages}{409--421}.
\newblock
\showISBNx{978-981-16-8484-5}


\bibitem[Jung et~al\mbox{.}(2020c)]%
        {DBLP:conf/kr/JungLPW20}
\bibfield{author}{\bibinfo{person}{Jean~Christoph Jung},
  \bibinfo{person}{Carsten Lutz}, \bibinfo{person}{Hadrien Pulcini}, {and}
  \bibinfo{person}{Frank Wolter}.} \bibinfo{year}{2020}\natexlab{c}.
\newblock \showarticletitle{Logical Separability of Incomplete Data under
  Ontologies}. In \bibinfo{booktitle}{\emph{Proceedings of {KR} 2020}},
  \bibfield{editor}{\bibinfo{person}{D.~Calvanese}, \bibinfo{person}{E.~Erdem},
  {and} \bibinfo{person}{M.~Thielscher}} (Eds.). \bibinfo{pages}{517--528}.
\newblock
\urldef\tempurl%
\url{https://doi.org/10.24963/kr.2020/52}
\showDOI{\tempurl}


\bibitem[Jung et~al\mbox{.}(2021)]%
        {DBLP:conf/kr/JungLPW21}
\bibfield{author}{\bibinfo{person}{Jean~Christoph Jung},
  \bibinfo{person}{Carsten Lutz}, \bibinfo{person}{Hadrien Pulcini}, {and}
  \bibinfo{person}{Frank Wolter}.} \bibinfo{year}{2021}\natexlab{}.
\newblock \showarticletitle{Separating Data Examples by Description Logic
  Concepts with Restricted Signatures}. In
  \bibinfo{booktitle}{\emph{Proceedings of {KR} 2021}},
  \bibfield{editor}{\bibinfo{person}{M.~Bienvenu},
  \bibinfo{person}{G.~Lakemeyer}, {and} \bibinfo{person}{E.~Erdem}} (Eds.).
  \bibinfo{pages}{390--399}.
\newblock
\urldef\tempurl%
\url{https://doi.org/10.24963/kr.2021/37}
\showDOI{\tempurl}


\bibitem[Jung et~al\mbox{.}(2020a)]%
        {DBLP:conf/aaai/JungLW20}
\bibfield{author}{\bibinfo{person}{Jean~Christoph Jung},
  \bibinfo{person}{Carsten Lutz}, {and} \bibinfo{person}{Frank Wolter}.}
  \bibinfo{year}{2020}\natexlab{a}.
\newblock \showarticletitle{Least General Generalizations in Description Logic:
  Verification and Existence}. In \bibinfo{booktitle}{\emph{Proceedings of
  {AAAI 2020}}}. \bibinfo{publisher}{{AAAI} Press},
  \bibinfo{pages}{2854--2861}.
\newblock


\bibitem[Jung et~al\mbox{.}(2020b)]%
        {Jung2020:least}
\bibfield{author}{\bibinfo{person}{Jean~Christoph Jung},
  \bibinfo{person}{Carsten Lutz}, {and} \bibinfo{person}{Frank Wolter}.}
  \bibinfo{year}{2020}\natexlab{b}.
\newblock \showarticletitle{Least General Generalizations in Description Logic:
  Verification and Existence}. In \bibinfo{booktitle}{\emph{Proceedings of the
  Thirty-Fourth {AAAI} Conference on Artificial Intelligence, {AAAI} 2020, New
  York, NY, USA, February 7-12, 2020}}. \bibinfo{pages}{2854--2861}.
\newblock
\urldef\tempurl%
\url{https://doi.org/10.1609/aaai.v34i03.5675}
\showDOI{\tempurl}


\bibitem[Kimelfeld and Ré(2018)]%
        {KimelfeldRe2018}
\bibfield{author}{\bibinfo{person}{Benny Kimelfeld} {and}
  \bibinfo{person}{Christopher Ré}.} \bibinfo{year}{2018}\natexlab{}.
\newblock \showarticletitle{A Relational Framework for Classifier Engineering}.
\newblock \bibinfo{journal}{\emph{ACM SIGMOD Record}}  \bibinfo{volume}{47}
  (\bibinfo{year}{2018}), \bibinfo{pages}{6--13}.
\newblock
\urldef\tempurl%
\url{https://doi.org/10.1145/3277006.3277009}
\showDOI{\tempurl}


\bibitem[Krisnadhi and Lutz(2007)]%
        {DBLP:conf/lpar/KrisnadhiL07}
\bibfield{author}{\bibinfo{person}{Adila Krisnadhi} {and}
  \bibinfo{person}{Carsten Lutz}.} \bibinfo{year}{2007}\natexlab{}.
\newblock \showarticletitle{Data Complexity in the \emph{EL} Family of
  Description Logics}. In \bibinfo{booktitle}{\emph{Proceedings of {LPAR
  2007}}} \emph{(\bibinfo{series}{Lecture Notes in Computer Science},
  Vol.~\bibinfo{volume}{4790})}. \bibinfo{publisher}{Springer},
  \bibinfo{pages}{333--347}.
\newblock


\bibitem[Larose et~al\mbox{.}(2007)]%
        {Larose2007:characterization}
\bibfield{author}{\bibinfo{person}{Benoit Larose}, \bibinfo{person}{Cynthia
  Loten}, {and} \bibinfo{person}{Claude Tardif}.}
  \bibinfo{year}{2007}\natexlab{}.
\newblock \showarticletitle{A Characterisation of First-Order Constraint
  Satisfaction Problems}.
\newblock \bibinfo{journal}{\emph{Log. Methods Comput. Sci.}}
  \bibinfo{volume}{3}, \bibinfo{number}{4} (\bibinfo{year}{2007}).
\newblock
\urldef\tempurl%
\url{https://doi.org/10.2168/LMCS-3(4:6)2007}
\showDOI{\tempurl}


\bibitem[Lehmann and Haase(2009)]%
        {DBLP:conf/ilp/LehmannH09}
\bibfield{author}{\bibinfo{person}{Jens Lehmann} {and}
  \bibinfo{person}{Christoph Haase}.} \bibinfo{year}{2009}\natexlab{}.
\newblock \showarticletitle{Ideal Downward Refinement in the $\mathcal{EL}$
  Description Logic}. In \bibinfo{booktitle}{\emph{Inductive Logic Programming,
  19th International Conference, {ILP} 2009, Leuven, Belgium, July 02-04, 2009.
  Revised Papers}} \emph{(\bibinfo{series}{Lecture Notes in Computer Science},
  Vol.~\bibinfo{volume}{5989})}, \bibfield{editor}{\bibinfo{person}{Luc~De
  Raedt}} (Ed.). \bibinfo{publisher}{Springer}, \bibinfo{pages}{73--87}.
\newblock
\urldef\tempurl%
\url{https://doi.org/10.1007/978-3-642-13840-9\_8}
\showDOI{\tempurl}


\bibitem[Lehmann and Hitzler(2010)]%
        {DBLP:journals/ml/LehmannH10}
\bibfield{author}{\bibinfo{person}{Jens Lehmann} {and} \bibinfo{person}{Pascal
  Hitzler}.} \bibinfo{year}{2010}\natexlab{}.
\newblock \showarticletitle{Concept learning in description logics using
  refinement operators}.
\newblock \bibinfo{journal}{\emph{Mach. Learn.}} \bibinfo{volume}{78},
  \bibinfo{number}{1-2} (\bibinfo{year}{2010}), \bibinfo{pages}{203--250}.
\newblock
\urldef\tempurl%
\url{https://doi.org/10.1007/s10994-009-5146-2}
\showDOI{\tempurl}


\bibitem[Li et~al\mbox{.}(2015)]%
        {Li2015:qfe}
\bibfield{author}{\bibinfo{person}{Hao Li}, \bibinfo{person}{Chee-Yong Chan},
  {and} \bibinfo{person}{David Maier}.} \bibinfo{year}{2015}\natexlab{}.
\newblock \showarticletitle{Query from Examples: An Iterative, Data-Driven
  Approach to Query Construction}.
\newblock \bibinfo{journal}{\emph{Proc. {VLDB} Endow.}} \bibinfo{volume}{8},
  \bibinfo{number}{13} (\bibinfo{year}{2015}), \bibinfo{pages}{2158--2169}.
\newblock
\showISSN{2150-8097}
\urldef\tempurl%
\url{https://doi.org/10.14778/2831360.2831369}
\showDOI{\tempurl}


\bibitem[Mitchell(1997)]%
        {Mitchell97}
\bibfield{author}{\bibinfo{person}{Tom~M. Mitchell}.}
  \bibinfo{year}{1997}\natexlab{}.
\newblock \bibinfo{booktitle}{\emph{Machine Learning}}.
\newblock \bibinfo{publisher}{McGraw-Hill}, \bibinfo{address}{New York}.
\newblock
\showISBNx{978-0-07-042807-2}


\bibitem[Muller and Schupp(1987)]%
        {DBLP:journals/tcs/MullerS87}
\bibfield{author}{\bibinfo{person}{David~E. Muller} {and}
  \bibinfo{person}{Paul~E. Schupp}.} \bibinfo{year}{1987}\natexlab{}.
\newblock \showarticletitle{Alternating Automata on Infinite Trees}.
\newblock \bibinfo{journal}{\emph{Theor. Comput. Sci.}}  \bibinfo{volume}{54}
  (\bibinfo{year}{1987}), \bibinfo{pages}{267--276}.
\newblock
\urldef\tempurl%
\url{https://doi.org/10.1016/0304-3975(87)90133-2}
\showDOI{\tempurl}


\bibitem[Nesetril and R{\"{o}}dl(1989)]%
        {NesetrilR89}
\bibfield{author}{\bibinfo{person}{Jaroslav Nesetril} {and}
  \bibinfo{person}{Vojtech R{\"{o}}dl}.} \bibinfo{year}{1989}\natexlab{}.
\newblock \showarticletitle{Chromatically optimal rigid graphs}.
\newblock \bibinfo{journal}{\emph{J. Comb. Theory, Ser. {B}}}
  \bibinfo{volume}{46}, \bibinfo{number}{2} (\bibinfo{year}{1989}),
  \bibinfo{pages}{133--141}.
\newblock
\urldef\tempurl%
\url{https://doi.org/10.1016/0095-8956(89)90039-7}
\showDOI{\tempurl}


\bibitem[Ne\v{s}et\v{r}il and de~Mendez(2008)]%
        {NesetrilOssona2008}
\bibfield{author}{\bibinfo{person}{Jaroslav Ne\v{s}et\v{r}il} {and}
  \bibinfo{person}{Patrice~Ossona de Mendez}.} \bibinfo{year}{2008}\natexlab{}.
\newblock \showarticletitle{Grad and classes with bounded expansion III.
  Restricted graph homomorphism dualities}.
\newblock \bibinfo{journal}{\emph{European Journal of Combinatorics}}
  \bibinfo{volume}{29}, \bibinfo{number}{4} (\bibinfo{year}{2008}),
  \bibinfo{pages}{1012 -- 1024}.
\newblock
\showISSN{0195-6698}
\urldef\tempurl%
\url{https://doi.org/10.1016/j.ejc.2007.11.019}
\showDOI{\tempurl}
\newblock
\shownote{Homomorphisms: Structure and Highlights}.


\bibitem[Ne\v{s}et\v{r}il and Ossona~de Mendez(2012)]%
        {nesetril2012sparsity}
\bibfield{author}{\bibinfo{person}{Jaroslav Ne\v{s}et\v{r}il} {and}
  \bibinfo{person}{Patrice Ossona~de Mendez}.} \bibinfo{year}{2012}\natexlab{}.
\newblock \bibinfo{booktitle}{\emph{Sparsity (Graphs, Structures, and
  Algorithms)}}. Vol.~\bibinfo{volume}{28}.
\newblock \bibinfo{publisher}{Springer}.
\newblock
\urldef\tempurl%
\url{https://doi.org/10.1007/978-3-642-27875-4}
\showDOI{\tempurl}


\bibitem[Ne\v{s}et\v{r}il and Tardif(2000)]%
        {NesetrilTardif2000}
\bibfield{author}{\bibinfo{person}{Jaroslav Ne\v{s}et\v{r}il} {and}
  \bibinfo{person}{Claude Tardif}.} \bibinfo{year}{2000}\natexlab{}.
\newblock \showarticletitle{Duality Theorems for Finite Structures
  (Characterising Gaps and Good Characterisations)}.
\newblock \bibinfo{journal}{\emph{Journal of Combinatorial Theory, Series B}}
  \bibinfo{volume}{80}, \bibinfo{number}{1} (\bibinfo{year}{2000}),
  \bibinfo{pages}{80 -- 97}.
\newblock
\showISSN{0095-8956}
\urldef\tempurl%
\url{https://doi.org/10.1006/jctb.2000.1970}
\showDOI{\tempurl}


\bibitem[Ne\v{s}et\v{r}il and Tardif(2005)]%
        {NesetrilTardif2005}
\bibfield{author}{\bibinfo{person}{Jaroslav Ne\v{s}et\v{r}il} {and}
  \bibinfo{person}{Claude Tardif}.} \bibinfo{year}{2005}\natexlab{}.
\newblock \showarticletitle{Short Answers to Exponentially Long Questions:
  Extremal Aspects of Homomorphism Duality}.
\newblock \bibinfo{journal}{\emph{SIAM J. Discret. Math.}}
  \bibinfo{volume}{19}, \bibinfo{number}{4} (\bibinfo{date}{Aug.}
  \bibinfo{year}{2005}), \bibinfo{pages}{914--920}.
\newblock
\showISSN{0895-4801}
\urldef\tempurl%
\url{https://doi.org/10.1137/S0895480104445630}
\showDOI{\tempurl}


\bibitem[Nienhuys-Cheng and de~Wolf(1997)]%
        {NienhuysCheng1997:Foundations}
\bibfield{author}{\bibinfo{person}{Shan-Hwei Nienhuys-Cheng} {and}
  \bibinfo{person}{Ronald de Wolf}.} \bibinfo{year}{1997}\natexlab{}.
\newblock \bibinfo{booktitle}{\emph{Foundations of Inductive Logic
  Programming}}. \bibinfo{series}{Lecture Notes in Computer Science},
  Vol.~\bibinfo{volume}{1228}.
\newblock \bibinfo{publisher}{Springer}.
\newblock


\bibitem[Plotkin(1971)]%
        {Plotkin1971:Automatic}
\bibfield{author}{\bibinfo{person}{G.~D. Plotkin}.}
  \bibinfo{year}{1971}\natexlab{}.
\newblock \emph{\bibinfo{title}{Automatic Methods of Inductive Inference}}.
\newblock \bibinfo{thesistype}{Ph.\,D. Dissertation}.
  \bibinfo{school}{University of Edinburgh, Department of Computer Science}.
\newblock
\newblock
\shownote{Ph.D. dissertation}.


\bibitem[Rizzo et~al\mbox{.}(2020)]%
        {DBLP:journals/fgcs/RizzoFd20}
\bibfield{author}{\bibinfo{person}{Giuseppe Rizzo}, \bibinfo{person}{Nicola
  Fanizzi}, {and} \bibinfo{person}{Claudia d'Amato}.}
  \bibinfo{year}{2020}\natexlab{}.
\newblock \showarticletitle{Class expression induction as concept space
  exploration: From {DL-FOIL} to {DL-FOCL}}.
\newblock \bibinfo{journal}{\emph{Future Gener. Comput. Syst.}}
  \bibinfo{volume}{108} (\bibinfo{year}{2020}), \bibinfo{pages}{256--272}.
\newblock
\urldef\tempurl%
\url{https://doi.org/10.1016/j.future.2020.02.071}
\showDOI{\tempurl}


\bibitem[Rothe(2003)]%
        {Rothe2001:exact}
\bibfield{author}{\bibinfo{person}{Jörg Rothe}.}
  \bibinfo{year}{2003}\natexlab{}.
\newblock \showarticletitle{Exact complexity of Exact-Four-Colorability}.
\newblock \bibinfo{journal}{\emph{Inform. Process. Lett.}}
  \bibinfo{volume}{87}, \bibinfo{number}{1} (\bibinfo{year}{2003}),
  \bibinfo{pages}{7--12}.
\newblock
\showISSN{0020-0190}
\urldef\tempurl%
\url{https://doi.org/10.1016/S0020-0190(03)00229-1}
\showDOI{\tempurl}


\bibitem[Tran et~al\mbox{.}(2014)]%
        {Tran2014:reverse}
\bibfield{author}{\bibinfo{person}{Quoc~Trung Tran}, \bibinfo{person}{Chee-Yong
  Chan}, {and} \bibinfo{person}{Srinivasan Parthasarathy}.}
  \bibinfo{year}{2014}\natexlab{}.
\newblock \showarticletitle{Query reverse engineering}.
\newblock \bibinfo{journal}{\emph{The {VLDB} Journal}} \bibinfo{volume}{23},
  \bibinfo{number}{5} (\bibinfo{year}{2014}), \bibinfo{pages}{721--746}.
\newblock
\urldef\tempurl%
\url{https://doi.org/10.1007/s00778-013-0349-3}
\showDOI{\tempurl}


\bibitem[Vardi(1998)]%
        {DBLP:conf/icalp/Vardi98}
\bibfield{author}{\bibinfo{person}{Moshe~Y. Vardi}.}
  \bibinfo{year}{1998}\natexlab{}.
\newblock \showarticletitle{Reasoning about The Past with Two-Way Automata}. In
  \bibinfo{booktitle}{\emph{Automata, Languages and Programming, 25th
  International Colloquium, ICALP'98, Aalborg, Denmark, July 13-17, 1998,
  Proceedings}} \emph{(\bibinfo{series}{Lecture Notes in Computer Science},
  Vol.~\bibinfo{volume}{1443})},
  \bibfield{editor}{\bibinfo{person}{Kim~Guldstrand Larsen},
  \bibinfo{person}{Sven Skyum}, {and} \bibinfo{person}{Glynn Winskel}} (Eds.).
  \bibinfo{publisher}{Springer}, \bibinfo{pages}{628--641}.
\newblock
\urldef\tempurl%
\url{https://doi.org/10.1007/BFb0055090}
\showDOI{\tempurl}


\bibitem[Willard(2010)]%
        {Willard10}
\bibfield{author}{\bibinfo{person}{Ross Willard}.}
  \bibinfo{year}{2010}\natexlab{}.
\newblock \showarticletitle{Testing Expressibility Is Hard}. In
  \bibinfo{booktitle}{\emph{Principles and Practice of Constraint Programming -
  {CP} 2010 - 16th International Conference, {CP} 2010, St. Andrews, Scotland,
  UK, September 6-10, 2010. Proceedings}} \emph{(\bibinfo{series}{Lecture Notes
  in Computer Science}, Vol.~\bibinfo{volume}{6308})},
  \bibfield{editor}{\bibinfo{person}{David Cohen}} (Ed.).
  \bibinfo{publisher}{Springer}, \bibinfo{pages}{9--23}.
\newblock
\urldef\tempurl%
\url{https://doi.org/10.1007/978-3-642-15396-9\_4}
\showDOI{\tempurl}


\end{thebibliography}

\newpage

\appendix

\section{Missing proofs from Section~\ref{sect:mostgenfitCQs}}
\label{app:wmg-existence}

\propencodable*

\begin{proof} (sketch)
    If $T$ is a proper $\Sigma$-labeled $d$-ary tree, then, it is clear from the definitions that
    $q_T$ satisfies (1) -- (4). Conversely, let $q(x_1, \ldots, x_k)$ be a $k$-ary query that satisfies 
    conditions (1) -- (4). Let $q_1, \ldots, q_n$ be the connected components of $q$
    (where $n\leq d$). Furthermore, choose arbitrarily one fact from each connected
    component. We will denote the chosen fact from $q_i$ by $f_i$.  We 
    construct a mapping $g$ from existential variables and facts of $q$ to sequences in $\{1, \ldots, d\}^*$ as follows, iteratively:
    \begin{itemize}
        \item $g(f_i)=\langle i \rangle$ for $i\leq n$.
        \item Let $f=R(z_1, \ldots, z_m)$ be a fact of $q_i$ (for some $i\leq n$) and suppose that $z_j$ is an existential variable (for some $i\leq m$). We say that $z_i$ is a \emph{parent variable} of $f$, if $z_i$ lies on the shortest path from $f_i$ to $f$ in the incidence graph of $q_i$. Note that, by c-acyclicity, there can be at most one such parent variable.
        If $z_j$ is \emph{not} a parent variable of $f$ and $g(f)=t$, then $g(z_j)=t\cdot j$. 
        \item Let $z$ be an existential variable of $q_i$ and let 
          $f'_1, \ldots, f'_m$ be the facts in which $z$ occurs (with $m\leq d+1$). 
          Exactly one of of these facts must lie on the shortest path from $f_i$ to $z$ 
          in the incidence graph of $q_i$, and we will will refer to it as the 
          \emph{parent fact} of $z$. For each fact $f'_j$ that is \emph{not} the parent fact of $z$,
            and $g(z)=t$, then
          we set $g(f'_j)=\pi \cdot j$.
    \end{itemize}
    The image of the map $g$ thus obtained, is a 
    non-empty prefix closed subset of $\{1, \ldots, d\}^*$ if we add to it also the empty
    sequence $\varepsilon$. We can expand this into a $\Sigma$-labeled $d$-ary tree $T$ by defining a 
    suitable labeling function $\lambda$. Specifically, we set $\lambda(x)=\nu$ if $x=\varepsilon$ or if $x\in\{1,\ldots,d\}^*$ is the $g$-image of an existential variable; if 
    $x\in\{1,\ldots,d\}^*$ is the $g$-image of a fact $R(z_1, \ldots, z_m)$, we set
    $\lambda(x)=(R,\langle dir_1, \ldots, dir_m\rangle)$, where $dir_j=\ans_i$ if $z_j=x_i$;
    $dir_j=\up$ if $z_j$ is the answer variable of $f$, and $dir_j=\down$ otherwise.
    See Figure~\ref{fig:example-tree-encoding} for an example.
    It is then easy to see that $q_T$ is isomorphic to $q$.
\end{proof}

\lemautomatonfrontier*

\begin{proof}
We will show how to construct the automaton in 
the case of a single data example. The general result then 
follows because we can construct $\mathfrak{A}^{\text{frontier}}_{E,i}$
as the union of the (polynomially many) automata
$\mathfrak{A}^{\text{frontier}}_{e,i}$ for all $e\in E$ (using non-determinism in the initial state transitions effectively to select the example).

Let $e=(I,\textbf{a})$ with $\textbf{a}=a_1, \ldots, a_k$.

The automaton $\mathfrak{A}^{\text{frontier}}_{e,i}$, intuitively, has to check
two things: (1) it has to check that subtree from the $i$-th child of the root
(which encodes the $i$-th connected component of the query), 
encodes a subquery $q'$ such that $F(q')$ fits $e$ as a positive example, and
(2) it has to check that, for every $j\neq i$, the subtree rooted at the 
$j$-th child of the root (if it exists), encodes a query that fits $e$ as a positive
example. For (2) we already showed in Lemma~\ref{lem:automaton-positive} how to do this, and 
in fact, we can include in our automaton a copy of the automaton from Lemma~\ref{lem:automaton-positive} to handle this part. Note that no
alternation is needed for this, as this is effectively a conjunction 
where each conjunct pertains to a different subtree of the input tree (i.e., a different child of the root).

Therefore, it suffices to focus on (1). 
Before we spell out the details of the automaton, we make two observations that 
provide the idea behind the automaton.
First of all, recall that $F(q')$ has an existential variable $u_{(y,f)}$ for every pair $(y,f)$, where $y$ is an existential variable of $q'$ and $f$ is a fact in which $y$ occurs. 
When we look at the tree encoding of the query, then we can see that 
each existential variable $y$ of $q'$ is a node in the tree encoding, 
and the facts in which $y$
occurs are precisely the children and the parent of $y$ in the tree encoding.

The second observation is that every existential variable of $q'$ may 
have up to $d+1$ many replicas in $F(q')$, and 
a homomorphism from $F(q')$ to $e$ is a map that, among other things, 
has to send each replica to a value in $\text{adom}(I)$. We want to set things
 up so that, from an accepting run of the automaton, we can obtain such a 
 homomorphism. To do this, we can create a different state of the automaton
 for every $d+1$-length vector of values from $\text{adom}(I)$. Such a state then encodes,
 for a given variable, the value in $\text{adom}(I)$ that each of its
 replicas gets mapped to. In addition, for each answer-variable $x$ of $q'$,
 $F(q')$ also includes an existential variable $u_x$ that has to be mapped to some
 value in $\text{adom}(I)$. We can include this information in the states as well 
 by further increasing the length of the vector by $k$.
 
Based on these ideas, the construction of the automaton is now a relatively
straightforward extension of the automaton given in the proof of 
Lemma~\ref{lem:automaton-positive}.

By a ``vector'' $v$ we will mean a $d+k+1$-length vector of values from $\text{adom}(I)$. 
We say that two vectors $v, v'$ are
\emph{compatible} if the last $k$ items of the vectors
are identical, i.e., $v[d+2,    \ldots, d+k+1] = 
v'[d+2, \ldots, d+k+1]$. (Recall that the last $k$
components of the vector are used to encode what 
$u_{x_i}$ gets mapped to, for each answer
variable $x_i$. By requiring all vectors to be 
compatible, we ensure that this choice is effectively 
made only once for the entire accepting run of the
automaton.)

The states of the automaton include all states of the automaton $\mathfrak{A}_e$ 
given in the proof of Lemma~\ref{lem:automaton-positive}, plus:
\begin{itemize}
    \item A state $q^{\text{exvar}}_v$ for every vector $v$,
    \item A state $q^{\text{root-fact}}_{R,\langle v_1, \ldots, v_n\rangle}$ where $n=\arity(R)$ and $v_1, \ldots, v_n$ are pairwise compatible vectors.
    \item A state $q^{\text{fact}}_{R,\langle v_1, \ldots, v_n\rangle,j,\ell}$ where $n=\arity(R)$ and $j\leq n$, $\ell\leq d$, and $v_1, \ldots, v_n$ are pairwise compatible vectors.
\end{itemize}

For a state of the form $q^{\text{fact}}_{R,\langle v_1, \ldots, v_n\rangle,j,\ell}$,
intuitively, $j$ represents the index at which the parent existential variable occurs in the fact, while $\ell$ will merely be used to keep track that the current node is going to be the $\ell$-th child of its parent in the tree.

The transitions include all the transitions of the automaton $\mathfrak{A}$, except
for those going to the root state.  In addition, we have:
\begin{itemize}
    \item $\langle q_1, \ldots, q_d\rangle\transition{\nu}q^{\text{root}}$ where
      $q_i$ is of the form $q^{\text{root-fact}}_{R,\langle v_1, \ldots, v_n\rangle}$,
      and where each $q_j$ for $j\neq i$ is either $\bot$ or is a state from
      $\mathfrak{A}_e$ of the form
      $q^{\text{root-fact}}_{R(b_1, \ldots, b_n)}$,
    \item $\langle q_1, \ldots, q_d\rangle\transition{\nu}q^{\text{root}}$ whenever
      $q_i=\bot$ (so that we accept whenever the root has no $i$-th child).
      \item $\langle q_1, \ldots, q_d \rangle \transition{\nu} q^{\text{exvar}}_v$ where
     each $q_j$ is either $\bot$ or of the form $q^{\text{fact}}_{R,\langle v_1, \ldots, v_n\rangle,j',j}$ with $v_{j'}=v$,
     \item $\langle q_1, \ldots, q_n, \bot, \ldots, \bot\rangle\transition{\langle R,dir_1, \ldots, dir_n\rangle}q^{\text{root-fact}}_{R,\langle v_1, \ldots, v_n\rangle}$ where (i) for each $j\leq n$, if $dir_j=\down$ then $q_j=q^{\text{exvar}}_{v_j}$, and (ii) $\langle R, dir_1, \ldots, dir_n\rangle$ is ``fulfilled'' by $\langle v_1, \ldots, v_n\rangle$ at $\ell=-1$,
     \item $\langle q_1, \ldots, q_n, \bot, \ldots, \bot\rangle\transition{\langle R,dir_1, \ldots, dir_n\rangle}q^{\text{fact}}_{R,\langle v_1, \ldots, v_n\rangle,j,\ell}$ where (i) $dir_j=\up$, (ii) for each $j'\leq n$, if $dir_{j'}=\down$ then $q_{j'}=q^{\text{exvar}}_{v_{j'}}$, and (iii) $\langle R, dir_1, \ldots, dir_n\rangle$ is ``fulfilled'' by $\langle v_1, \ldots, v_n\rangle$ at $\ell$.
\end{itemize}

The above definition of the transitions refer to the notion of ``being fulfilled'',
which we define now. This definition naturally reflects the frontier construction in 
Definition~\ref{def:F}. A node label
$\sigma=\langle R,dir_1, \ldots, dir_n\rangle\in\Sigma$ is ``fulfilled'' by 
$n$-tuple of vectors $(v_1, \ldots, v_n)$ at $\ell$, if 
every ``acceptable instance'' of $\pi$ relative to $(v_1, \ldots, v_n)$ and $\ell$ is a tuple in $R$. Here, 
an \emph{acceptable instance} of $\sigma$ relative to $(v_1, \ldots, v_n)$ and $\ell$
is a tuple $(y_1, \ldots, y_n)$ where
\begin{itemize}
    \item for each $i\leq n$, either $dir_i\in\{\down,\up\}$ and $y_i\in v_i[1\ldots d+1]$, or $dir_i=\ans_j$ and $y_i\in \{a_j, v_i[d+1+j]\}$, and
    \item for some $i\leq n$, either (i) $dir_i=\down$ and $y_i\in v_i[2,\ldots, d+1]$,
    or (ii) $dir_i=\up$ and $y_i= v_i[m+1]$ for some $m\in\{1,\ldots, d+1\}$ with $m \neq \ell+1$  or (iii) $dir_i=\ans_j$ and $y_i = v_i[d+1+j]$.
\end{itemize}
Note that, the first entry in the vector corresponds to the replica
$u_{(x,f)}$ where $f$ is the parent fact of $x$, while the $i+1$-th 
entry in the vector (for $i\leq d$) corresponds to the replica
$u_{(x,f)}$ where $f$ is the $i$-th child fact of $x$.

If the automaton has an accepting run on input $T$, then, it already follows
from the above root transition and the proof of Lemma~\ref{lem:automaton-positive},
that every connected component of $q_T$ other than the $i$-th one, 
admits a homomorphism to $e$. It should also be clear from the above
discussion that if we denote the $i$-th connected component by $q'$, then
$F(q')$ has a homomorphism to $e$. The converse holds as well, and therefore
the automaton accepts $T$ if and only if $F^i(q_T)$ admits a homomorphism to $e$.
\end{proof}

\corwmgwithoutunp*

\begin{proof}
We will restrict
ourselves here to giving a high-level explanation of the changes required for this.
As a concrete example, let us consider the $3$-ary CQ
\[ q(x_1,x_2,x_3) \colondash R(x_1,x_2,x_3), P_1(x_1), P_2(x_2), P_3(x_3), (x_1=x_2)\]

Note that we use here equalities in the body of the CQ, but the same query could  be expressed equivalently using repeated occurrences of variables in the head.

By an \emph{equality-type} we will mean an equivalence relation over the set of answer variables $\{x_1, x_2, x_3\}$. The equality type $\equiv_q$ of the above query $q$ is the equivalence relation that identifies $x_1, x_2$ with each other but not with $x_3$.

A frontier for such a query can be obtained by taking the set $F$ of all instances that can be obtained in one of the following two ways:
\begin{enumerate}
    \item Compute a CQ $q'$ of lower arity by replacing every equivalence class of answer variable by a single representative of that class. By construction, 
    $q'$ has the UNP. We take all queries belonging to the frontier of $q'$, and, finally, add equalities to obtain queries of the original arity. Specifically, in the case of our example query $q$ above, $q'$ is the 
    2-ary query $q'(x_1, x_3) \colondash R(x_1,x_1,x_3), P_1(x_1), P_2(x_1), P_3(x_3)$. We then take each CQ belonging to the frontier of $q'$, and extend it with a conjunct $x_2=x_1$ to make that CQ ternary again, and add the result to $F$.
    \item Let $\equiv$ be the equality type of the query at hand. A \emph{minimal weakening} of $\equiv$ is an equivalence relation in which some equivalence class is divided in two. In our example, the only minimal weakening of $\equiv_q$ is the
    equality type $\equiv'$ that does not identify any answer variables with each other. For each such weakening (i.e., in the case of our example, $\equiv'$), 
    we then construct another CQ $q'$ where we drop from $q$ all equalities and
    replace them with the equalities $(x_i=x_j)$ for $(x_i,x_j)\in\equiv'$. We
    add $q'$ to our set $F$. In our specific example, $q'$ ends up being identical 
    to $q$ except without the $x_1=x_2$ conjunct.
\end{enumerate}
It follows from~\cite[Proof of Theorem 3.8]{tCD2022:conjunctive} that $F$, thus constructed, 
is a frontier for $q$.

All of the above can be implemented by a tree automaton. First of all, this
requires a minor modification to the tree representation of c-acyclic CQs:
we will store the equality type of the query in the root label of the tree. 
Next, with a non-deterministic root transitions, we guess whether item
1 or 2 as described above, applies. In first case, we reuse (with minimal 
modification) the automata we constructed earlier. In the second case, 
we guess a minimally weaker equality type. It is not hard to write an automaton
that accepts a tree-encoding of a c-acyclic CQ if and only its corresponding
minimal weakening fits the labeled examples. We omit the details here. 
\end{proof}

\section{Verification and Construction of Relativized homomorphism dualities (Theorem~\ref{thm:relative-dualities})}
\label{app:relative-dualities}
This section is dedicated to the proof of:

\thmrelativedualities*

\newcommand{\ba}{\operatorname{{\bf a}}}
\newcommand{\bp}{\operatorname{{\bf p}}}
\newcommand{\be}{\operatorname{{\bf e}}}
\newcommand{\bi}{\operatorname{{\bf i}}}

\newcommand{\bb}{\operatorname{{\bf b}}}
\newcommand{\bx}{\operatorname{{\bf x}}}

\newcommand{\dom}{\text{adom}}
\newcommand{\diag}{\text{diag}}

To prove item (1) we proceed in two steps: we first reduce to the case where $D$ is a single example, 
and then we provide a characterization for this case, which leads to an criterion
that can be evaluated in NP.  As a byproduct of the proof of the above-mentioned criterion we obtain the upper bound on the size of the elements in $F$ stated in (2), from which a 2Exptime algorithm is immediately derived by brute-force search.


\subsection*{Reduction to the case where $D$ is a single example}

It will be convenient to regard equivalence relations $\alpha$ on some set $A$, as subsets of $A^2$. In this way we can write $\alpha\subseteq\beta$ to indicate that $\alpha$ refines $\beta$. We shall
use $\wedge$ and $\vee$ to indicate the meet and join of equivalence relations.

 Let be $(I,\ba)$ be a pointed instance and let $\alpha$ be an equivalence relation on $\dom(I)$. For every array $\bb$ containing values from $\dom(I)$ we define $\bb_{\alpha}$ to
be the tuple obtained by replacing every value in $\bb$ by its $\alpha$-class. Also, we define 
$(I,\ba)_{\alpha}$ to be $(I_{\alpha},\ba_{\alpha})$ where the facts of $I_{\alpha}$ are $\{R(\bb_{\alpha}) \mid R(\bb)\in I\}$. 

For this part of the proof it will be convenient to generalize the disjoint union of a pair of pointed instances $q_1=(I,\ba_1)$ and $q_2=(I,\ba_2)$ which, so far, has only been defined under the assumption that $q_1$ and $q_2$ have the UNP. In its full generality we define $q_1\uplus q_2$ as $(I_1\cup I_2,\ba_1)_{\alpha}$ where $\alpha$ is the more refined equivalence such that $\ba_1[i]$ and $\ba_2[i]$ are $\alpha$-related for every $i\in [k]$.
Note that  $(I_1,\ba_1)\uplus (I_2,\ba_2)$ is defined so that for every pointed instance $(I,\ba)$ the following is equivalent:
\begin{enumerate}
\item $(I_i,\ba_i)\rightarrow (I,\ba)$ for $i=1,2$
\item $(I_1,\ba_1)\uplus (I_2,\ba_2)\rightarrow (I,\ba)$. 
\end{enumerate}

Recall that $(I,\ba)$ is {\em connected} if it cannot be expressed
as the disjoint union $(I_1,\ba_1)\uplus (I_2,\ba_1)$ where both $I_1$ and $I_2$ are non-empty.

We say that an example $e\in D$ is {\em strictly subsumed} relative to $D$ and $p$ if, for some $e'\in D$, 
$p\times e\rightarrow e'$ and $p\times e'\nrightarrow e$.
We say that an example $e\in D$ is \emph{non-subsumed} if it is not strictly subsumed
by any $e'\in D$.
We note that a pair $(F,D)$ is a homomorphism duality 
relative to $p$ if and only if $(F,D')$ is a homomorphism duality relative to $p$, where $D'$ is the set of non-subsumed examples in $D$.

The next lemma reduces our problem to the case where $D$ contains a single example.

\begin{lemma}\label{le:reductiontosingleexample}
For every finite set of data examples $D$ and data example $p$,
the following are equivalent:
\begin{enumerate}
\item 
There exists a finite set $F$ such that $(F,D)$ is a generalized duality relative to $p$.
\item For every non-subsumed $e\in D$ there exists a finite set $F_e$ such that $(F_e,\{e\})$ is a generalized duality relative to $p$.
\end{enumerate}
\end{lemma}
\begin{proof} 
Let $D'=\{e_1,\dots,e_n\}$ be the set of non-subsumed examples in $D$.

$(2)\Rightarrow(1)$ It can easily verified that $(F,D')$ (and hence $(F,D)$) is a generalized duality relative to $p$ 
when $F=\{q_1\uplus\cdots\uplus q_n \mid q_i\in F_{e_i}, i\in [n]\}$. 

$(1)\Rightarrow(2)$ Let $e_i\in D'$. For every pointed instance $x=(X,\bx)$ with $k$ distinguished elements we shall use 
$\gamma(x)$ to denote the equivalence relation on $[k]$ where $i,j\in[k]$ are related
if $\bx[i]=\bx[j]$.  

Let $\delta=\gamma(e)\wedge\gamma(p)$.
For every pointed instance $x=(X,\bx)$ satisfying $\delta\subseteq\gamma(x)$ we shall use 
$x'$ to denote the pointed instance $x'=(X,\bx')$ where $\bx'$ is obtained from $\bx$ by removing elements that belong to the same $\delta$-class. Formally, fix a representative $i_1,\dots,i_r$ for each one of the $\delta$-classes and define $\bx'[j]$ to be $\bx[i_j]$. 

Let $m$ be the maximum domain size of any pointed instance in $F$ and let $F'_{i}$ be the set
of all pointed instances with at most $m$ values that are not homomorphic to $e_i'$. We shall
show that $(F'_i,\{e'_i\})$ is a generalized duality relative to $p'$. This completes the proof
of $(1)\Rightarrow(2)$ as it is easily verified
that $e_i$ and $p$ satisfy Lemma~\ref{le:characterization}(1) if and only if $e'_i$ and $p$ satisfy it as well.

Let us then show that $(F'_i,\{e'_i\})$ is a generalized duality relative to $p'$. Let $y$ be satisfying $y\nrightarrow e_i'$ and $y\rightarrow p'$. We note that $y=x'$
for some $x=(X,\bx)$ satisfying $x\nrightarrow e_i$, $x\rightarrow p$, and $\delta\subseteq\gamma(x)$.
We can assume that $\gamma(x)\subseteq\gamma(e_i)$ since otherwise, $(\emptyset,\bx')$ (which belongs to $F'_{e_i}$) is homomorphic to $x'$ and we are done.

We note that $\delta=\gamma(p\times e_i)$. Also, since $\gamma(x)\subseteq\gamma(e_i)$
and $\gamma(x)\subseteq\gamma(p)$ it follows that $\gamma(x)\subseteq\delta$. Hence, we have $\gamma(x)=\delta$. 

Consider $x\uplus (p\times e_i)$. We note that in general we cannot assume that the disjoint union $y_1\uplus y_2$ of two pointed instances contains $y_1$ (or $y_2$) as a subinstance since some values can be identified while computing the disjoint union. However, since $\gamma(x)=\delta=\gamma(p\times e_i)$ it follows that $x\uplus (p\times e_i)$ contains $x$ as a subinstance. This fact will be necessary later in the proof.

We claim
that $x\uplus (p\times e_i)\nrightarrow e_j$ for every $j\in [n]$. The case $j=i$ follows
from $x\nrightarrow e_i$. The case $j\neq i$ follows from the fact
that $e_i$ is non-subsumed. Also, since $x\rightarrow p$ and $(p\times e_i)\rightarrow p$ we have that
$x\uplus (p\times e_i)\rightarrow p$. By  (1) it follows that there exists  $q$ satisfying $q\rightarrow x\uplus (p\times e_i)$ and $q\nrightarrow e_j$ for every $j\in [n]$.

Let $h\colon q\rightarrow x\uplus (p\times e_i)$. We can assume that $h$ is injective since otherwise
we could replace $q$ by $q_{\alpha}$ where $\alpha$ is the equivalence relation that partitions $\dom(q)$ according to the image of $h$. This implies, in particular, that $\gamma(q)=\delta$. To simplify notation we will assume further that $q$ is a subinstance of $x\uplus (p\times e_i)$, i.e, $q$ has been just obtained by (possibly) removing facts from $x\uplus (p\times e_i)$. 

Let $q_j, j\in J$ be the connected components of $q$ and let $t=\biguplus_{j\in J'} q_j$ where
$j\in J'$ if $q_j\nrightarrow e_i$. We note that by definition $t\nrightarrow e_i$.

For every $j\in J'$, $q_j\nrightarrow (p\times e_i)$ and, since $q_j$ is connected, we can conclude that $q_j$ is a subinstance of $x$ (here we are using implicitly $x\uplus(p\times e_i)$ contains $x$ as a subinstance). Hence, $t\rightarrow x$.

Finally, we have $\delta\subseteq\gamma(t)$. Consequently $t'\rightarrow x'$ and $t'\in F'_e$, completing the proof.
\end{proof}

\subsection*{The case where $D$ contains a single example}

We shall now deal with the case where $D$ contains a single example. To this end, we adapt the techniques introduced in \cite{Briceno2021:dismantlability} to a broader setting, since the setup in \cite{Briceno2021:dismantlability} did not consider distinguished elements and, more importantly, did not include the relativized version considered here. In addition, the proof given here is more streamlined as, unlike in \cite{Briceno2021:dismantlability}, it does not go via mixing properties.

 We shall need to introduce a few extra definitions. Let $A$ be an instance. We define a {\em walk} in  $A$ as a walk in its incidence graph that starts and finishes at values from $\dom(A)$. That is, a walk $\rho$ in $A$ is a sequence
$$a_0,R_1(\ba_1),a_1,\dots,a_{n-1},R_n(\ba_n),a_n$$
for some $n \geq 0$, such that, for all $1 \leq \ell \leq n$,
\begin{itemize}
\item $R_\ell(\ba_{\ell})$ is a fact of $A$, and 
\item $a_{\ell-1},a_{\ell}\in\{\ba_{\ell}\}$.
\end{itemize}

In this case, we will say that $a_0$ and $a_n$ are the starting and ending point of $\rho$, and that the {\em length} of the walk $\rho$ is $n$. The {\em distance} between two values is defined to be the smallest length among all the walks that join them. 
The \emph{diameter} of a connected instance is the maximum distance of any pair of its values,
while the diameter of an instance with multiple connected components is the maximum of the diameters of its components.

Let $a,b$ be values from $\dom(A)$. We say that $b$ {\em dominates} $a$ (in $A$) if for every fact $R(a_1,\dots,a_r)$
in $A$ and for every $i\in[r]$ with $a_i = a$, we also have that the fact $R(a_1,\dots,a_{i-1},b,a_{i+1},\dots,a_r)$ belongs to $A$. Additionally, if $a\neq b$, then the instance $A'$ obtained from $A$ by removing $a$ and all the facts in which $a$ participates is said to be obtained from $A$ {\em by folding} $a$. 

A sequence of instances $A_0,\dots,A_{\ell}$ is a {\em dismantling sequence} if for every $0\leq j<\ell$, $A_{j+1}$ has been obtained from $A_j$ by folding some value $a_j$ dominated in $A_j$. In this case, we say that $A_0$ {\em dismantles to} $A_{\ell}$.

In what follows, let $(P,\bp)$, $(E,\be)$ be pointed instances over a
common schema $\sigma$. Consider the new schema
$\sigma\subseteq\overline{\sigma}$ containing, in addition, a new
unary relation symbol
$R_p$ for each $p\in\dom(P)$ and consider the instances $\overline{P}$ and $\overline{E}$ over schema $\overline{\sigma}$ defined as 
$\overline{P}=P\cup\{R_p(p) \mid p\in\dom(P)\}$ and $\overline{E}=E\cup\{R_p(e) \mid p\in\dom(P), e\in\dom(E)\}$.

Let $u=\langle (p_1,d_1),(p_2,d_2)\rangle\in\dom(P\times E)^2$. We shall use $\pi_i u$ to denote the $i$th projection $(p_i,d_i)$, $i=1,2$ of $u$. We say
that $u$ is $P$-{\em diagonal} if $p_1=p_2$. If additionally, $a_1=a_2$ then we say that $u$ is {\em diagonal}. The \emph{symmetric pair} of $u$ is the value
$\langle (p_2,d_2),(p_1,d_1)\rangle$. 

For every subinstance $\overline{I}$ of 
$(\overline{P}\times \overline{E})^2$ we shall use $\diag_P(\overline{I})$ (resp. $\diag(\overline{I})$) to denote the subinstance of $\overline{I}$
induced by its $P$-diagonal (resp.\ diagonal) values.

An \emph{endomorphism} is a homomorphism from a instance
into itself. 
A \emph{retraction} is an endomorphism $h$ with the property
that $h(x)=x$ for every $x$ in the range of $h$. 

We will say that a pointed instance $(A,\ba)$ is a {\em critical obstruction of $(E,\be)$ relative to $(P,\bp)$}, if $(A,\ba)\rightarrow (P,\bp)$, $(A,\ba)\nrightarrow (E,\be)$, and $(A'\ba)\rightarrow (E,\be)$ for
any $A'\subsetneq A$. It is easy to see that critical obstructions are always connected.

\begin{lemma}\label{le:characterization}
Let $(P,\bp)$ and $(E,\be)$ be pointed instances.
The following are equivalent:
\begin{enumerate}
\item 
There exists a retraction from $(\overline{P},\bp)\times (\overline{E},\be)$ to some subinstance $(\overline{I},\bi)$ such that $\diag_P(\overline{I}^2)$ dismantles to $\diag(\overline{I}^2)$. 
\item Every critical obstruction $(A,\ba)$ for $(E,\be)$ relative to $(P,\bp)$ has diameter at most $m= |\dom(P)|\cdot|\dom(E)|^2+2$.
\item There are finitely many (modulo isomorphism) critical obstructions of $(E,\be)$ relative to $(P,\bp)$. 
More specifically,
every critical obstruction has size
at most $2^{O(|\dom(P)|\cdot|\dom(E)|^2\cdot \log(|\dom(E)|))}$.
\item 
There exists $F$ such that
$(F,\{(E,\be)\})$ is a generalized duality relative to $(P,\bp)$.
\end{enumerate}
\end{lemma}

\begin{proof}
 $(3)\Leftrightarrow (4)$: The equivalence $(3)\Leftrightarrow (4)$ is immediate. Indeed, in the direction
from $(3)$ to $(4)$, we can set $F$ to be the set of critical obstructions, while
in the direction from $(4)$ to $(3)$, it is easy to see that the size of each 
critical obstruction is bounded by the size of the instances in $F$.

$(2)\Rightarrow (3)$. This proof is an adaptation of \cite{Larose2007:characterization}. Let $(A,\ba)$ be a critical obstruction of $(E,\be)$ relative to $(P,\bp)$. We shall
use the sparse incomparability lemma (SIL) \cite{NesetrilR89} (see also Theorem 5 in \cite{FV98}). However, since SIL was originally only proved for instances without constants we need to do some adjustments. By the \emph{girth} of an instance $A$, we will mean the
length of the shortest cycle in the incidence graph of $A$, where
the length is measured by the number of facts the lie on the cycle. If the incidence graph is acyclic, the instance is said to have 
girth $\infty$.
Let $\ba=(a_1,\dots,a_k)$ and associate to $(A,\ba)$ the instance $A'$ obtained from $A$ by adding facts $R_i(a_i)$ where $R_i,i\in [k]$ are new relation symbols. We define similarly $E'$ and $P'$. Note that 
$A'\rightarrow P'$ and $A'\nrightarrow E'$. Then, according to SIL there is an instance $B'$ with girth greater than $m$ satisfying $B'\rightarrow A'$ and $B'\nrightarrow E'$. 

Consider the pointed instance $(B,\bb)$ obtained from $B'$ in the following way. For every $i\in [k]$ we remove all facts with relation symbol $R_i$ and glue all the values occurring in them into a single value, which then we place in the $i$th coordinate of $\bb$.  
Clearly, we have 
$(B,\bb)\rightarrow (A,\ba)$ (and, hence, $(B,\bb)\rightarrow (P,\bp)$) and $(B,\bb)\nrightarrow (E,\be)$. Further, remove facts and values from $(B,\bb)$ until becomes a critical obstruction of $(E,\be)$ relative to $(P,\bp)$. Note that by assumption $B$ has diameter at most $m$. 

Now, consider the subinstance, $C$, of $B$ induced by its non-distinguished elements.  By the minimality of $B$, $C$ must be connected.
Note that $C$ is a subinstance of $B'$ as well and, hence, it has  girth larger than $m$. Since $B$ has diameter at most $m$ it follows that $C$ is a tree (i.e., an acyclic connected instance). We claim that every value $c$ of $C$ appears in at most $n_E$ facts where $n_E$ is the number of values in $E$. 
Indeed, let $f_1,\dots,f_r$ be the facts in which $c$ participates. For every 
$I\subseteq [r]$, let $C_I$ be the maximal subinstance of $C$ containing all
the facts $f_i,i\in I$ and none of the facts $f_i,i\in[r]\setminus I$, let $B_I$ be the subinstance of $B$ induced by $C_I\cup\{\bb\}$ and let 
$$S_I=\{h(c) \mid h \text{ is an homomorphism from $(B_I,\bb)$ to $(E,\be)$}\}$$

We have that $S_{[1]},S_{[2]},\dots,S_{[k]}$ are all
different since, otherwise, say $S_{[i-1]}=S_{[i]}$, then $S_{[k]\setminus i}=\emptyset$ contradicting the minimality of $(B,\bb)$.

Since both the diameter and the branching of $C$ are bounded it follows that there is a bound (depending only on $m$ and $e$) on the number of values of $C$ (and, hence, of $B$). 
Indeed, $|\dom(B)| \leq 2^{O(|\dom(P)|\cdot|\dom(E)|^2\cdot \log(|\dom(E)|))}$.
Since $(A,\ba)$ is critical and admits an homomorphism from $(B,\bb)$ it follows that the size of the domain of $A$ is not larger than the size of the domain of $B$.

$(1)\Rightarrow (2)$
Assume that (1) holds. Let $\overline{J}=\diag_P(\overline{I}^2)$ and 
let $\overline{J}_0,\overline{J}_1,\dots,\overline{J}_n$ be a dismantling sequence where $\overline{J}_0=\overline{J}$ and $\overline{J}_n=\diag(\overline{J})$. For every $i\in[n]$ there is a natural retraction $s_i\colon \overline{J}_{i-1}\rightarrow \overline{J}_i$ that sends the folded value $a\in \dom(\overline{J}_{i-1})\setminus\dom(\overline{J}_i)$ to a value in $\dom(\overline{J}_i)$ that dominates it. 

We shall use $I$, $J$, $J_0,\dots,J_n$ to denote be the instance on schema  $\sigma$ 
obtained by removing all facts with relation symbol $R_p$, $p\in P$ from $\overline{I}$, $\overline{J}$, $\overline{J}_0,\dots,\overline{J}_n$ respectively.

We claim that for every critical obstruction $(A,\ba)$ for $(E,\be)$ relative to $(P,\bp)$ the diameter of $A$ is bounded above by $m=n+2\leq |\dom(P)|\cdot|\dom(E)|^2+2$.

Towards a contradiction, let $(A,\ba)$ contradicting the claim. Since the diameter of $A$ is larger than $2n+2$ 
it follows that there exists two facts $f_1=R_1(\ba_1),f_2=R(\ba_2)$ in $A$ such that $N_n(\{\ba_1\})\cap N_n(\{\ba_2\})=\emptyset$
where for every $X\subseteq\dom(A)$ and $i\geq 0$, $N_i(X)$ 
denotes the set of all values from $A$ that are at distance at most $i$ from some value in $X$.

Let $A_i$, $i=1,2$ be the instance obtained removing fact $f_i$ from $A$. It follows that there are homomorphisms
$g_i\colon(A_i,\ba)\rightarrow (E,\be)$, $i=1,2$. 
Let $v\colon(A,\ba)\rightarrow (P,\bp)$. Hence, for every $i=1,2$, mapping $a\mapsto(v(a),g_i(a))$ defines an homomorphism from $A_i$ to $P\times E$.
 If we let $h_i=u\cdot (v,g_i)$ where $u\colon (\overline{P},\bp)\times (\overline{E},\be)\rightarrow(\overline{I},\bi)$ is the retraction guaranteed to exist from (1), we have $h_i\colon A_i\rightarrow I$.
 
 Let $B$ be the
subinstance of $A$ obtained by removing both $f_1$ and $f_2$ and let $h=(h_1,h_2)\colon B\rightarrow I^2$.
Also, 
note that due to the facts with relation symbol $R_p, p\in\dom(P)$ added in $\overline{P}\times\overline{E}$ it follows that the image of $h$ is necessarily in $\dom(J)$, and, hence $h\colon B\rightarrow J$. Since $g_1(\ba)=g_2(\ba)=\be$ it follows that $h(\ba)$ contains only diagonal values.

Let $r_0,\dots,r_n \colon \dom(B)\rightarrow\dom(J)$ be the sequence of mappings where $r_0=h$ and $r_i(b)$, $b\in \dom(B),i\in[n]$, is defined as follows:
$$r_i(b)=\left\{
\begin{array}{ll}
r_{i-1}(b) & \text{if } b\in N_i(\{\ba_1\})\cup N_i(\{\ba_2\}) \\
s_i\cdot r_{i-1}(b) & \text{otherwise}
\end{array}
\right.
$$

It follows that $r_i\colon B\rightarrow J$ for every $i\leq n$. 
Further, since $N_n(\{\ba_1\})\cap N_n(\{\ba_2\})=\emptyset$ it follows that there is no fact in $B$
 that contains at the same time at least one value from $N_{n-1}(\{\ba_1\})$ and at least one value from $N_{n-1}(\{\ba_2\})$. In addition $r_n(b)$ is a diagonal value whenever
$b\not\in N_n(\{\ba_1\})\cup N_n(\{\ba_2\})$. Consequently the mapping $z$, defined as 
$$z(b)=\left\{\begin{array}{ll}
\pi_1\cdot r_n(b) & \text{if } b\in N_{n-1}(\{\ba_1\}) \\
\pi_2\cdot r_n(b) & \text{otherwise} 
\end{array}
\right.$$ 
is an homomorphism from $B$ to $P\times E$. Note that $r_i$, $i\leq n$ agrees with $h$ on $\{\ba_1\}\cup\{\ba_2\}$. This implies that $z$ agrees with
$h_i$ on $\{\ba_i\}$ for $i=1,2$. In consequence, mapping $z$ preserves facts $f_1$ and $f_2$ as well. Further, since $h(\ba)$ contains only diagonal values it follows that $r_n$ agrees with $h$ on $\ba$, and, hence $z(\ba)=\be$.
Putting all together we have $z\colon(A,\ba)\rightarrow (E,\bb)$, contradicting our assumptions on $(A,\ba)$.

$(3)\Rightarrow (1)$
Assume that (1) does not hold. Let $(\overline{I},\bi)$ be the core of $(\overline{P},\bp)\times (\overline{E},\be)$
and let $\overline{J}=\diag_P(\overline{I}^2)$. We define $\overline{K}$ be any instance obtained from a distmantling sequence starting at $\overline{J}$ until no further distmantling if possible. Since $\overline{J}$ does not distmantle to its diagonal it $\dom(\overline{K})$ contains some non-diagonal value.
It is easy to see (see \cite{Briceno2021:dismantlability}, Remark 3.8)  that such distmantling sequence can be done by folding symmetric pairs
so that $\overline{K}$ contains the symmetric pair of any of its values.

Define $I$, $J$ and $K$ to be obtained by removing all the facts with
relation symbol $R_p, p\in\dom(P)$ from $\overline{I}$, $\overline{J}$, and $\overline{K}$ respectively. We shall show that for every $m\geq 0$ there is a critical obstruction of $(E,\be)$ relative to $(P,\bp)$ with at least $m$ values.

Let $k_0$ be a non-diagonal value in $K$. We construct an instance $T$ (without distinguished elements) in the following way. The domain of $T$ consists of all the walks in $K$ of length $m+1$ starting at $k_0$. For every fact $f=R(k_1,\dots,k_r)$ in $K$, for every $i\in[r]$, and for every walk $\rho$ ending at $k_i$, we include in $T$ the fact $R(\rho_1,\dots,\rho_{i-1},\rho,\rho_{i+1},\dots,\rho_r)$, where $\rho_j$, $j \neq i$, is the walk obtained from $\rho$ by extending it with $f,k_j$. The {\em root} of $T$ is the walk $\rho_0$ of length $0$ consisting only of $k_0$ and the {\em leafs} of $T$ are the walks of length $m+1$.

It is immediate that the map $u$ sending every walk $\rho$ to its last node is an homomorphism from $T$ to $(P\times E)$. Also let $v\colon J\rightarrow P$ be the mapping sending value $\langle (p,a_1),(p,a_2)\rangle$ to $p$.

\begin{claim}
\label{cl:tree}
Let $g\colon T\rightarrow K$ be any homomorphism that agrees with $u$ on the leaves of $T$. Further, assume that $v\circ g=v\circ u$. Then we have that $g=u$. 
\end{claim}
\begin{proof}[Proof of claim]
This follows by induction by showing that for every $\ell\in[m+1]$, if $g$ agrees with $u$ for every walk of length $\ell$ it also agrees for every walk  of length $\ell-1$. In particular, let $\rho\in T$ of length $\ell-1$. It follows from the definition of $T$ and the inductive hypothesis that $g(\rho)$ dominates $u(\rho)$ in $K$. Additionally, since $v\circ g(\rho)=v\circ u(\rho)$ it follows that $g(\rho)$ dominates $u(\rho)$ also in $\overline{K}$. Since $\overline{K}$ cannot be further dismantled it follows that $g(\rho)=u(\rho)$ as desired. 
\end{proof}

Let $(A_i,\bi_{\alpha}),i=1,2$ be the pointed instances $(T\cup I,\bi)_{\alpha}$ where
$\alpha$ can be informally described as gluing every leaf $\rho\in \dom(T)$ with value $\pi_i\circ u(\rho)$ in $\dom(I)$. We also glue the root $\rho_0$ to some leaf $\rho'$ such that $u(\rho_0)$ and
$u(\rho')$ are symmetric (here we use the fact that  $\dom(\overline{K})$ is closed under symmetric pairs).

We note that all the values in $\dom(I)$ have been glued to some value in $T$. This allows to simplify a bit our notation as we can naturally extend $u$ to $\dom(A_1)$ (and $\dom(A_2)$) by defining $u(a)$ as $u(\rho)$ for any walk $\rho$ contained in $\alpha$-class $a$.

\begin{claim}
$(A_1,\bi_{\alpha})\nrightarrow (E,\be)$ or $(A_2,\bi_{\alpha})\nrightarrow (E,\be)$.
\end{claim}
\begin{proof}[Proof of claim]
Assume towards a contradiction that there are homomorphisms $f_i\colon(A_i,\bi_{\alpha})\rightarrow (E,\be)$ for $i=1,2$. Then, mapping $g_1(a)=(v\cdot u(a),f_1(a))$ defines an homomorphism from $(A_1,\bi_{\alpha})$
to $(P,\bp)\times (E,\be)$. Let 
$z\colon (\overline{P},\bp)\times (\overline{E},\be)\rightarrow (\overline{I},\bi)$ be any retraction and let $h_1=z^n \circ g_1$ for some $n\geq 1$ to be chosen later. Note that $h_1 \colon (A_1,\bi_{\alpha})\rightarrow (I,\bi)$. Note that map $i\mapsto i_{\alpha}$ defines an isomorphism from 
$(I,\bi)$ to 
$(A_1,\bi_{\alpha})$. We might abuse slightly notation and refer to the copy or $(I,\bi)$ in $(A_1,\bi_{\alpha})$
simply as $(I,\bi)$.
Then,
the restriction of $h_1$ to $\dom(I)$ must be a bijection since otherwise this would contradict the fact that $(\overline{I},\bi)$ is a core. Hence we can choose $n$ so that
$h_1$ acts as the identity on $(I,\bi)$. Note that $v\cdot h_1=v\cdot u$

We can similarly show that there exists $h_2\colon (A_2,\bi_{\alpha})\rightarrow (\overline{I},\bi)$ that acts as the identity on $(\overline{I},\bi)$ and $v\cdot h_2=v\cdot u$. It follows that the mapping $h(\rho)=(h_1(\rho_{\alpha}),h_2(\rho_{\alpha}))$ defines
an homomorphism from $T$ to $J$ satisfying $v\cdot h=v\cdot u$. Note also that $h$ acts as the identity on the leaves of $T$, which implies that, indeed, $h\colon T\rightarrow K$. Hence
Claim~\ref{cl:tree} implies that $h=u$.

Since $\alpha_0$ and $\alpha'$ have been glued in both $A_1$ and $A_2$ it follows that $h$ agrees on $\alpha_0$ and $\alpha'$.
However this is impossible since we have chosen $u(\alpha_0)$ to be non-diagonal and $u(\alpha')$ and 
$u(\alpha_0)$ are symmetric pairs.
\end{proof}

Assume that $(A_1,\bi_{\alpha})\nrightarrow (E,\be)$ (the case $(A_2,\bi_{\alpha})\nrightarrow (E,\be)$ is analogous). Then there exists some $B\subseteq A_1$
such that $(B,\bi_{\alpha})$ is a critical obstruction for $(E,\be)$ relative to $(P,\bp)$. To conclude our proof it is only necessary to note that $B$ has at least $m$ values as a consequence of the following claim.

\begin{claim}
For every $\ell=1,\dots,m$, $B$ contains at least one value of $T$ of level $\ell$.
\end{claim}
\begin{proof}[Proof of claim]
Assume that some $\ell$ falsifies the claim. Then the mapping $g$ defined as follows is an homomorphism from $(B,\bi_{\alpha})$ to $(E,\be)$, a contradiction. Let $b\in B$ (recall that $b$ is a $\alpha$-class). 
If $b$ contains some value $(a,e)$ in $I$ then define $g(b)$ to be $e$. If $b$ contains some $\rho$ in $T$ we define $g(\rho)$ in the following way: Let $i$ be the length of $\rho$ and let $u(\rho)=\langle (a,e_1),(a,e_2)\rangle$ be its last value. Then $g(\rho)$ is defined to be $e_2$ if $i<\ell$ and $e_1$ if $i>\ell$.  It is immediate to see that $g$ is well defined and that defines an homomorphism.
\end{proof}

This concludes the proof of Lemma~\ref{le:characterization}.
\end{proof}

\subsection*{Putting everything together}

\begin{proof}[Proof of Theorem~\ref{thm:relative-dualities}]

(1)
By combining Lemmas~\ref{le:reductiontosingleexample} and~\ref{le:characterization} we obtain an NP algorithm to decide whether there exists $F$ such that $(F,D)$ is a generalized duality relative to $p$:
for every $e\in D$, we non-deterministically verify that there is some homomorphism $(p\times e)\rightarrow e'$ for some $e'\in D$ different than $e$ or that condition (1) in Lemma~\ref{le:characterization} is satisfied for $e$ and $p$. This
condition can be tested in NP by guessing the retraction.

The NP lower bound holds already in the non-relativized case without designated elements (i.e., where $k=0$ and $(P,\bp)$ is the pointed instance 
containing all possible facts over a single-element domain)
\cite{Larose2007:characterization}.

(2) By Lemma~\ref{le:characterization}, in the single-example case, when the set of critical obstructions is finite, then, in fact, each critical obstruction has domain size at most $2^{O(|\dom(P)|\cdot|\dom(E)|^2\cdot \log(|\dom(E)|))}$. Hence, $F$ can be constructed to 
consist of instances of this size. In the general case with a set of examples $D$, inspection of 
the proof of Lemma~\ref{le:reductiontosingleexample} shows that we take 
 $F=\{e'_1\uplus\cdots\uplus e'_n \mid e'_i\in F_{e_i}, i\in [n]\}$, where $\{e_1,\dots,e_n\}$ be the set of all non-subsumed examples in $D$. It follows that each member of $F$ 
 has size at most $2^{O(|\dom(P)|\cdot||\dom(D)||^2\cdot \log(||\dom(D)||))}$,
 as claimed.
\end{proof}

\section{Missing proofs from section~\ref{sec:tree-cq}}



%
%

\subsection{Arbitrary Fitting Tree CQs}
\label{app:tree-degree-bound}

\lemdegreebnd*
\begin{proof}
  Take any tree CQ $q(x_0)$ that is a fitting for $(E^+,E^-)$.  We
  consider the answer variable $x_0$ to be the root of $q$, imposing a
  direction on it and allowing us to speak about successors,
  predecessors, etc. %

  Assume that $q$ contains a variable $x$ with more than $||E^-||$
  successors $y_1,\dots,y_m$. For $1 \leq i \leq m$, let $q|_x^i$
  denote the tree CQ obtained from the subquery of $q$ rooted at $x$
  by deleting all successors $y_i,\dots,y_m$ and the subtrees below
  them. We define $S_i$ to be the set of all values $a$ such that for
  some $e \in E^-$, there is a homomorphism $h$ from $q|_x^i$ to $e$
  with $h(x)=a$. Clearly, $S_1 \supseteq S_2 \cdots$. Consequently,
  $S_j=S_{j+1}$ for some $j \leq ||E^-||$. Let $q'$ be obtained from
  $q$ by removing the successor $y_{j+1}$ of $x$ and the
  subtree below it.
  
  We show in the following that $q'$ is a fitting for $(E^+,E^-)$. It
  must then be weakly/strongly most-general resp.\ unique if $q$ is
  since $q'$ may only be more general than $q$ (and $q'$ must be
  equivalent to $q$ if $q$ is a unique fitting). Likewise, $q'$ cannot
  have more variables than $q$ and thus if $q$ had a minimal number
  of variables, then so does $q'$. In fact, $q'$ has less variables
  than $q$, so showing that $q'$ is a fitting for $(E^+,E^-)$
  establishes a contradiction to our assumption that $q$ has
  degree exceeding $||E^-||$.

  \smallskip It is also clear that $q'$ fits all positive
  examples. For the negative examples, assume to the contrary that
  there is an $(I,c) \in E^-$ and a homomorphism $h'$ from $q'$ to $I$
  with $h'(x_0)=c$. We can construct from $h'$ a homomorphism $h$ from
  $q$ to $I$ with $h(x_0)=c$, yielding a contradiction. Clearly, $h'$
  is also a homomorphism from $q|^{j}_x$ to $I$ and since
  $S_j=S_{j+1}$, this means that we also find a homomorphism $g$ from
  $q|^{j+1}_x$ to $I$ such that $h'(x)=g(x)$. We can plug $g$ into $h$
  in an obvious way to find the desired homomorphism $h$ from $q$ to
  $I$ with $h(x_0)=c$.
  
  If we apply the above argument repeatedly, we thus find a fitting of
  degree at most $||E^-||$ and this fitting is weakly/strongly
  most-general and has a minimal number of variables if this was the
  case for the original fitting.
\end{proof}

\subsection{Most-Specific Fitting Tree CQ}
\label{app:ms-fit-tree-cq}

\lemdegreebndtwo*
\begin{proof}
  The case where we disregard the number of variables is immediate.
  Proposition~\ref{prop:most-specific-tree-initial-piece} tells us that if
  there exists a most-specific fitting $q$, then some initial piece
  $p$ of the unraveling of $\Pi_{e \in E^+}(e)$ is a fitting of the
  same kind. Clearly, the degree of $p$ is bounded by $2^{||E^+||}$
  and we are done.

  Let us now add the requirement that the number of variables be
  minimal. It suffices to show that $q(x)$ has an injective
  homomorphism $h$ to $p(x)$ with $h(x)=x$ because this means that $q$
  is actually a subquery of $p$ (can be obtained from it by dropping
  atoms), and thus the degree of $q$ cannot be larger than that of
  $p$. Since $p$ and $q$ are simulation equivalent, we find
  homomorphisms $h_1$ from $p$ to $q$ and $h_2$ from $q$ to $p$, both
  the identity on $x$. The composition $h_1 \circ h_2$ is a
  homomorphism from $q$ to $q$ and must be surjective as otherwise it
  identifies a strict subquery of $q$ (with fewer variables!) that is
  homomorphically equivalent to $q$, and thus simulation equivalent.
  But the composition can only be surjective if $h_2$ is injective, so
  $h_2$ is the desired injective homomorphism from $q$ to $p$.

  For the `moreover part', we note that we have already shown above that
  a most-specific fitting $q$ with a minimal number of variables is
  isomorphic to a subtree $p'$ of some initial piece $p$ of the
  unraveling $U$ of $\Pi_{e \in E^+}(e)$, but that subtree $p'$ is an
  initial piece of $U$ itself, and thus it remains to prove that $p'$
  is complete. However, any incomplete initial piece $p'$ of $U$ does
  not admit a simulation from $U$ as otherwise the simulation would
  witness completeness of $p'$. Consequently, $p'$ must be complete as
  by Proposition~\ref{prop:most-specific} it is simulation equivalent
  to $U$.
\end{proof}

\subsection{Weakly Most-General Fitting Tree CQs}
\label{app:wmg-tree-cq}

\thmwmgexistencetree*

Let $(E^+,E^-)$ be a collection of labeled examples over
schema~$\mathcal{S}$ and set $m := ||E^-||$. We aim to construct a
TWAPA $\mathfrak{A}$ with polynomially many states that accepts
exactly the weakly most-general tree CQ fittings for $(E^+,E^-)$ which
have degree at most $m$.  The latter restriction is justified by
Lemma~\ref{lem:degreebnd}. To prove Theorem~\ref{thm:wmg-existence-tree}, it then
remains to invoke (once again) Point~(1) and~(5) of
Theorem~\ref{thm:TWAPAstuff}.

By Proposition~\ref{prop:weakly-most-general-tree-cq}, we may
construct $\mathfrak{A}$ as the intersection of two TWAPAs
$\mathfrak{A}_1$ and $\mathfrak{A}_2$ where $\mathfrak{A}_1$ verifies
that the $q$ fits $(E^+,E^-)$ and $\mathfrak{A}_2$ that every element
of the frontier $\mathcal{F}$ for $q$ w.r.t.\ tree CQs simulates to an
example in $E^-$. For $\mathfrak{A}_1$, we can use the TWAPA from
the proof of Theorem~\ref{thm:sizetreeCQarbupper}.

We next describe the frontier construction. %

\medskip
\noindent
\textbf{Step~1: Generalize.}  For each variable $x$ in $q$,
define a set $\mathcal{F}_0(x)$ that contains all tree CQs which can be
obtained by starting with the subquery of $q$ rooted at $x$ and then doing one of the
following:
\begin{enumerate}

\item choose an atom $A(x)$ and remove it;

\item choose a successor $y$ of $x$, with $R(x,y) \in q_x$, and then
  \begin{enumerate}
  \item 
  remove $R(x,y)$ and the subtree 
  rooted at $y$ and

\item for each $q'(y) \in \mathcal{F}_0(y)$, add a disjoint copy
  $\widehat q'$ of $q'$ and the role atom $R(x,y'')$ with $y''$ the
  copy of $y$ in $\widehat q'$.
  
  \end{enumerate}
  \end{enumerate}
  Every variable $x$ in the resulting tree CQs may be associated in an obvious
  way with a variable from $q$ that it derives from. We denote that original
  variable with  $x^\downarrow$.

  \smallskip
\noindent\textbf{Step~2: Compensate.}
We construct the frontier $\mathcal{F}$ of $q(x_0)$ by including, for
each $p \in \mathcal{F}_0(x_0)$, the tree CQ obtained from $p$ by
adding, for every atom $R(x,y)$ in $p$ directed away from the root, an
atom $R(z,y)$, $z$ a fresh variable, as well as a disjoint copy
$\widehat q$ of $q$ and glue the copy of $x^\downarrow$ in
$\widehat q$ to~$z$.

It was shown in \cite{tCD2022:conjunctive} that $\mathcal{F}$ is indeed a frontier of $q$ w.r.t.\ all tree CQs.

\begin{lemma}[Theorem 3.18 in \cite{tCD2022:conjunctive}]
 $\mathcal{F}$ is a frontier of $q$ w.r.t.\ all tree CQs.
\end{lemma}

\begin{example}
Consider the schema that consists of a binary relation $R$ and two unary relations $P$ and $Q$.
Let $q$ be the tree CQ $q(x)\colondash P(x) \land R(x, y) \land Q(y)$. Then $\mathcal{F}_0(x)$ consists of the tree CQs $q_1(x)\colondash  R(x, y) \land Q(y)$ and $q_2(x)\colondash P(x) \land R(x, y)$. Consequently,
$\mathcal{F}$ contains the two tree CQs $q'_1(x) \coloneq R(x, y) \land Q(y) \land R(x', y) \land P(x') \land R(x', y') \land Q(y')$ and $q'_2(x)\colondash P(x) \land R(x, y) \land R(x', y) \land P(x') \land R(x', y') \land Q(y')$. 
\end{example}

\medskip

We now describe how $\mathfrak{A}_2$ can be constructed.

Let $E^-=\{(I_1,\widehat c_1),\dots,(I_{n},\widehat c_{n})\}$. We assume w.l.o.g.\ that
the negative counterexamples have pairwise disjoint domains.  Set
$I=I_1 \cup \cdots \cup I_{n}$ and $\mn{adom}=\mn{adom}(I)$.
The TWAPA $\mathfrak{A}_2$ starts in state $s_0$ and universally
branches over all queries in the frontier, choosing for each of them a
negative example that it simulates into. Since the queries in the
frontier are tree CQs, we can actually verify the existence of a
homomorphism in place of a simulation (which we consider slightly
more intuitive).

Assume that the input tree represents the tree CQ $q(x_0)$. Then by
construction, $\mathcal{F}$ contains a query $q_A$ for every atom
$A(x_0) \in q$ and a query $q_y$ for every successor $y$ of $x$ in
$q$. We set for all $\sigma \in \Gamma$:
 $$
 \delta(s_0,\sigma)=\bigwedge_{A \in \sigma}  \bigvee_{1 \leq \ell
   \leq n} s^A_{\widehat c_\ell}
 \wedge
 \bigwedge_{1 \leq i \leq m} ([i] \bot \vee \bigvee_{1 \leq \ell \leq n} s^i_{\widehat c_\ell}).
 $$
 So the TWAPA is now located at the root of the input tree, in state
 $s^A_{\widehat c_\ell}$ to verify that the query $q_A$ in
 $\mathcal{F}$ maps to ${\widehat c_\ell}$, and in state
 $s^i_{\widehat c_\ell}$ to verify that the query $q_y$ in
 $\mathcal{F}$, where $y$ is the variable in $q$ represented by the
 $i$-th successor of the root in the input tree, maps to $\widehat c_\ell$.

The queries $q_A$ are easy to deal with. We can essentially use the
same TWAPA as in the proof of Theorem~\ref{thm:sizetreeCQarbupper},
except that we must ignore the atom $A(x_0)$ and also take into account the
subqueries added in the compensate step. 
More precisely, for all $c \in \mn{adom}$ and $\sigma \in \Gamma$
we set
$$
\delta(s^A_c,\sigma) = \displaystyle\bigwedge_{1 \leq i \leq m} \ \bigvee_{R \ \mathcal{S}\text{-role and }
  R(c,d) \in I} [i] s_{d,R}
$$
if $B \in \sigma \setminus \{ A\}$ implies $B(c) \in I$, and
$\delta(s^A_c,\sigma) = \text{false}$ otherwise. For all $c \in \mn{adom}$, $\mathcal{S}$-roles
$R$, and $\sigma \in \Gamma$ we also set
  $$
\delta(s_{c,R},\sigma) = \displaystyle t^0_{c,R} \wedge \bigwedge_{1 \leq i \leq m} \ \bigvee_{R' \ \mathcal{S}\text{-role and }
  R'(c,d) \in I} [i] s_{d,R'}
$$
if $R \in \sigma$ and $A \in \sigma$ implies $A(c) \in I$, and
$\delta(s_{c,R},\sigma) = \text{false}$ otherwise. The state $t^0_{c,R}$
is for verifying the subqueries added in the compensate step. For all
$\sigma \in \Gamma$, set 
$$
\delta(t^0_{c,R},\sigma) = \bigvee_{R' \ \mathcal{S}\text{-role and }
  R(d,c) \in I} \langle -1 \rangle t_{d,R'}
$$
and for all $c \in \mn{adom}$, $\mathcal{S}$-roles $R$, and $\sigma \in
\Gamma$, set
$$
\delta(t_{c,R},\sigma) = s_{c,R} \wedge \bigvee_{R' \ \mathcal{S}\text{-role and }
  R(d,c) \in I} [-1] t_{d,R'} 
$$
if $R \in \sigma$ and $A \in \sigma$ implies $A(c) \in I$, and
$\delta(s_{c,R},\sigma) = \text{false}$ otherwise. 

It remains to deal with the states $s^i_c$. Remember that their
purpose is to verify that the query $q_y$ in $\mathcal{F}$, where $y$
is the variable in $q$ represented by the $i$-th successor of the root
in the input tree, maps to $\widehat c_\ell$. Also recall that in the
generalize step of the construction of $q_y$, the successor $y$ of
$x_0$ is replaced with one successor for each query in
$\mathcal{F}_0(y)$. For $1 \leq i \leq m$ and all $c \in
\mn{adom}$, set
$$
\delta(s^i_c,\sigma)= \displaystyle \bigvee_{R \ \mathcal{S}\text{-role}}
   \langle i \rangle u_{c,R} \wedge
  \mathop{\bigwedge_{1 \leq j \leq m}}_{j \neq i}
  \bigvee_{R \ \mathcal{S}\text{-role and }
  R(c,d) \in I}
[ j ] s_{d,R}$$
State $u_{c,R}$ expresses that variable $y$ of the input tree CQ
$q$ that the TWAPA is currently visiting is replaced with each of the
queries in $\mathcal{F}_0(y)$, and that it is an $R$-successor of
its predecessor, which is mapped to $c$. There is one such query for each 
atom $A(y)$ in $q$ and every successor of $y$ in $q$. We are thus
in a very similar, though not identical, situation as in the
beginning. For all $c \in \mn{adom}$, $\mathcal{S}$-roles $R$, and
$\sigma \in \Gamma$ set
 $$
 \delta(u_{c,R},\sigma)=\bigwedge_{A \in \sigma}  \bigvee_{R(c,d) \in I} s^A_{d}
 \wedge
 \bigwedge_{1 \leq i \leq m} ([i] \bot \vee \bigvee_{R(c,d) \in I} s^i_{d}).
 $$
 if $R \in \sigma$ and  $\delta(u_{c,R},\sigma)=\bot$ otherwise.
 This finishes the construction of the automaton.

\subsection{Basis of Most-General Fitting Tree CQ}
\label{app:bmg-tree-cq}

\lemcritdeg*
\begin{proof}
  The proof is similar to that of Lemma~\ref{lem:degreebnd}. 
  Let $q(x_0)$ be a critical fitting tree CQ for $(E^+,E^-)$.  Assume
  to the contrary of what we want to show that $q$ contains a variable
  $x$ with more than $||E^-||$ successors $y_1,\dots,y_m$. For
  $1 \leq i \leq m$, let $q|_x^i$ denote the tree CQ obtained from the
  subquery of $q$ rooted at $x$ by deleting all successors
  $y_i,\dots,y_m$ and the subtrees below them. We define $S_i$ to be
  the set of all values $a$ such that for some $e \in E^-$, there is a
  homomorphism $h$ from $q|_x^i$ to $e$ with $h(x)=a$. Clearly,
  $S_1 \supseteq S_2 \cdots$. Consequently, $S_j=S_{j+1}$ for some
  $j \leq ||E^-||$. Let $q'$ be obtained from $q$ by removing the
  successor $y_{j+1}$ of $x$ and the subtree below it.
    We can show as in Lemma~\ref{lem:degreebnd} that $q'$ is fitting for
    $(E^+,E^-)$. This, however, contradicts the fact that $q$ is critical.
\end{proof}

\section{The Product Simulation Problem into Trees is ExpTime-hard (Theorem~\ref{thm:simtreeexp})}
\label{app:product-sim-problem}


\thmsimtreeexp*

The proof is by reduction from the product $\downarrow$-simulation
problem, proved ExpTime-hard in~\cite{Funk2019:when}. 
Actually, an inspection of the proof shows that~\cite{Funk2019:when}
achieves something stronger: there is a fixed domain $D$ such
that given as input instances $I_1,\dots,I_n, I_t$ with
$\text{adom}(I_t)=D$, a value $\bar c_s \in \text{adom}(\prod_{1 \leq i \leq n}
I_i)$, and a value $c_t \in \text{adom}(I_t)$, it is 
ExpTime-hard to decide whether $(\prod_{1 \leq i \leq n} I_i,\bar c_s)\preceq^{\downarrow} (I_t,c_t)$.
Note that the target instance $I_t$ need not be a tree and that the
schema is not fixed. Also note that this problem is defined in
terms of $\downarrow$-simulations while the product simulation problem we aim to reduce
to is defined in terms of $\updownarrows$-simulations.

Assume that we are given an input $I_1,\dots,I_n, I_t, \bar c_s,
c_t$ as above. We refer to $I_1,\dots,I_n$ as the source instances and to
$\bar c_s$ as the source value, and to $I_t$ and $c_t$ as the target instance and value. Let $\Sigma$ be the set of relation symbols
used in $I_1,\dots,I_n, I_t$.

Let $\Gamma$ be the schema that contains all unary relations from~$\Sigma$, a
single fresh binary relation $R$, the fresh unary relations $\mn{Out}$ and
$\mn{Impossible}$, as well as fresh unary relations ${\sf From}_c$ and ${\sf
To}_c$ for all
$c \in D$ and fresh unary relations $A_S$ for every binary
relation $S \in \Sigma$. Note that $\Gamma$ is not fixed as
it depends on $\Sigma$ and thus on the input. At the very
end of the proof, we will explain how to rectify this. 

We convert every source instance $I_i$ with $1 \leq i \leq n$, into a
$\Gamma$-instance $I'_i$ by replacing every fact $S(c,c') \in I_i$
with a gadget as shown in Figure~\ref{fig:prodsim-first}. More
precisely, the gadget contains the following facts for all $d \in D$
where all values except $c$ and $c'$ are fresh:
\begin{itemize}

    \item $R(a_{c,c',S,d,1},c)$, $R(a_{c,c',S,d,2},a_{c,c',S,d,1})$,
    $R(a_{c,c',S,d,3},a_{c,c',S,d,2})$, $R(a_{c,c',S,d,3},a_{c,c',S,d,4})$;
    
  \item ${\sf Out}(a_{c,c',S,d,1})$, $A_S(a_{c,c',S,d,3})$,
    ${\sf From}_d(a_{c,c',S,d,4})$;

    \item for all $d' \in D$:
    
      \begin{itemize}
      \item $R(a_{c,c',S,d,4}, a_{c,c',S,d,d',5})$, $R(a_{c,c',S,d,d',5},c')$;

      \item ${\sf To}_{d'}(a_{c,c',S,d,d',5})$;

      \item ${\sf Impossible}(a_{c,c',S,d,d',5})$ if $S(d,d') \notin I_t$.

    \end{itemize}
\end{itemize}
This is illustrated in Figure~\ref{fig:prodsim-first} where we 
show only the `$i$' component of
values $a_{c,c',S,d,i}$ and $a_{c,c',S,d,d',i}$.  Informally, the
labels ${\sf From}_d$ and ${\sf To}_{d'}$ identify the fact $S(d,d')$
in $I_t$ that $S(\bar c,\bar c')$ is mapped to and the predicate
\mn{Impossible} marks impossible such choices. We explain this in more
detail later on. The ${\sf Out}$ labels serve to deal
with the issue that we are reducing from a problem defined in terms of
$\downarrow$-simulations to a problem defined in terms of
$\updownarrows$-simulations. We refer to this as the
\emph{$\downarrow$/$\updownarrows$-issue}.

\begin{figure}
    \centering
    \begin{tikzpicture}[on grid, node distance = 0.8cm and 0.8cm, label distance = -0.1cm]
\tikzset{numbered/.style={font=\footnotesize}}
\tikzset{every label/.style={align=center}}
\node (c) {$c$};
\node (c') [below = 5.1cm of c] {$c'$};

\node(1a) [numbered, below left = of c, label = left:$\textsf{Out}$]{$1$};
\node(1b) [numbered, below right = of c, label = right:$\textsf{Out}$]{$1$};
\node(2a) [numbered, below left = of 1a]{$2$};
\node(2b) [numbered, below right = of 1b]{$2$};
\node(3a) [numbered, below left = of 2a, label = left:$A_S$]{$3$};
\node(3b) [numbered, below right = of 2b, label = right:$A_S$]{$3$};
\node(4a) [numbered, below = of 3a, label = left:$\textsf{From}_{d_1}$]{$4$};
\node(4b) [numbered, below = of 3b, label = left:$\textsf{From}_{d_n}$]{$4$};
\node(5a) [numbered, below left = of 4a, label = left:{$\textsf{To}_{d_1}$ \\ ($\textsf{Imp}$)}]{$5$};
\node(5b) [numbered, below right = of 4a, label = right:{$\textsf{To}_{d_n}$ \\ ($\textsf{Imp}$)}]{$5$};
\node(5c) [numbered, below left = of 4b, label = left:{$\textsf{To}_{d_1}$ \\ ($\textsf{Imp}$)}]{$5$};
\node(5d) [numbered, below right = of 4b, label = right:{$\textsf{To}_{d_n}$ \\ ($\textsf{Imp}$)}]{$5$};
\node(dots5a) [below = of 4a] {$\cdots$};
\node(dots5b) [below = of 4b] {$\cdots$};
\node(dots2) [below =  of c] {$\ldots$};
\node(dots4) [below =2.4cm of c] {$\ldots$};

\draw[->] (1a) to (c);
\draw[->] (1b) to (c);
\draw[->] (2a) to (1a);
\draw[->] (2b) to (1b);
\draw[->] (3a) to (2a);
\draw[->] (3b) to (2b);
\draw[->] (3a) to (4a);
\draw[->] (3b) to (4b);
\draw[->] (4a) to (5a);
\draw[->] (4a) to (5b);
\draw[->] (4b) to (5c);
\draw[->] (4b) to (5d);
\draw[->, bend right] (5a) to (c');
\draw[->, bend right, out=-45] (5b) to (c');
\draw[->, bend left, out=45] (5c) to (c');
\draw[->, bend left] (5d) to (c');

    \end{tikzpicture}
    \caption{Gadget in $I_i'$ replacing  $S(c, c') \in I_i$.}
    \label{fig:prodsim-first}
\end{figure}

The reader may verify that every edge $S(\bar c,\bar c')$ in the
product $\prod_{1 \leq i \leq n} I_i$ is replaced in the product
$\prod_{1 \leq i \leq n} I'_i$ also by a gadget of the form shown in
Figure~\ref{fig:prodsim-first}, with the starting value $c$ now
actually being a tuple $\bar c$ and the ending value $c'$ being a
tuple $\bar c'$. There are in fact some additional paths in the
product gadget, not shown in the figure, that carry no
${\sf From}_{d}$ label and/or no ${\sf To}_{d'}$ label, but these map
homomorphically into the properly labeled paths and can be
disregarded. There is another issue here, namely that two values
$\bar c, \bar c'$ in $\prod_{1 \leq i \leq n} I'_i$ may be connected
by a weakening of the gadget even if there is no corresponding fact
$S(\bar c,\bar c')$ in $\prod_{1 \leq i \leq n} I_i$. In particular,
this happens when for every $i$, the $i$-th component of $\bar c$ is
connected to the $i$-th component of $\bar c'$ in $I_i$, but via different
relations from $\Sigma$.  Note that these weakened gadgets can be
distinguished from the one in Figure~\ref{fig:prodsim-first} as they
will not carry any $A_S$ label. We refer to this as the \emph{ghost
  gadget issue}.

For any instance $I'$ and $c' \in D$, it follows from
Lemma~\ref{lem:unravbasics} that
$(\prod_{1 \leq i \leq n} I'_i,\bar c_s)\preceq^{\updownarrows}
(I',c')$ if and only if there is a homomorphism $h'$ from the unraveling $U'$ of
$\prod_{1 \leq i \leq n} I'_i$ at $\bar c_s$ to $I'$ with
$h(\bar c_s)=c'$. We prefer to think in terms of this latter
presentation.  Let us discuss the shape of $U'$.  Recall that the
values in unravelings are paths. We are most interested in the paths
$p \in \mn{adom}(U')$ that end in a value
$\bar d \in \mn{adom}(\prod_{1 \leq i \leq n} I_i)$.  This
is the case for the root of $U'$, that is, the path of length~1 that
consists of only the source value $\bar c_s$. Let $p$ be a path that
ends in a value $\bar c \in \mn{adom}(\prod_{1 \leq i \leq n}
I_i)$. Then for every fact $S(\bar c,\bar c')$ in
$\prod_{1 \leq i \leq n} I_i$, we find in $U'$ a subtree rooted at $p$
that can be obtained from the gadget in Figure~\ref{fig:prodsim-first}
by duplicating the point $\bar c'$ sufficiently many times so that a
tree is obtained. This subtree contains, for all $d,d' \in I_t$, a
path of length~6 that starts at $p$ and ends at a
$p' \in \mn{adom}(U')$ that in turn ends with a value $\bar c'$. The
first value on the path is labeled with ${\sf Out}$, the fourth with
${\sf From}_d$, and the fifth with ${\sf To}_{d'}$, and
possibly with ${\sf Impossible}$.  Informally, each of the paths
represents the choice for $h'$ to map edge $S(\bar c,\bar c')$ to
$S(d,d') \in I_t$.  Note that there is an \emph{and/or-issue} arising
here: $h'$ needs to map $S(\bar c,\bar c')$ only to a \emph{single}
$S(d,d') \in I_t$, but we have paths for \emph{all} possible choices.

Unsurprisingly, the $\downarrow$/$\updownarrows$-issue also shows up
in $U'$.  There may in fact be other successors of $p$ in~$U'$ than
the ones described above: for every fact $S(\bar c',\bar c)$ in
$\prod_{1 \leq i \leq n} I_i$, there is a gadget in
$\prod_{1 \leq i \leq n} I'_i$ of the form shown in
Figure~\ref{fig:prodsim-first} with $\bar c'$ and $\bar c$ swapped
(that is, the gadget `starts' in $\bar c'$ and `ends' in $\bar c$
rather than the other way around). This leads to undesired successors
of $p$ in $U'$ that are not labeled with $\mn{Out}$, as is the case for the
`desired' successors. The ghost edge issue leads to further undesired successors
of $p$.

We next define a tree $\Gamma$-instance $I'_t$ that replaces $I_t$.
We start with a tree of depth three that branches only at the root
$b_0$ and has one leaf for every pair of values in $I_t$. It contains
the following facts for all $c,c' \in D$:
\begin{itemize}

  \item $A_S(b_0)$ for all binary relations $S \in \Sigma$;
  
    \item 
$
  R(b_0,b_{c,c',1}),R(b_{c,c',1},b_{c,c',2}),R(b_{c,c',2},b_{c,c',3})$;
  
  \item ${\sf From}_c(b_{c,c',1}), {\sf To}_{c'}(b_{c,c',2}), \mn{Out}(b_{c, c', 2})$;
  
  \item $P(b_{c,c',3})$ for all $P(c') \in I$.
  
  \end{itemize}
To address the $\downarrow$/$\updownarrows$-issue, the ghost gadget
issue, and the and/or issue, we include in $I'_t$ additional
gadgets. Note that all these issues pertain to additional, undesired
successors in $U'$. The additional gadgets, which we refer to as
\emph{sinks}, can accommodate the surplus successors and the subtrees
below them. There are sinks of four types:
\begin{itemize}

\item[I.] when $S(\bar c,\bar c')$ is mapped to $S(c,c') \in I_t$,
a sink that takes paths labeled ${\sf From}_e$ with $e \neq c$ (and/or
issue);

\item[II.] when $S(\bar c,\bar c')$ is mapped to $S(c,c') \in I_t$,
a sink that takes paths labeled ${\sf From}_c$ and ${\sf To}_e$ with
$e \neq c'$ (and/or
issue);

\item[III.] sinks that deal with additional successors in $U'$
due to the $\downarrow$/$\updownarrows$-issue.

\item[IV.] sinks that deal with additional successors in $U'$
due to the ghost gadget issue.

\end{itemize}
Each sink takes the form of a path. For a word $w=\sigma_1 \cdots \sigma_k \in \{R,R^-\}^*$, 
a \emph{full $w$-path} is a path in which the $i$-th edge is a forward $R$-edge if $\sigma_i=R$ and a backward $R$-edge if $\sigma_i=R^-$. Moreover,
every value on the path except the starting value is labeled
with all unary relation symbols from $\Gamma$.

Let $c,c' \in D$.
The sink of Type~I is attached to value $b_{c,c',1}$. We add the following facts:
\begin{itemize}
    
    \item $R(s_{c,c',I,3},b_{c,c',1}),R(s_{c,c',I,3},s_{c,c',I,4})$;
    
    \item $A_S(s_{c, c', I, 3})$ for all $S \in \Sigma$;

    \item ${\sf From}_e(s_{c,c',I,4})$ for all $e \in D \setminus \{ c' \}$;
    
    \item a full $RRR^-R^-R^-RRR$-path attached to $s_{c,c',I,4}$.
    
\end{itemize}
The sink of Type~II is also attached to value $b_{c,c',1}$. We add the following facts:
\begin{itemize}
    
    \item $R(b_{c,c',1},s_{c,c',II,5})$;
    
    \item ${\sf To}_e(s_{c,c',II,5})$ for all $e \in D \setminus \{ c' \}$ and ${\sf Impossible}(s_{c,c',II,5})$;
    
    \item a full $RR^-R^-R^-RRR$-path attached to $s_{c,c',II,5}$.
    
\end{itemize}
The sink of Type~III is attached to value $b_{c,c',3}$. We add the following facts:
\begin{itemize}
    
    \item $R(s_{c,c',III,1},b_{c,c',3})$;
    
    \item ${\sf To}_e(s_{c,c',III,1})$ for all $e \in D$ and $\mn{Impossible}(s_{c, c', III, 1})$;
    
    \item a full $R^-R^-RRRR^-R^-R^-RRR$-path attached to $s_{c,c',III,1}$.
    
\end{itemize}
The sink of Type~IV is again attached to value $b_{c,c',1}$. We add the following facts:
\begin{itemize}
    
    \item $R(s_{c,c',IV,1},b_{c,c',1})$;
      
    \item a full $RRRR^-R^-R^-RRR$-path attached to $s_{c,c',IV,1}$.
    
\end{itemize}

This finishes the construction of $I'_t$, which is illustrated in
Figure~\ref{fig:prodsim-fourth}. We make a few remarks on sinks.  It is
important to keep in mind which labels are \emph{missing} on sink
nodes. For sinks of Type~I, the missing label is $\mn{From}_c$ on the
second node, for sinks of Type~II it is ${\sf To}_{c'}$ on the first
node, for sinks of Type~III it is ${\sf Out}$ on the first node, and
for sinks of Type~IV it is all labels $A_S$ on the first node. The
latter, for example, is important to make sure that sinks of Type~IV
will only accommodate the ghost gadgets (which are missing the $A_S$ labels), but not any of the gadgets representing real edges in $\prod_{1 \leq i \leq n} I_i$.

The length and shape of the full paths that we attach to the sinks
is determined by which nodes of the gadget in
Figure~\ref{fig:prodsim-first} we intend to map to the first node in
the sink. These are nodes of depth~3 for sinks of Type~I and~IV, nodes of
depth~5 for sinks of Type~II, nodes of depth~1 for sinks of Type~III.
A path of length 6 is not enough since, as described, there is
a node on each sink that misses some labels. But note that all
attached full paths end in a full path of length 6 with the pattern
$R^-R^-R^-RRR$. These are actual sinks in the sense that
any subtree of $U'$ has a $\updownarrows$-simulation into them.
We also note that the \mn{Impossible} label is present \emph{only} in sinks (of Type~II). Consequently, any
element of $U'$ that is labeled with \mn{Impossible} must be mapped to
a sink which intuitively ensures that edges $S(\bar c,\bar c')$ in
$\prod_{1 \leq i \leq n} I_i$ are only mapped to edges $S(c,c')$ in
$I_t$ that actually exist.

\begin{figure}
    \centering
    \begin{tikzpicture}[on grid, node distance = 0.8cm and 0.8cm, label distance = -0.0cm]
\tikzset{numbered/.style={font=\footnotesize}}
\tikzset{every label/.style={align=center}}
\tikzset{sink/.style={font=\footnotesize, fill, circle, inner sep=1pt}}

    \node (b) [numbered, label=right:{$A_S$ for all $S \in \Sigma$}]{$b_0$};
    \node (b1a) [numbered, below left = of b, label=left:$\textsf{From}_c$]{$b_{c, c', 1}$};
    \node (b1b) [numbered, below right = of b]{$1$};
    \node (b2a) [numbered, below left = of b1a, label=left:{$\textsf{To}_{c'}$ $\textsf{Out}$}]{$b_{c, c', 2}$};
    \node (b2b) [numbered, below right = of b1b]{$2$};
    \node (b3a) [numbered, below left = of b2a]{$b_{c, c', 3}$};
    \node (b3b) [numbered, below right = of b2b]{$3$};

    \node (t1) [sink, below left = 2.6cm and 1.2cm of b1a, label=left:{I}, label=right:$A_S$] {};
    \node (t1a) [sink,fill,circle,inner sep=1pt, below = of t1, label = right:{$\textsf{From}_e$\\($e \neq c'$)}] {};
    \node (t1b) [below = of t1a] {$\vdots$};
    \node (t2) [sink, below right = 2.6cm and 0.8cm of b1a, label=left:{II}] {};
    \node (t2label) [below right = 0.1cm and 0.7cm of t2, align=left] {$\textsf{To}_e$\\$(e \neq c')$ \\ $\textsf{Imp}$};
    \node (t2a) [below =  of t2] {$\vdots$};
    \node (t3) [sink, below left = 2.6cm and 2.5cm of b1a, label=left:{III}] {};
    \node (t3label) [below right = 0cm and 0.40cm of t3, align=left] {$\textsf{To}_e$ \\ $\textsf{Imp}$};
    \node (t3a) [below = of t3] {$\vdots$};
    \node (t4) [sink, below right = 2.5cm and 3.0cm of b1a, label=left:{IV}] {};
    \node (t4a) [below =  of t4] {$\vdots$};

    \draw[->, dashed](t1) to [bend right=10] (b1a);
    \draw[->, dashed](t1) to  (t1a);
    \draw[->, dashed](t1a) to  (t1b);
    \draw[->, dashed](b1a) to [bend left=10] (t2);
    \draw[->, dashed](t2) to  (t2a);
    \draw[->, dashed](t3.north) to [bend left=40] (b3a);
    \draw[->, dashed](t3a) to  (t3);
    \draw[->, dashed](t4) to [bend right=25] (b1a);
    \draw[->, dashed](t4) to  (t4a);

\draw[->] (b) to (b1a);
\draw[->] (b) to (b1b);
\draw[->] (b1a) to (b2a);
\draw[->] (b1b) to (b2b);
\draw[->] (b2a) to (b3a);
\draw[->] (b2b) to (b3b);

\node(dots1) [below = of b] {$\ldots$};

    \end{tikzpicture}
    \caption{The instance $I'_t$ %
    and the four types of sinks.}
    \label{fig:prodsim-fourth}
\end{figure}

We also note that the structure of $I_t$ is not reflected by $I'_t$, but
rather by (the labels in) $\prod_{1 \leq i \leq n} I'_i$. Instead,
$I'_t$ is merely a navigation gadget that is traversed by any $\updownarrows$-simulation
of $\prod_{1 \leq i \leq n} I'_i$ in a systematic way. This allows $I'_t$ to be a tree, as desired.
\begin{lemma}
  $(\prod_{1 \leq i \leq n} I_i,\bar c_s)\preceq^{\downarrow} (I_t,c_t)$
  if and only if
    $(\prod_{1 \leq i \leq n} I'_i,\bar c_s)\preceq^{\updownarrows} (I'_t,b_{c_t,c_t,3})$.\footnote{We could use any $b_{c,c_t,3}$ in place
    of $b_{c_t,c_t,3}$.}
\end{lemma}
\begin{proof}
  ``if''. Let $\mathcal{S}'$ be a $\updownarrows$-simulation that witnesses
  $(\prod_{1 \leq i \leq n} I'_i,\bar c_s)\preceq^{\updownarrows} (I'_t,b_{c_t,c_t,3})$. Define a relation~$\mathcal{S}$ by setting
  $$
  \begin{array}{r@{\;}c@{\;}l}
    \mathcal{S} &:=& \{ (\bar c,c) \in {\sf adom}(\prod_{1 \leq i \leq n} I_i)
     \times D \mid (\bar c,b_{d,c,3}) \in \mathcal{S}' \text{ for some } d \}.
  \end{array}
  $$
  We argue that $\mathcal{S}$ is a $\downarrow$-simulation from
  $\prod_{1 \leq i \leq n} I_i$ to~$I_t$. In fact, it is easy to use the
  definitions of the instances $I'_i$ and $I'_t$ to verify that
  Condition~1 of $\downarrow$-simulations is satisfied.

  For Condition~2, take an edge $R(\bar c,\bar c')$ in
  $\prod_{1 \leq i \leq n} I_i$ and let $(\bar c,d) \in \mathcal{S}$. Then
  $(\bar c,b_{e,d,3}) \in \mathcal{S}'$ for some~$e$.  The edge
  $R(\bar c,\bar c')$ gives rise to a corresponding gadget in
  $\prod_{1 \leq i \leq n} I'_i$ that starts at $\bar c$ and ends
  at $\bar c'$, as shown in Figure~\ref{fig:prodsim-first}. The
  gadget has multiple paths that branch at the end, one for each
  value in $D$. Consider the path associated with
  $d \in D$, meaning it is labeled with $\mn{From}_d$, and let us analyze the values in $I'_t$ that
  $\mathcal{S}'$ simulates this path into, starting from the root that $\mathcal{S}'$
  simulates into $b_{e,d,3}$.

  The first node on the path is labeled \mn{Out} and $\mathcal{S}'$ cannot
  simulate it into the sink of Type~III attached in $I'_t$ to
  $b_{e,d,3}$ since the first node in that sink is missing the
  \mn{Out} label. 
  Thus, the first node on the path is simulated into $b_{e,d,2}$.  The
  second node on the path must then be simulated into $b_{e,d,1}$.  The
  third and fourth node, the first labeled $A_S$ and the latter labeled ${\sf From}_d$, are best
  considered together. They cannot be simulated into the sink of
  Type~I attached to $b_{e,d,1}$ because the fourth node is labeled
  ${\sf From}_d$, but the second node in the sink is
  not. They also cannot be simulated into the sink of Type~IV attached to 
  $b_{e,d,1}$ since the first node of this sink does not carry any $A_S$ label.
  Consequently, the third node is simulated into $b_0$ and the
  fourth one into the value $b_{d,f,1}$ for some $f \in D$ 
  as only such nodes are labeled ${\sf From}_d$.

  At the fourth node, the path that we are following branches. We
  are interested in further following the branch on which the fifth
  node is labeled ${\sf To}_{f}$. This node cannot be simulated into
  the sink of Type~II attached to $b_{d,f,1}$ because the first node
  in that sink is not labeled ${\sf To}_{f}$. It must consequently be
  mapped to $b_{d,f,2}$ which leaves $\mathcal{S}'$ with the only option of
  simulating the ending node  $\bar c'$ of the gadget to
  $b_{d,f,3}$. The definition of $\mathcal{S}$ thus yields
  $(\bar c',f) \in \mathcal{S}$.  Since $S'$ simulates the node labeled  ${\sf To}_{f}$
  to $b_{d,f,2}$, which does not carry an $\mn{Impossible}$
  label, also the ${\sf To}_{f}$ node does not carry that label. 
  By construction of the
  source instances $I'_1,\dots,I'_n$ we therefore know that $R(d,f) \in
  I_t$. We have thus shown that Condition~2 of $\downarrow$-simulations
  is satisfied.
  
  \medskip ``only if''.  Assume that
  $(\prod_{1 \leq i \leq n} I_i,\bar c_s)\preceq^{\downarrow}
  (I_t,c_t)$ and let $\mathcal{S}$ be a witnessing $\downarrow$-simulation. To
  prove that
  $(\prod_{1 \leq i \leq n} I'_i,\bar c_s)\preceq^{\updownarrows}
  (I'_t,b_{c_t,c_t,3})$, it suffices to show that there is a
  homomorphism $h'$ from the unraveling of
  $\prod_{1 \leq i \leq n} I'_i$ at~$\bar c_s$ to $I'_t$ with
  $h'(\bar c_s)=b_{c_t,c_t,3}$.  To avoid unnecessary case
  distinctions, we actually work with a slightly more careful version
  of unravelings. Recall that the unraveling of an instance $I$ is defined based on paths
  which are sequences $p=a_1R_1\cdots R_{k-1}a_k$, $k \geq 1$, where
  $a_1,\dots,a_k \in \text{adom}(I)$ and $R_1,\dots,R_{k-1}$ are roles
  such that $R_i(a_i,a_{i+1}) \in I$ for $1 \leq i< k$. Here, we
  additionally assume that $a_{i+1} \neq a_{i-1}$ for $1 < i < k$.
  It is easy to verify that Lemma~\ref{lem:unravbasics} still holds.

  We define $h'$ step by step, obtaining the desired homomorphism in
  the limit.  Start with setting $h'(\bar c_s)=b_{c_t,c_t,3}$ and note that
  $b_{c_t,c_t,3}$ satisfies the same unary relations in $I'_t$ that $c_t$
  satisfies in $I_t$.

  We call a path $p \in \mn{adom}(U')$ a \emph{frontier point} if $h'(p)$ is
  defined and $U'$ contains an edge $R(p,pR\bar a)$ with $h'(pR\bar a)$
  undefined---the construction of $h'$ will ensure that then
  $h'(pR\bar a)$
  is undefined for \emph{all} edges $R(p,pR\bar a)$ in $U'$.  We 
  maintain the invariant that if $p$ is a frontier point, then
  \begin{enumerate}

  \item[($*$)] $p$  ends in a value
    $\bar c \in \mn{adom}(\prod_{1 \leq i \leq n} I_i)$ and there is
    an $e \in D$ such that $h'(p)=b_{f,e,3}$ (for some $f$) and
    \mbox{$(\bar c,e) \in \mathcal{S}$}.

  \end{enumerate}
  We extend $h'$ by repeatedly selecting a frontier point $p$ whose
  length as a path is as short as possible (to achieve that all
  frontier points will eventually be treated). By
  Invariant~($*$), $p$ ends with a value
  $\bar c \in \mn{adom}(\prod_{1 \leq i \leq n} I_i)$ and there is an
  $e \in D$ such that $h'(p)=b_{f,e,3}$ (for some $f$) and
  $(\bar c,e) \in S$.

  Now consider (independently of each other) all
  $p_1=pR_1\bar a_1 \in \mn{adom}(U')$ with $h'(pR_1\bar a_1)$
  undefined. Each component of $\bar a_1$, say that $i$-th, is reached
  from the $i$-th component of $\bar c$ by traversing to a 1-point or
  to a 5-point in the gadget shown in
  Figure~\ref{fig:prodsim-first}. If any transition is to a 5-point,
  then $p_1$ is not labeled \mn{Out} in $U'$ and we can
  map $p_1$ as well as the entire subtree of $U'$ below
  it to the sink of Type~III attached to $h'(p)=b_{f,e,3}$.

  Now assume that this is not the case. Note that $p_1$ has a
  single successor $p_2=p_1R_2 \bar a_2$ (consisting only of
  2-points from the gadget in Figure~\ref{fig:prodsim-first}) which
  has a single successor $p_3 = p_2 R_3 \bar a_3$ that
  consists solely of 3-points.\footnote{This relies on our modified
    version of unravelings. Otherwise, there would be multiple
    successors.}  We set $h'(p_1)=b_{f,e,2}$ and
  $h'(p_2)=b_{f,e,1}$.  If
  $p_3$ is not labeled with any
  $A_S$, then we can map   $p_3$
  and the entire subtree below it to the sink of Type~IV attached to
  $b_{f,e,1}$.

  Now assume that this is not the case.  Since the 3-point of the
  gadget in Figure~\ref{fig:prodsim-first} has two successors, 
  $p_3$ may also have multiple successors in $U'$: in each component, we
  may follow the gadget upwards or downwards, reaching a 2-point or a
  4-point. Consider (independently) any successor
  $p_4=p_3R_3 \bar a_4$ of $p_3$. If $p_4$ is not labeled with
  $\mn{From}_e$, then we can map $p_3$ and the subtree below it to the
  sink of Type~I attached to $b_{f,e,1}$. Otherwise we set
  $h'(p_3)=b_0$. But how to map $p_4$? Since $p_4$ is labeled
  $\mn{From}_e$, each component of $\bar a_4$, say the $i$-th, is a
  4-point. Moreover, it is on a path in $I'_i$ that starts at the
  $i$-th component of $\bar c$ and whose 3-point is labeled with
  $A_S$, for the same $S$ across all $i$. Thus, all 4-points in
  $\bar a_4$ must belong to the same gadget (as in
  Figure~\ref{fig:prodsim-first}) in $I'_i$. The gadget starts at the
  $i$-the component $c_i$ of $\bar c$ and ends at a value $c_i'$. By
  construction of $I'_i$, we have $S(c_i,c_i') \in I_i$.  For
  $\bar c' = (c_1',\dots,c_n')$, we then have
  $S(\bar c, \bar c') \in \prod_{1 \leq i \leq n} I_i$. Since
  $(\bar c, e) \in \mathcal{S}$, we find an $e' \in D$ with
  $S(e,e') \in I_t$ and $(\bar c',e') \in \mathcal{S}$. We are now ready to
  map $p_4$: set $h'(p_4)=b_{e,e',1}$.

  At $p_4$, there is another branching in $U'$. Consider
  (independently) any successor $p_5=p_4 R_5 \bar a_5$ of $p_4$.  If
  $p_5$ is not labeled $\mn{To}_{e'}$, then we can map $p_5$ as well
  as the subtree below it to the sink of Type~II attached to
  $b_{e,e',1}$. Otherwise, we set $h'(p_5)=b_{e,e',2}$. Note that this
    is a valid target: since $S(e,e') \in I_t$, $p_5$ is not labeled
    with \mn{Impossible} (which is missing at $b_{e,e',2}$).

    $p_5$ has a single successor $p_6=p_5 R_6 \bar a_6$ in $U'$. We map it to
    $b_{e,e',2}$ and leave $p_6$ as a frontier point to treat in a subsequent
    round. Note that the Invariant~($*$) holds for $p_6$.

\medskip
It remains to explain how to get to a fixed schema $\Gamma$. Note that only the unary relation symbols in $\Gamma$ depend on $\Sigma$, and thus on the input. Let these symbols be $A_1,\dots,A_k$. We can introduce a fresh binary relation symbol $S$ and a fresh unary relation symbol $A$ and replace each fact $A_j(x)$, both in our
construction of the instances $I'_i$ and $I'_t$, with an $S$-path
of length $j$ that starts at $x$ and whose final node is labeled
with $A$. It is easy to verify that this simple modification does not compromise the correctness of the reduction. 
\end{proof}

\section{Lower Bounds for Tree CQs (Theorems~\ref{thm:mostspecific-exptime-hard-trees} and~\ref{thm:mauriceexpapp})}
\label{app:lowerboundstreecqs}

\thmmostspecificexptimehardtrees*

\begin{proof}
  We  reduce from the product simulation problem into
  trees.  Assume that we are given
  finite pointed instances $(I_1,a_1),\dots,(I_n,a_n)$ and $(J,b)$
  with $J$ a tree, and that we are to decide whether
  $\Pi_{1 \leq i \leq n} (I_i,a_i) \preceq (J,b)$. 
  
  We construct a collection of labeled examples $E=(E^+, E^-)$, as
  follows. Assume w.l.o.g.\ that
  $\text{adom}(I_i) \cap \text{adom}(J) = \emptyset$ for
  $1 \leq i \leq n$. Let $R$ be a fresh binary relation symbol and
  $A_1,A_2,B_1,B_2$ fresh unary relation symbols. $E^+$ contains
  instances $(I'_1,a'_1),\dots,(I'_n,a'_n)$ where $(I'_i,a'_i)$ is
  obtained by starting with $(I_i,a_i)$ and adding the following
  facts:
  \begin{enumerate}
      
      \item $R(a'_i,a_i)$;
      
      \item  $R(a'_i,b)$ and all facts from $J$ (we refer to this as the copy of $J$ in $I'_i$);
      
      \item $R(a_i,c_1),R(a_i,c_3),R(c_2,c_1),R(c_2,c_3),
        A_1(a_i),B_1(c_1),B_2(c_3),A_2(c_2)$ with $c_1,c_2,c_3$ fresh
        values;

      \item $R(b,d), A_1(b), A_2(b), B_1(d),B_2(d)$ with $d$ a fresh value.
        
   \end{enumerate}
   The structure of $(I'_i, a'_i)$ is shown in Figure~\ref{fig:most-specific-reduction}.
   The purpose of Point~3 above is to create, in the unraveling of
   $I_i$ at $a_i$, an infinite path that starts at~$a_i$. The path
   is obtained by traveling $a_i$, $c_2$, $c_4$, $c_3$, $a_i$, ad
   infinitum. It alternates between forward and backwards $R$-edges
   and the labeling of its nodes with $A_1,B_1,A_2,B_2,A_1,\dots$
   ensures that the path does not simulate into a finite prefix of
   itself. Also note that there is a simulation of the subinstance
   created in Point~3 in the subinstance created in Point~4 that maps $a_i$ to $b$.
  
  \begin{figure}   
    \centering
    \begin{tikzpicture}[on grid, node distance = 1cm and 1cm, label distance = -0.1cm]
      \node (aip) at (0, 0) {$a_i'$};
      \node (ai) [below left = of aip, label=right:{$A_1$}] {$a_i$};
      \node (b) [below right = of aip, label=45:{$A_1, A_2$}] {$b$};
      \node (d) [below right = 0.5cm and 1cm of b, label=right:{$B_1, B_2$}] {$d$};
      \node (c1) [above left = 0.5cm and 1cm of ai, label=above:{$B_1$}] {$c_1$};
      \node (c3) [below left = 0.5cm and 1cm of ai, label=below:{$B_2$}] {$c_3$};
      \node (c2) [below left = 0cm and 2cm of ai, label=left:{$A_2$}] {$c_2$};
      
      \node[draw, isosceles triangle, shape border rotate=90] (Ii) [below = of ai] {$I_i$};
      \node[draw, isosceles triangle, shape border rotate=90] (J) [below = of b] {$J$};

      \draw[->] (aip) to node [above] {} (ai);
      \draw[->] (aip) to node [above] {} (b);
      \draw[->] (b) to node [below] {} (d);
      \draw[->] (ai) to node [above] {} (c1);
      \draw[->] (ai) to node [above] {} (c3);
      \draw[->] (c2) to node [above] {} (c1);
      \draw[->] (c2) to node [above] {} (c3);
    \end{tikzpicture}
    \caption{An instance $I'_i$.}
    \label{fig:most-specific-reduction}
  \end{figure}

   Since $(J,b)$ is a tree, we may view it as a tree CQ $q(x)$.
   It was shown in \cite{tCD2022:conjunctive} that every tree CQ
   has a frontier w.r.t.\ tree CQs which can be computed in polynomial time. We may thus compute such a frontier
   $\mathcal{F} = \{ p_1(x),\dots,p_k(x)\}$ for $q$. $E^-$ contains instances $(L_1,x'),\dots,(L_k,x')$ 
  where $(L_i,x')$ is obtained by starting with $(p_i,x)$ and
   adding the fact  $R(x',x)$.
     %
      
      
      
   %
   
   Let $q'(x')$ be the tree CQ obtained from $q(x)$ by making $x'$ the
   answer variable and  adding the atoms
   \begin{enumerate}

   \item $R(x',x)$;

   \item  $R(x,y), A_1(x), A_2(x), B_1(y),B_2(y)$ with $y$ a fresh variable.

   \end{enumerate}
   Note that Point~2 above creates the same gadget as Point~4 in the
  definition of $I'_i$.

   Observe that $q'$
   is a fitting for $E$. In fact, it is clear by construction of $E^+$
   that $q' \preceq (I'_i,a'_i)$ for all $(I'_i,a'_i) \in
   E^+$. Moreover, since $q \not\preceq p$ for any $p \in \mathcal{F}$
   we also have $q' \not\preceq (L_i,x')$ for all
   $(L_i,x') \in E^-$.
   
    To establish the theorem, it remains to show the following:
    \begin{enumerate}

         \item[(a)] If $\Pi_{1 \leq i \leq n} (I_i,a_i) \preceq
           (J,b)$, then $q'$ is a  most-specific fitting for $E$;
         
         \item[(b)] If $\Pi_{1 \leq i \leq n} (I_i,a_i) \not\preceq (J,b)$, then $E$ has no most-specific fitting.
        
    \end{enumerate}
    For Point~(a), assume that
    $\Pi_{1 \leq i \leq n} (I_i,a_i) \preceq (J,b)$ and let
    $\widehat q$ be a fitting for $E$.  We have to show that
    $\widehat q \preceq q'$.
    We first argue that
    $\Pi_{1 \leq i \leq n} (I'_i,a'_i) \preceq q'$. Let $S$ be a
    simulation witnessing
    $\Pi_{1 \leq i \leq n} (I_i,a_i) \preceq (J,b)$. We obtain $S'$
    from $S$ as follows:
    \begin{enumerate}

    \item add $(\bar a',x')$ for $\bar a' =
          (a'_1,\dots,a'_n)$;

        \item add $(\bar a,x)$ for all $\bar a \in 
          \text{adom}(\Pi_{1 \leq i \leq n} (I'_i,a'_i))$ that
          contain only values $a_i$, $b$, and $c_2$;

        \item add $(\bar a,y)$ for all $\bar a \in 
          \text{adom}(\Pi_{1 \leq i \leq n} (I'_i,a'_i))$ that
          contain only values $c_1$, $c_3$, and $d$;
        
        \item for all
          $\bar a \in \text{adom}(\Pi_{1 \leq i \leq n} (I'_i,a'_i))$
          that contain an element $c$ of $\text{adom}(J)$, add
          $(\bar a,c)$.
        
    \end{enumerate}
    It can be verified that $S'$ is a simulation of
    $\Pi_{1 \leq i \leq n} (I'_i,a'_i)$ in $q'$. In particular, every
    tuple $\bar a \in \text{adom}(\Pi_{1 \leq i \leq n} (I'_i,a'_i))$
    that is reachable in $\Pi_{1 \leq i \leq n} (I'_i,a'_i)$ from
    $(a'_1,\dots,a'_n)$ and contains any of the values
    $c_1,c_2,c_3,d$ must be of one of the forms treated
    in Points~2 and~3 above. We have thus shown that
    $\Pi_{1 \leq i \leq n} (I'_i,a'_i) \preceq q'$.
    Together with 
    $\widehat q \preceq \Pi_{1 \leq i \leq n} (I'_i,a'_i)$, which
    holds since $\widehat q$ is a fitting of $E$, 
    we obtain     $\widehat q \preceq q'$, as desired.
    
    \smallskip
    
    For Point~(b), assume that
    $\Pi_{1 \leq i \leq n} (I_i,a_i) \not\preceq (J,b)$. By
    Lemma~\ref{lem:unravbasics}, there is then also an $m \geq 1$ such
    that $(U_m,\bar a) \not\preceq (J,b)$ with $U_m$ the $m$-unraveling
    of $\Pi_{1 \leq i \leq n} I_i$ at
    $\bar a =(a_1,\dots,a_n)$. Let $U'_m$ be $U_m$ extended with fact
    $R(x,\bar a)$ and let $\widehat q(x)$ be $U'_m$ viewed as a tree
    CQ.  By construction of $E^+$, $\widehat q \preceq (I'_i,a'_i)$
    for all $(I'_i,a'_i) \in E^+$. Moreover,
    $\widehat q \not\preceq (L_i,x')$ for all $(L_i,x') \in E^-$
    because otherwise from the composition of simulations witnessing
    $\widehat q \preceq (L_i,x')$ and $p_i \preceq (J,b)$ we may
    obtain a simulation witnessing $(U_m,\bar a) \preceq (J,b)$, a
    contradiction. Thus, $\widehat q$ is a fitting for $E$. For every
    $i \geq 1$, let $\widehat q_i$ be $\widehat q$ extended with a path on variables $\bar a, x_1,\dots,x_i$ that alternates between forwards and backwards $R$-edges
    (starting with forwards) and is additionally labeled
    with atoms
    $$A_1(\bar a),B_1(x_1),A_2(x_2),B_2(x_3),A_1(x_4)\dots
    $$
    to achieve that it does not map into a finite prefix of itself, as
    described above. It is easy to verify that $\widehat q_i$ is a
    fitting for all $i \geq 1$. Clearly,
    $\widehat q_i \prec \widehat q_{i+1}$ for all $i \geq 1$. To
    finish the proof, it thus remains to show that there is no fitting
    $p$ for $E$ such that $\widehat q_i \preceq p$ for all $i \geq
    1$. Assume to the contrary that there is such a $p(x)$.  Then
    $p \preceq (I'_1,a'_1)$. Let $h$ be a homomorphism from $p$ to
    $(I'_1,a'_1)$ with $h(x)=a'_1$.  We must find a $R(x,y) \in p$ such that for
    infinitely many $i$, there is a homomorphism $h_i$ from
    $\widehat q_i$ to $p$ with $h_i(x)=x$ and $h_i(\bar a)=y$. Then
    $h(y)=a_i$ as the only other option $h(y)=b$ implies that
    $(U_m,\bar a) \preceq (J,b)$, which is not the case. Consequently,
    each $h_i \circ h$ maps the path $\bar a, x_1, \dots,x_i$ in
    $\widehat q_i$ to the subinstance of $I'_1$ induced by the
    values $a_i,c_1,c_2,c_3$. But clearly there is no (finite!) tree instance
    (resp.\ tree CQ) that admits a homomorphism from all paths
    $\bar a, x_1, \dots,x_i$, $i \geq 1$, and also a homomorphism
    to this subinstance. 
\end{proof}

\thmmauriceexpapp*

\begin{proof}
It is well known that there is a fixed alternating, linear space bounded Turing
  machine (TM)  whose
  word problem is ExpTime-complete~\cite{chandra81}.  Given a
  word $w$, we thus construct a collection of labeled examples
  $(E^+, E^-)$ such that $(E^+, E^-)$ permits a weakly most-general
  fitting CQ if and only if $M$ does not accept $w$.  The weakly most-general
  fitting CQ of $(E^+, E^-)$ is always a tree CQ, therefore this
  reduction shows hardness of the existence problem for CQs as well as
  of the existence problem for tree CQs.

  For our purposes, an \emph{alternating Turing machine (ATM)} $M$ is
  a tuple
  $M=(\Gamma, Q_\forall, Q_\exists, \mapsto, q_0, F_{\mathrm{acc}},
  F_{\mathrm{rej}})$ consisting of a finite set of tape
  symbols~$\Gamma$, a set of universal states $Q_\forall$, a set of
  existential states $Q_\exists$, a set of accepting states
  $F_{\mathrm{acc}}$, a set of rejecting states $F_{\mathrm{rej}}$,
  which together form the set of all states $Q = Q_\forall \cup Q_\exists \cup
  F_{\mathrm{acc}} \cup F_{\mathrm{rej}}$, as well as an
  initial state $q_0 \in Q_\forall$ and a transition relation
  $\mapsto \subseteq Q \times \Gamma \times Q \times \Gamma \times
  \{-1, 0, +1\}$. The last component of the transition relation that
  is either $-1, 0$ or $+1$ indicates the head of the TM moving to the
  left, staying at the same tape cell, and moving to the right,
  respectively.  The sets $Q_\forall$, $Q_\exists$,
  $F_{\mathrm{acc}}$, $F_{\mathrm{rej}}$ partition $Q$ and we refer to
  states in $F_{\mathrm{acc}} \cup F_{\mathrm{rej}}$ as \emph{final}
  states.  A configuration of $M$ is \emph{universal} if its state is
  universal, and likewise for \emph{existential} configurations
  and \emph{final} configurations.  In our model of alternation, every
  existential or universal configuration has exactly two successor
  configurations and every final configuration has no successor
  configurations.  Hence, we write
  $(q, a) \mapsto ((q_\ell, b_\ell, \Delta_\ell), (q_r, b_r,
  \Delta_r))$ to indicate that when $M$ is in state
  $q \in Q_\forall \cup Q_\exists$ and reading symbol $a$, it branches to
  ``the left'' with $(q_\ell, b_\ell, \Delta_\ell)$ and to ``the
  right'' with $(q_r, b_r, \Delta_r)$. These directions are not
  related to the movement of the head on the tape.  Furthermore, we
  assume that $\mapsto$ alternates between existential states and
  universal states, that $q_0$ is a universal state, and that $M$
  always reaches a final state.

  With each configuration that is reached by an alternating TM $M$ on
  an input $w$, we associate an acceptance value of $1$ or $0$ as
  follows.  Final configurations with an accepting state have
  acceptance value $1$ and final configurations with a rejecting state
  have acceptance value~$0$.  The acceptance value of a universal
  configuration is the minimum of the acceptance value of its two
  successors. The acceptance value of an existential configuration is
  the maximum of the acceptance value of its two successors. An
  alternating TM \emph{accepts} input $w$ if the initial configuration
  $q_0w$ of $M$ on $w$ has acceptance value $1$ and \emph{rejects}
  $w$ otherwise. 

  \smallskip

  Let
  $M = (\Gamma, Q_\forall, Q_\exists, \mapsto, q_0, F_{\mathrm{acc}},
  F_{\mathrm{rej}})$ be a fixed alternating TM with linear space bound
  $s(n)$.  Given a word $w$ with $|w| = n$, we construct pointed
  instances $(I_i, c_i)$ for all $i$ with $1 \leq i \leq s(n)$ to be
  used as positive examples and a pointed instance $(J, c)$ to be used
  as the only negative example.  As the schema, we use the unary
  relation symbols $\mathsf{Reject}$, $\mathsf{Accept}$ and the binary
  relation symbols $r_{q, a, i}$ and $\ell_{q, a, i}$ for all
  $q \in Q$, $a \in \Gamma$ and $i$ with $1 \leq i \leq s(n)$. What we
  want to achieve is that
  \begin{enumerate}

  \item if $M$ accepts $w$, then
    $(\prod_{1 \leq i \leq s(n)} I_i, c_1\dots c_{s(n)}) \to (J, c)$
    and thus there is no fitting CQ;

  \item if $M$ rejects $w$, then the computation tree of $M$ on $w$,
    defined in the usual way, describes a fitting tree CQ $q$;
    moreover, we can extract from $q$ a weakly most-general CQ
    by dropping subtrees.

  \end{enumerate}
  We start with the pointed instances $(I_i, c_i)$.
    Each $I_i$ uses the values $a$ and $(q, a)$ for
    all $q \in Q$ and $a \in \Gamma$ to represent the $i$-th tape cell of $M$.
    The value $a$ represents that the head of $M$ is not on cell $i$ and that
    cell $i$ contains the symbol $a$. The value $(q, a)$ represents that the
    head of $M$ is on cell $i$, that $M$ is in state $q$ and that the cell $i$
    contains the symbol $a$.
    The facts in each $I_i$ ensure that  $r_{q, a, i}(e, e')$ is true in the
    part of $\prod_{1 \leq i \leq s(n)} I_i$ that is reachable from
    the value $c_1\dots
    c_{s(n)}$ if and only if in state $q$, reading symbol $a$ and head at tape cell $i$,
    $M$ branches right from the configuration represented by $e$ to the
    configuration represented by $e'$. The same is true for the facts $\ell_{q, a,
    i}(e, e')$ and branching left.

    For each transition $(q, a) \mapsto ((q_\ell, b_\ell, \Delta_\ell), (q_r, b_r,
    \Delta_r))$ of $M$, $I_i$ contains the following facts:
    \begin{enumerate}
        \item Facts that correspond to the head moving away from cell $i$:
            \begin{align*}
                &\ell_{q, a, i}((q, a), b_\ell)\ \text{if}\ \Delta_\ell \neq 0, \\
                &r_{q, a, i}((q, a), b_r)\ \text{if}\ \Delta_r \neq 0.
            \end{align*}
        \item Facts that correspond to the head staying on cell $i$:
            \begin{align*}
                &\ell_{q, a, i}((q, a), (q_\ell, b_\ell))\ \text{if}\ \Delta_\ell = 0,\\
                &r_{q, a, i}((q, a), (q_r, b_r))\ \text{if}\ \Delta_r = 0.
            \end{align*}
        \item Facts that correspond to the head moving onto cell $i$ from
            cell $i-1$ or $i + 1$. For all $b \in \Gamma$:
            \begin{align*}
                &\ell_{q, a, i - 1}(b, (q_\ell, b))\ \text{if}\ \Delta_\ell = +1,\\
                &r_{q, a, i - 1}(b, (q_r, b))\ \text{if}\ \Delta_r = +1,\\
                &\ell_{q, a, i + 1}(b, (q_\ell, b))\ \text{if}\ \Delta_\ell = -1,\\
                &r_{q, a, i + 1}(b, (q_r, b))\ \text{if}\ \Delta_r = -1.
            \end{align*}
        \item Facts that correspond to the transition not modifying the 
            cell~$i$. For all $j \neq i$ with $1 \leq j \leq s(n)$:
            \begin{align*}
                &\ell_{q, a, j}(b, b)\ \text{if}\ \Delta_\ell = +1\ \text{and}\ j \neq i - 1,\\
                &r_{q, a, j}(b, b)\ \text{if}\ \Delta_r = +1\ \text{and}\ j \neq i - 1,\\
                &\ell_{q, a, j}(b, b)\ \text{if}\ \Delta_\ell = -1\ \text{and}\ j \neq i + 1,\\
                &r_{q, a, j}(b, b)\ \text{if}\ \Delta_r = -1\ \text{and}\ j \neq i + 1,\\
                &\ell_{q, a, j}(b, b)\ \text{if}\ \Delta_\ell = 0,\\
                &r_{q, a, j}(b, b)\ \text{if}\ \Delta_r = 0.
            \end{align*}
    \end{enumerate}
    Additionally, $I_i$ includes the following unary facts for all $a \in
    \Gamma$ to mark accepting and rejecting final configurations:
    \begin{align*}
        &\textsf{Reject}((q, a))\text{, for all}\ q \in F_{\mathrm{rej}},\\
        &\textsf{Reject}(a),\\
        &\textsf{Accept}((q, a))\text{, for all}\ q \in F_{\mathrm{acc}},\\
        &\textsf{Accept}(a).
    \end{align*}
    Note that $\prod_{1 \leq i \leq s(n)} I_i$ contains the fact
    $\textsf{Accept}(e)$ if and only if $e$ represents a configuration in an
    accepting state, similarly for $\textsf{Reject}(e)$ and rejecting
    states.  We do not treat the cases $i = 1$ and $i = s(n)$ in a
    special way since we can assume that $M$ does not move its head
    beyond tape cell $1$ or $s(n)$.  This completes the description of
    the instances $I_i$. We choose the values $c_i$ such that the
    value
    $c_1\dots c_{s(n)} \in \text{adom}(\prod_{1 \leq i \leq s(n)}
    I_i)$ represents the initial configuration of $M$ on $w$.  For
    input $w = a_1\dots a_n$ and all $i$ with $1 \leq i \leq s(n)$ we
    choose
    \[
        c_i = \begin{cases}
            (q_0, a_1)&\text{if}\ i = 1\\
            a_i & \text{if}\ 2 \leq i \leq n\\
            \beta & \text{otherwise}
        \end{cases}
    \]
    where $\beta \in \Gamma$ is the symbol for an empty tape cell.

    Next, we describe the negative example $(J, c)$. Informally, the
    instance $J$ together with the choice of $c \in \text{adom}(J)$
    encodes that a computation is accepting.  For that,
    $\text{adom}(J)$ contains the two values $0$ and $1$ for final
    configurations, the values $(\forall, 0, 0, 0)$,
    $(\forall, 1, 0, 0)$, $(\forall, 0, 1, 0)$, $(\forall, 1, 1, 1)$
    $(\exists, 0, 0, 0)$, $(\exists, 1, 0, 1)$, $(\exists, 0, 1, 1)$,
    $(\exists, 1, 1, 1)$ as well as the two ``sink''-values $s_1$ and
    $s_2$.  A value of the form $(\forall, \ell, r, v)$ represents
    that a configuration is in a universal state, that the left
    successor configuration has acceptance value $\ell$ and that the
    right successor configuration has acceptance value $r$ and the
    configuration hence has acceptance value $v$, and similarly for
    values of the form $(\exists, \ell, r, v)$. Reflecting this
    intuition, $J$ includes the following facts for all $q \in Q$,
    $a \in \Gamma$, $i$ with $1 \leq i \leq s(n)$, and
    $(*, \ell, r, v), (*', \ell', r', v') \in \text{adom}(J)$ with
    $* \neq *'$:
    \begin{itemize}
    \item $r_{q, a, i}((*, \ell, r, v), (*', \ell', r', v'))$ if $v' = r$, and
    \item $\ell_{q, a, i}((*, \ell, r, v), (*', \ell', r', v'))$ if $v' = \ell$.
    \end{itemize}
    Reflecting the acceptance behavior of final configurations, $J$
    additionally includes the facts
    $$\textsf{Reject}(0),
    \textsf{Accept}(1)$$
    as well as the following facts, for all $q\in Q$, $a \in \Gamma$, $i$
    with $1 \leq i \leq s(n)$, and
    $(*, \ell, r, v) \in \text{adom}(J)$:
    $$r_{q, a, i}((*, \ell, r, v), r) \text{ and }
    \ell_{q, a, i}((*, \ell, r, v), \ell).$$
    At this point, we have completely described the computational
    behavior of $M$. If we stopped here, however, then a weakly
    most-general fitting CQ would \emph{never} exist, no matter
    whether $M$ accepts $w$ or not. We thus extend $J$ with
    the following facts which ensure that if there is a fitting CQ at
    all, then there is a weakly most-general fitting CQ:
    \begin{itemize}
        \item $r_{q, a, i}(s_1, e)$ and $\ell_{q, a, i}(s_1, e)$ for all $e \in \text{adom}(J)$,
        \item $r_{q, a, i}(e, s_2)$ and $\ell_{q, a, i}(e, s_2)$ for all $e \in \text{adom}(J) \setminus \{s_1, s_2\}$,
        \item $r_{q, a, i}((\exists, 1, 0, 1), s_1)$, $\ell_{q, a, i}((\exists, 0, 1, 1), s_1)$, 
        \item $r_{q, a, i}((\exists, 1, 0, 1), (\forall, 1, 1, 1))$, $\ell_{q, a, i}((\exists, 0, 1, 1), (\forall, 1, 1, 1))$, and
        \item $\textsf{Accept}(s_1)$, $\textsf{Reject}(s_1)$
    \end{itemize}
    for all $q \in Q$, $a \in \Gamma$ and $i$ with $1 \leq i \leq s(n)$.
    This completes the construction of $J$. We choose $c = (\forall, 1, 1, 1)$
    and set $E = (E^+, E^-)$ with $E^+ = \{ (I_i, c_i) \mid
    1 \leq i \leq s(n)\}$ and $E^- = \{ (J, c)\}$.

    It remains to show that the reduction is correct, that
    is:
    \begin{center}
    $M$ rejects $w$ if and only if $E$ has a weakly most-general fitting CQ.
    \end{center}
    
    First, assume that $M$ accepts $w$. To show that $E$ has
    no weakly most-general fitting CQ, it suffices to show that
    $(\prod_{1 \leq i \leq s(n)} I_i, c_1\dots c_{s(n)}) \to (J, c)$. 

    Let $I = \prod_{1 \leq i \leq s(n)} I_i$ and $\bar c = c_1\dots c_{s(n)}$.
    We define a homomorphism $h$ from $I$ to $J$ with $h(\bar c) = c$.
    If $e \in \text{adom}(I)$ is not reachable from $\bar c$, set $h(e) = s_1$.
    If $e \in \text{adom}(I)$ is reachable from $\bar c$, then it
    describes a configuration
    of $M$ that appears in the computation of $M$ on $w$. In the
    following,
    we will not distinguish values reachable from $\bar c$ in $I$ and configurations of $M$.
    Let $v$ be the acceptance value associated with $e$.
    If $e$ is a final configuration, set $h(e) = v$.
    If $e$ is an existential or a universal configuration, then $e$ must have a left successor and a right successor.
    Let $\ell$ be the acceptance value of the left
    successor and $r$ the acceptance value of the right successor of $e$. Set
    $h(e) = (\forall, \ell, r, v)$ if $e$ is universal and $h(e) = (\exists,
    \ell, r, v)$ if $e$ is existential.

    To verify that $h$ is as required, first note that
    $h(\bar c) = (\forall, 1, 1, 1) = c$ as $M$ accepts $w$ and
    $\bar c$ is a universal configuration.  Then, let
    $r_{q, a, i}(e, e')$ be a fact in $I$ that is reachable from
    $\bar c$ with $e$ a universal configuration. The case of facts
    $\ell_{q, a, i}$ and of existential configurations is similar. Then
    $h(e) = (\forall, \ell, r, v)$ and
    $h(e') = (\exists, \ell', r', v')$ for some
    $\ell, r, v, \ell', r', v'$ with $v' = r$ by definition of
    computations of $M$ and definition of $h$. Hence,
    $r_{q, a, i}(h(e), h(e')) \in J$ by construction.  Thus, $h$ is a
    homomorphism as required.

\smallskip
    
    For the other direction, assume that $M$ rejects $w$. From
    the computation of $M$ on $w$ we construct a CQ $q$ that is a
    weakly most-general fitting of $E$.
    Informally, $q$ will be a smallest subset of the unraveling of the
    computation of $M$ on $w$ that still witnesses that $M$ rejects $w$.

    For defining $q$ formally, we first introduce the notion of a minimal path
    of the computation of $M$ on $w$.
    A \emph{path} of the computation of $M$ on $w$ is a sequence
    $p = e_1 d_1 \dots d_{n - 1} e_n$ where $e_1$ is the initial
    configuration of $M$ on $w$ and for all $i$, $d_i = r$ if $e_i$ has right successor $e_{i + 1}$ and $d_i =
    l$ if $e_i$ has left successor $e_{i + 1}$. 
    We define $\mn{tail}(e_1 d_1 \dots d_{n - 1} e_n) = e_n$.
    A path in the computation of $M$ on $w$ is \emph{minimal} if for all $i$, $e_i$ has
    acceptance value $0$ and if $e_i$ is a universal configuration and has a
    left successor with acceptance value $0$, then $d_i = \ell$.
    
    Now, let $q(e_1)$ be the unary CQ that contains the following atoms for all
    minimal paths $p, p'$ of the computation of $M$ on $w$:
    \begin{itemize}
        \item $r_{q, a, i}(p, p')$ if $p' = pre$ and $\text{tail}(p)$ is a configuration with
            state $q$ and head at tape cell $i$, which  contains $a$.
        \item $\ell_{q, a, i}(p, p')$ if $p' = ple$ and $\text{tail}(p)$ is a configuration with
            state $q$ and head at tape cell $i$, which  contains $a$.
        \item $\mathsf{Reject}(p)$ if $\text{tail}(p)$ is a configuration in a rejecting state
    \end{itemize}
    Note that $q$ is finite due to the assumption
    that $M$ always terminates. By construction, it is a tree CQ and connected.
    By Definition~\ref{def:F} and Proposition~\ref{prop:frontier}, $q$
    therefore has a frontier consisting of a single query $F(q)$.  To
    prove that $q(e_1)$ is a weakly most-general fitting CQ for $E$,
    it thus remains to show that $q(e_1) \to (I, \bar c)$,
    $q(e_1) \not \to (J, c)$, and $F(q) \to (J, c)$.

    We begin with $q(e_1) \to (I, \bar c)$. Recall that by construction of
    $I$, the values that are reachable from $\bar c$ represent
    configurations of the computation of $M$ on $w$ and that $\bar c$
    corresponds to the initial configuration of $M$ on $w$. We can thus
    construct a homomorphism $h$ from $q$ to $I$ with $h(e_1) = \bar c$ by
    setting $h(p) = \text{tail}(p)$ for all $p \in \text{var}(q)$.

    Next, we show that $q(e_1) \nrightarrow (J, c)$. Recall that
    $c = (\forall, 1, 1, 1)$. For all $p \in \text{var}(q)$, we use
    $q_p(p)$ to denote the restriction of $q$ to all paths that start
    with $p$ and has answer variable $p$. We show that
    $q_p(p) \nrightarrow (J, c)$ for all $p \in \text{var}(q)$ if
    $\mathrm{tail}(p)$ is universal or final, by induction on the
    depth of tree CQ $q_p(p)$.  The desired
    result $q(e_1) \nrightarrow (J, c)$ then follows for $p = e_1$.  In the induction
    start, let $q_p(p)$ be of depth $0$. Then $\mathrm{tail}(p)$ must
    be  final by construction of $q$ and hence
    $q_p(p) = \textsf{Reject}(p)$. It follows that
    $q_p(p) \nrightarrow (J, c)$.
    
    Next, let $q_p(p)$ have depth $> 0$, with $\mn{tail}(p)$
    universal, and assume that the statement holds for all $q_p'(p')$
    of smaller depth. By construction of $q$ $\mathrm{tail}(p) = e$
    has acceptance value~$0$.  Thus, there must be an atom
    $r_{q, a, i}(p, pre')$ or $\ell_{q, a, i}(p, ple')$ in $q_p$ and
    $e'$ must be existential or final.  Assume that
    $r_{q, a, i}(p, pre')$ is in $q_p$, the other case is similar.  If
    $e'$ is final, it must be rejecting and hence $q_p$ contains
    $\textsf{Reject}(pre')$, implying that $q_p(p) \nrightarrow(J, c)$.  If
    $e'$ is in an existential state, then by construction $q_p$ must
    contain both atoms $r_{q, a, i}(pre', p'_r)$ for $p'_r = pre're''$
    and $\ell_{q, a, i}(pre', p'_\ell)$ for $p'_\ell = pre'le'''$.
    For both $p' = p'_r$ and $p' = p'_\ell$, we have
    $q_{p'} \nrightarrow (J, 1)$, $q_{p'} \not \to (J, s_2)$ since $p'$ must either
    be rejecting or must have successors that do not exist for $1$ and
    $s_2$. Additionally, $q_{p'} \nrightarrow (J, c)$ by the induction
    hypothesis. Consequently,
    $q_{pre'}(pre') \nrightarrow (J, (\exists, 1, 1, 1))$,
    $q_{pre'}(pre') \nrightarrow (J, (\exists, 1, 0, 1))$, and
    $q_{pre'}(pre') \nrightarrow (J, (\exists, 0, 1, 1))$, implying that
    $q_p(p) \nrightarrow (J, (\forall, 1, 1, 1))$, as required.

    It remains to show $F(q) \to (J, c)$. For that, recall that by
    Definition~\ref{def:F}, the answer variable of $F(q)$ is $e_1$ and
    the existential variables of $F(q)$ are $u_{e_1}$ and $u_{(p, f)}$
    for all minimal paths $p$ and atoms $f \in q$ such that $p$ occurs
    in $f$.  A variable of the latter kind is a \emph{replica} of
    $p$. We construct a homomorphism $h$ from $F(q)$ to $J$ with
    $h(e_1) = c = (\forall, 1, 1, 1)$.  Start by setting
    $h(e_1) = (\forall, 1, 1, 1)$ and $h(u_{e_1}) = s_1$.  Now let
    $u_{(p, f)}$ be a replica of the variable $p$ of $q$.

    If $\text{tail}(p)$ is rejecting, then $p$
    occurs in exactly two atoms in~$q$: $f_1 = d_{q, a, i}(p', p)$ for some $d
    \in \{r, \ell\}$ and $f_2 = \mathsf{Reject}(p)$.
    Set $h(u_{(p, f_1)}) = 0$ and $h(u_{(p, f_2)}) = s_2$.

    If $\text{tail}(p)$ is existential, then
    $p$ occurs in exactly three atoms in~$q$: $f_1 = d_{q, a, i}(p', p)$
    for some $d \in \{r, \ell\}$, $f_2 = r_{q', a', i'}(p, pre_r)$ and $f_3 =
    \ell_{q', a', i'}(p, ple_\ell)$.
    Set $h(u_{(p, f_1)}) = s_1$, $h(u_{(p, f_2)}) = (\exists, 0, 1, 1)$ and
    $h(u_{(p, f_3)}) = (\exists, 1, 0, 1)$.

    If $\text{tail}(p)$ is universal and not
    $e_1$, then $p$ occurs in exactly two atoms in $q$: $f_1 = d_{q, a,
    i}(p', p)$ for some $d \in \{r, \ell\}$ and $f_2 = d'_{q', a', i'}(p,
    pd'e)$ for some $d' \in \{r, \ell\}$.
    Set $h(u_{(p, f_1)}) = s_1$ and $h(u_{(p, f_2)}) = (\forall, 1, 1, 1)$.
   
    To verify that $h$ is a homomorphism, consider an atom $r_{q, a,
    i}(p, p')$ in $q$, let $u_{(p, f)}$ be a replica of $p$ in $F(q)$ and let
    $u_{(p', f')}$ be a replica of $p'$ in $F(q)$.
    The case for $\ell_{q, a, i}$ atoms is symmetrical.

    If $\text{tail}(p)$ is a universal configuration, then by definition of $h$, 
    $h(u_{(p, f)}) \in \{s_1, (\forall, 1, 1, 1)\}$ and
    $h(u_{(p', f')}) \in \{0, s_1, s_2, (\exists, 1, 0, 1), (\exists, 0, 1,
    1)\}$, since $\text{tail}(p')$ must be  existential or final. 
    By construction of $J$, 
    $r_{q, a, i}(h(u_{(p, f)}), h(u_{(p', f')})) \notin J$ implies that 
    $h(u_{(p, f)}) = (\forall, 1, 1, 1)$ and $h(u_{(p', f')}) = s_1$ or $h(u_{(p', f')}) = 0$.
    In both cases the definitions of $h$ and $q$ imply $f = f'$ and hence
    $r_{q, a, i}(u_{(p, f)}, u_{(p', f')}) \notin F(q)$ by construction of
    $F(q)$.
    Note that this case also applies to $p = e_1$, where $u_{(p, f)}$ is either
    $p$ or $u_p$ and the fact $f$ is uniquely determined.

    If $\text{tail}(p)$ is an existential configuration, then by definition of
    $h$, $h(u_{(p, f)}) \in \{s_1, (\exists, 1, 0, 1), (\exists, 0, 1, 1)\}$
    and $h(u_{(p', f')}) \in \{0, s_1, s_2, (\forall, 1, 1, 1) \}$, since
    $\text{tail}(p')$ must be universal or final.
    By construction of $J$, 
    $r_{q, a, i}(h(u_{(p, f)}), h(u_{(p', f')})) \notin J$ implies that 
    $h(u_{(p, f)}) = (\exists, 0, 1, 1)$ and $h(u_{(p', f')}) = s_1$ or $h(u_{(p', f')}) = 0$.
    In both cases the definitions of $h$ and $q$ imply $f = f'$ and hence
    $r_{q, a, i}(u_{(p, f)}, u_{(p', f')}) \notin F(q)$ by construction of
    $F(q)$.
    
    Hence, $h$ is a homomorphism as required.
\end{proof}

\end{document}